\documentclass{ws-ijmpe}

\usepackage{slashed}

\usepackage{color}
\usepackage{feynmf}

\usepackage{bbm}

\newcommand{\bhalflife}{T^{0\nu}_{1/2}}

\def\ra{\rightarrow}
\def\be{\begin{equation}}
\def\ee{\end{equation}}
\def\gs{\mathrel{
   \rlap{\raise 0.511ex \hbox{$>$}}{\lower 0.511ex \hbox{$\sim$}}}}
\def\ls{\mathrel{
   \rlap{\raise 0.511ex \hbox{$<$}}{\lower 0.511ex \hbox{$\sim$}}}}
\newcommand{\obb}{0\mbox{$\nu\beta\beta$}}
\newcommand{\zbb}{2\mbox{$\nu\beta\beta$}}
\newcommand{\onbb}{neutrino-less double beta decay}
\newcommand{\ba}{\begin{array}{c}}
\newcommand{\baz}{\begin{array}{cc}}
\newcommand{\bad}{\begin{array}{ccc}}
\newcommand{\bav}{\begin{array}{cccc}}
\newcommand{\bea}{\begin{equation} \begin{array}{c}}
\newcommand{\eea}{ \end{array} \end{equation}}
\newcommand{\ea}{\end{array}}
\newcommand{\D}{\displaystyle}
\newcommand{\dms}{\mbox{$\Delta m^2_{\odot}$}}
\newcommand{\dma}{\mbox{$\Delta m^2_{\rm A}$}}
\newcommand{\meff}{\mbox{$\langle m_{ee} \rangle$}}

\newcommand{\sss}{\sin^2 \theta_{12}}
\newcommand{\sch}{\sin^2 \theta_{13}}

\newcommand{\imeff}{\mbox{$\langle \frac 1 m \rangle$}}

\begin{document}

\markboth{}{}

\catchline{}{}{}{}{}

\title{NEUTRINO-LESS DOUBLE BETA DECAY AND PARTICLE PHYSICS
}
\author{\footnotesize WERNER RODEJOHANN}

\address{Max--Planck--Institut f\"ur Kernphysik\\ 
Postfach 103980, D--69029 Heidelberg\\
Germany\\
werner.rodejohann@mpi-hd.mpg.de}
\maketitle


\begin{abstract}
We review the particle physics aspects of neutrino-less double beta
decay. This process can be mediated by light massive Majorana
neutrinos ({\it standard interpretation}) or by something else
({\it non-standard interpretations}). The physics potential of both
interpretations is summarized and the consequences of future
measurements or improved limits on the half-life of neutrino-less double beta
decay are discussed.  We try to cover all proposed alternative
realizations of the decay, including light sterile neutrinos, 
supersymmetric or left-right symmetric theories, Majorons, and other exotic
possibilities. Ways to distinguish the mechanisms from one another are
discussed. Experimental and nuclear physics aspects are also briefly 
touched, alternative processes to double beta decay are discussed,  
and an extensive list of references is provided.

\end{abstract}

\newpage
\tableofcontents

\markboth{W.~Rodejohann}{\obb~and Particle Physics}
\section{\label{sec:intro}Introduction: General Aspects of Double Beta Decay and
Lepton Number Violation}

Neutrino-less double beta decay (\obb) is a process of fundamental
importance for particle physics. It is defined as the transition of 
a nucleus into a nucleus with proton number larger by two units, and the
emission of two electrons\cite{furry}: 
\be \label{eq:main}
(A,Z) \ra (A,Z+2) + 2 \, e^- ~~~~(\obb)\, . 
\ee
There are no leptons in the initial state, but two in the final
state. Observation of \obb~would therefore show that lepton number, an
accidental and classical symmetry of the Standard Model (SM) of particle physics, is
violated by Nature. The process therefore stands on equal footing with
baryon
number violation, i.e.~proton decay. 
For this reason a huge amount of experimental and theoretical activity
is pursued in order to detect and predict the 
process\cite{haxton,doi,EV,Tomoda:1990rs,Gnu1,vergados,APS,AEE,nme_rev,barabash,andrea}.

As well known, the main motivation to search for \obb~is the fact that neutrinos are, in
contrast to the prediction of the SM, massive particles and that
basically all theories beyond the SM predict them to be
Majorana\cite{Majorana} particles. However, as we will discuss in this review,
there are many other well-motivated particle physics scenarios and 
frameworks that allow for \obb. Before
discussing these aspects, let us first give some general comments on
lepton number violation.

Why should we look for Lepton Number Violation (LNV)? The conservation
of lepton (and baryon) number in the SM is an accidental
one at the classical level only. In fact, chiral anomalies actually 
violate this conservation
law, and it can be shown that the currents associated with baryon and
lepton number have non-vanishing divergences: $\partial^\mu 
J^{B,L}_\mu= c \, G_{\mu \nu} \, \tilde G^{\mu \nu} \neq 0$. Here $G_{\mu
\nu}$ is the electroweak gauge field strength and $J^{B}_\mu = \sum 
\overline{q_i} \, \gamma_\mu \, q_i$, $J^{L}_\mu = \sum 
\overline{\ell_i} \, \gamma_\mu \, \ell_i$. Though this LNV is not the
one related to \obb~or Majorana neutrinos, and the rates of processes associated to it are
negligible at low temperatures, it shows that lepton number is nothing
sacred, not even in the Standard Model. Extending the picture from the
SM to Grand Unified Theories (GUTs), quarks and leptons live 
together in multiplets, and hence both $B$ and $L$ are not expected to be
conserved quantities. The combination $B-L$, which is conserved in the
SM both at the classical and quantum level, often plays an important
role in GUTs, and is broken at some stage.   In the spirit of baryogenesis,
one needs to require that baryon number is
violated, and hence lepton number should be violated
too. The search for baryon number violation proceeds in proton decay, 
or neutron--anti-neutron oscillation experiments. Lepton number
violation is investigated in neutrino-less double beta decay
experiments, and should be treated on the same level as baryon number
violation. An observation of LNV would be far more 
fundamental than a ``simple'' measurement of neutrino properties,
which is often quoted as the main goal of \obb-searches. Its implications
are far beyond that.

\begin{figure}[t]
\centerline{\psfig{file=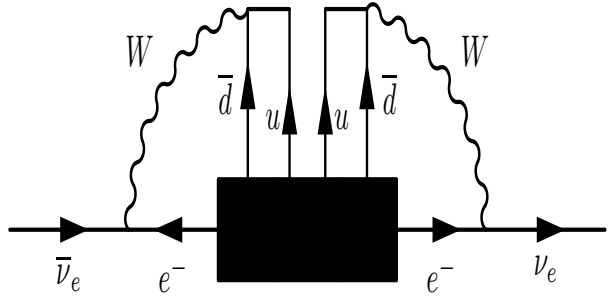,width=8cm,height=4.5cm}
}
\vspace*{8pt}
\caption{\label{fig:blackbox}Black-box illustration of 
neutrino-less double beta decay. }
\end{figure}

In this review we wish to summarize the particle physics aspects of
limits and possible measurements of this process. 
A large number of theories and mechanisms to violate lepton number exists, 
and the most often considered light Majorana neutrino exchange 
(though well-motivated) is only one possibility. We should note that via the black-box, 
or Schechter-Valle, theorem\cite{SV} (see also\cite{SV1}), all realizations of 
Eq.~(\ref{eq:main}) are connected to a Majorana neutrino mass. Crossing the process 
on the quark level gives from $d \, d \ra u \, u \, e^-  e^-$ the relation  
$0 \to u \bar d \, u \bar d \, e^-  e^-$, and with the only input of 
$SU(2)_L$ gauge theory one can couple each $u \bar d$ pair via a $W$ 
to the electrons, as illustrated in Fig.~\ref{fig:blackbox}. 
The result is a $\bar{\nu}_e$--$\nu_e$ transition, which is nothing but a Majorana mass term. 
Needless to say, this is a tiny mass generated at the 4-loop level, and too small to 
explain the neutrino mass scale (or its splitting) as observed in oscillation 
experiments. Naively, one can estimate the contribution to neutrino
mass as 
\be \label{eq:SV}
(m_\nu)_{ee} \ls \frac{1}{(16 \pi^2)^4} \frac{{\rm MeV}^5}{m_W^4}
\simeq 10^{-23} ~ {\rm eV} \, , 
\ee
where we inserted a factor $1/(16 \pi^2)$ for each loop, put $m_W^{-2}$ for
each of the two $W$ in the loop, and MeV is the typical mass of the
involved electron, up- and down-quark. An explicit calculation of the 
4-loop diagram with an effective operator as the source of \obb~yields
a very similar number\cite{DLM}. 
Note that this tiny mass is much smaller than the Planck-scale contribution to
the Majorana neutrino mass, which is $v^2/M_{\rm Pl} \simeq 10^{-5}$
eV. There are now two main possibilities: 
\begin{itemize}
\item[(i)] 
the mechanism leading to \obb~is connected to neutrino
oscillation. Here there are again two possibilities: 
\begin{itemize}
\item[(ia)] there is a direct connection to neutrino oscillation. This is
the standard mechanism of light neutrino exchange; 
\item[(ib)] there is an indirect connection to neutrino
oscillation. Examples would be heavy neutrino exchange, where the
heavy neutrinos are responsible for light neutrino masses via the 
seesaw mechanism. Another case would be
$R$-parity violating SUSY particles generating \obb, where via loops
the same particles generate light neutrino
masses; 
\end{itemize}
\item[(ii)] the mechanism leading to \obb~is {\it not} connected to neutrino 
oscillation. The underlying physics of \obb~could be either: 
\begin{itemize}
\item[(iia)] giving a sub-leading contribution to neutrino mass, maybe
$R$-parity violating SUSY being responsible for only one of the neutrino masses; 
\item[(iib)] giving no contribution to neutrino mass,
for instance a right-handed Higgs triplet in the absence of a Dirac mass
matrix for neutrinos. Hence only the Schechter-Valle term from
Eq.~(\ref{eq:SV}) can generate a neutrino mass. 
\end{itemize}
In both cases (iia) and (iib) we would need another source 
for neutrino mass and oscillation\cite{DLM}. 
\end{itemize}

As already mentioned, the assumption that massive Majorana neutrinos
generate \obb~is presumably the best motivated one, though there are
many more. We can thus
classify the possible interpretations of \obb~as follows: 

\begin{enumerate}
\item {\it Standard Interpretation}: 

neutrino-less double beta decay is mediated by light and massive
Majorana neutrinos (the ones which oscillate) and all other mechanisms
potentially leading to \obb~give negligible or no contribution;

\item {\it Non-Standard Interpretations}:

neutrino-less double beta decay is mediated by some other lepton number
violating physics, and light massive
Majorana neutrinos (the ones which oscillate) potentially leading 
to \obb~give negligible or no contribution. 
\end{enumerate}

In this review we will consider both cases and aim to
discuss all possible realizations of \obb. In Sections \ref{sec:exp} and
\ref{sec:nme} we will deal with experimental 
and nuclear physics aspects of \obb, respectively. The standard 
interpretation of light neutrino exchange is discussed in Section
\ref{sec:meff}, where we summarize in detail 
our current understanding of neutrino physics and its many aspects 
which can be tested with \obb. 
Section \ref{sec:non-standard} is devoted to the various non-standard
interpretations, such as  left-right symmetric
theories, $R$-parity violating supersymmetry, Majorons, and other
proposals. Section \ref{sec:distinguish} deals with possibilities to
distinguish the mechanisms from one another, and Section \ref{sec:alt}
is concerned with alternative processes to \obb. A 
summary is presented in Section \ref{sec:concl}. For all aspects we
provide an extensive list of references.

\begin{table}[b]
\tbl{\label{tab:Gnu}$Q$-value, natural abundance and phase space factor 
$G$ (standard mechanism) for all isotopes with $Q \ge 2$ MeV using $r_0=1.2$ fm. 
Values taken from Table 6 of\protect\cite{Gnu1} and scaled to
$g_A=1.25$. Note that there is a misprint in Ref.\protect\cite{Gnu1}, which
quotes $G^{0\nu}$ for $^{100}$Mo as $11.3 \times 10^{-14}$ yrs$^{-1}$.}
{\begin{tabular}{@{}cccc@{}} \toprule
Isotope		&	$G$ [10$^{-14}$ yrs$^{-1}$]&
$Q$ [keV] & nat.~abund.~[\%]\\ \colrule
$^{48}$Ca	&	6.35    &  4273.7 & 0.187 \\
$^{76}$Ge	&	0.623   &  2039.1 & 7.8 \\
$^{82}$Se	&	2.70    &  2995.5 & 9.2 \\
$^{96}$Zr	&	5.63    &  3347.7 & 2.8\\
$^{100}$Mo	&	4.36    &  3035.0 & 9.6\\
$^{110}$Pd	&	1.40    &  2004.0 & 11.8 \\
$^{116}$Cd	&	4.62    &  2809.1 & 7.6\\
$^{124}$Sn	&	2.55    &  2287.7 & 5.6\\
$^{130}$Te	&	4.09    &  2530.3 &  34.5 \\
$^{136}$Xe	&	4.31    &  2461.9 & 8.9 \\
$^{150}$Nd	&	19.2    &  3367.3 & 5.6 \\
\botrule
\end{tabular}}
\end{table}

\section{\label{sec:exp}Experimental aspects}
\begin{figure}[t]
\centerline{\psfig{file=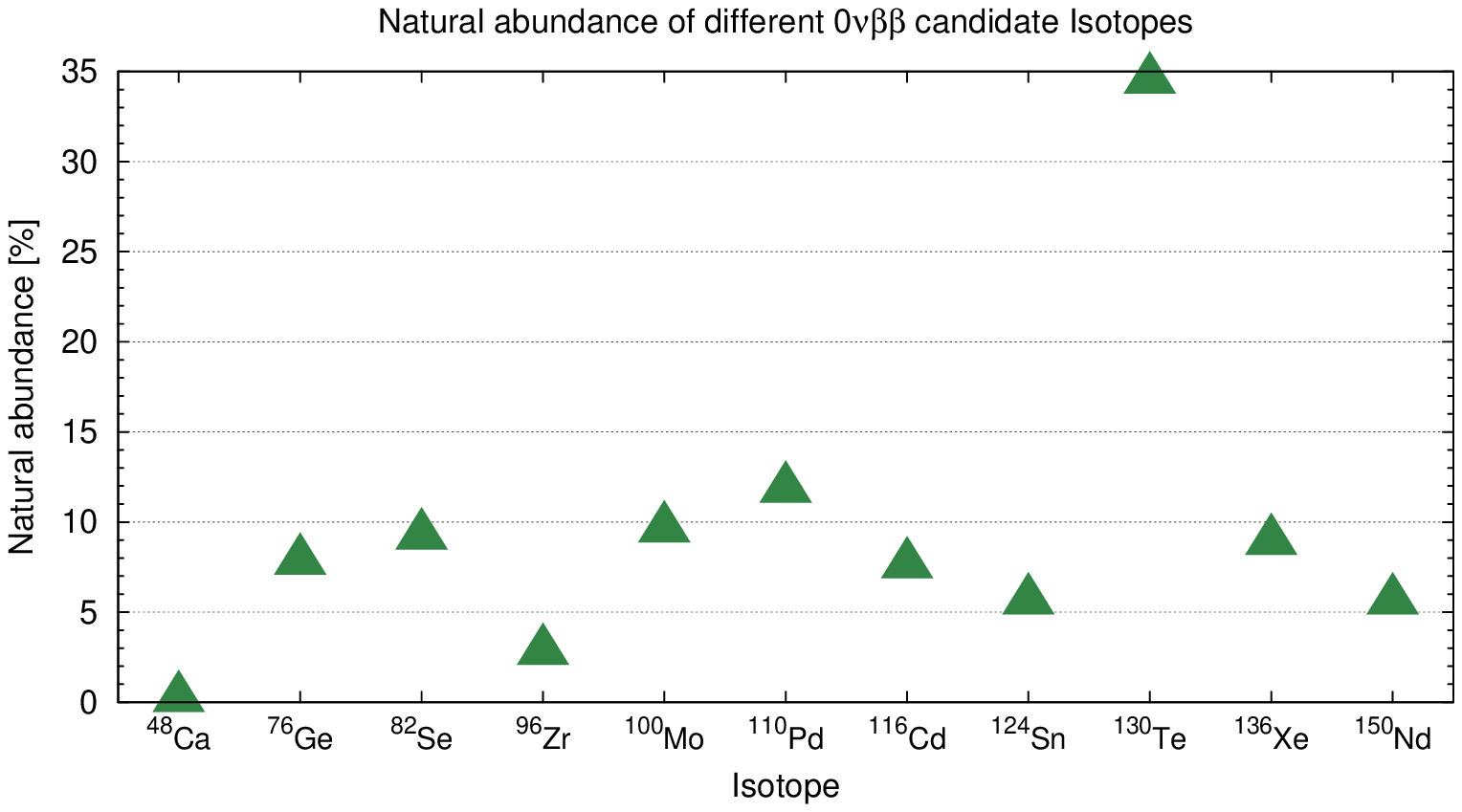,width=6cm,height=5cm}
\psfig{file=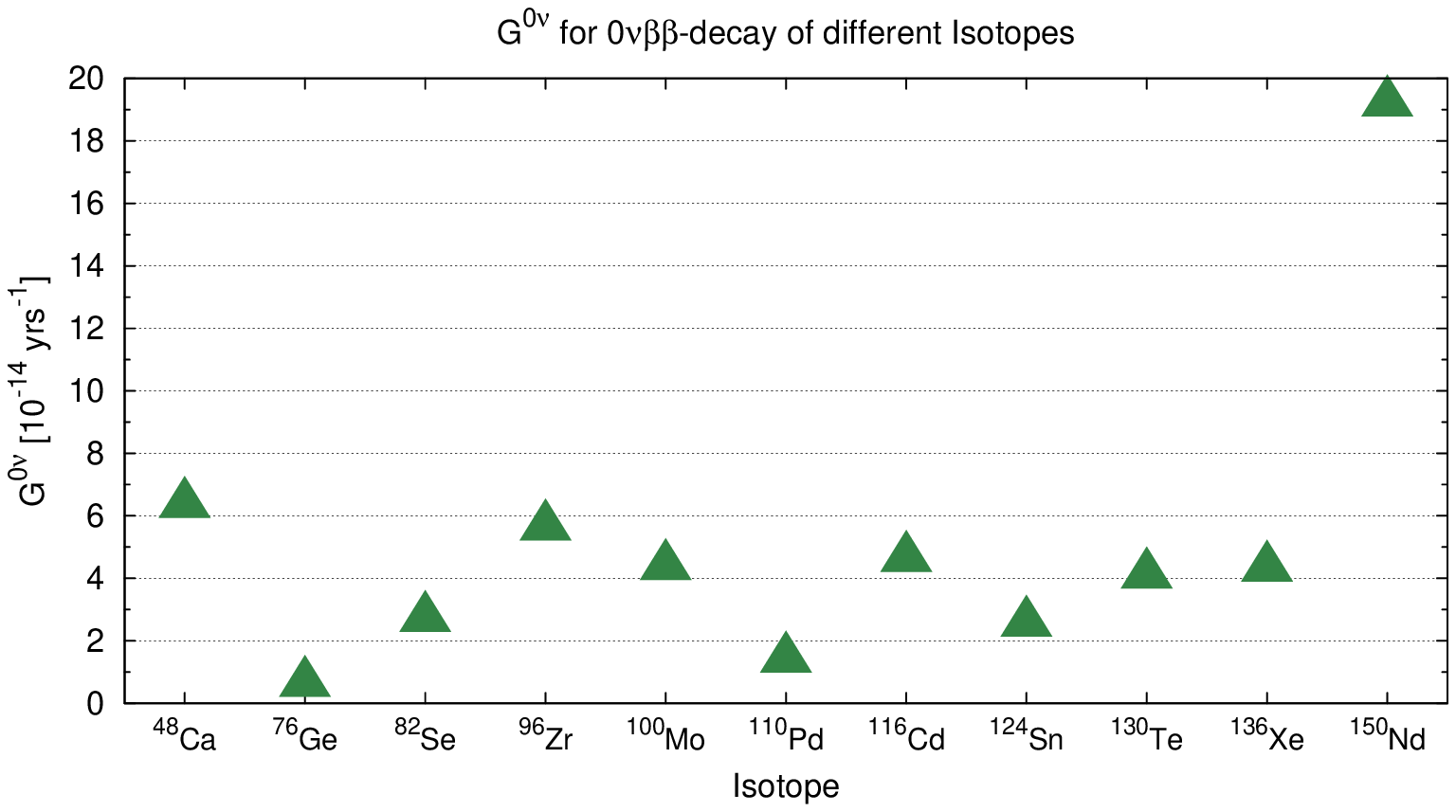,width=6cm,height=5cm}
}
\vspace*{8pt}
\caption{\label{fig:ovbb_para}Natural abundance and phase space factor
  for all 11 $\obb$-isotopes above $Q = 2$
  MeV. }
\end{figure}

Neutrino-less double beta decay can only be observed if the usual beta
decay is energetically forbidden. This is the case for some even-even
nuclei (i.e.~even proton and neutron numbers), whose ground states are
energetically lower than their odd-odd neighbors. 
If the nucleus with atomic number higher by one unit has a smaller binding
energy (preventing beta decay from occurring), and the nucleus with 
atomic number higher by two units has a larger binding energy, the double beta decay
process is allowed. In principle 35 nuclei can undergo \obb, though
realistically only nine emerge as interesting candidates and are under
investigation in competitive experiments, namely 
$^{48}$Ca, $^{76}$Ge, $^{82}$Se, $^{96}$Zr, $^{100}$Mo, $^{116}$Cd, 
$^{130}$Te, $^{136}$Xe, $^{150}$Nd. There is no ``super isotope'', one
has to find compromises between the natural abundance, 
reasonably priced enrichment, the association with a well controlled 
experimental technique or the $Q$-value, because the decay rate for \obb~goes 
with $Q^5$ (except for Majoron emission, see Section
\ref{sec:Majorons}). Table \ref{tab:Gnu} and Fig.~\ref{fig:ovbb_para} give the relevant
parameters of all 11 isotopes with a $Q$-value above 2 MeV, including
the nine most studied isotopes given above. 
The experimental signal is the sum of energy of the two emitted
electrons, which should equal the known $Q$-value. 
The neutrino-less mode has to be distinguished from 2 neutrino
double beta decay\cite{GM} 
\be \label{eq:2nubb}
(A,Z) \ra (A,Z+2) + 2 \, e^- + 2 \, \bar{\nu}_e  ~~~~(\zbb) \, , 
\ee 
which experimentally can be an irreducible background for
the neutrino-less mode. The half-life of \zbb~is typically around
$10^{19}$--$10^{21}$ yrs (it is important to note that the process is
allowed in the SM), and has been observed for a number of isotopes
already, see\cite{barabash} for a list of results. 
Obviously, the countless
peaks arising from natural radioactivity, cosmic ray reactions
etc.~need to be understood and/or the experiments have to be ultrapure
and/or heavily shielded. 
The energy release $Q$ should also be large due to the background of 
natural radioactivity, which drops significantly beyond 2.614 MeV,
which is the highest significant $\gamma$-line in the natural decay
chains of Uranium and Thorium.  
In general, the decay rate for \obb~can be factorized as 
\be \label{eq:fact}
\Gamma^{0\nu} = G_x(Q,Z) \, |{\cal M}_x(A,Z) \, \eta_x|^2 \, ,
\ee
where $\eta_x$ is a function of the particle physics parameters 
responsible for the decay. The nuclear matrix element (NME)
${\cal M}_x(A,Z)$ depends on the mechanism and the nucleus. 
The term ${\cal M}_x(A,Z) \,
\eta_x$ can in fact be a sum of several terms, therefore including the
possibility of destructive or constructive interference, a situation
we will deal with in Section \ref{sec:distinguish_simul}. Finally, 
$G_x(Q,Z)$ is a phase space factor which can have dependence on 
the particle physics. For most of the processes in which only two electrons
are emitted, the phase space factor can be considered almost 
independent of the mechanism. The biggest effect for $G$ occurs in double beta decay
with Majoron emission, in which the final state contains one or two
additional particles, see Section \ref{sec:Majorons}. 
 Table \ref{tab:limits} summarizes 
the current best limits on the half-life. 
Neutrino-less double beta decay is definitely a rare process. 
\begin{table}[b]
\tbl{\label{tab:limits}Experimental limits at 90\% C.L.~on the most
interesting isotopes for \obb. Using the nuclear matrix element ranges from Table
\ref{tab:nme} we also give the maximal and minimal limits on 
\meff. }
{\begin{tabular}{@{}ccccc@{}} \toprule
Isotope & $T_{1/2}^{0\nu}$ [yrs]  & Experiment  & $\meff^{\rm
lim}_{\rm min}$ [eV] & $\meff^{\rm lim}_{\rm max}$ [eV]  \\ \colrule
$^{48}$Ca & $5.8 \times 10^{22}$ & CANDLES\cite{Umehara:2008ru} & 3.55  &   9.91\\ 
$^{76}$Ge & $1.9 \times 10^{25} $ & HDM\cite{KlapdorKleingrothaus:2000sn} &   0.21  &   0.53\\ 
 & $1.6 \times 10^{25} $ & IGEX\cite{Aalseth:2002rf} & 0.25    &  0.63 \\ 
$^{82}$Se & $3.2 \times 10^{23} $ & NEMO-3\cite{Arnold:2005rz} & 0.85
&   2.08 \\ 
$^{96}$Zr & $9.2 \times 10^{21} $ & NEMO-3\cite{Argyriades:2009ph}&   3.97  &  14.39\\ 
$^{100}$Mo & $1.0 \times 10^{24} $& NEMO-3\cite{Arnold:2005rz} &   0.31  &   0.79\\ 
$^{116}$Cd & $1.7 \times 10^{23} $ & SOLOTVINO\cite{Danevich:2003ef}&   1.22  &   2.30\\ 
$^{130}$Te &  $2.8 \times 10^{24} $ &
CUORICINO\cite{Arnaboldi:2005cg} & 0.27 & 0.57 \\ 
$^{136}$Xe &  $5.0 \times 10^{23} $& DAMA\cite{Bernabei:2002bn}&   0.83  &   2.04\\  
$^{150}$Nd & $1.8 \times 10^{22} $ & NEMO-3\cite{Argyriades:2008pr} &   2.35  &   5.08\\ \botrule
\end{tabular}}
\begin{tabnote}
The limits on $T_{1/2}^{0\nu}$ from NEMO-3 measurements assume the standard light
neutrino mechanism. 
\end{tabnote}
\end{table}
In Table \ref{tab:limits} we already quote the limits on the 
effective mass (the particle physics parameter in the standard
interpretation) from the respective experiments, for which we used a
compilation of nuclear matrix elements discussed later.

In the past the search relied mainly on geo- and radiochemical 
measurements, which are insensitive to the mode of double beta decay,
but led to the first observation that two neutrino double beta
decay occurs in nature\cite{kirsten}. Here the approach is to identify
accumulation of the decay isotope (in particular if it is a noble gas,
for which mass spectroscopy can be done very precisely) during geological time periods in
samples which are rich in a double beta decay isotope. 
In principle, this method can be
used to test the time-dependence of the parameters associated with the
mechanism of double beta decay.

\begin{table}[b]
\tbl{\label{tab:expts}Planned experiments categorized according
to\protect\cite{andrea} and the isotope(s) under consideration.}
{\begin{tabular}{@{}cccccc@{}} \toprule
Name & Isotope & \multicolumn{3}{c}{source $=$ detector; calorimetric with} 
& source $\neq$ detector with \\ 
 &      & high energy res. & low energy res.  & sensit.~to event
topology & sensit.~to event topology\\ \colrule
CANDLES\cite{candles} & $^{48}$Ca & -- & \checkmark & -- & -- \\ 
 COBRA\cite{cobra}& $^{116}$Cd (and $^{130}$Te)  & -- &  -- & \checkmark &  -- \\ 
CUORE\cite{cuore} & $^{130}$Te  &\checkmark  &  -- &  -- &  -- \\
DCBA\cite{dbca}  & $^{150}$Nd    &  -- &  -- &  -- & \checkmark \\ 
EXO\cite{exo} & $^{136}$Xe  &  -- &  -- & \checkmark &  -- \\ 
GERDA\cite{gerda}& $^{76}$Ge & \checkmark &  -- &  -- &  -- \\ 
KamLAND-Zen\cite{kamland} & $^{136}$Xe &  -- & \checkmark &  -- &  -- \\ 
LUCIFER\cite{lucifer} & $^{82}$Se or $^{100}$Mo or $^{116}$Cd  & 
\checkmark &  -- &  -- & -- \\ 
 MAJORANA\cite{majorana}& $^{76}$Ge & \checkmark &  -- &  -- &  -- \\ 
MOON\cite{moon} & $^{82}$Se or $^{100}$Mo or $^{150}$Nd 
 &  -- &  -- &  -- & \checkmark \\ 
NEXT\cite{next} & $^{136}$Xe &  -- &  -- & \checkmark &  -- \\ 
SNO+\cite{sno+} & $^{150}$Nd  &  -- & \checkmark &  -- &  -- \\ 
SuperNEMO\cite{supernemo}& $^{82}$Se or $^{150}$Nd  &  -- &  -- &  -- & \checkmark \\ 
XMASS\cite{xmass} & $^{136}$Xe  &  -- & \checkmark &  -- &  -- \\ \botrule
\end{tabular}}
\end{table}

Nowadays 
only direct methods are applied, based mainly on the observation of
the two electrons in the form of measuring their total sum energy, which
should equal the $Q$-value of the decay. Some experiments have the
possibility of tracking the individual electrons. 
There are a number of recent reviews on the experimental
situation in double beta decay, to which we refer for more 
details\cite{APS,AEE,barabash,andrea}. The number of expected events
in an experiment can be written as 
\be \label{eq:Nbb}
N = \ln 2 \, a \, M \, t \, N_A \, (T_{1/2}^{0\nu})^{-1} \, , 
\ee
where $a$ is the abundance of the isotope, $M$ the used mass,
$t$ the time of measurement and $N_A$ is Avogadro's number. 
The half-life sensitivity depends on whether there is background or
not\cite{Avignone:2005cs}: 
\be \label{eq:Texp}
(T_{1/2}^{0\nu})^{-1} \propto \left\{ 
\baz \D 
a \, M \, \varepsilon \, t & \mbox{without background,} \\ \D 
a \, \varepsilon \, \sqrt{\frac{M \, t}{B \, \Delta E}} &
\mbox{with background.} 
\ea 
\right. 
\ee
Here $B$ is the background index with natural
units of counts/(keV kg yr) and $\Delta E$ the energy resolution at the
peak.  
In Table \ref{tab:expts} we follow the classification proposed 
by A.~Guiliani\cite{andrea} and list some properties of the main up-coming experiments. 
Table \ref{tab:fut_exp} lists the most developed
experiments according to\cite{barabash}.  
Roughly speaking, at present the transition from 10 kg yrs to 100 kg
yrs experiments is being made, background levels below $10^{-2}$ counts/(keV kg yr) are
planned, and half-life sensitivities above $10^{26}$ yrs are 
foreseen.

\begin{table}[pt]
\tbl{\label{tab:fut_exp} 
Sensitivity at 90\% C.L.~of the seven most developed projects 
for about three (phase II of GERDA and MAJORANA, KamLAND, SNO+) 
five (EXO, SuperNEMO and CUORE) and ten (full-scale GERDA plus MAJORANA) 
years of measurements. Taken from\protect\cite{barabash}.  }
{\begin{tabular}{ccccccc} 
\toprule 
Experiment & Isotope & Mass of & Sensitivity 
& 
& Status & Start of  \\
& & Isotope [kg] & $T_{1/2}^{0\nu}$ [yrs] & 
& &
data-taking \\ \hline
GERDA & $^{76}$Ge & 18 & $3\times10^{25}$ 
& 
& running & $\sim$ 2011 \\
  &  & 40 & $2\times10^{26}$ 
& 
& in progress & $\sim$ 2012 \\
& & 1000 & $6\times10^{27}$ & 
& R\&D & $\sim$ 2015\\ 
CUORE & $^{130}$Te & 200 & $6.5\times10^{26}$$^{*}$ & 
& in progress & $\sim$ 2013 \\ 
& & & $2.1\times10^{26}$$^{**}$ & 
& \\
MAJORANA & $^{76}$Ge & 30-60 & $(1-2)\times10^{26}$ & 
& in progress & $\sim$ 2013 \\
& & 1000 & $6\times10^{27}$ & 
& R\&D & $\sim$ 2015\\ 
EXO 
& $^{136}$Xe & 200 & $6.4\times10^{25}$ & 
& in progress & $\sim$ 2011 \\
& & 1000 & $8\times10^{26}$ & 
& R\&D & $\sim$ 2015\\ 
SuperNEMO & $^{82}$Se & 100-200 & $(1-2)\times10^{26}$ & 
& R\&D & $\sim$ 2013-2015\\
& & & & &\\
KamLAND-Zen & $^{136}$Xe & 400 & $4\times10^{26}$ & 
& in progress & $\sim$ 2011 \\
& & 1000 & $10^{27}$ & 
& R\&D & $\sim$ 2013-2015\\
SNO+ & $^{150}$Nd & 56 & $4.5\times10^{24}$ & 
& in progress & $\sim$ 2012 \\
& &  500 & $3\times10^{25}$ & 
&        R\&D  & $\sim$ 2015\\
\botrule 
\end{tabular}}
\begin{tabnote}$^{*}$ For a background of $10^{-3}$/keV/kg/yr; 
$^{**}$ for a background of $10^{-2}$/keV/kg/yr. 
\end{tabnote}
\end{table}

The current best values come from the 
Heidelberg-Moscow\cite{KlapdorKleingrothaus:2000sn} experiment, using
 $^{76}$Ge enriched Germanium calorimetric detectors. As is well known, 
part of the collaboration claims observation of the
process\cite{claim}, at the level of about $2 \times
10^{25}$ yrs, with a 99.73\% C.L.~range of $(0.7 - 4.2) \times 10^{25}$
yrs.  This has been criticized by a large part of the
community\cite{anticlaim1}, but eventually needs to be tested experimentally. In the later part of
this review we will discuss limits on lepton number violating
parameters from \obb. As the limit on the half-life of $^{76}$Ge corresponds roughly
to the claimed signal, one could easily translate the limits of the
lepton number violating parameters into their values, in case the
claim is actually valid.\\ 

A bolometric experiment, also run in Gran Sasso, was 
CUORICINO\cite{Arnaboldi:2005cg}, using $^{130}$Te in the form of
TeO$_2$ crystals. Similar limits to Heidelberg-Moscow could be
reached. An experiment with source $\neq$ detector was NEMO-3\cite{Arnold:2005rz}, using
foils of several potential \obb-emitters in a magnetized tracking
volume. Here the main point is measuring the energy of the individual
energies and their angular distribution. This approach is of
interest in testing different mechanisms for \obb, as we will discuss 
later. Again, limits of the order of Heidelberg-Moscow were obtained.

We shortly discuss presently running and upcoming experiments. 
Basically all of them will use enriched material, and all are located
in underground laboratories. 
GERDA\cite{gerda} and MAJORANA\cite{majorana} will use $^{76}$Ge, in
the case of GERDA operated in liquid Argon. 
Phase I consists of 18 kg previously used by IGEX and
Heidelberg-Moscow and will test in the very near future (till 2013) 
the Heidelberg-Moscow claim, which
unambiguously will only be possible with the same
isotope. Phase II will work with 40 kg and depending on the outcome, a
phase III, probably joined with MAJORANA ($2\times60$ kg), is possible. 
Inauguration of GERDA was in November 2010. CUORE\cite{cuore}
extends CUORICINO to several towers of material, aiming at 200 kg
$^{130}$Te and a start of data taking in 2013. 
EXO\cite{exo}, whose prototype with 200 kg of liquid Xenon enriched to
80\% is in commissioning, will apply liquid or gaseous
Xenon; by using a time-projection chamber there is sensitivity to event
topology. It will be attempted in a later phase to laser-tag the $^{136}$Ba$^{++}$ ion,
which is the decay product of the isotope under investigation, $^{136}$Xe.  
SuperNEMO\cite{supernemo} uses the NEMO-3 approach and will work with
about 100 kg of $^{82}$Se or $^{150}$Nd. 
SNO+\cite{sno+} wishes to fill the large SNO detector with a total of
44 kg of $^{150}$Nd. KamLAND-Zen\cite{kamland} pursues a similar approach with
the KamLAND experiment, using $^{136}$Xe.
CANDLES\cite{candles} will investigate CaF$_2$ scintillators, and is
currently analyzing enrichment options for later phases.
COBRA\cite{cobra} will be an array of CdZnTe room temperature semiconductors, mainly
sensitive to \obb~of $^{116}$Cd, but to other decay modes as
well. LUCIFER\cite{lucifer} proposes to use scintillating bolometers
at low temperature. 
MOON\cite{moon} wants to use scintillators in between source foils, 
DCBA\cite{dbca} aims at putting source foils with $^{150}$Nd in a magnetized drift
chamber. 
XMASS\cite{xmass} proposes liquid scintillating Xenon,  
NEXT\cite{next} a gaseous Xenon TPC. Some of the experiments can also
be used as solar neutrino or dark matter experiments, such as XMASS,
NEXT or MOON. More details on the experiments can be found in the
respective publications and the reviews\cite{APS,AEE,barabash,andrea}.

It is encouraging that different experimental techniques will be
pursued, and that different isotopes are under study. 
Eventually, a multi-isotope determination of
\obb~would be preferable, to make it more unlikely that 
a peak coming from an unidentified background process mimics the
signal. This is the {\bf first reason for multi-isotope
determination}. \\

We will focus in this review on \onbb. However, there are similar
processes called neutrino-less double beta$^+$ decay 
$(0\nu\beta^+\beta^+)$, or beta$^+$-decay electron capture
$(0\nu\beta^+\rm EC)$, or double electron capture $(0\nu\rm
ECEC)$ of bound state electrons $e^-_b$, which can also be searched for: 
\begin{eqnarray}
(A,Z) \ra (A,Z-2) + 2 \, e^+  & ~~~~(0\nu\beta^+\beta^+)\, , \label{eq:b+b+}\\
e_b^- + (A,Z) \ra (A,Z-2) + e^+ & ~~~~(0\nu\beta^+\rm EC)\,,\label{eq:b+EC}\\
2 \, e_b^- + (A,Z) \ra (A,Z-2)^\ast & ~~~~(0\nu\rm ECEC)\,. \label{eq:ECEC}
\end{eqnarray} 
Observation of one of those processes would also imply the
non-conservation of lepton number. The rates depend on
the particle physics parameters in the same way as \obb~does. 
The creation of two positrons reduces the phase space and renders
rates for $0\nu\beta^+\beta^+$ very low. Somewhat less suppressed are
$(0\nu\beta^+\rm EC)$ processes. In $0\nu\rm ECEC$ the final atom (and 
sometimes the nucleus) are in excited states and generate photons (and $\gamma$
rays). The rate is low, unless a resonance can be met. This occurs\cite{ECECenh} for
certain $0\nu\rm ECEC$ modes, if the initial and final 
states of the system are degenerate in energy. 
Here the $^{152}$Gd--$^{152}$Sm transition has, via Penning-trap
mass-ratio measurements, recently been identified as an
interesting candidate for neutrino-less double electron
capture\cite{klaus1}, though it is currently unclear if an experiment
competitive to \obb-searches can be realized. 
The current limits of the reactions (\ref{eq:b+b+},\ref{eq:b+EC},\ref{eq:ECEC}) 
are summarized in\cite{barabash}. The main focus of future experiments 
is on the standard process \obb~in Eq.~(\ref{eq:main}). 
The other reactions could however be used to distinguish different
\obb-mechanisms from each other, see Section \ref{sec:distinguish}.

\section{\label{sec:nme}Nuclear physics aspects}
Nuclear physics is (unfortunately) an almost irreducible difficulty in making
interpretations of neutrino-less double beta decay. Observation of the process means
of course the proof of lepton number violation, but more precise
particle physics interpretations suffer from any nuclear physics
uncertainty. 
The calculation of the Nuclear Matrix Element (NME) ${\cal M}$ 
is a complicated many body nuclear physics problem as old as \obb. It basically describes the overlap of
the nuclear wave functions of the initial and final states. 
A nuclear model typically has a set of single-particle states with a number of possible
wave function configurations, and diagonalizes a Hamiltonian in a 
mean background field. 
A general property of solving Hamiltonians is that the energy levels are rather
stable in what regards small modifications. Wave functions, and
hence overlap, are however very sensitive to small modifications of the
Hamiltonian, and this is the origin of the uncertainty in the values
of NMEs.

\subsection{Standard mechanism}

Most theoretical work has been invested into the study of the standard mechanism of 
light neutrino exchange, on which we will focus in the following
discussion. The process is evaluated as
two pointlike Fermi vertices and the exchange of a light neutrino with 
momentum of about $q \simeq 0.1$ GeV, corresponding to the average
distance $r\simeq 1/q \simeq 1$ fm 
between the two decaying nuclei. Since the neutrino is very
light with respect to the energy scale one denotes the situation as a 
``long-range process''.

The expression for the decay rate is 
\be \label{eq:Gamma_mass}
\Gamma^{0\nu} = G^{0\nu}(Q,Z) \, |{\cal M}^{0\nu}|^2 \, \frac{\meff^2}{m_e^2} \, , 
\ee
with the phase space factor $G^{0\nu}(Q,Z)$, $\meff$ the particle
physics parameter\footnote{Sometimes one includes the electron mass 
$m_e^2$ in the phase space factor.} 
in case of light neutrino exchange (to be defined in
Section \ref{sec:meff}), and ${\cal M}^{0\nu}$ the NME. The quantity \meff~is usually called the
``effective mass'', or the ``effective electron neutrino mass''. 
The 5 main approaches to tackle
the problem are the Quasi-particle Random Phase Approximation 
(QRPA, including its many variants and evolution
steps)\cite{tuebingen2009,jyvaskyla2009}, 
the Nuclear Shell Model (NSM)\cite{NSM}, the Interacting Boson Model (IBM)\cite{IBM}, 
the Generating Coordinate Method (GCM)\cite{GCM}, and 
the projected Hartree-Fock-Bogoliubov model (pHFB)\cite{phfb}. 
We will not go into
comparing in detail the different procedures, and refer the reader
to the cited papers and the reviews\cite{haxton,doi,EV,Tomoda:1990rs,Gnu1,vergados,APS,AEE,nme_rev}. One example
on how the approaches differ is to note that QRPA calculations can
 take into account a huge number of single particle states but only a
limited set of configurations, whereas in the NSM the situation is
essentially the opposite. The issue of which
method should be used is far from settled. 
Another point are short-range correlations (SRC),  
since the main contribution to NMEs comes from 
internucleon distances $r \ls (2 - 3)$ fm\cite{tue08:src}, and the
nucleons tend to overlap. SRC take the hard core repulsion into
account. There are different proposals on how to treat SRC, namely via a 
Jastrow--like function, Unitary Correlation Operator Method (UCOM), 
or Coupled Cluster Method (CCM).
The Jastrow method leads typically to a reduction of NMEs by about 
20\% while UCOM and CCM both reduce NME by about 5\% as compared 
to calculations without SRC\cite{tue08:src,tuebingen2009}.

In contrast to \zbb, which involves only Gamov-Teller transitions
through intermediate $1^+$ states (because of low momentum transfer), 
\obb~involves all multipolarities in the intermediate odd-odd 
$(A,Z+1)$ nucleus, and contains a Fermi and a Gamov-Teller part 
(plus a negligible tensor contribution from higher order currents): 
\be \label{eq:NME}
{\cal M}^{0\nu} = \left(\frac{g_A}{1.25} \right)^2 \left( 
{\cal M}_{\rm GT}^{0\nu} - \frac{g_V^2}{g_A^2} {\cal
M}_{\rm F}^{0\nu} \right) \, . 
\ee
The matrix elements for the final and initial states $|f \rangle$ and
$|i \rangle$ can be written as 
\bea \label{eq:M0nu}
{\cal M}_{\rm GT}^{0\nu} = 
\langle f | \sum\limits_{l k} \sigma_l \, \sigma_k  \, \tau_l^- \,
\tau_k^- \, H_{\rm GT}(r_{lk}, E_a) | i \rangle \, , \\
{\cal M}_{\rm F}^{0\nu} = 
\langle f | \sum\limits_{l k} \tau_l^- \,
\tau_k^- \, H_{\rm F}(r_{lk}, E_a) | i \rangle \, ,
\eea
where $r_{lk} \simeq 1/q \simeq 1/(0.1~{\rm GeV})$ 
is the distance between the two decaying neutrons and
$E_a$ is an average energy (closure approximation due to 
the large momentum of the virtual neutrino). The ``neutrino
potential''  
\be \label{eq:nupot}
H(x,y) \propto \frac{1}{x} \int\limits_0^\infty dq \frac{\sin q x}{x +
y - (E_i + E_f)/2} 
\ee
integrates over the virtual neutrino momenta. The two emitted
electrons are usually 
described in $s$-wave form because one focusses on $0^+_{\rm g.s.} \to 0^+_{\rm g.s.}$
transitions. $p$-wave emission, which
would lead to transitions to excited states, is suppressed in the standard neutrino 
exchange mechanism\cite{doi}, see Section \ref{sec:distinguish_exp}. The \zbb~matrix elements
can be written as (note the different energy dependence in
comparison with the NMEs for \obb)
\bea \label{eq:M2nu} \D 
{\cal M}_{\rm GT}^{2\nu} = \sum\limits_{n} 
\frac{\langle f | \sum\limits_{a} \sigma_a \, \tau_a^- |n\rangle
\langle n| \sum\limits_b \sigma_b \, \tau_b^- |i\rangle}{E_n - (M_i -
M_f)/2}\, , \\ \D 
{\cal M}_{\rm F}^{2\nu} = \sum\limits_{n} 
\frac{\langle f | \sum\limits_{a}  \tau_a^- |n\rangle
\langle n| \sum\limits_b \tau_b^- |i\rangle}{E_n - (M_i -
M_f)/2}\, ,
\eea
where the sum over $n$ includes only $1^+$ states. This is the reason 
why \zbb~gives only indirect information on \obb.


\begin{table}[t]
\tbl{\label{tab:nme}Dimensionless NMEs calculated in different
frameworks, normalized to 
$r_0 = 1.2$ fm and $g_A = 1.25$. 
 The method used to take into account short range
correlations is indicated in brackets. There is also a pseudo-$SU(3)$
model\protect\cite{Hirsch:1994fw} for the highly deformed nucleus 
$^{150}$Nd, with a matrix element 1.00. Taken from\protect\cite{DRZ}.} 
{\begin{tabular}{@{}lcccccc@{}}
    &     NSM   &     T\"ubingen  &  Jyv\"askyl\"a   & IBM    & GCM  &
pHFB \\
Isotope    &     (UCOM)\cite{NSM}  &     (CCM)\cite{tuebingen2009} &
(UCOM)\cite{jyvaskyla2009}   &  (Jastrow)\cite{IBM}   &
(UCOM)\cite{GCM} & (mixed)\cite{phfb} \\ \colrule
$^{48}$Ca  &        0.85       &       -       &       -         &       -         &      2.37      &       -        \\
$^{76}$Ge  &        2.81       &  4.44 - 7.24  &  4.195 - 5.355  &  4.636 - 5.465  &      4.6      &       -        \\
$^{82}$Se  &        2.64       &  3.85 - 6.46  &  2.942 - 3.722  &  3.805 - 4.412  &      4.22      &       -        \\
$^{96}$Zr  &          -        &  1.56 - 2.31  &  2.764 - 3.117  &       -         &      5.65      &  2.24 - 3.46 \\
$^{100}$Mo &          -        &  3.17 - 6.07  &  3.103 - 3.931  &  3.732 - 4.217  &      5.08      &  4.71 - 7.77 \\
$^{110}$Pd &          -        &       -       &       -         &       -         &       -       &  5.33 - 8.91 \\
$^{116}$Cd &          -        &  2.51 - 4.52  &  2.996 - 3.935  &       -         &      4.72      &       -        \\
$^{124}$Sn &        2.62       &       -       &       -         &       -         &      4.81      &       -        \\
$^{130}$Te &        2.65       &  3.19 - 5.50  &  3.483 - 4.221  &  3.372 - 4.059  &      5.13      &  2.99 - 5.12 \\
$^{136}$Xe &        2.19       &  1.71 - 3.53  &  2.38 - 2.802  &       -         &      4.2      &       -        \\
$^{150}$Nd &          -        &      3.45     &       -         &
2.321 - 2.888  &      1.71      &  1.98 - 3.7 \\  \botrule
\end{tabular}}
\end{table}

\begin{figure}[t]
\centerline{\psfig{file=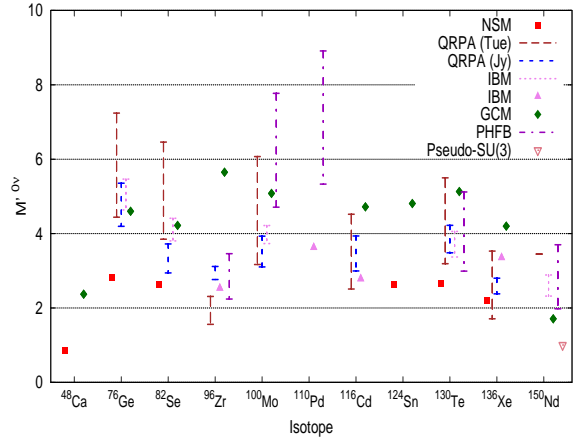,width=8cm,height=6cm}}
\vspace*{8pt}
\caption{\label{fig:nme}Nuclear Matrix Elements for \obb, different
isotopes and calculational approaches. 'Tue' and 'Jy' are both QRPA results.}
\end{figure}

We would like to stress here
that care has to be taken when different calculations are compared\cite{care,DRZ}: 
for instance, NMEs are made dimensionless by putting a factor $1/R_A^2 =
1/(r_0 \, A^{\frac 13})^2$ in the phase space factor (in the
convention of Eq.~(\ref{eq:NME}) the phase space 
becomes independent of $g_A$), where in the
nuclear radius $R_A$ the parameter $r_0$ is sometimes chosen as 
$1.1$ fm or $1.2$ fm. 
The axial-vector coupling $g_A$ is often chosen to be $1.25$ or $1.0$. 
In addition, it is often overlooked (see the discussion in\cite{care,TuBa}) 
that the phase space factors $G^{0\nu}(Q,Z)$ can differ by up 
to order 10\%, for instance when one compares the results from\cite{Gnu1} or\cite{Gnu2}. 
The results from\cite{Gnu1} are given in Table \ref{tab:Gnu}. 
In Table \ref{tab:nme} and Fig.~\ref{fig:nme} we give a compilation\cite{DRZ} of NME values from
different calculations (see Refs.\cite{comp1,comp2,Bilenky:2011tr} for
similar recent compilations).  For definiteness, we will often apply the
values of this table in what follows. 
Main features of the current status are that NMEs using QRPA seem to agree
with each other and also with IBM calculations. NSM evaluations are
consistently smaller and show little dependence on $Z$. However, it is
encouraging that conceptually different approaches give results 
in the same ballpark. 
All in all, there has been some improvement in recent years, in
particular the number of groups and approaches, as well as
experimental support, has increased. However, full understanding
and/or consensus has not been reached yet. Currently, one has to take
the uncertainty at face value and keep it in mind when interpreting the results of
\obb~experiments. Table \ref{tab:limits} gives the
current limits on the effective mass \meff~obtained with the NME compilation
from Table \ref{tab:nme} and the phase space factors from Table
\ref{tab:Gnu}. The best limit on \meff~is provided by $^{76}$Ge, but, 
as we will see, for some other mechanisms stronger limits stem from
other isotopes, in particular $^{130}$Te.

\begin{figure}[b]
\centerline{\psfig{file=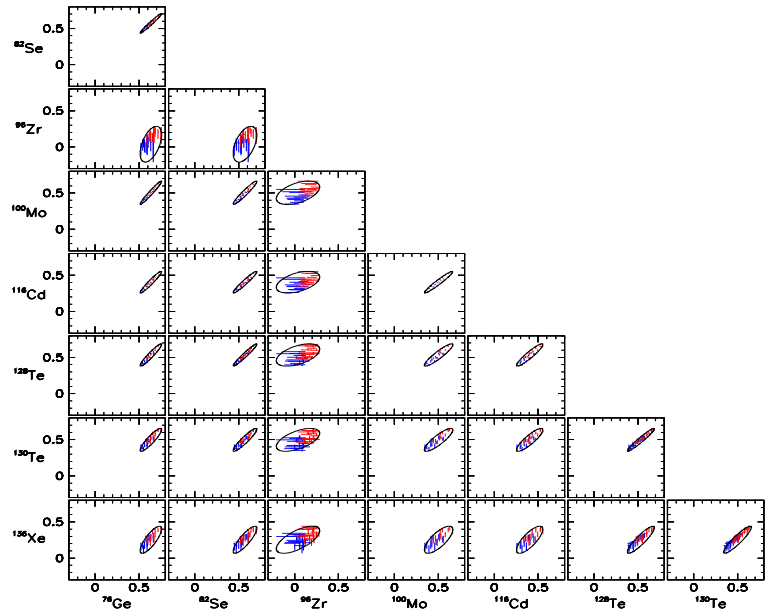,width=10cm,height=8cm}}
\vspace*{8pt}
\caption{\label{fig:nme_error}
$1\sigma$ error ellipses (logarithms of the NMEs) within QRPA calculations. The major axis
corresponds to variation of the short-range correlation model (blue is
Jastrow, red is UCOM) and $g_A$, while the minor axis corresponds 
to variations of $g_{pp}$. 
Taken from\protect\cite{errors}.}
\end{figure}
If available, we include uncertainties of the NMEs in Table \ref{tab:nme}
and Fig.~\ref{fig:nme}. 
However, not all authors provide errors in their calculations. 
Those theoretical uncertainties can arise from
varying $g_A$, $g_{pp}$ (the particle-particle strength parameter in
QRPA models), or other model details. 
One can distinguish
here between correlated (e.g.~the use of SRC or the value of $g_A$ or
$g_{pp}$) and uncorrelated errors (e.g.~the model space of the single
particle base)\cite{errors}.  Eventually, a multi-isotope determination of
\obb~would be preferable, to disentangle the different types of
errors. This is the {\bf second reason for multi-isotope
determination}. Ideally, if there was one adjustable parameter $x$ in the
 calculations, then two measurements would suffice to fix \meff~and 
$x$ (up to degeneracies). A third result would overconstrain the
system\cite{BP} and allow for cross checks. This requires analyses 
of degeneracies and realistic estimates of theoretical
errors, an effort which has recently started\cite{errors}.
With the factorization in Eq.~(\ref{eq:fact}) it is clear
that the ratios of two measured half-lifes are 
\be
\frac{T_{1/2}^{0\nu}(A_1,Z_1)}{T_{1/2}^{0\nu}(A_2,Z_2)} 
= \frac{G(Q_2,Z_2) \, |{\cal M}(A_2,Z_2)|^2}
{G(Q_1,Z_1) \, |{\cal M}(A_1,Z_1)|^2} \, , 
\ee
i.e.~the particle physics parameter drops out. The ratio is sensitive
to the NME calculation\cite{BP}, and systematic errors are expected to
cancel. Fig.~\ref{fig:nme_error}, taken from\cite{errors}, shows the error ellipses of matrix elements
within a QRPA analysis.\\

Free parameters of an Ansatz can in principle also be fixed or tested by other means,
for instance the particle-particle strength parameter $g_{pp}$ in QRPA
models can be adjusted to reproduce the \zbb\cite{fix1}, single beta
decay\cite{fix2} or electron capture rates. Overconstraining the parameters is
possible if data on all these processes is available\cite{Faessler:2007hu}. 

In recent years the uncertainty {\it of the individual approaches} to NMEs, 
which was somewhat overestimated in the past\cite{toomuch}, has been
reduced. An experimental program to support the
calculations with as much information as possible was
launched\cite{kai_nme}, including charge exchange
reactions\cite{frekers} to determine Gamov-Teller transition strengths. 
The latter are directly related to \zbb~matrix elements;  
applying the results to \obb~requires theoretical input, see
e.g.\cite{measure?}. Occupation numbers of neutron valence orbits in 
the initial and final nuclei are not known very well, and 
measurements\cite{schiffer} via nucleon transfer reactions 
are helpful for all NME approaches. 
Muon capture rates can also be useful\cite{muon}, because the
momentum transfer is of order $m_\mu \simeq 100$ MeV, i.e.~of the same order as for
\obb. Determinations of $Q$-values\cite{Q} with precision 
spectroscopy\cite{klaus} are also ongoing.  This is particularly
helpful for experiments in which the energy resolution is comparable
to the current uncertainty of the $Q$-value. Another motivation for
precise $Q$-value determinations is the 
identification of candidates for $0\nu$ECEC which show resonance behavior, as
mentioned above. 

The QRPA particle-particle strength parameter
$g_{pp}$ can be fixed by the measured \zbb~rates and
used as input for \obb~predictions. However, using this $g_{pp}$ value
for calculating the rates of beta-decay or electron capture 
of the intermediate double beta decay isotope sometimes fails.  
One particular observational approach to these issues is the TITAN-EC 
experiment\cite{titan}, which aims at testing with ion traps the badly known electron
capture rates of the intermediate odd-odd state of double beta decay
via observing the de-excitation $X$-rays. \\

It is expected that the future the uncertainty in the NMEs will
further decrease, though (owing to the enormous complexity of the
problem) the precision by the end of the decade will probably not be better than
20\%. 

\subsection{\label{sec:NME_NSI}Non-standard mechanisms}
The evaluation of NMEs in non-standard mechanisms is a less well
developed field, with less calculations available, and often only
within one particular nuclear physics approach. 
In general the NMEs can obtain now contributions from Fermi, Gamov-Teller, 
pseudo-scalar, tensor, etc.~contributions, and the realization of \obb~can differ from the standard
mechanism in 
\begin{itemize}
\item[(i)] the Lorentz structure of the currents (e.g.~right-handed currents);
\item[(ii)] the mass scale of the exchanged particle (e.g.~exchange
of heavy SUSY particles); 
\item[(iii)] the number of particles in the final states (e.g.~modes
with additional Majoron emission); 
\end{itemize}
A frequent feature here is that the scale of lepton number
violating physics is larger than the momentum transfer or nuclear energies, in
which case one speaks of a ``short-range process''. Non-standard physics
including light neutrino exchange, hence long-range, is however 
also possible. Within QRPA, a general Lorentz-invariant parametrization of the
\obb~decay rate has been developed for long-\cite{super_long} and 
short-range\cite{super_short} processes 
(these papers are in fact the only entries on SPIRES with the word
``superformula'' in the title). In those works the
most general Lagrangian for \obb~was written down and each  
term includes an individual prefactor $\epsilon_i$. 
These $\epsilon_i$
can in principle via Fierz-transformations 
be translated (see also\cite{Prezeau:2003xn}) into the particle physics
parameters of the alternative realizations of \obb~which we will
discuss in Section \ref{sec:non-standard}.

It is possible that the different Lorentz structure leads to
additional contributions to the NMEs, which are not present in other realizations. 
Different Lorentz structure implies also that the energies of the
individual electrons, and their angular distribution, may differ from the
standard mechanism\cite{doi,angular1,angular2}. Further potential differences are
modified relations between the rate of \obb~and $0\nu\beta^+\rm EC$\cite{bbvsEC}, or with
the decay rate to excited states\cite{excited0,excited0a,excited2,Simkovic:2001ft}. 
Some details will be
discussed in Section \ref{sec:distinguish_exp}. 
Short distance physics implies that the heavy particles with mass $M_X$ can be
integrated out and the pure particle physics amplitude is inversely 
proportional to $M_X$ or $M_X^2$, 
depending on whether it is a fermion or boson. A nuclear 
aspect of short distance physics is that the inner structure of the
nucleons becomes relevant, which is taken into account by multiplying
the weak nucleon vertices with (dipole) form factors\cite{formfactor}
\be
g_A(q^2) = \frac{g_A}{(1 - q^2 /M_A^2)^2} \, , 
\ee
with a mass 
parameter $M_A^2 \simeq (0.9 \, \rm GeV)^2$. This 
introduces e.g.~for heavy neutrino exchange a dependence proportional
to $M_A^2$, after an integration over the neutrino momenta has
been performed in the potential Eq.~(\ref{eq:nupot}). Here the form
factor avoids the otherwise exponential suppression of the amplitude
due to the repulsion of the nuclei. 
Finally, if additional particles are emitted in addition to the two
electrons, such as in Majoron modes (Section \ref{sec:Majorons}), 
significant phase space effects can be expected. 

Another aspect of heavy particle exchange is that pion exchange can
dominate\cite{Vergados:1981bm,Faessler:1996ph}. This means that the
pions which are present in the nuclear medium undergo transitions like
$\pi^- \to \pi^+ \, e^- e^-$, i.e.~the hadronization procedure of the
quark level diagram differs from the 2 nucleon mode discussed so far. 
Though the probability of finding pions in the 
nuclear soup is less than 1, this can be compensated by the fact that
the suppression due to the short-range nature of the usual 2 nucleon
mode is absent, because low mass pions can mediate between more
distant nucleons. One or both of the two initial quarks can be placed
into a pion. 
In fact, $R$-parity violating SUSY contributions 
(Section \ref{sec:SUSY}), turn out to be dominated by pion NMEs.

We will discuss aspects relevant to particular non-standard mechanisms
and means to distinguish them, from one another and from the standard
one, in the later sections which deal with the respective mechanisms. Different
realizations of the decay influence the NMEs in a way which depends on
the isotope and on particle physics. Therefore, eventually a multi-isotope determination of
\obb~would be preferable, in order to disentangle the different 
mechanisms\cite{diff_nme1,diff_nme2,diff_nme3,Faessler:2011rv,Faessler:2011qw}, see Section
\ref{sec:distinguish}. 
This is the {\bf third reason for multi-isotope determination}.

\section{\label{sec:meff}Standard Interpretation}

In this Section we will discuss the standard mechanism of
neutrino-less double beta decay, let us repeat for convenience the
definition:

{\it 
Neutrino-less double beta decay is mediated by light and massive
Majorana neutrinos (the ones which oscillate) and all other mechanisms
potentially leading to \obb~give negligible or no contribution.}

We will first summarize the current status of our understanding of 
lepton mixing and neutrino mass, before discussing the amount of
information encoded in \obb~combined with the standard
interpretation. Readers who are very familiar with neutrino physics
can go directly to Section \ref{sec:SI} and skip the summary of
neutrino physics in  Section \ref{sec:neutrinos}.

\subsection{\label{sec:neutrinos}Neutrino physics}
Most part of the review will deal with the standard, and presumably
best motivated, interpretation of \obb, light massive Majorana
neutrino exchange. We will first review the current theoretical and 
phenomenological status of neutrino physics.

\subsubsection{Neutrino mass and mixing: theoretical origin}

The theory behind neutrino mass and lepton mixing has been reviewed 
in several places\cite{Bilenky:1987ty,King:2003jb,Mohapatra:2005wg,Mohapatra:2006gs}. 
Lepton mixing is rather different from quark
mixing. In addition, the mass of the two lepton partners in an $SU(2)_L$
doublet (e.g.~$\nu_e$ and $e$) is extremely hierarchical, in sharp
contrast to the partners in quark doublets (e.g.~$u$ and $d$, with
$m_u = {\cal O}(m_d)$). It is natural to believe that these
discrepancies are related to special properties of the neutrinos. 
Indeed, most, if not all, theorists believe that neutrinos are Majorana
particles. This is the case in basically all Grand Unified Theories
(GUTs), and also from an effective theory point of view, in
which non-renormalizable higher dimensional operators invariant under the SM gauge group are
constructed. The lowest dimensional (Weinberg) operator is unique, and 
reads\cite{Weinberg}
\be \label{eq:Leff}
{\cal L}_{\rm eff} = \frac 12 \frac{h_{\alpha\beta}}{\Lambda}
\overline{L_{\alpha}^c}  \, \tilde \Phi \, \tilde{\Phi}^T \, L_{\beta} 
 \stackrel{\rm EWSB}{\longrightarrow} \frac 12 \, (m_\nu)_{\alpha \beta} \,
\overline{\nu_{\alpha}^c} \, \nu_{\beta} \, ,    
\ee
Here the superscript '$^c$' denotes the charge-conjugated spinor, 
$L_\alpha = (\nu_\alpha, \alpha)^T$ are the lepton doublets of
flavor $\alpha = e, \mu, \tau$ and $\Phi$ is the Higgs doublet with 
vacuum expectation value $v = 174$ GeV. A Majorana neutrino mass 
matrix is induced by this operator, given by $m_\nu = h \, v^2 /\Lambda$. 
With the typical mass scale of $m_\nu \simeq 0.05$ eV, it follows that $\Lambda \simeq 10^{15}$ GeV,
tantalizingly close to the GUT scale. This is one of the main reasons
why neutrino physics is popular: large scales are probed by small
neutrino masses. 
It has been shown that within the minimal standard
electroweak gauge model, there are only three tree-level
realizations\cite{Ma} of the Weinberg operator. 
One is the canonical type I seesaw mechanism\cite{I} with right-handed
neutrinos. Another approach is introducing a scalar Higgs triplet (type II, or
triplet seesaw\cite{II}), and  
the third one involves hypercharge-less fermion triplets (type III
seesaw\cite{III}). In Table \ref{tab:seesaws} we summarize the main
approaches for generating small neutrino mass. 

\begin{table}[t]
\tbl{\label{tab:seesaws}Tree-level approaches to small neutrino mass
classified according to the ingredient which has to be added to the SM, and the electroweak
quantum numbers of the new particles. LNV denotes Lepton Number Violation.}
{\begin{tabular}{cccccc} \toprule
approach  & ingredient & $\ba SU(2)_L \times U(1)_Y \\ \mbox{quantum number} 
\\ \mbox{of messenger} \ea $ 
& ${\cal L}$ & $m_\nu$ & scale  \\ \colrule \hline 
$\ba \mbox{``SM''} 
\\ \mbox{\bf (Dirac mass)} \ea $ & RH $\nu$
& ${ N_R} \sim (1,0)$ 
& $h \overline{N_R}  \Phi  { L}$ & 
${ h v}$ 
 & $h = {\cal O}(10^{-12})$
\\ \hline\hline 
$\ba \mbox{``effective''} 
\\ \mbox{\bf (dim 5 operator)} \ea $ 
& $ \ba \mbox{ new scale} \\ \mbox{+ LNV} \ea $
& -- 
& $ h \, \overline{L^c}  \, \Phi \, \Phi \, L $ 
& $ { \frac{\D h \, v^2}{\D \Lambda}}$  
&  $\Lambda = \frac 1h \left(\frac{\D \rm 0.1~eV}{\D m_\nu}\right) 
 10^{14} $ GeV
\\ \hline 
$\ba \mbox{``direct''} \\ 
\mbox{\bf (type II seesaw)} \ea $ 
& $\ba \mbox{ Higgs triplet} \\ \mbox{+ LNV} \ea $ 
& ${ \Delta} \sim (3,2)$  
& $ h  \overline{L^c}  { \Delta}  L + \mu \Phi \Phi { \Delta}$  
& ${ h  v_T}$ 
& $\Lambda = \frac{\D  1}{\D h  \mu} M_\Delta^2 $
\\ \hline 
$\ba \mbox{``indirect 1''}
\\ \mbox{\bf(type I seesaw)} \ea $ 
& $\ba \mbox{ RH $\nu$} \\ \mbox{ + LNV} \ea $
&  ${ N_R} \sim (1,0)$ 
& $h \overline{N_R}  \Phi  { L} + 
\overline{ N_R}  M_R  { N_R^c} $ & 
${ \frac{\D (h  v)^2}{\D M_R} } $  
& $\Lambda = \frac{\D 1}{\D h}  M_R$ 
\\ \hline 
$\ba \mbox{``indirect 2''} 
\\ \mbox{\hspace{-.432cm}\bf(type III seesaw)\hspace{-.432cm}} \ea $  
&  $\ba \mbox{\hspace{-.432cm} fermion triplets\hspace{-.432cm}} \\ \mbox{+ LNV} \ea $ 
& ${ \Sigma} \sim (3,0)$  
& $h \overline{\Sigma} 
\, { L}  \Phi + {\rm Tr} 
\overline{ \Sigma}  M_\Sigma  { \Sigma} $ & 
${ \frac{\D (h  v)^2}{\D M_\Sigma} } $  
& $\Lambda = \frac{\D 1}{\D h}  M_\Sigma$ \\ \botrule
\end{tabular}}
\end{table}      

Taking first the standard type I seesaw as an example, one introduces 3 
(actually 2 would suffice) Majorana neutrinos $N_{R, i}$, which have a
Majorana mass matrix $M_R$. After electroweak symmetry breaking a
Dirac mass term with the SM neutrinos is present and the full
Lagrangian for neutrino masses is 
\be \label{eq:Lseesaw}
{\cal L} = -\frac 12 M_R \, \overline{N_R} \, N_R^c - m_D \, \overline{N_R }
\, \nu_L 
= -\frac 12  \, \overline{(\nu_L^c , \, N_R)} 
\left(
\bad 
0 & m_D^T \\
m_D & M_R 
\ea 
\right) \left(\ba \nu_L \\ N_R^c \ea \right)
\, , 
\ee
with $N_R = (N_{R 1},N_{R 2},N_{R 3})$ and  $\nu_L = (\nu_e, \nu_\mu, 
\nu_\tau)_L$.  We see that the combination of a Dirac and a Majorana
mass term is a Majorana mass term, no matter how small the Majorana mass is.  
Being SM singlets, the scale of $M_R$ is not connected
to the only energy scale of the SM (the Higgs vacuum expectation
value), and hence can be arbitrarily high. Integrating out the heavy
states, or block-diagonalizing\footnote{The
condition for this is that the eigenvalues of $M_R$ are much heavier
than the entries of $m_D$.} the mass matrix in Eq.~(\ref{eq:Lseesaw})
gives a Majorana mass term for the light neutrinos, 
\be\label{eq:mnuseesaw}
m_\nu = - m_D^T \, M_R^{-1} \, m_D \, , 
\ee
plus terms of order $m_D^4/M_R^3$. The states for which this mass
matrix is valid are the initial $\nu_L$ plus a contribution of 
$N_R^c$, which is however suppressed by $m_D/M_R$. We see that the
Weinberg operator is realized with $\Lambda \simeq M_R$. 

Often one considers a triplet term for neutrino masses, generated by 
an $SU(2)_L$ triplet scalar with non-zero vev $v_L$ of its neutral
component. The coupling of the triplet to two lepton doublets with the
Yukawa coupling matrix $h$ gives a neutrino mass 
$m_\nu = M_L$, i.e.~a direct contribution (see Section \ref{sec:Delta}). 
Of course, both the type I and the type II
term could be present. In this case the zero in the upper left entry 
of Eq.~(\ref{eq:Lseesaw}) is filled with a term $M_L$. 
The neutrino mass matrix in this case reads 
\be\label{eq:mnuseesawI+II}
m_\nu = M_L - m_D^T \, M_R^{-1} \, m_D \, . 
\ee
Finally, type III seesaw introduces 3 hypercharge-less fermion triplets
(one for each massive light neutrino), 
whose neutral components play the role of the $N_{R i}$ of type I seesaw. 

What about production of low scale seesaw messengers at colliders? A
recent review on the situation can be found in\cite{seesaw_LHC}. While
Majorana neutrino production proves difficult because of the
constraint of its small mixing with SM particles, Higgs and fermion triplets have
gauge quantum numbers and can be observed up to TeV masses. Note that
in left-right symmetric models, or models with gauged $B-L$, the
right-handed neutrinos can have gauge interactions and can be produced more easily at
colliders. The lepton number violation associated with the seesaw messengers 
can lead to their identification and spectacular {\it like-sign lepton
events}. For this to be realized one needs to bring the seesaw scale
down to TeV, see\cite{Chen:2011de} for a recent review on how to
achieve this.

It is clear that, either way, neutrinos are Majorana particles, i.e.  
\be
\nu_i^c = C \, \bar{\nu}_i^T = \nu_i \, . 
\ee
Here we have chosen a convention in which there 
is no phase in the above relation. 

In general the mass matrices for neutrinos ($m_\nu$) and for charged
leptons ($m_\ell$) are non-trivial. 
Diagonalizing those matrices with unitary\footnote{Strictly speaking
the matrix $U_\nu$ is not unitary in type I seesaw, due to 
mixing of the leptons with the heavy neutrinos. This is 
however usually a very small effect $|U_\nu U_\nu^\dagger - \mathbbm
1| \sim (m_D/M_R)^2$ and phenomenologically constrained to be less
than a permille effect\cite{SA}.}
$U_\nu$ and $U_\ell$, respectively, results in the charged current
term in the appearance of the Pontecorvo-Maki-Nakagawa-Sakata (PMNS) 
matrix $U = U_\ell^\dagger \, U_\nu$: 
\be
{\cal L}_{CC} = -\frac{g}{\sqrt{2}} \, \overline{\ell}_\alpha \,
\gamma^\mu \, U_{\alpha i} \, \nu_i  \, W^-_\mu \, . 
\ee
In the basis in which the charged leptons are real and diagonal, the
neutrino mass matrix is diagonalized by $U$. It is useful for our
purposes to stay in this basis, in which  
the mass matrix for Majorana neutrinos can be written as 
\be \label{eq:mnu}
m_\nu = U^\ast \, m_\nu^{\rm diag} \, U^\dagger  ~,\mbox{ where } 
m_\nu^{\rm diag} = {\rm diag}(m_1, m_2, m_3) \, . 
\ee
The mass matrix is complex and symmetric;  after rephasing of three
phases there are 9 physical parameters. Because the mass term goes as 
$\overline{\nu^c} \, \nu \propto \nu^T \, \nu $, the Lagrangian is not
invariant under a global transformation $\nu \to e^{i \phi} \nu$. The
charge associated with this transformation, lepton number, is
therefore not a conserved quantity and $L$ is violated by two 
units\footnote{It should be noted that there are alternatives to the seesaw
mechanism, see\cite{seesaw?} for a discussion. Examples are radiative
mechanisms, supersymmetric scenarios, or extra-dimensional
approaches. It can happen that lepton number is conserved in such
frameworks and \obb~cannot take place, but this is clearly a rare exception rather than the
rule.}. This is exactly what is required for the presence of \onbb.

Another appealing prediction of seesaw is the possible 
generation of the baryon asymmetry of the Universe via
leptogenesis\cite{lepto}. Here the heavy seesaw messengers decay out 
of equilibrium (Sakharov condition I) in the early Universe and, due to CP violating
phases (condition II), create a lepton asymmetry which subsequently is 
transfered into a baryon asymmetry via $B+L$ violating (condition III) non-perturbative SM
processes. 
In this context, proving the Majorana nature of neutrinos
and the presence of CP violation in the lepton sector would strengthen
our belief in this already very appealing mechanism.  This remains
true even though a model-independent connection between the low energy
CP phases 
and the necessary CP violation for leptogenesis cannot be
established\cite{lepto}. In general, taking the standard type I seesaw as an
example, there are in total six CP phases, three of which get lost
when the heavy Majorana neutrino mass matrix is integrated out to
obtain $m_\nu$ (see Eq.~(\ref{eq:mnuseesaw})). In principle, one could
construct models in which the ``low energy phases'' take CP conserving
values, while the remaining three phases are responsible for
leptogenesis\footnote{Note that the opposite case is also
possible.}. Leaving this seemingly unnatural possibility aside, one
expects that CP violation in the lepton sector at low energy is 
present if there is ``high energy'' CP violation responsible for
leptogenesis. One should however not expect that the Majorana phases are 
``more connected'' to leptogenesis than the Dirac phase. At the
fundamental (seesaw) scale, there are six CP phases and the three low energy
phases will be some complicated function of these phases and the other
seesaw parameters. From this point of view, the low energy Dirac and Majorana
phases are not different from each other. 

A final remark necessary here
is that a link between \obb~and the baryon asymmetry is {\it not 
guaranteed}. The often-made and popular statement that \obb-experiments probe the
origin of matter in the Universe is not true. 
For instance, if neutrino mass is simply generated by a Higgs triplet,
then this triplet alone cannot generate a baryon asymmetry, but
\obb~is very well possible.

\subsubsection{\label{sec:mnu_obs}Neutrino mass and mixing: observational status}

Neutrino oscillations have been observed with solar, atmospheric and
man-made (reactor, accelerator) neutrinos,
see\cite{concha_review,StrumiaVissani} for extensive reviews on the
status of neutrino physics. 
This implies that in the 
charged current term of electroweak interactions the neutrino flavor
states $\nu_{e}, \nu_\mu$ and $\nu_\tau$ are superpositions of
neutrino mass states: 
\be
\nu_\alpha = U_{\alpha i}^\ast \, \nu_i \, , 
\ee
where $\alpha = e, \mu, \tau$ and $i = 1,2,3$. The PMNS mixing
 matrix $U$ is unitary and can
be written in its standard parametrization as 
\be \label{eq:U}
U = \left( \bad 
c_{12}   c_{13} 
& s_{12}  c_{13} 
& s_{13}  e^{-i \delta}  \\ 
-s_{12}  c_{23} 
- c_{12}  s_{23} \, 
s_{13}   e^{i \delta} 
& c_{12}  c_{23} - 
s_{12}  s_{23}  s_{13} 
\, e^{i \delta} 
& s_{23}   c_{13}  \\ 
s_{12}    s_{23} - c_{12} 
 c_{23}  s_{13}  e^{i \delta} & 
- c_{12}  s_{23} 
- s_{12}  c_{23} \, 
s_{13}  e^{i \delta} 
& c_{23}   c_{13}  
\ea   
\right) P \,,
\ee
where $s_{ij} = \sin \theta_{ij}$, $c_{ij} = \cos \theta_{ij}$ and
$\delta$ is the ``Dirac phase'' responsible for CP violation in
neutrino oscillation experiments. This phase is expressible in a parametrization
independent form as a Jarlskog invariant: 
\be \label{eq:J}
J_{\rm CP} = {\rm Im}\left\{ U_{e1}^\ast \, U_{\mu 3}^\ast \, U_{e3} \,
U_{\mu 1} \right\} = 
\frac 18  \sin 2 \theta_{12} \, \sin 2 \theta_{23} \, \sin 2
\theta_{13} \, \cos \theta_{13} \, \sin \delta \, , 
\ee
where we have given it in its explicit form for the standard
parameterization. 
In Eq.~(\ref{eq:U}) we
have included a diagonal phase matrix $P$, containing the two ``Majorana
phases'' $\alpha$ and $\beta$: 
\be
P = {\rm diag}(1,e^{i \alpha}, e^{i (\beta + \delta)}) \, . 
\ee
These phases are physical\cite{MajPha}
if neutrinos are Majorana particles. Note that
we have included $\delta$ in $P$, in which case the first row of the
PMNS matrix is independent of $\delta$. 
For three neutrinos we have therefore 9 physical parameters, three
masses $m_{1,2,3}$, three mixing angles $\theta_{12}, \theta_{13},
\theta_{23}$ and three phases $\delta, \alpha, \beta$.

One can also define invariants for the Majorana phases\cite{MajPhainv}, for instance 
$S_1 = {\rm Im}\left\{ U_{e 1}  \, U_{e 2}^\ast \right\} = 
-c_{12} \, s_{12} \, c_{13}^2 \sin \alpha $ and 
$S_2 = {\rm Im}\left\{ U_{e 2}  \, U_{e 3}^\ast \right\}
= s_{12} \, c_{13} \, s_{13} \, \sin(\delta - \beta)$. 
Note that CP violation due to the Majorana phases is present only if, 
in addition to $S_1 = {\rm Im}\left\{ U_{e 1}  \, U_{e 2}^\ast
\right\} \neq 0$, ${\rm Re}\left\{ U_{e 1}  \, U_{e 2}^\ast
\right\} \neq 0$ also holds. The reason for this is that the cases $\alpha, \beta =
\pi/2$ correspond to the CP parities of the Majorana fields\cite{CPparities}, which can be
either positive or negative. Majorana phases are present because the
mass term in the Lagrangian is proportional to 
$(m_\nu)_{\alpha\beta} \, \nu_{\alpha}^T \, \nu_\beta$ and a
rephasing of the spinors $\nu_\alpha$ can eliminate fewer phases than
in the Dirac case, where the mass term is $(m_\nu)_{\alpha\beta} \,
\bar{\nu}_{\alpha} \, \nu_\beta$. For $N$
Majorana neutrinos, there are $N-1$ Majorana phases in addition to 
$\frac 12 \, (N - 2) \, (N-1)$ Dirac phases and $\frac 12 N \, (N-1)$
mixing angles\footnote{This counting is valid for active neutrinos
only. For $N$ massive families including  $0 \neq N_s = N-3$ massive sterile neutrinos,
one has $N - 1 = N_s + 2$ Majorana phases, $3 \, (N-2) = 3 \, (N_s + 1)$ mixing angles and 
$2 \, N - 5 = 2 \, N_s + 1$ Dirac phases. The number of angles and Dirac
phases is less because the 
$\frac 12 \, N_s \, (N_s - 1)$ rotations between sterile states are unphysical. }. 

\begin{figure}[t]
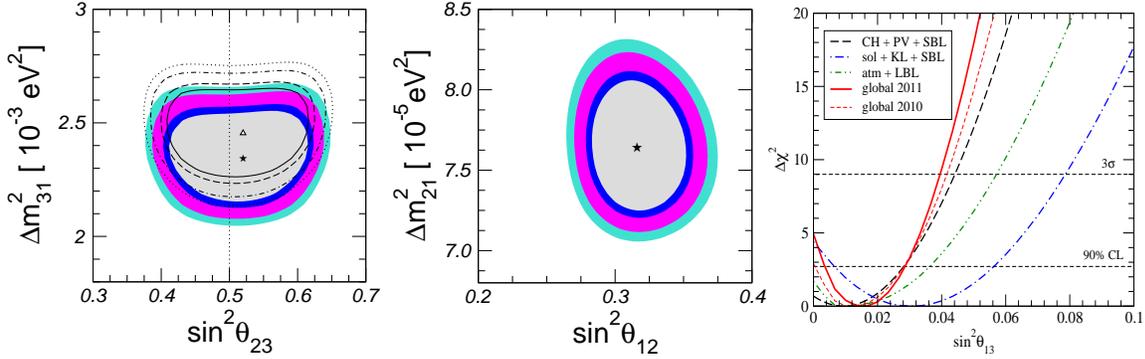

\centerline{\psfig{file=comp-atm+minos-H-2.eps,width=5cm}
\psfig{file=plot-sol+kaml+sbl.eps,width=5cm}\psfig{file=chisq-th13-glob-NH.eps,width=5cm,height=4.65cm}
}
\vspace*{8pt}
\caption{\label{fig:osc}Left plot: allowed ranges (lines are for
NH, colored regions for IH) of $|\Delta m^2_{31}|$ 
and $\sin^2 \theta_{23}$ at 90\%, 95\%, 99\% and 99.73\%. Middle
plot: allowed ranges $\Delta m^2_{21}$ and $\sin^2 \theta_{12}$. Right
plot: constraints on $\sin^2 \theta_{13}$ from various experiments. 
Taken from\protect\cite{STV}.}
\end{figure}

For three fermion families, neutrino oscillation experiments are sensitive to the
three mixing angles, the two independent mass-squared differences
(including their sign), and
the Dirac phase $\delta$. The general formula for oscillation
probabilities is 
\bea
P(\nu_\alpha \to \nu_\beta) = \delta_{\alpha\beta} - 4 \sum\limits_{i>j}{\rm
  Re} \left\{  U_{\alpha i}^\ast \, U_{\beta j}^\ast \, U_{\beta i} \,
  U_{\alpha j} \right\} \sin^2 \frac{\Delta m^2_{ij} \, L}{4 \, E} \\
 + 2 \sum\limits_{i>j}{\rm
  Im} \left\{  U_{\alpha i}^\ast \, U_{\beta j}^\ast \, U_{\beta i} \,
  U_{\alpha j} \right\} \sin \frac{\Delta m^2_{ij} \, L}{2 \, E} \, , 
\eea
with $E$ the neutrino energy and $L$ the baseline. 
To leading order, using the hierarchy of $\dms \equiv \Delta m^2_{21} \ll |\Delta
m^2_{31}| \simeq |\Delta m^2_{32} | \equiv \dma$, this formula usually
breaks down to two neutrino oscillation formulas. The angle 
$\theta_{12}$ and $\Delta m^2_{21} \equiv \dms$ are 
responsible for solar neutrino (suitably modified with matter
effects) and long-baseline reactor neutrino oscillations. Atmospheric neutrinos
are governed by $\theta_{23}$ and $\Delta m^2_{32}$, the same
parameters which long-baseline accelerator neutrinos are sensitive
to. Finally, $\theta_{13}$ (if non-zero) and $\Delta m^2_{31}$ 
are responsible for short-baseline reactor neutrino and 
long-baseline $\nu_\mu \to \nu_e$ oscillations. 
Non-zero $\theta_{13}$ also provides a link between the solar
and atmospheric sector and is intensively searched for, as leptonic
CP violation in oscillations would be absent if it was zero. 
Our current knowledge of the oscillation quantities 
is summarized in Fig.~\ref{fig:osc} and Table \ref{tab:nudata}, taken from\cite{STV}. 
It is noteworthy that the sign of the (atmospheric) mass-squared
difference is unknown, as are the three CP phases, thus including the
Majorana phases. The hint\cite{t13_hint} towards non-zero
$\theta_{13}$ recently exceeded the $3\sigma$
level\cite{Fogli:2011qn}, after the T2K long-baseline experiment provided evidence
for electron neutrino appearance\cite{Abe:2011sj}.

\begin{table}
\tbl{\label{tab:nudata}Current values from global fits to the world's
neutrino oscillation experiments. Taken from\protect\cite{STV}. The
values in brackets are for the inverted ordering.}
{\begin{tabular}{@{}cccc@{}} \toprule
parameter    	&	best-fit$^{+1\sigma}_{-1\sigma}$ & $2\sigma$ &
$3\sigma$ \\ \colrule
$\Delta m^2_{21}\, \left[ 10^{-5}\, {\rm eV}^2 \right]$  &
$7.64^{+0.19}_{-0.18} $ & 7.27 -- 8.03 & 7.12 -- 8.23 \\ \hline 
$|\Delta m^2_{31}| \, \left[ 10^{-3}\, {\rm eV}^2 \right]$ & 
$2.45^{+0.09}_{-0.09} $ & 2.28 -- 2.64 & 2.18 -- 2.73 \\
& $\left(2.34^{+0.10}_{-0.09}\right) $ & (2.17 -- 2.54) & (2.08 --
2.64) \\  \hline 
$\sin^2 \theta_{12}$  & $0.316_{-0.016}^{+0.016}$ & 0.29 -- 0.35 & 0.27
-- 0.37 \\ \hline 
$\sin^2 \theta_{23}$ & $0.51_{-0.06}^{+0.06}$ & 0.41 -- 0.61 & 0.39 --
0.64 \\ 
& $\left(0.52^{+0.06}_{-0.06}\right) $ & (0.42 -- 0.61)  & (0.39 -- 0.64)  \\ \hline 
$\sin^2 \theta_{13}$ & $0.017^{+0.007}_{-0.009} $ & $\le 0.031$ & $\le
0.040$ \\
& $\left(0.020^{+0.008}_{-0.009}\right) $ & $(\le 0.036)$  & $(\le 0.044)$  \\
 \botrule
\end{tabular}}
\end{table}

\begin{figure}[t]
\centerline{\psfig{file=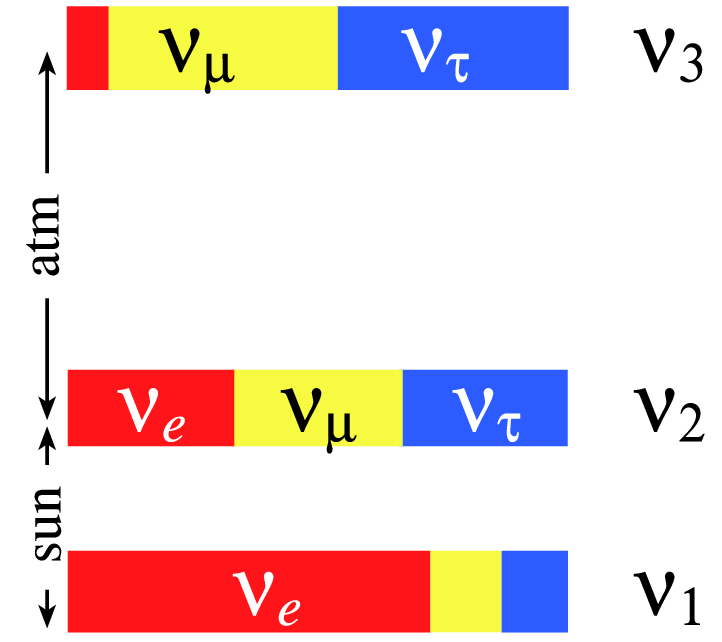,width=4cm,height=4cm}\hspace{2cm}
\psfig{file=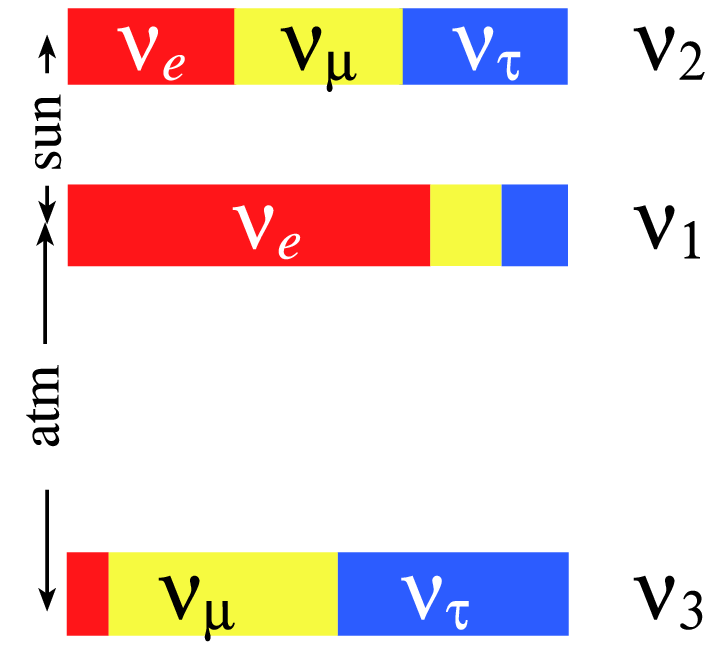,width=4cm,height=4cm}}
\vspace*{8pt}
\caption{\label{fig:NHIH}Normal (left) vs.~inverted (right) mass ordering. The red
  area denotes the electron content $|U_{ei}|^2$ in the mass state
  $\nu_i$. Accordingly, the yellow and blue areas denote the muon and
  tau contents. Taken from\protect\cite{StrumiaVissani}.}
\end{figure}

A recent review on the details of current and future
determinations of the parameters can be found in
Ref.\cite{concha_review}. An extensive program to improve the
precision on $\theta_{13}$, $\theta_{23}$ and $\Delta m^2_{31}$
(including its sign) has been launched, while improvement in the
precision of $\theta_{12}$ and 
$\Delta m^2_{21}$ is not on top of the neutrino community's agenda,
mostly because they are currently the best-known parameters. As we
will see below, precision determination of the solar parameters may be required
if the inverted mass ordering is to be tested with \obb. 
The unknown sign of the atmospheric mass-squared difference defines
the mass ordering (see Fig.~\ref{fig:NHIH}): normal for $\dma>0$, inverted for $\dma<0$. 
The two larger masses for each ordering are given in terms of 
the smallest mass and the mass squared differences as 
\be
\label{eq:masses}
\bad
\text{normal:}  &  m_2 = \sqrt{m_1^{2}+\dms} ~,~~  m_3 =
\sqrt{m_1^{2}+\dma} \, ,\\
\text{inverted: }~&  m_2 = \sqrt{m_3^{2}+\dms+\dma} ~;~~m_1 =
\sqrt{m_3^{2} + \dma}  \, .
\ea
\ee
Note that the oscillation data and the possible mass spectra and
orderings are independent on whether neutrinos are Dirac or Majorana
particles. 
The two possible mass orderings are shown in Fig.~\ref{fig:NHIH}. 
Of special interest are the following three extreme cases:  
\be
\baz  
\mbox{ normal hierarchy~(NH): }  &  
   m_3 \simeq \sqrt{\dma} \gg m_{2} \simeq \sqrt{\dms} \gg m_1\,,\\[0.3cm]
\mbox{ inverted hierarchy~(IH): } &  
 m_2 \simeq m_1 \simeq \sqrt{\dma} \gg m_{3} \, ,\\[0.3cm]
\mbox{ quasi-degeneracy~(QD): } &  
  m_0^2 \equiv m_1^2 \simeq m_2^2 \simeq m_3^2  \gg \dma \,.
\label{eq:mass}
\ea
\ee 
As can be seen from Fig.~\ref{fig:osc} and Table \ref{tab:nudata}, 
the current data is well described by so-called tri-bimaximal mixing\cite{tbm}, 
corresponding to 
$\sin^2 \theta_{13} = 0 \times \cos^2 \theta_{13} $, 
$\sin^2 \theta_{12} = \frac 12 \times \cos^2 \theta_{12} $ and 
$\sin^2 \theta_{23} = 1 \times \cos^2 \theta_{23} $: 
\be \label{eq:tbm}
U = \left(\bad
\sqrt{\frac 23} & \sqrt{\frac 13} & 0 \\
-\sqrt{\frac 16} & \sqrt{\frac 13} & \sqrt{\frac 12} \\
\sqrt{\frac 16} & -\sqrt{\frac 13} & \sqrt{\frac 12} 
\ea \right) . 
\ee
The application of  
flavor symmetries to the fermion sector, in order to obtain this and
other possible mixing schemes is a very active field of
research. For references and an overview of flavor symmetry models,
see\cite{flavsym}.\\  

A longstanding issue in oscillation physics is the indication of the
presence of sterile neutrinos.  The LSND experiment\cite{lsnd} found
evidence for $\overline{\nu}_\mu \to \overline{\nu}_e$ 
transitions which, when interpreted in terms of oscillations, are 
described by a $\Delta m^2 \sim $ eV$^2$ and small mixing $\ls 0.1$. 
These values survive even when combined\cite{lsnd+karmen,MaltoniSchwetz,carlo} with the negative
results from the KARMEN experiment\cite{karmen}. 
This mass scale cannot be compatible with solar and atmospheric
oscillation and hence a fourth, sterile neutrino needs to be
introduced. So-called 1+3 (3+1) scenarios would then be realized, in
which one sterile neutrino is heavier (lighter) than the three active
ones, separated by a mass gap of order eV. The MiniBooNE experiment was
designed to test the LSND scale with different $L$ and $E$, but very
similar $L/E$. The results\cite{MB} could not rule out the LSND
parameters, and are also compatible with the presence of 2 sterile
neutrinos\cite{2st}. In fact, the difference between MiniBooNE's
neutrino and anti-neutrino results can be explained by two additional
eV-like $\Delta m^2$ plus CP violation. Here one could envisage 2+3 or 3+2 scenarios, in
which 2 sterile neutrinos lie above or below the three active ones, 
or 1+3+1 scenarios\cite{131}, in which one sterile neutrino is heavier than the
three active ones and the other sterile neutrino is lighter. Recently,
reactor neutrino fluxes have been re-evaluated and an underestimation of
3\% with respect to previous results has been
found\cite{thierry}. The null results of previous very short-baseline reactor
experiments can now be interpreted as in fact being a deficit of
neutrinos, which again is compatible with oscillations corresponding
to $\Delta m^2 \sim $ eV$^2$
and small mixing $\ls 0.1$. A recent analysis of short-baseline
neutrino oscillation data in a framework with one or two sterile
neutrinos can be found in Ref.\cite{Kopp:2011qd}. The global fit
improves considerably when the existence of two sterile neutrinos is
assumed. \\

We will discuss the situation on neutrino mass from now on. 
Neutrino mass can be measured in three and complementary different
ways\footnote{Alternatives such as time-of-flight measurements of supernova
neutrinos cannot give comparable limits. Other
ideas\cite{other_numass} are presumably not realizable.}: \\

{\bf 1) Kurie-plot experiments,}\newline
in which the non-zero
neutrino mass influences the energy distribution of electrons in beta
decays close to the kinematical endpoint of the spectrum. As long as
the energy resolution is larger than the mass splitting, the spectrum 
is described by a function 
\be
(E_e - Q) \, \sqrt{(E_e - Q)^2 - m_\beta^2} \, , 
\ee
where the observable neutrino mass parameter is 
\be \label{eq:mbeta}
m_\beta \equiv \sqrt{\sum |U_{ei}|^2 m_i^2 } \, . 
\ee
The current limit to this quantity from spectrometer approaches
is 2.3 eV at 95\% C.L., obtained from the Mainz\cite{mainz} and
Troitsk\cite{troitsk} collaborations. The KATRIN experiment\cite{katrin,fre_bas} has a design sensitivity of 
$m_\beta = 0.2$ eV at (90\% C.L.) and a discovery potential of
$m_\beta = 0.35$ eV with $5\sigma$ significance. It represents the
ultimate spectrometer experiment for neutrino mass, in which an
external source of beta emitters (tritium) is used. Further
improvement of the limits must e.g.~come from calorimeter approaches, 
where the source is identical to the detector. The MARE\cite{MARE}
proposal will use $^{187}$Re modular crystal bolometers. A history of neutrino mass limits
from beta decays and reviews of upcoming experiments can be found
in\cite{mbetaRev}. A different
Ansatz called Project 8 aims to detect the coherent cyclotron radiation
emitted by mildly relativistic electrons (like those in tritium 
decay) in a magnetic field. The relativistic shift of the
cyclotron frequency allows to extract the electron energy from the
emitted radiation\cite{project8}. In principle, MARE and Project 8 can
reach limits of 0.1 eV. 
Investigation of beta spectra is usually considered to be the
least model-dependent Ansatz to probe neutrino mass. For instance,
Refs.\cite{RHC_KATRIN} have shown that admixture of right-handed
currents can be not more than a 10\% effect in KATRIN's determination
of $m_\beta$;\\  

{\bf 2) Cosmological and astrophysical observations}\newline 
are sensitive to
neutrino mass, see\cite{steen_rev} for a review. In particular,
effects of neutrinos in cosmic 
structure formation are used to extract limits on neutrino masses. 
The quantity which is
constrained by such efforts is 
\be \label{eq:msum}
\Sigma = \sum m_i \, , 
\ee
familiar from the contribution of neutrinos to hot dark
matter\footnote{While one usually considers light neutrinos as 
a sub-leading part of dark matter, arguments in favor of neutrino hot dark
matter are given in\cite{TMN}.},
$\Omega_\nu \, h^2 = \Sigma/(94.57 \, \rm eV)$. Finite neutrino masses
suppress the matter power spectrum on scales smaller than the
free-streaming scale $k_{\rm FS} \simeq 0.8 \, h \, m_i/{\rm eV}$
Mpc$^{-1}$. However, neutrino mass is highly degenerate with other
cosmological parameters, for instance\cite{steen_omega} with the dark energy equation of
state parameter $\omega$, so that one needs to break the degeneracies
with different and complementary data sets.  
Besides cosmic microwave background (CMB) experiments, one
can use the Hubble constant (H0) measurements, high-redshift Type-I
supernovae (SN) results, information from large scale structure (LSS) surveys, the LSS matter
power spectrum (LSSPS) and baryon acoustic oscillations (BAO). The
impact on the neutrino mass limit is shown in Table \ref{tab:cosmo},
taken from\cite{concha_cosmo}, in which a fit to a cosmological model allowing 
for neutrino mass, non-vanishing curvature, dark energy with equation of state $\omega \neq -1$, 
and the presence of new particle physics whose effect on the
present cosmological observations can be parameterized in terms of additional relativistic
degrees of freedom $\Delta N_{\rm rel}$, has been performed. As can be
seen, depending on the data sets, 
the limit on $\Sigma$ varies by a factor of 3. Future cosmological probes will 
add additional information, a summary of expectations for this is
shown in Table \ref{tab:cosmo_fut}.

\begin{table}
\tbl{\label{tab:cosmo}95\% C.L.~upper bound on the sum of the neutrino masses 
from different cosmological analyses. Taken from\protect\cite{concha_cosmo}.}
{\begin{tabular}{ccc}  \toprule
Model & Observables & $\sum m_i$ [eV] \\ 
\hline
$
o\omega{\rm CDM} 
+\Delta N_{\rm rel}+m_\nu
$ 
& {\footnotesize CMB+HO+SN+BAO} & 
$\leq 1.5$ \\
$
o\omega{\rm CDM} 
+\Delta N_{\rm rel}+m_\nu
$ 
& {\footnotesize CMB+HO+SN+LSSPS} & 
$\leq 0.76$ \\
$\Lambda{\rm CDM} +m_\nu$
& {\footnotesize CMB+H0+SN+BAO} & 
$\leq 0.61$ 
\\
$\Lambda{\rm CDM} +m_\nu$
& {\footnotesize CMB+H0+SN+LSSPS} & 
$\leq 0.36$  \\
$\Lambda{\rm CDM} +m_\nu$
& {\footnotesize CMB (+SN)} & 
$\leq 1.2$ \\
$\Lambda{\rm CDM} +m_\nu$
& {\footnotesize CMB+BAO} & 
$\leq 0.75$  \\
$\Lambda{\rm CDM} +m_\nu$
& {\footnotesize CMB+LSSPS} & 
$\leq 0.55$ \\
$\Lambda{\rm CDM} +m_\nu$
& {\footnotesize CMB+H0} & 
$\leq 0.45$ \\
\botrule
\end{tabular} } 
\end{table}

\begin{table}
\tbl{\label{tab:cosmo_fut}Future probes of neutrino mass, with
their projected sensitivity.
Sensitivity in the short term means within the next few years, 
while long term means by the end of the decade. Taken from\protect\cite{steen_rev}.}
{\begin{tabular}{ccc}
\toprule 
Probe &  Potential sensitivity [eV] & Potential sensitivity [eV] \\
& (short term) & (long term) \\ 
\hline
CMB & 0.4--0.6 & 0.4  \\
CMB with lensing & 0.1--0.15 & 0.04 \\
CMB + Galaxy Distribution & 0.2 & 0.05--0.1 \\
CMB + Lensing of Galaxies & 0.1 & 0.03--0.04 \\
CMB + Lyman-$\alpha$ & 0.1--0.2 & Unknown \\
CMB + Galaxy Clusters & -- & 0.05 \\
CMB + 21 cm & -- & 0.0003--0.1 \\
\botrule 
\end{tabular} } 
\end{table}
\begin{figure}[t]
\centerline{\psfig{file=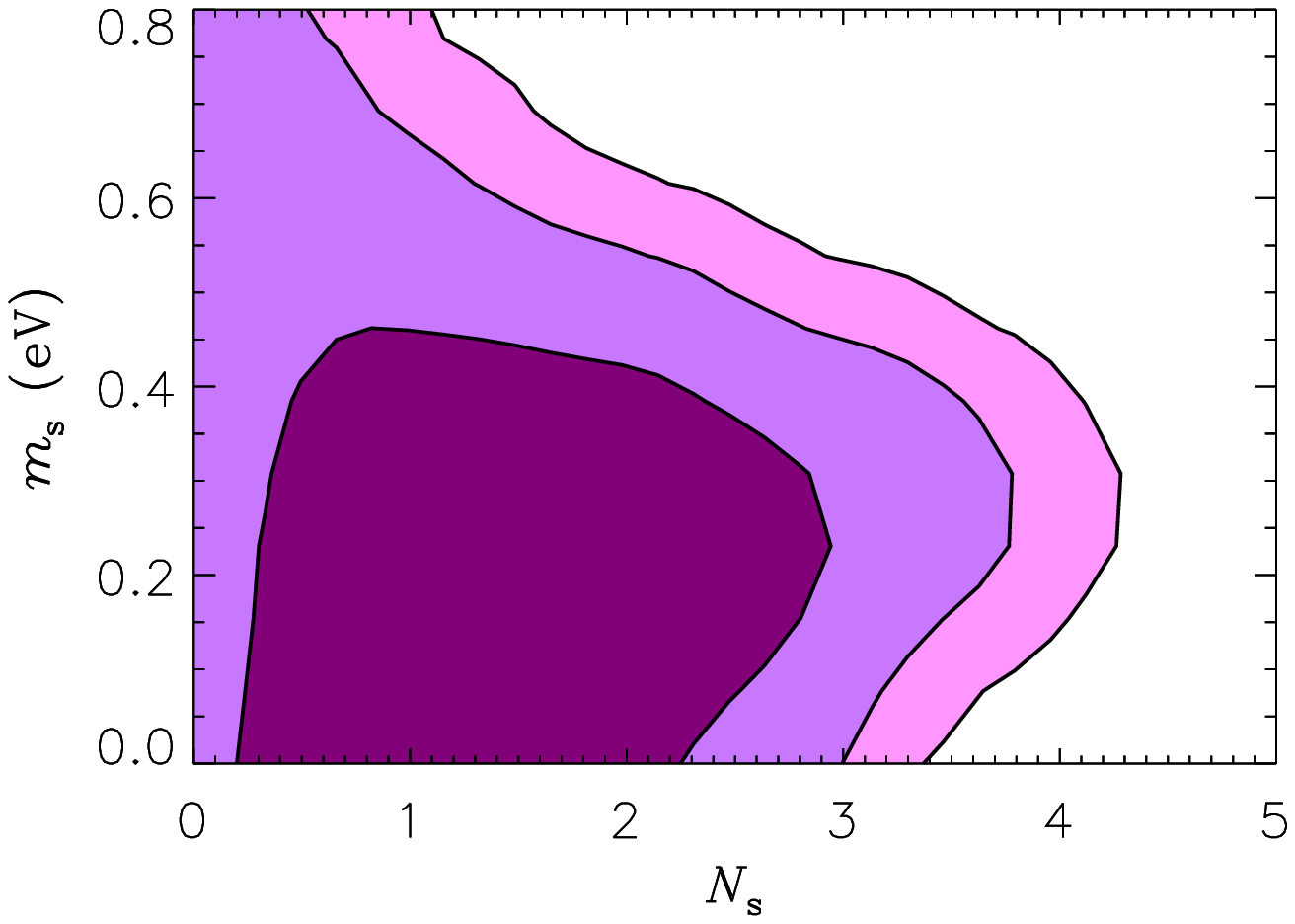,width=5cm,height=4cm}\hspace{.4cm}\psfig{file=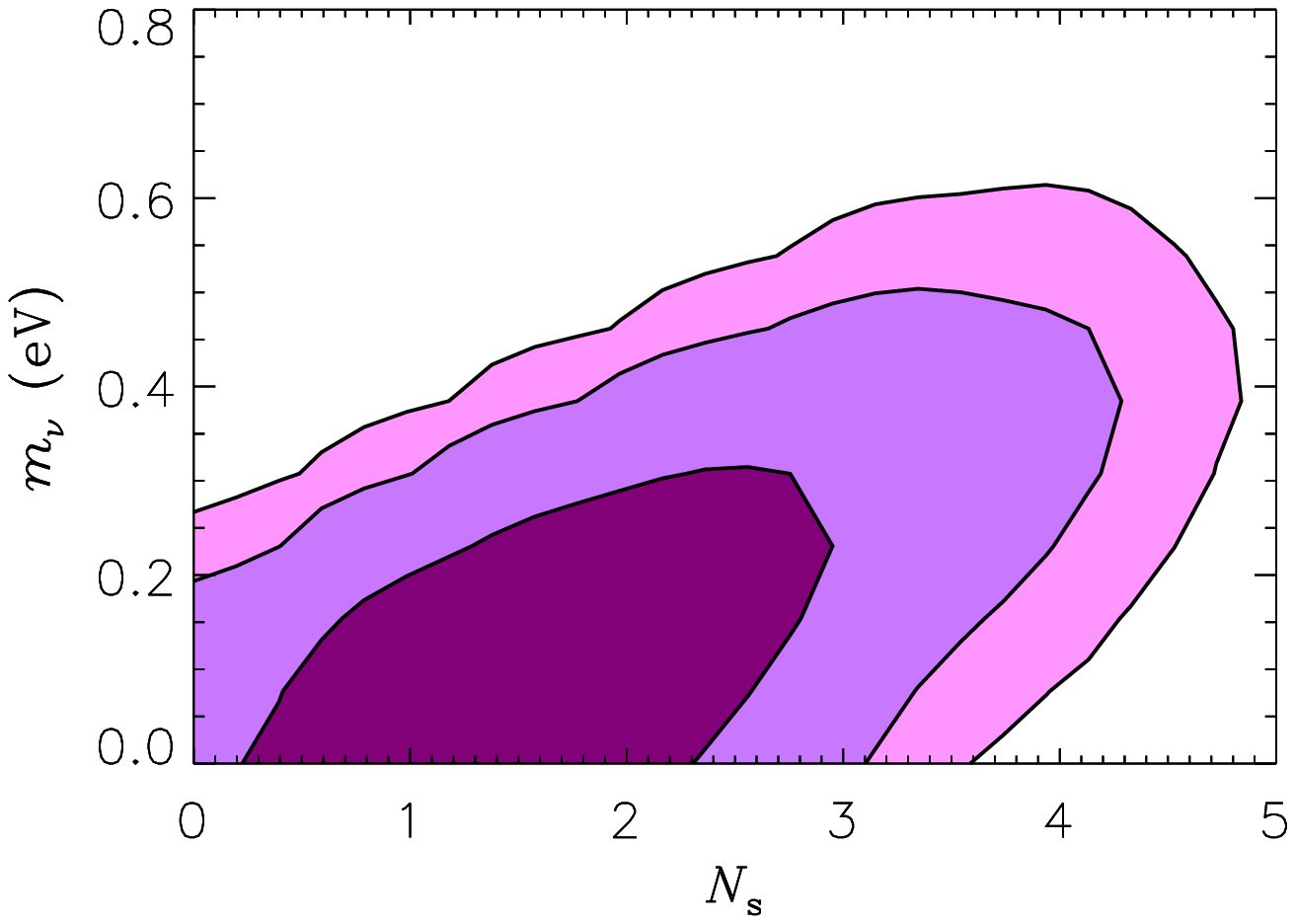,width=5cm,height=4cm}}
\vspace*{8pt}
\caption{
68\%, 95\% and 99\% C.L.~regions for
  the neutrino mass and thermally excited number of degrees of freedom
  $N_{\rm s}$. The left plot is the $N_{\rm s} + 3$ scheme, in which ordinary
  neutrinos have $m_\nu=0$, while sterile states have a common mass
  scale $m_{\rm s}$, hence $\Sigma \simeq N_{\rm s} \, m_{\rm s}$. The
right plot is for the $3 + N_{\rm s}$ scheme, where
  the sterile states are taken to be massless $m_{\rm s}=0$, and 3.046
  species of ordinary neutrinos have a common mass
  $m_\nu$, hence $\Sigma \simeq 3  \, m_{\nu}$. 
Taken from\protect\cite{cosmo_sterile?}. \label{fig:contours}}
\end{figure}

At the present stage it is worth noting that precision cosmology and
Big Bang nucleosynthesis mildly favor extra radiation in the Universe
beyond photons and ordinary neutrinos. While this
could be any relativistic degree of freedom, the interpretation in
terms of additional sterile neutrino species is 
straightforward (recall the discussion on sterile neutrinos from
above). Fit results very well compatible with
more radiation than the SM value have been found e.g.~by the 
WMAP collaboration\cite{Komatsu:2010fb} or in\cite{Hamann:2010pw}. This is supported by the recently
reported higher $^4$He abundance\cite{bbn_spec}, which in the
framework of Big Bang Nucleosynthesis can be accommodated by additional relativistic degrees
of freedom, as this leads to earlier freeze-out of the weak reactions,
resulting in a higher neutron-to-proton ratio. In
Fig.~\ref{fig:contours} the result of a recent
fit\cite{cosmo_sterile?} to cosmological data is shown, in which two
situations are analyzed: massless active neutrinos plus $N_{\rm s}$
massive sterile states ($N_s + 3$ scheme); and $N_{\rm s}$ massless sterile states plus 3
massive active states ($3 + N_s $). The Planck satellite, with a projected
sensitivity of $\pm 0.2 $ to the number of extra degrees of freedom, 
will be decisive in order to test this presence of additional radiation. It is
rather interesting that hints for the presence of sterile neutrinos
are given by fundamentally different probes: neutrino oscillations,
Big Bang Nucleosynthesis and CMB + LSS.

Cosmological mass limits can be considered robust 
with respect to reasonable modifications of the $\Lambda$CDM
model\cite{Seattle}, in particular if different and complementary data
sets are applied. However, several non-standard cosmologies exist for which no detailed
study on the effect on the $\Sigma$ bound has been performed yet, 
for instance coupled dark energy scenarios.  
Nevertheless, it is fair to say that neutrino
masses heavier than $\Sigma \simeq 2$ eV or so would be rather
surprising, and correspond to very unusual scenarios. Note however that 
any information about the neutrino mass can be obtained only by way 
of statistical inference from the observational data after a parametric
model has been chosen as the basis for the
analysis\cite{Seattle}. This is a difference to the investigation of
energy spectra in single or double beta decay experiments; \\

{\bf 3) Neutrino-less Double Beta Decay}\newline 
This possibility to test neutrino mass will be dealt with in Section
\ref{sec:meff_mass} in some detail. In the ideal case, results from two or all three
approaches to neutrino mass are present, and we will discuss this
interesting case too. Neutrino
mass limits from \onbb~need to assume that neutrinos are Majorana
particles, and that no mechanism other than light neutrino
exchange is responsible for the process. Let us note here that from 
\obb~limits one can extract two different ``masses''. First, we can
extract the physical masses, i.e.~the eigenvalues of the mass
matrix. These quantities are the ones tested in the other approaches. 
Second, \obb~tests directly the quantity $(m_\nu)_{ee}$, i.e.~the $ee$ element of
the neutrino mass matrix in the charged lepton basis: 
\be
\sqrt{\frac{1}{T^{0\nu}_{1/2}}} \propto |(m_\nu)_{ee}|~\mbox{ with  }
(m_\nu)_{ee} = \frac{h_{ee} \, v^2}{\Lambda} \mbox{ in } 
{\cal L}_{\rm eff} = \frac 12 \frac{h_{\alpha\beta}}{\Lambda}
\overline{L_{\alpha}^c}  \, \tilde \Phi \, \tilde{\Phi}^T L_{\beta} 
  \, . 
\ee 
Thus, the decay width is directly proportional to the fundamental 
quantity which originates at the fundamental large (seesaw) scale, without any
diagonalization procedure dependent on known and unknown
parameters. Note that in order to extract neutrino mass limits from
\obb~one needs to assume the neutrinos are Majorana particles, and
that no other mechanism contributes. We will comment later on the
complementarity of the three neutrino mass observables.

\subsection{\label{sec:SI}Standard three neutrino picture and \obb}
In this Section we will summarize the standard analysis of
neutrino-less double beta decay with the standard three neutrino
framework. Several works have been devoted in the literature to
this\cite{Petcov:1993rk,KlapdorKleingrothaus:2000gr,Farzan:2001cj,Pascoli:2002xq,Pascoli:2005zb,Choubey:2005rq,Lindner:2005kr,Bilenky:2011tr},
an earlier review can be found in\cite{Petcov:2005yq}. 

\begin{figure}[t]
 \centerline{
\psfig{file=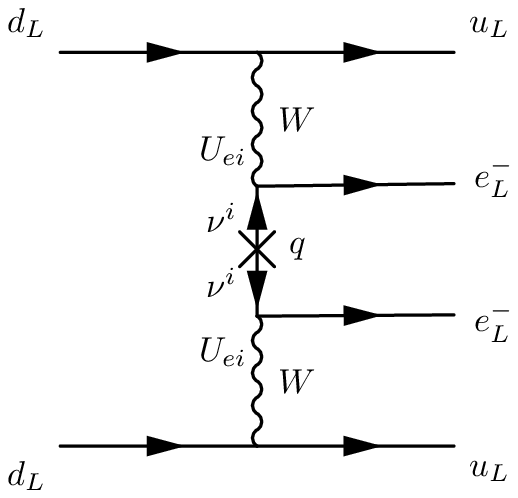,width=6cm,height=4cm}\quad
\psfig{file=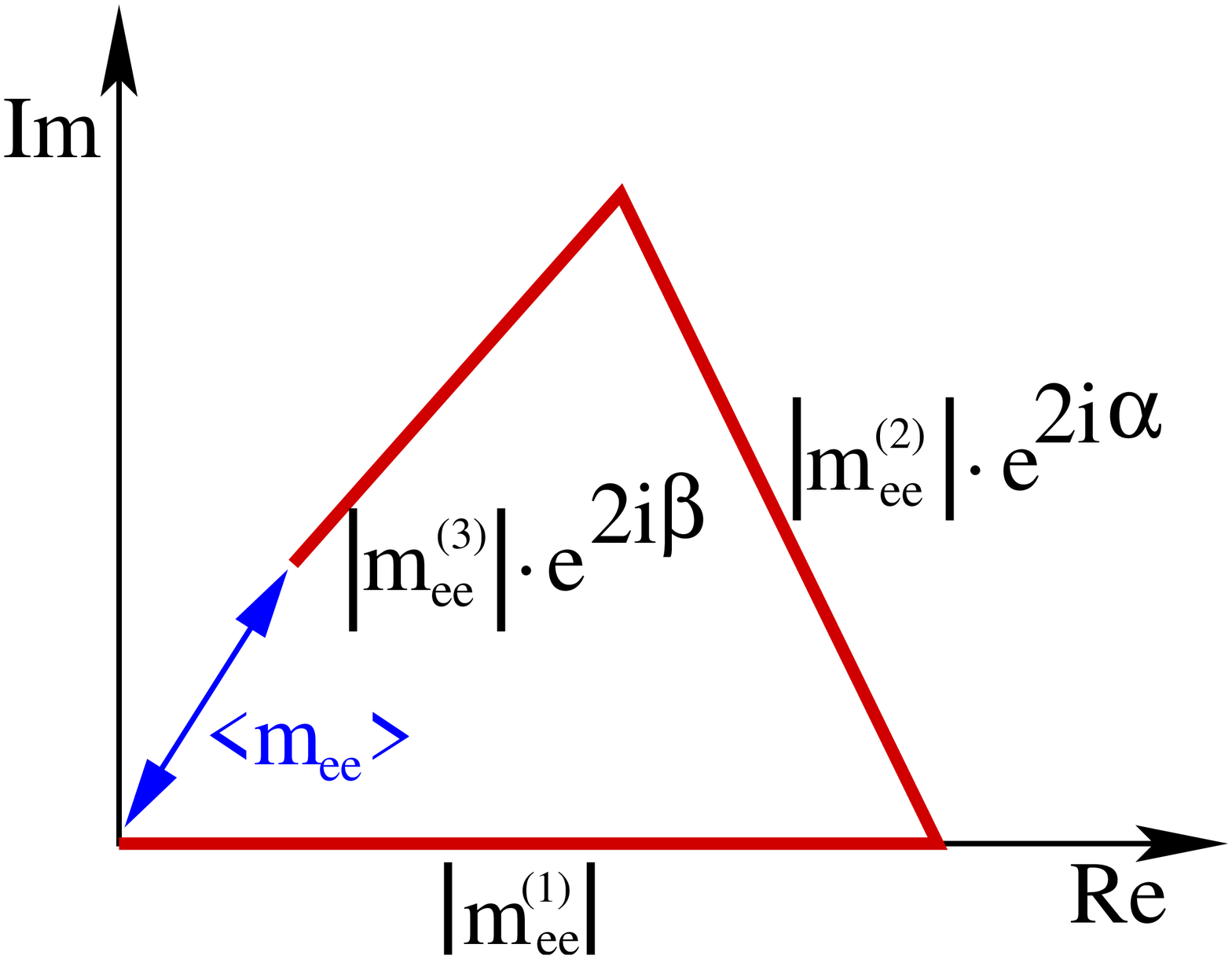,width=5cm}
}
\vspace*{8pt}
\caption{\label{fig:FD_mass_mech}Left: quark level Feynman diagram for the
standard interpretation of \onbb. Right: geometrical visualization of the effective
mass. }
\end{figure}

\begin{figure}[t]
\centerline{
\psfig{file=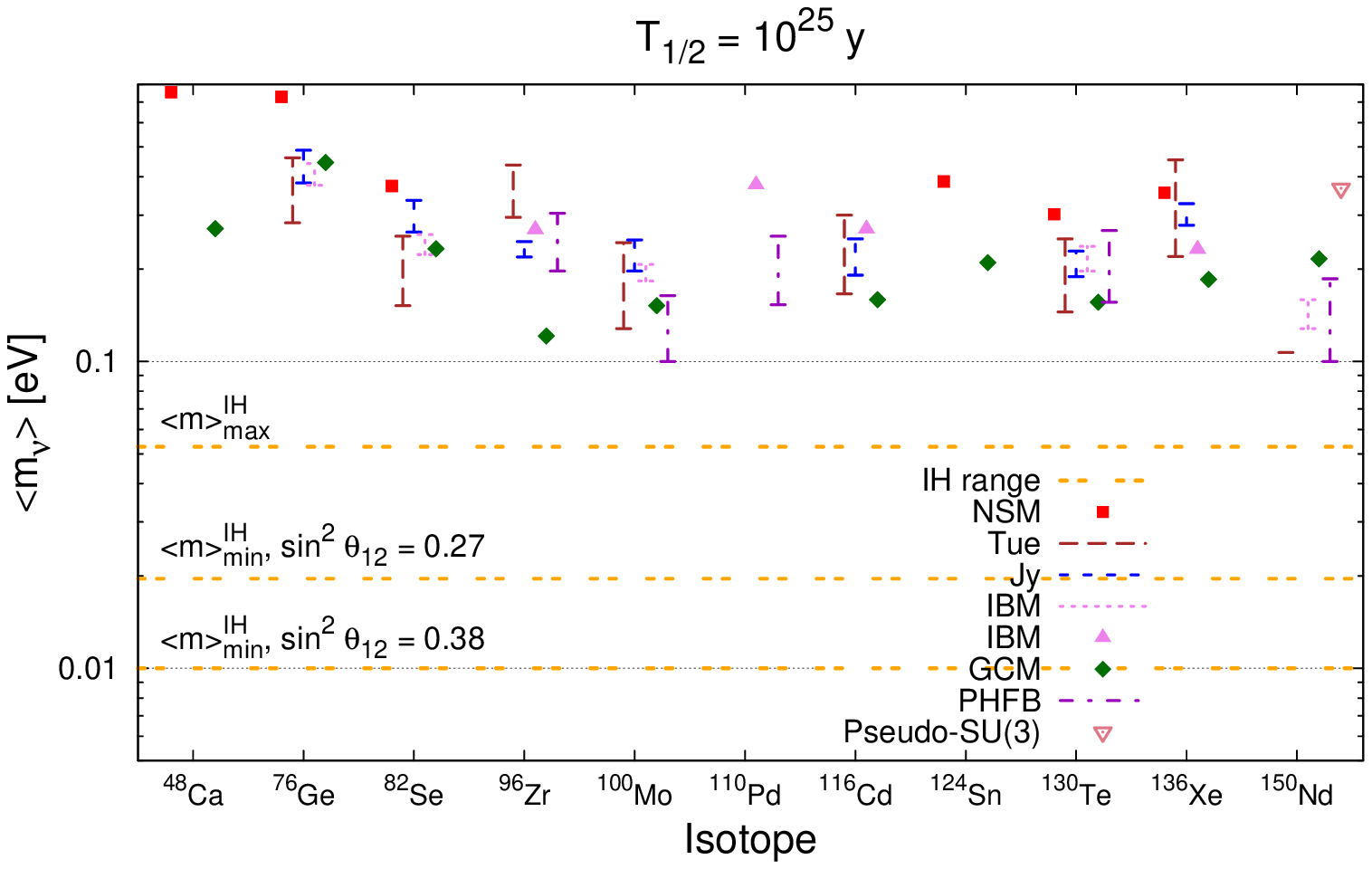,width=7cm,height=5cm} \quad
\psfig{file=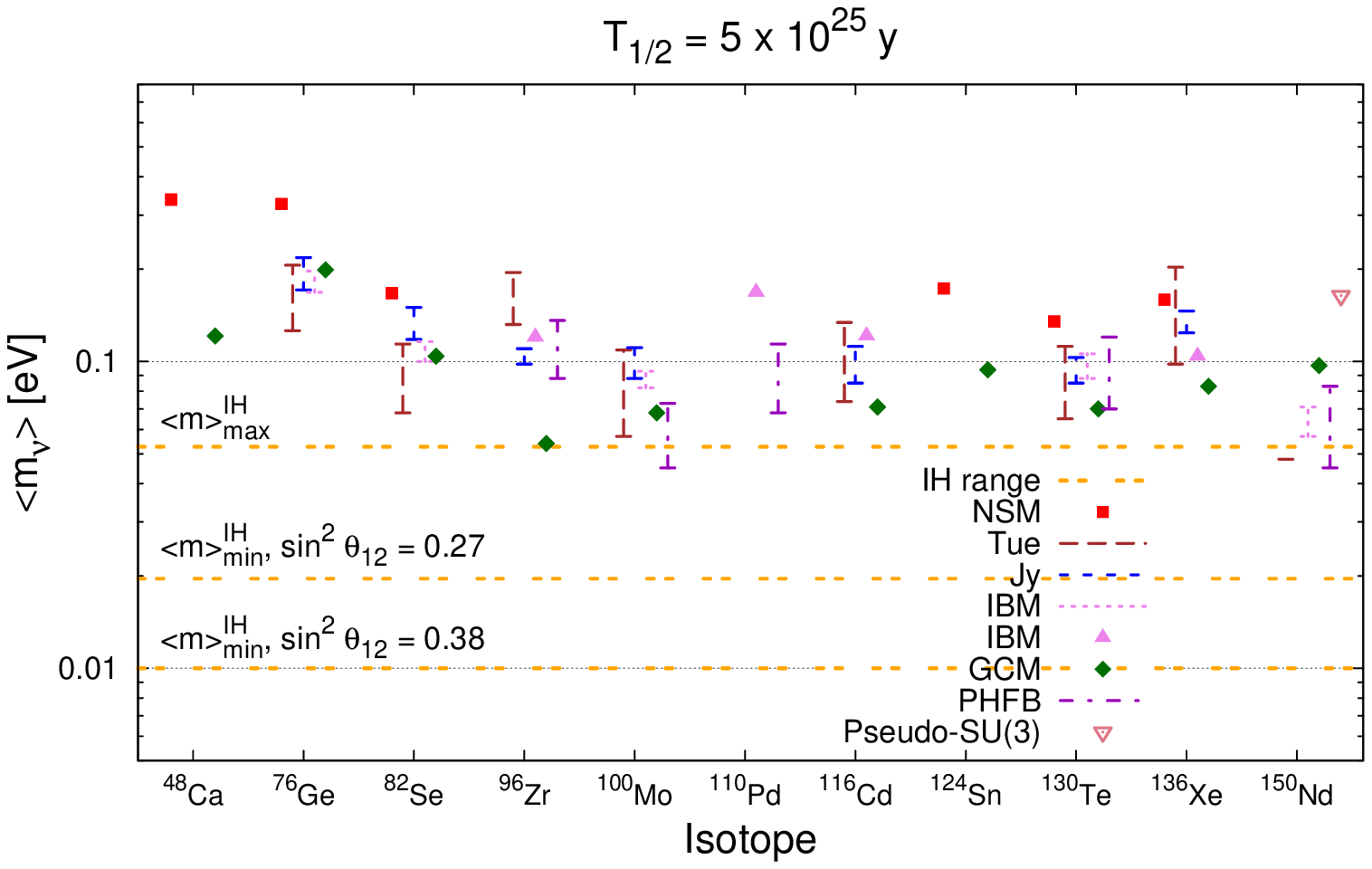,width=7cm,height=5cm} }

\centerline{
\psfig{file=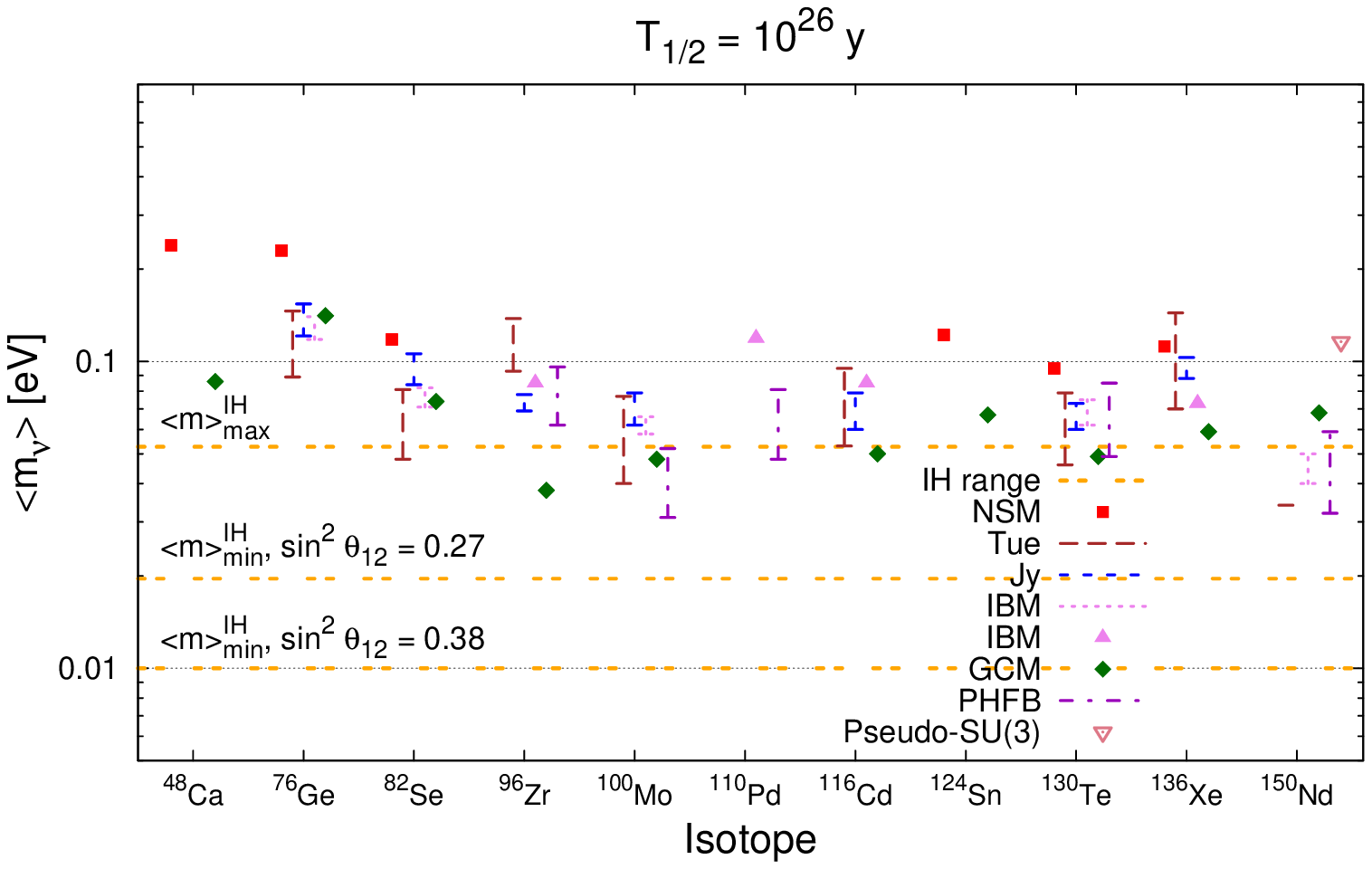,width=7cm,height=5cm}  \quad
\psfig{file=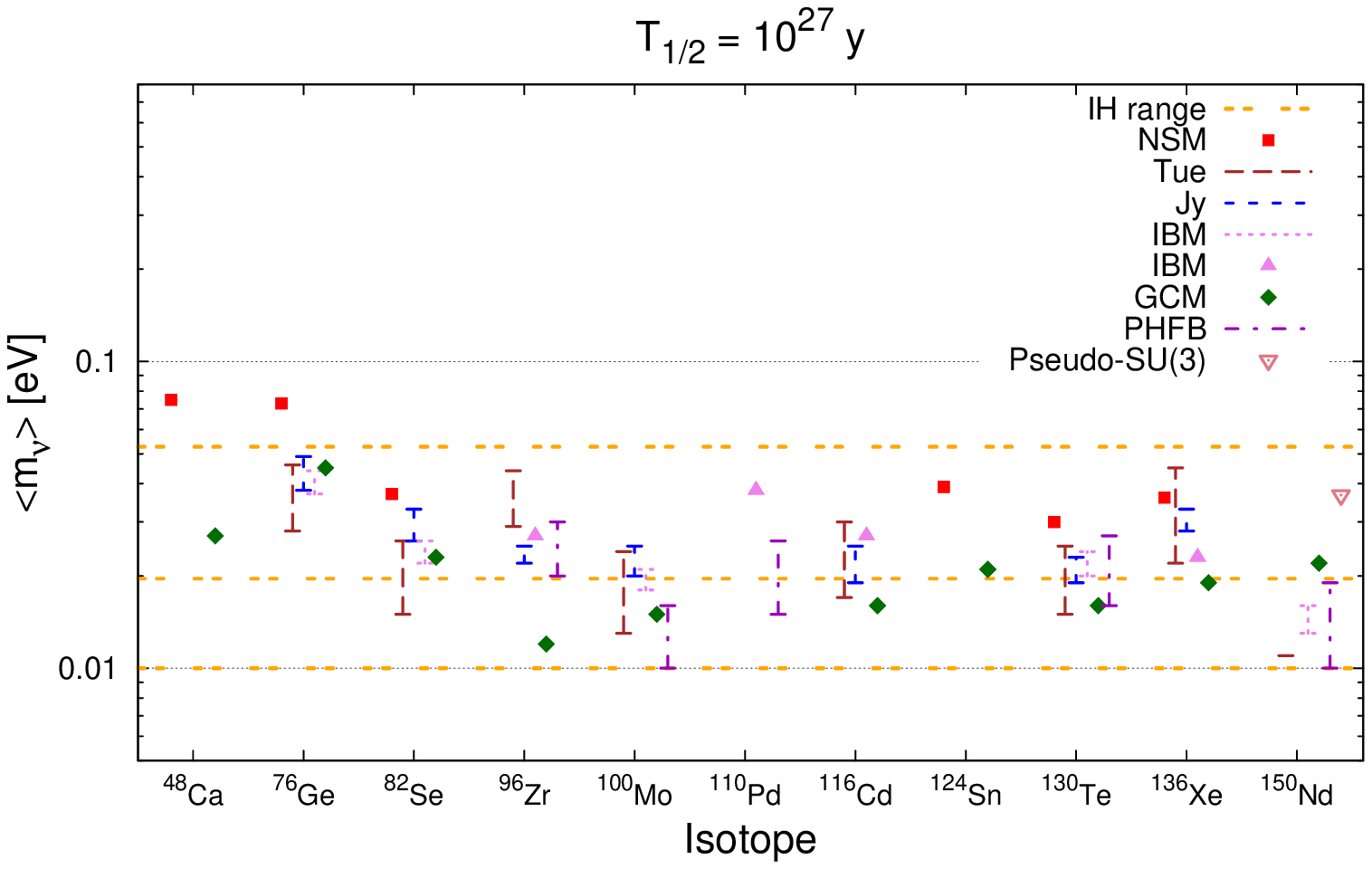,width=7cm,height=5cm} 
}
\vspace*{8pt}
\caption{\label{fig:fut_lim}Limits on the effective mass for different
half-life limits. The horizontal lines are the maximal and minimal
values of \meff~in the inverted mass ordering.}
\end{figure}

The Feynman diagram for \obb~on the quark level in this
interpretation is shown in Fig.~\ref{fig:FD_mass_mech}.  
Due to the typical structure of the process it
 is sometimes called ``lobster diagram''. The amplitude
of the process is for the $V - A$ interaction of the SM proportional to 
\bea \D  \label{eq:SA_calc}
\sum G_F^2 \, U_{ei}^2 \, \gamma_\mu \, 
\gamma_+ \frac{\slashed{q} + m_i}{q^2 - m_i^2} \, \gamma_\nu \, \gamma_-
= \sum G_F^2 \, U_{ei}^2 \frac{m_i}{q^2 - m_i^2} \gamma_\mu \,
\gamma_+ \, \gamma_\nu \\ \D 
\simeq  \sum G_F^2 \, U_{ei}^2 \frac{m_i}{q^2} \gamma_\mu \, \gamma_+ \,
\gamma_\nu \, , 
\eea
where $\gamma_{\pm} = \frac12 (1 \pm \gamma_5)$, $m_i$ is the neutrino
mass, $q \simeq 100$ MeV is the typical neutrino momentum, and $U_{ei}$ an
element of the first row of the PMNS matrix. 
The linear dependence on the neutrino mass is expected from the
requirement of a spin-flip, as the neutrino can be though of being emitted as a
right-handed state and absorbed as a left-handed state. In case the
interactions are not left-handed at one of the vertices, the linear
dependence on $m_i$ will be absent; we will consider these cases later in
Section \ref{sec:LR}. If both interactions are right-handed, the same
linear dependence on $m_i$ appears. 
Note that the amplitude is proportional to a coherent sum, which implies the possibility of
cancellations. 
The decay width is proportional to the square of the so-called effective mass 
\be \label{eq:meff}
\meff = \left| \sum U_{ei}^2 \, m_i \right| 
= \left| |m_{ee}^{(1)}| + |m_{ee}^{(2)}| \, e^{2i\alpha} + 
|m_{ee}^{(3)}| \, e^{2i\beta} \right|  , 
\ee
which is visualized in Fig.~\ref{fig:FD_mass_mech} as the sum of three 
complex vectors $m_{ee}^{(1,2,3)}$. If one cannot form a triangle 
with the $m_{ee}^{(1,2,3)}$, then the effective mass is non-zero. 
The Majorana phases $2\alpha$ and 
$2\beta$ correspond to the relative orientation of the three
vectors. The standard analysis of the effective mass is the geometry
of the three vectors expressed in terms of neutrino parameters. In the
standard parametrization of the PMNS matrix we have 
\begin{eqnarray}
|m_{ee}^{(1)}| &=& m_1 \, |U_{e1}|^{2} = m_1 \, c_{12}^{2} \, c_{13}^{2} \,,
\nonumber\\
|m_{ee}^{(2)}| &=& m_2 \, |U_{e2}|^{2} = m_2 \, s_{12}^{2} \, c_{13}^{2} \,,\\
|m_{ee}^{(3)}| &=& m_3 \, |U_{e3}|^{2} = m_3 \, s_{13}^{2} \, .\nonumber
\end{eqnarray}
The individual masses can, using Eq.~(\ref{eq:masses}), be expressed in
terms of the smallest mass and the mass-squared differences, whose
currently allowed ranges, as well as those of the mixing angles,
are given in Table \ref{tab:nudata}. From Table \ref{tab:limits} we can
read off the current limit on the effective mass: 
\be \label{eq:limit_meff}
\meff \ls 0.5 ~{\rm eV} \, . 
\ee
For later use we define
the standard amplitude for light Majorana neutrino exchange: 
\be\label{eq:am_SI}
{\cal A}_{\rm l} \propto G_F^2 \, \frac{\meff}{q^2} \simeq 7 \times
10^{-18} \left(\frac{\meff}{0.5~\rm eV} \right) 
\, {\rm GeV}^{-5} \, . 
\ee 
Fig.~\ref{fig:fut_lim} shows the future limits on the effective mass
for different isotopes and half-life limits (see also\cite{comp1,Bilenky:2011tr}). We have again used the
NME compilation from Table \ref{tab:nme}. 

The effective mass depends on 7 out of the 9 physical parameters of 
low energy neutrino physics (only $\theta_{23}$ and $\delta$ do not
appear), hence contains an enormous amount of
information. It is the only realistic observable in which the two
Majorana phases appear. For the other five quantities there will be
complementary information from oscillation experiments or other
experiments probing neutrino mass. 
It is also noteworthy that \meff~is the $ee$ element of the
neutrino mass matrix $m_\nu$, see Eq.~(\ref{eq:mnu}), which is a
fundamental object in the low energy Lagrangian. In terms of the
origin of neutrino mass, \meff~is $h_{ee} \, v^2/\Lambda$, see
Eq.~(\ref{eq:Leff}) and the realizations of $\Lambda$ in terms of
fundamental mass scales in Table \ref{tab:seesaws}. 

\begin{figure}[t]
\centerline{\psfig{file=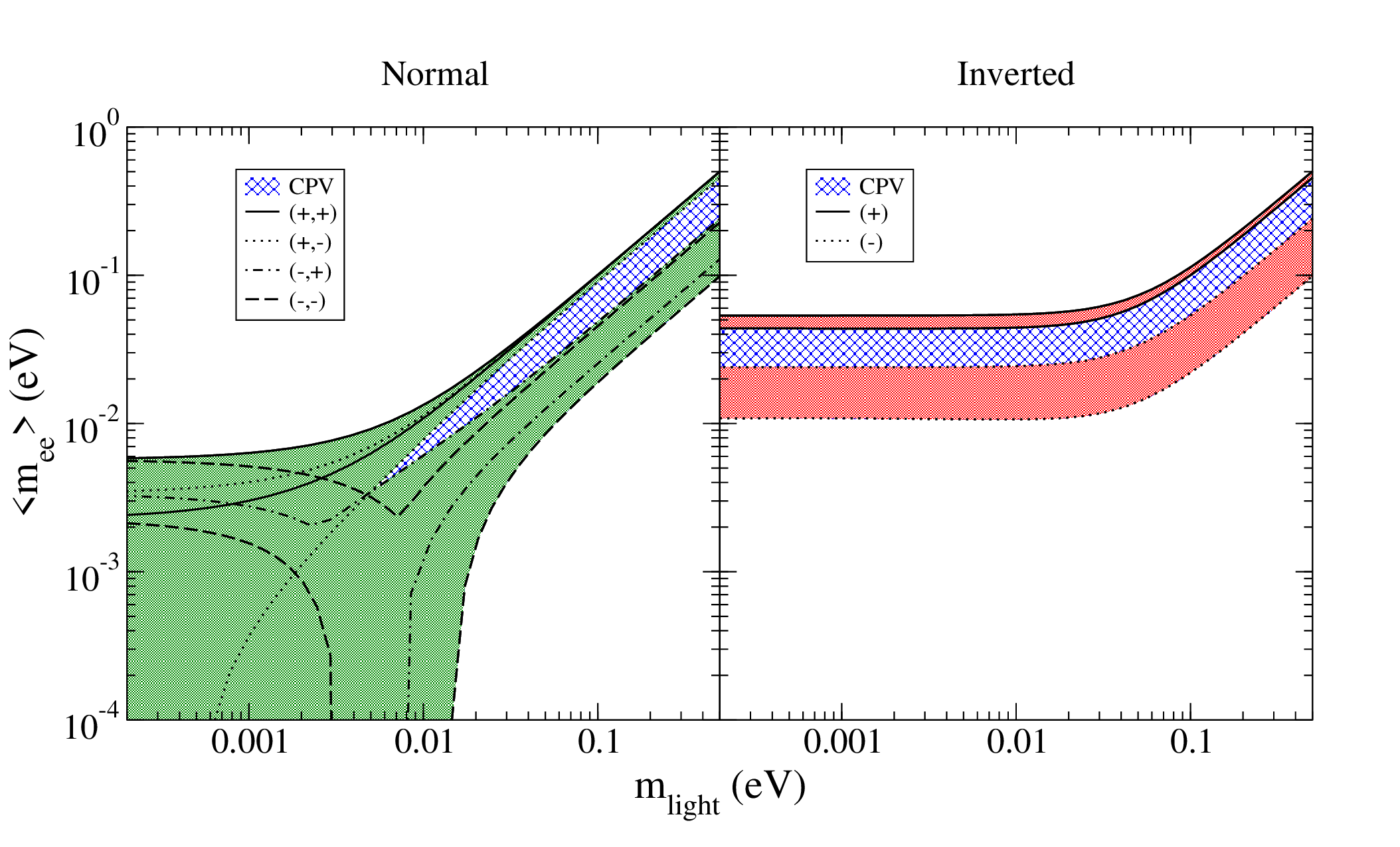,width=7cm,angle=270}}
\centerline{
\psfig{file=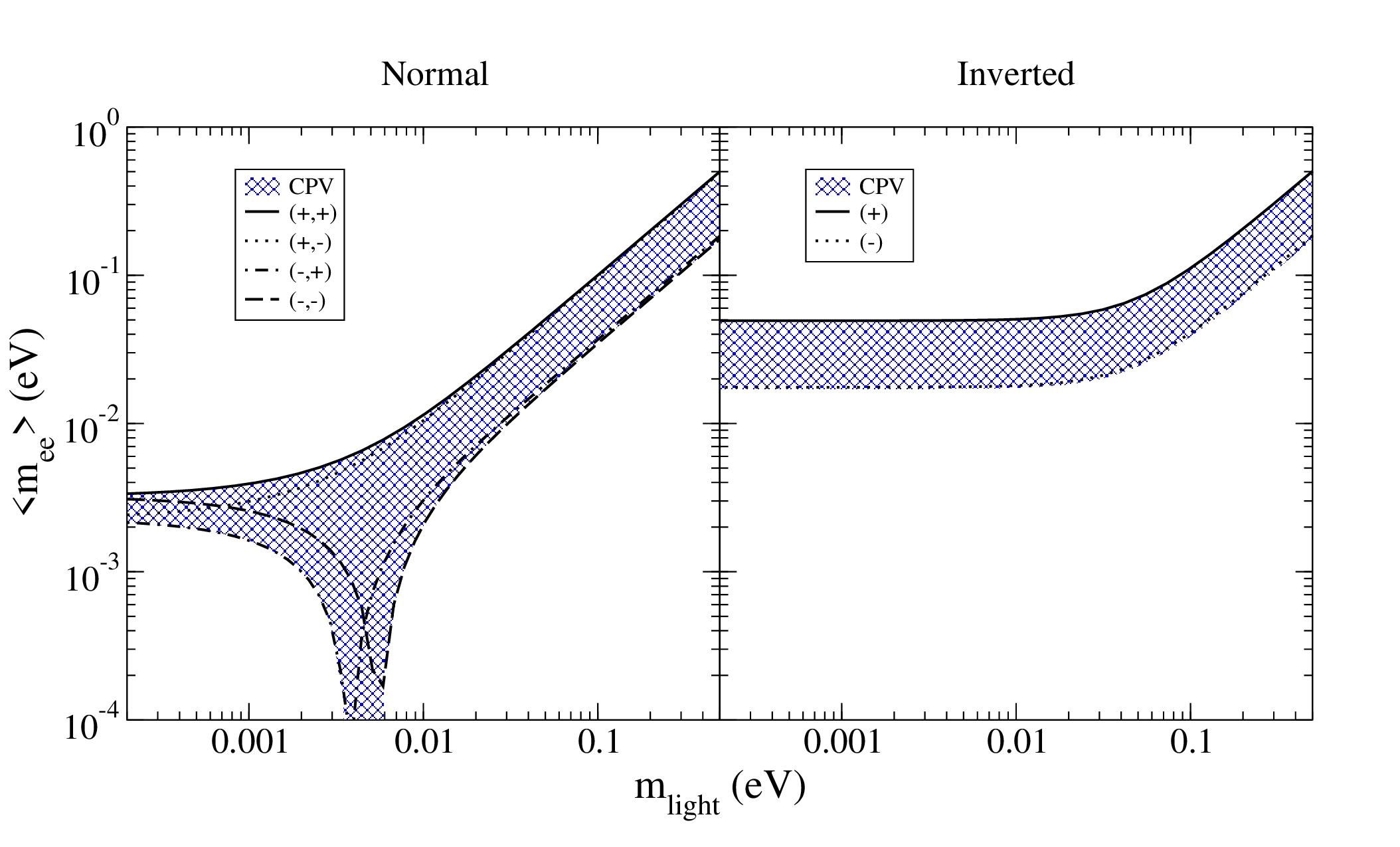,width=7cm,angle=270}
}
\vspace*{8pt}
\caption{\label{fig:meff_mass}Effective mass against the smallest
neutrino mass for the $3\sigma$ ranges (top) and best-fit values
(bottom) of the oscillation parameters. CP conserving and 
violating areas are indicated.}
\end{figure}

\begin{figure}[th]
\centerline{
\psfig{file=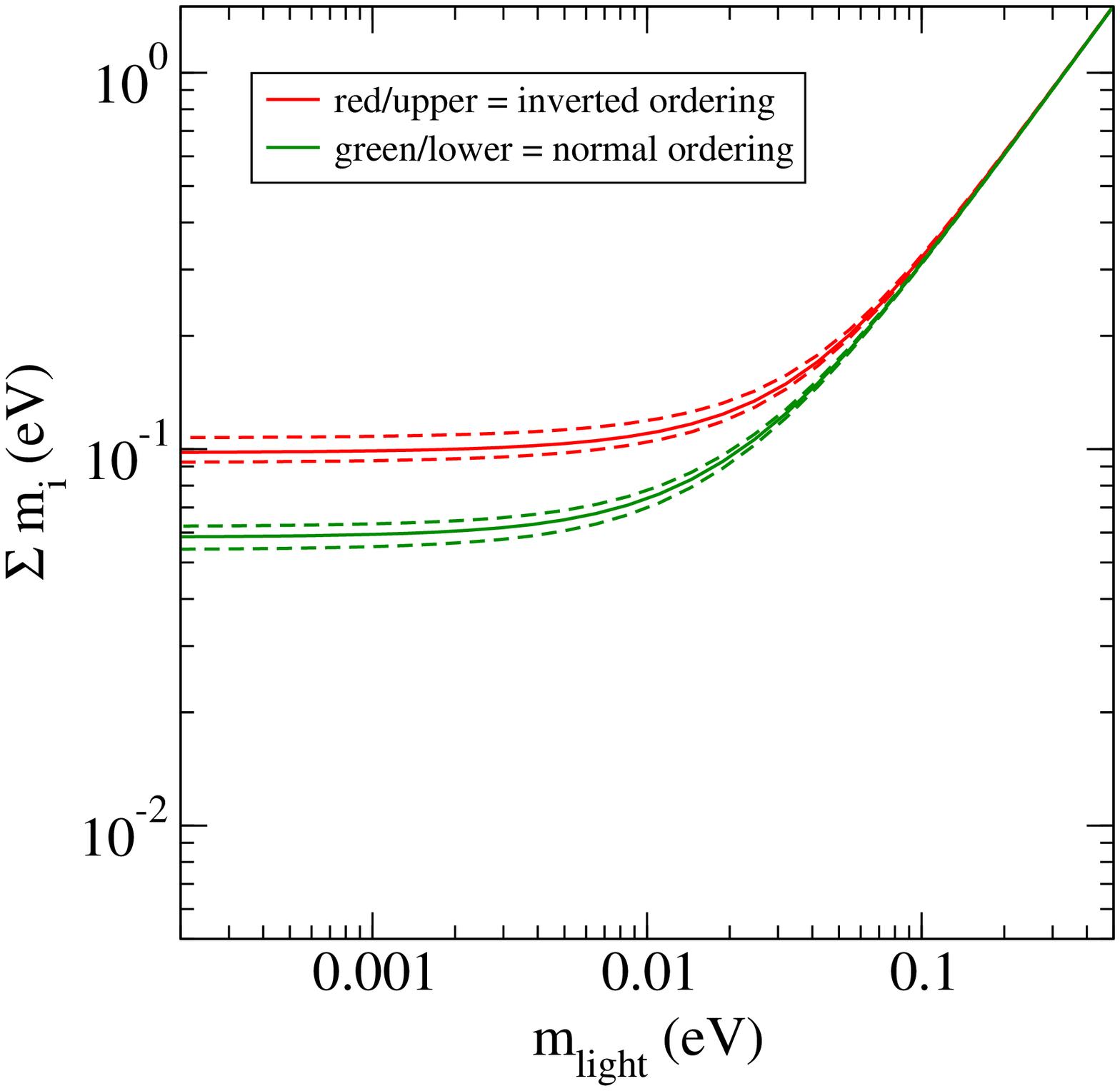,width=7cm,height=6cm}
\psfig{file=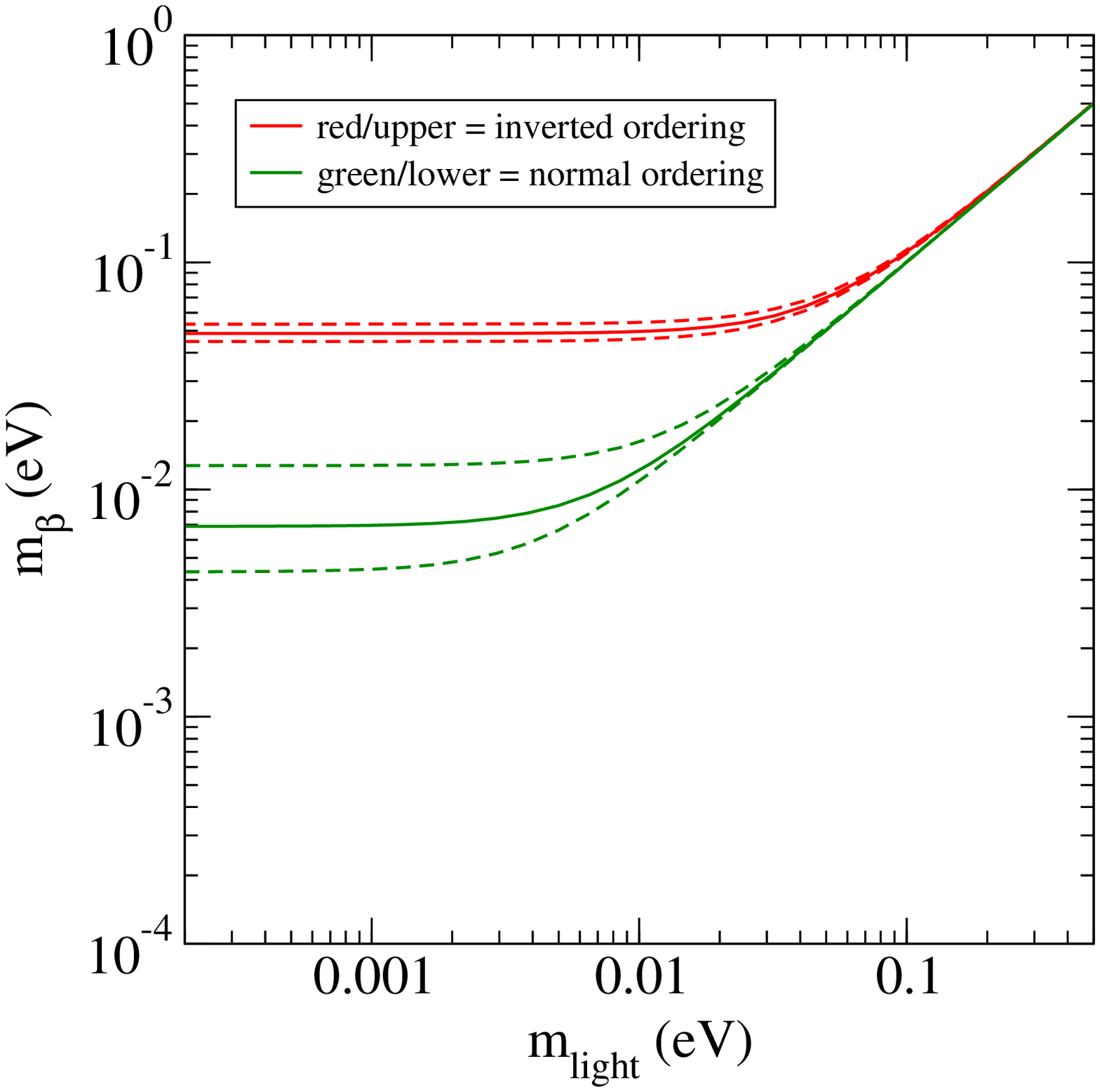,width=7cm,height=6cm}
}
\vspace*{8pt}
\caption{\label{fig:other_mass}Sum of masses $\Sigma$ and kinematic
neutrino mass $m_\beta$ against the smallest neutrino mass.}
\centerline{\psfig{file=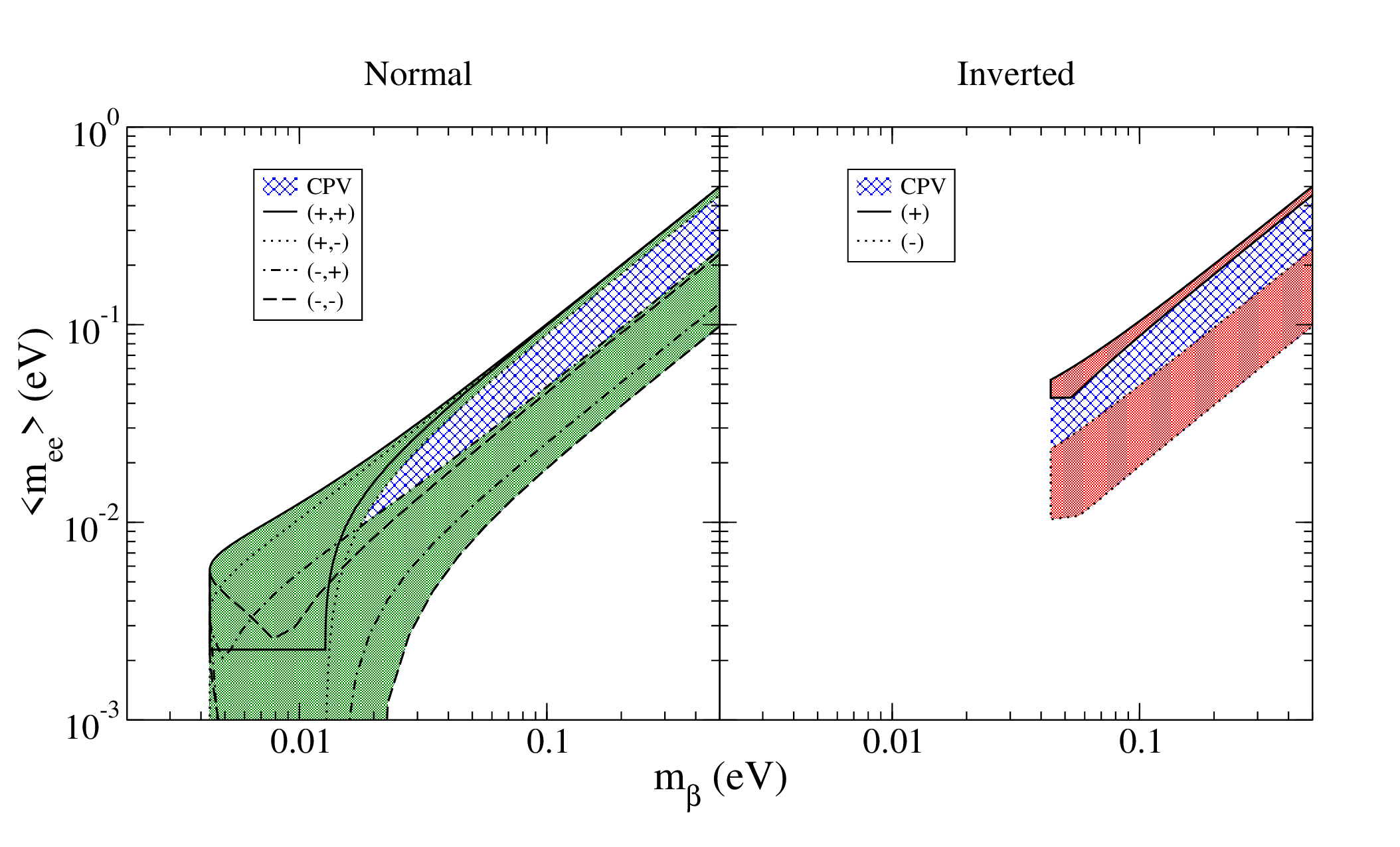,height=7cm,width=7cm,angle=270}
\psfig{file=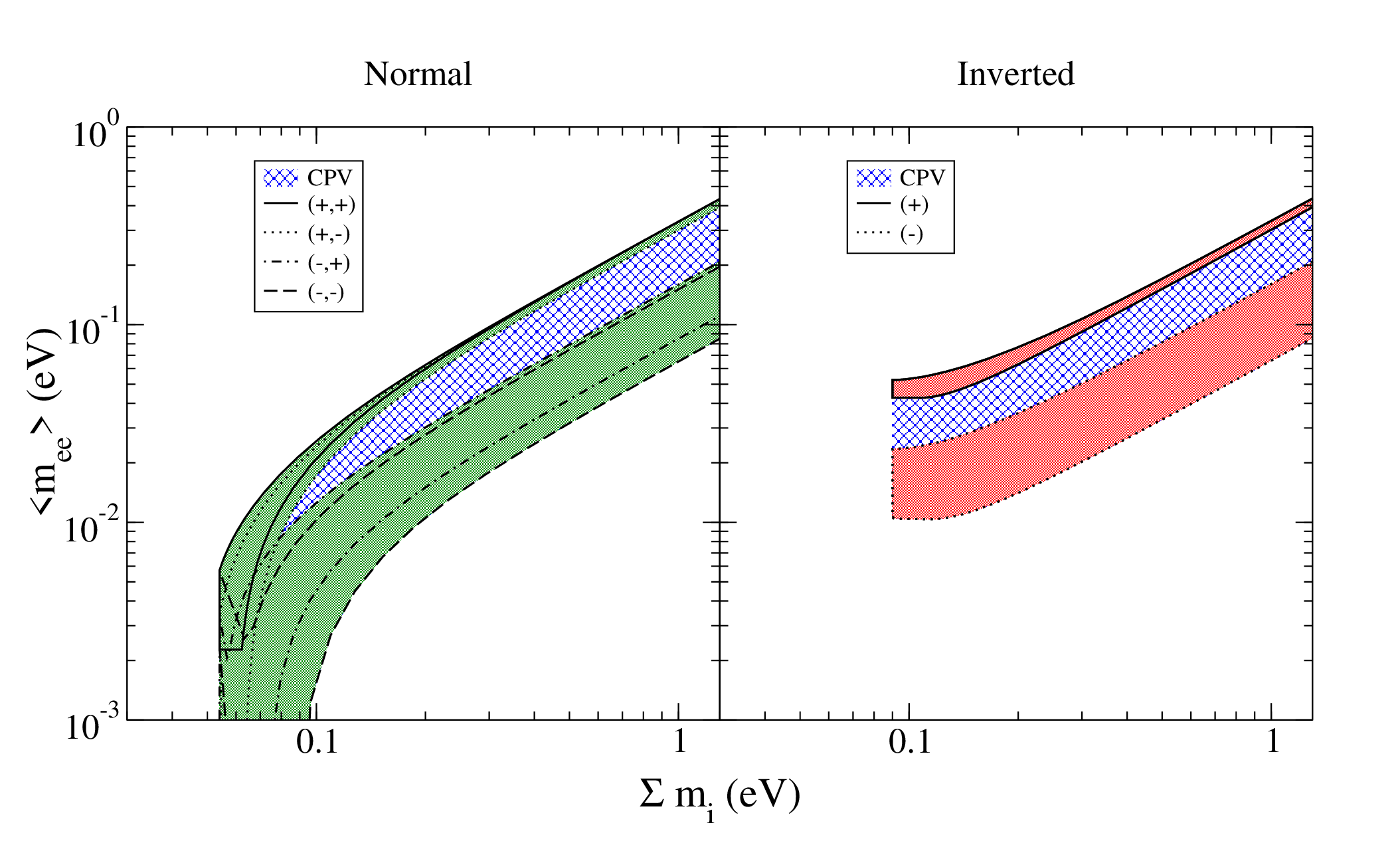,width=7cm,height=7cm,angle=270}}
\vspace*{8pt}
\caption{\label{fig:meff_obs}Effective mass against sum of masses $\Sigma$ and kinematic
neutrino mass $m_\beta$ for the $3\sigma$ ranges of the oscillation
parameters. CP conserving and violating areas are indicated. }
\end{figure}

\begin{figure}[t]
\begin{center}
\input{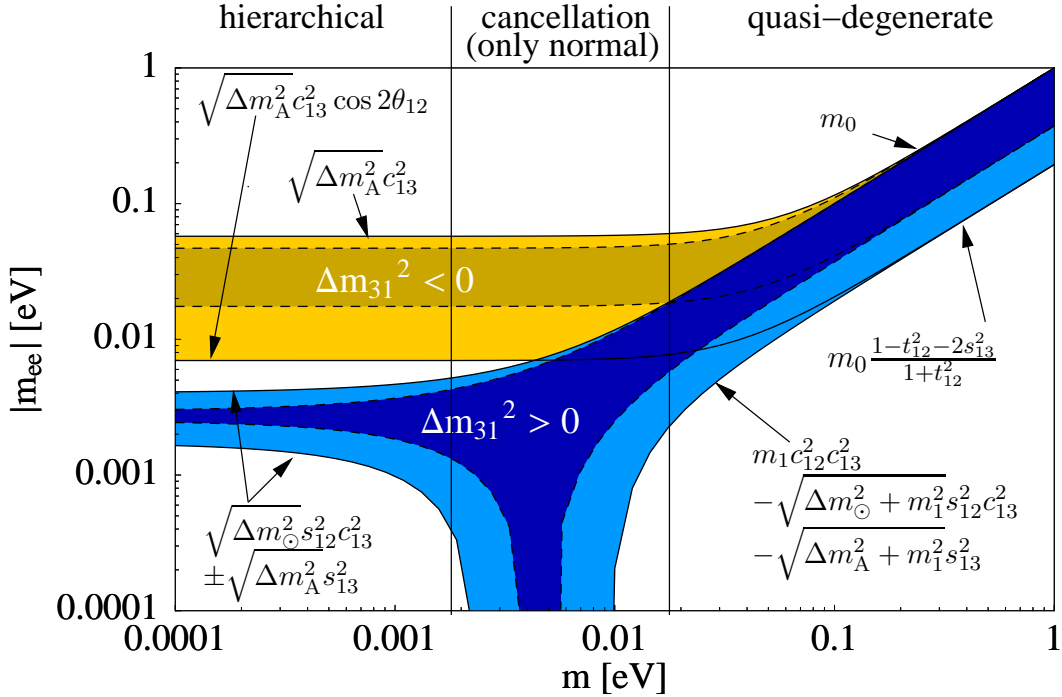}
\vspace*{8pt}
\caption{\label{fig:lovely_isn't_it?}The main properties 
of the effective mass as function of the smallest neutrino mass. 
We indicate the relevant formulae and the three important regimes: 
hierarchical, cancellation (only possible for normal mass ordering) and 
quasi-degeneracy. $c_{ij} = \cos \theta_{ij}$, $s_{ij} = \sin
\theta_{ij}$ and $t_{ij} = \tan \theta_{ij}$. 
Taken from\protect\cite{Lindner:2005kr}.}
\end{center}
\end{figure}

A typical analysis of the effective mass would plot it against the smallest
neutrino mass, while varying the Majorana phases and/or the
oscillation parameters. This results in Fig.~\ref{fig:meff_mass}, for
which the best-fit values and 3$\sigma$ ranges of the oscillation parameters have been
used. The blue shaded area is of interest because it can only be
covered if the CP phases are non-trivial, i.e.~if $\alpha, \beta \neq
0,\pi/2$. The values $\alpha, \beta = 0,\pi/2$ correspond to CP
conserving situations, associated with positive or negative signs of
the neutrino masses, and the resulting span of \meff~is also indicated
in the figure. 
For comparison, the other mass-related observables $\Sigma$ and $m_\beta$
are shown as a function of the smallest neutrino mass in
Fig.~\ref{fig:other_mass}. 
Actually, the smallest neutrino mass is not really an observable, so
it is interesting to plot the effective mass against 
$\Sigma$ and $m_\beta$, which is shown in Fig.~\ref{fig:meff_obs}. 
The analytical expressions for the effective mass in certain extreme
cases are given in Fig.~\ref{fig:lovely_isn't_it?}.

\subsubsection{\label{sec:meff_normal}Normal mass ordering}
Let us begin with the normal mass ordering. The effective mass is 
\be
\meff^{\rm nor} = \left| m_1 \, c_{12}^{2} \, 
c_{13}^{2} + \sqrt{m_1^{2} + \dms} \, s_{12}^{2} \, c_{13}^{2}\, 
e^{2i\alpha} + \sqrt{m_1^{2} + \dma} \, s_{13}^{2} \, e^{2i\beta}
\right| .
\label{eq:mee_norm}
\ee
The maximum of the effective mass is obtained when 
the Majorana phases are given by $\alpha=\beta=0$. 
The case of small $m_1$, which 
corresponds to a normal hierarchy (NH) defines 
 the ``hierarchical regime'' in Fig.~\ref{fig:lovely_isn't_it?}. Neglecting $m_1$ gives 
\be
\meff^{\rm NH} = \left|\sqrt{\dms} \, s_{12}^{2} \, c_{13}^{2} + 
\sqrt{\dma} \, s_{13}^{2} \, e^{2i(\alpha-\beta)}
\right| , 
\ee
where both terms can be of comparable magnitude. If $\theta_{13} = 0$
one has 
\be
\meff = \left|m_1 \, c_{12}^{2} \, e^{2i\alpha}  + \sqrt{\dms + m_1^2} \, s_{12}^{2}   \right| , 
\ee
where both terms can again be comparable. Note that in the last two
expressions, as well as for other situations with small $m_1$, 
the effective mass can vanish, a special case we will deal with in
Section \ref{sec:meff0}. If the smallest mass $m_1$ is much larger
than the mass-squared differences, the effective mass for
quasi-degenerate neutrinos is obtained: 
\be
\meff^{\rm QD} = m_0  \left|  c_{12}^{2} \, 
c_{13}^{2} + s_{12}^{2} \, c_{13}^{2}\, 
e^{2i\alpha} + s_{13}^{2} \, e^{2i\beta}
\right| .
\ee
Recall that $m_0$ denotes the common neutrino mass for QD neutrinos. 
The third term is now much smaller than the minimal combination of the
first two terms, $m_0 ( c_{12}^{2} - s_{12}^2)$, because
$\theta_{12}$ lies below $\pi/4$. Therefore, the effective mass cannot
vanish for quasi-degenerate neutrinos. The estimate for the effective
mass in case of quasi-degenerate neutrinos is 
\be \label{eq:meff_NH}
\cos 2 \theta_{12} \, m_0 \ls \meff^{\rm QD} \ls m_0 
\, . 
\ee
This corresponds, for $\meff^{\rm QD} \simeq 0.1$ eV, to half-lifes in the
regime of $10^{25}$ to $10^{26}$ yrs. Current experiments are testing
this regime, thus $\ls 100$ kg yrs facilities with $10^{-2}$ or less 
background counts are sufficient for the QD regime.

A rough estimate for the effective mass in terms of a normal hierarchy is 
\be \label{eq:meff_NH_scale}
\meff^{\rm NH} \sim \left\{ \baz 
\sqrt{\dms} \, \sin^2 \theta_{12} & \simeq 0.003 \, \rm eV \, , \\ 
\left(
\mbox{ or } \sqrt{\dma} \, \sin^2 \theta_{13} \right. 
&
\left. 
\ls  0.003 \, \rm eV \right) . 
\ea \right.
\ee
The meV scale of the effective mass should be the final goal of
experiments, but the possibility of strong or even complete
cancellation has to be kept in mind. The half-lifes corresponding to
meV effective masses are $10^{28}$ to $10^{29}$ yrs. Multi-ton scale
experiments are necessary for such extremely low numbers, 
with background levels below $10^{-4}$.  
It has been argued that if single electron events cannot be distinguished from double electron
events, the elastic $\nu_e e$ scattering of solar neutrinos represents
an irreducible background for \obb-experiments probing the NH regime\cite{kai_barros}.  

In Figs.~\ref{fig:other_mass} and \ref{fig:meff_obs} we see that in
case of NH $m_\beta$ lies below KATRIN's sensitivity of about 0.2 eV for the
normal hierarchy regime. With $\Sigma \simeq \sqrt{\dma} \simeq 0.05$ eV, only very
optimistic or far future cosmological observations can test this value. 
If the quasi-degenerate scenario is realized we find 
$\Sigma \simeq 3 \,m_\beta \simeq 3 \, \meff^{\rm max} \simeq 3\, m_0$.  
Table \ref{tab:mass_obs} shows the mass observables for the NH and QD
schemes. The interplay of the mass observables assumes unitarity of
the PMNS matrix; corrections due to possible non-unitarity
have been discussed in Ref.\cite{Rodejohann:2009ve} and found to be
negligible.

\subsubsection{\label{sec:meff_inverted}Inverted mass ordering}
For the inverted mass ordering, the smallest neutrino mass is denoted $m_3$ 
and the mass matrix element is given by 
\be
\meff^{\rm inv} = \left| \sqrt{m_3^{2}+\dma} \,  c_{12}^{2} \, 
c_{13}^{2} + \sqrt{m_3^{2} + \dms + \dma} \,  s_{12}^{2} \, c_{13}^{2} \, 
e^{2i\alpha} + m_3 \,  s_{13}^{2} \, e^{2i\beta} \right|  .
\label{eq:meeIH}
\ee
The maximal effective mass is -- as for the normal mass ordering -- 
obtained when $\alpha = \beta = 0$. The minimal value is   
\be
\meff^{\rm inv}_{\rm min} 
=\sqrt{m_3^{2}+\dma}\, c_{12}^{2} \, c_{13}^{2} - 
\sqrt{m_3^{2} + \dms + \dma}\, s_{12}^{2} \, c_{13}^{2} - m_3\,
s_{13}^{2} \, .
\label{eq:meeIHmin}
\ee
The third term of \meff~is usually 
negligible because
$\theta_{13}$ is small and $m_3$ is the smallest mass. 
In this case: 
\bea
\meff^{\rm IH} \simeq \sqrt{\dma}\,c_{13}^{2}  
\left| c_{12}^{2} + s_{12}^{2}  \, e^{2i\alpha} \right| \\[0.2cm]
\mbox{ and } 
\meff^{\rm IH}_{\rm max} \equiv \sqrt{\dma}\,c_{13}^{2} \le 
\meff^{\rm IH} \le \sqrt{\dma}\,c_{13}^{2} \cos 2 \theta_{12} 
\equiv \meff^{\rm inv}_{\rm min} \, . 
\label{eq:meeIHleft}
\eea
It is important to note that owing to the non-maximal value of
$\theta_{12}$ the minimal value of the effective mass is
non-vanishing\cite{Pascoli:2002xq}. Therefore, if limits below the
minimal value 
\begin{equation}
\meff^{\text{inv}}_{\text{min}} = \meff^{\text{IH}}_{\text{min}} = 
\left(1 - |U_{e3}|^2 \right)  \sqrt{\dma} \left(1 - 2 \, \sin^2
\theta_{12} \right) , \label{eqn:meffminih}
\end{equation}
are reached by an experiment, the inverted mass ordering is ruled out
if neutrinos are Majorana particles. If we knew by independent
evidence that the mass ordering is inverted (by a long-baseline
oscillation experiment or a galactic supernova observation) then we
would rule out the Majorana nature of neutrinos. Of course, one has to
assume here that no other lepton number violating mechanism
interferes. The two scales of
\meff~corresponding to the minimal and maximal value of \meff~in case
of the inverted hierarchy, given in Eq.~(\ref{eq:meeIHleft}), should be the intermediate or long-term goal
of future experiments.  

\begin{table}[t]
\tbl{\label{tab:mass_obs}Approximate analytical expressions for the neutrino mass
observables for the extreme cases of the mass ordering. For \obb~the
typical (isotope-dependent) half-lifes are also given.}
{\begin{tabular}{cccc}\hline 
 &  $\Sigma$ &  $m_\beta$ 
&  \meff \\ \toprule 
 NH &   $\sqrt{\dma}$ 
&   $\sqrt{\sss \dms + |U_{e3}|^2  \dma}  $ 
&  $\left| \sss  \sqrt{\dms} + |U_{e3}|^2 \sqrt{\dma} 
e^{2 i (\alpha - \beta)} \right| $ \\ 
& $\simeq 0.05$ eV & $\simeq 0.01$ eV & $\sim 0.003$ eV $\Rightarrow
T^{0\nu}_{1/2} \gs 10^{28 - 29}$ yrs \\ \hline 
 IH &   $2  \sqrt{\dma} $  
&   $\sqrt{\dma} $ &   $\sqrt{\dma}  
\sqrt{1 - \sin^2 2 \theta_{12} \sin^2 \alpha}$  \\ 
& $\simeq 0.1$ eV & $\simeq 0.05$ eV & $\sim 0.03$ eV $\Rightarrow T^{0\nu}_{1/2} \gs 10^{26 - 27}$ yrs\\
\hline 
 QD &   $3  m_0$  &   $m_0$ 
&   $m_0 \sqrt{1 - \sin^2 2 \theta_{12}  
\sin^2 \alpha}$  \\ 
& & & $\gs 0.1$ eV $\Rightarrow T^{0\nu}_{1/2} \gs 10^{25 - 26}$ yrs\\ \botrule 
\end{tabular} }
\end{table}

The typical effective mass values of order $\simeq 0.03$
eV are one order of magnitude
larger than for the normal hierarchy and
roughly one order of magnitude smaller than for quasi-degenerate
neutrinos. They implies half-lifes of order 
$10^{26}$ to $10^{27}$ yrs. A few 100 kg yrs of data taking with
background levels below $10^{-2}$ counts will be necessary. 
Upcoming next generation experiments will test
the inverted ordering, but cannot completely rule it out.

A more detailed analysis, focussing on
the important dependence on $\theta_{12}$, will be
summarized in Section \ref{sec:meff_NHIH}.

The transition to the quasi-degenerate regime takes place when 
$m_3 \gs 0.03$ eV. If the smallest mass assumes such values, the normal and inverted mass 
ordering generate identical predictions for the effective mass. 
The results in this case are therefore identical to the ones for the normal mass 
ordering treated above in Section~\ref{sec:meff_normal}. 

For the inverted hierarchy case $m_\beta$ is again below KATRIN's
sensitivity, and $\Sigma \simeq 2 \sqrt{\dma} \simeq 0.1$ eV is in the
range of future limits. Table \ref{tab:mass_obs} shows the mass
observables for the IH scheme. 
The values of the effective mass for the various special cases are
displayed in Fig.~\ref{fig:lovely_isn't_it?}. In Table
\ref{tab:mass_obs_compl} it is attempted to illustrate the
complementarity of neutrino mass observables. Prospective sensitivity
values of $m_\beta = 0.2$ eV, $\meff = 0.02$ eV, and $\Sigma = 0.1$ eV
are assumed and the interpretation of positive and/or negative results
in all 3 approaches is given.

\begin{table}[t]
\tbl{\label{tab:mass_obs_compl}``Neutrino mass matrix'' for the
present decade. It is assumed that KATRIN will reach its sensitivity
limit of $m_\beta = 0.2$ eV, that \obb-experiments can obtain values
down to $\meff = 0.02$ eV, and that cosmology can probe the sum of
masses down to $\Sigma = 0.1$ eV. N-SI denotes non-standard interpretation of
\obb, N-SC is non-standard cosmology. }
{\begin{tabular}{cccccccc}
& & \multicolumn{2}{c}{\bf KATRIN} & \multicolumn{2}{c}{$\boldsymbol{\hspace{-.3cm}0\nu\beta\beta}$} &
\multicolumn{2}{c}{\hspace{.6cm}\bf cosmology} \\ \hline 
& & yes & no & yes & no & yes & no \\ \toprule 
\bf KATRIN & $\ba \mbox{yes} \\ \mbox{no} \ea $ & $ \ba-\\- 
\ea  $ & $ \ba-\\-\ea  $ & $ \ba \mbox{QD + Majorana}
\\ \mbox{N-SI} \ea $  & $ \ba \mbox{QD + Dirac} \\ \mbox{low IH or NH
or Dirac} \ea $ & $\ba \mbox{QD} \\ m_\nu \ls 0.1 \, \mbox{eV or N-SC}
\ea $ & $\ba \mbox{N-SC} \\ \mbox{NH} \ea $ \\ \hline
 
$\boldsymbol{0\nu\beta\beta}$ & $\ba \mbox{yes} \\ \mbox{no} \ea $ 
& $ \ba \bullet \\ \bullet \ea  $ & $  \ba \bullet \\ \bullet \ea  $
& $ \ba-\\- 
\ea  $ & $ \ba-\\-\ea  $  
& $\ba \mbox{(IH or QD) + Majorana} \\ \mbox{low IH or (QD + Dirac)} \ea $ 
& $\ba \mbox{N-SC or N-SI} \\ \mbox{NH} \ea $ \\ \hline 

\bf cosmology & $\ba \mbox{yes} \\ \mbox{no} \ea $ & $ \ba \bullet \\
\bullet \ea  $ & $ \ba \bullet \\ \bullet \ea  $ & $ \ba \bullet \\
\bullet \ea  $ & $ \ba \bullet \\ \bullet \ea  $ & $ \ba-\\
- \ea  $ & $ \ba- 
\\-\ea  $ \\ \botrule 
\end{tabular} }
\end{table}

\subsubsection{\label{sec:meff_mass}Mass scale}
\begin{figure}[t]
\centerline{\psfig{file=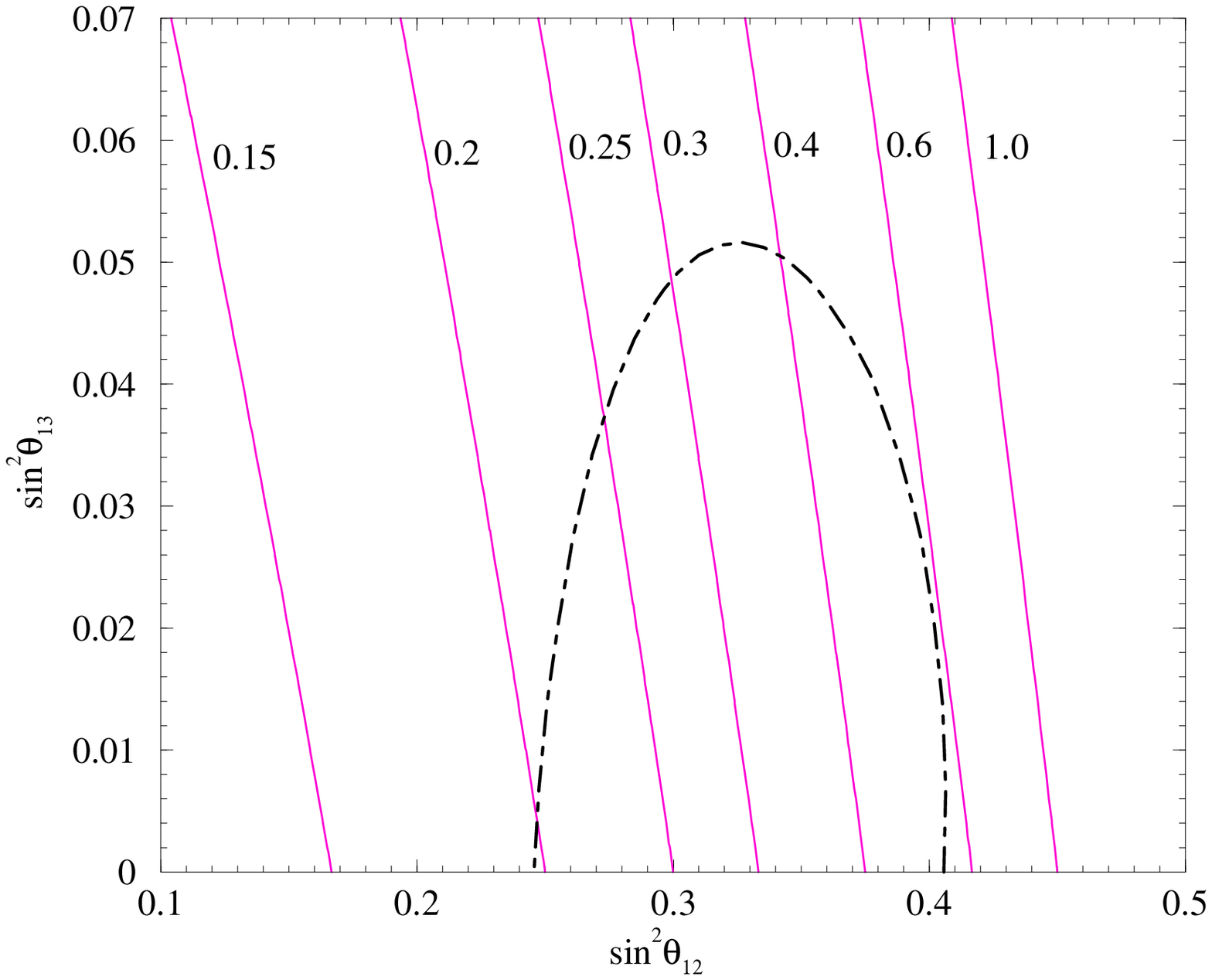,width=5cm,angle=0}}
\vspace*{8pt}
\caption{\label{fig:limitm0}Lines of constant $m_0$ in the 
$\sss-\sch$ plane, predicted 
for a QD mass spectrum with $\meff = 0.1$ eV. Also shown is an allowed 
$3\sigma$ region in $\sss$ and $\sch$. }
\end{figure}

As mentioned above, from the fundamental quantity \meff~one can also
extract information on the masses of the individual neutrinos. We focus
here on the quasi-degenerate regime, which is the easiest, though
still non-trivial, case. The smallest effective mass can be written as 
\be  
\meff^{\rm QD}_{\min} = m_0 \, 
\left( 
|U_{e1}|^2 - |U_{e2}|^2 - |U_{e3}|^2 
\right) = m_0 \,  \frac{1 - \tan^2 \theta_{12} - 2 \, |U_{e3}|^2 }
{1 + \tan^2 \theta_{12}} \, . 
\ee
We show in Fig.~\ref{fig:limitm0} iso-contours\cite{Choubey:2005rq} of $m_0$ in the plane
spanned by $\sin^2\theta_{12}$ and $|U_{e3}|^2$. 
With a limit $\meff_{\min}^{\rm exp}$ on the effective mass at hand,
one can translate this into a limit on the neutrino mass, which reads 
\be \label{eq:m0_lim}
m_0 \le  \meff_{\min}^{\rm exp} \,  \frac{1 + \tan^2 \theta_{12}}
{1 - \tan^2 \theta_{12} - 2 \, |U_{e3}|^2 } 
\equiv \meff_{\min}^{\rm exp} \,  f(\theta_{12}, \theta_{13}) \, .
\ee 
This function $f(\theta_{12}, \theta_{13})$ varies from 2.57 to 3.29
at $1\sigma$ and from 2.17 to 4.77 at $3\sigma$. The limit on the
effective mass is about 0.5 eV (see Table \ref{tab:limits}), and hence
$m_0 \le 1.6$ eV and 2.4 eV, respectively. Therefore, the current limit
on $m_0$ from \obb~is very similar to the one from the Mainz experiment. 

\begin{figure}[t]
\centerline{
\psfig{file=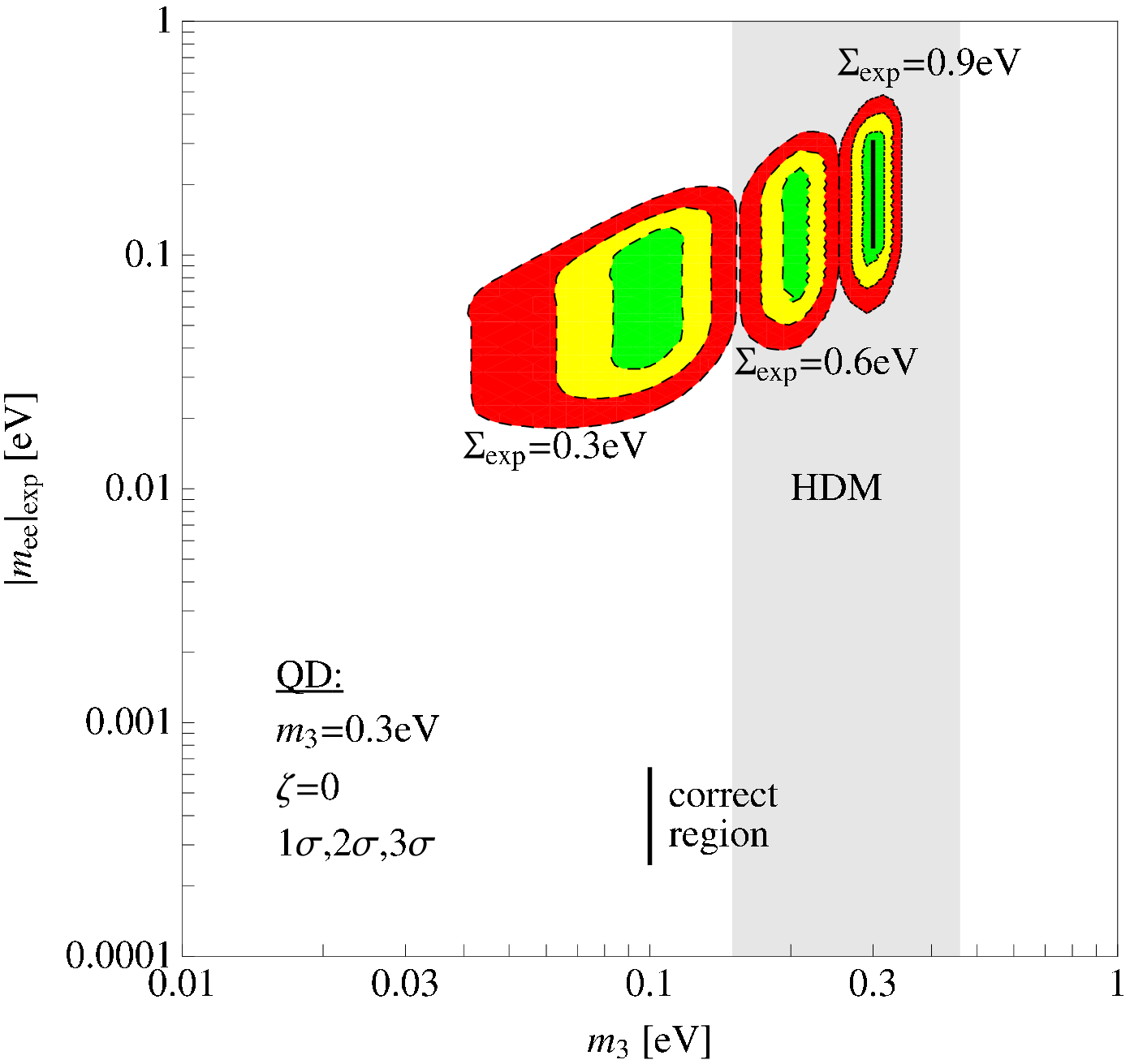,height=5cm,width=6cm}
\psfig{file=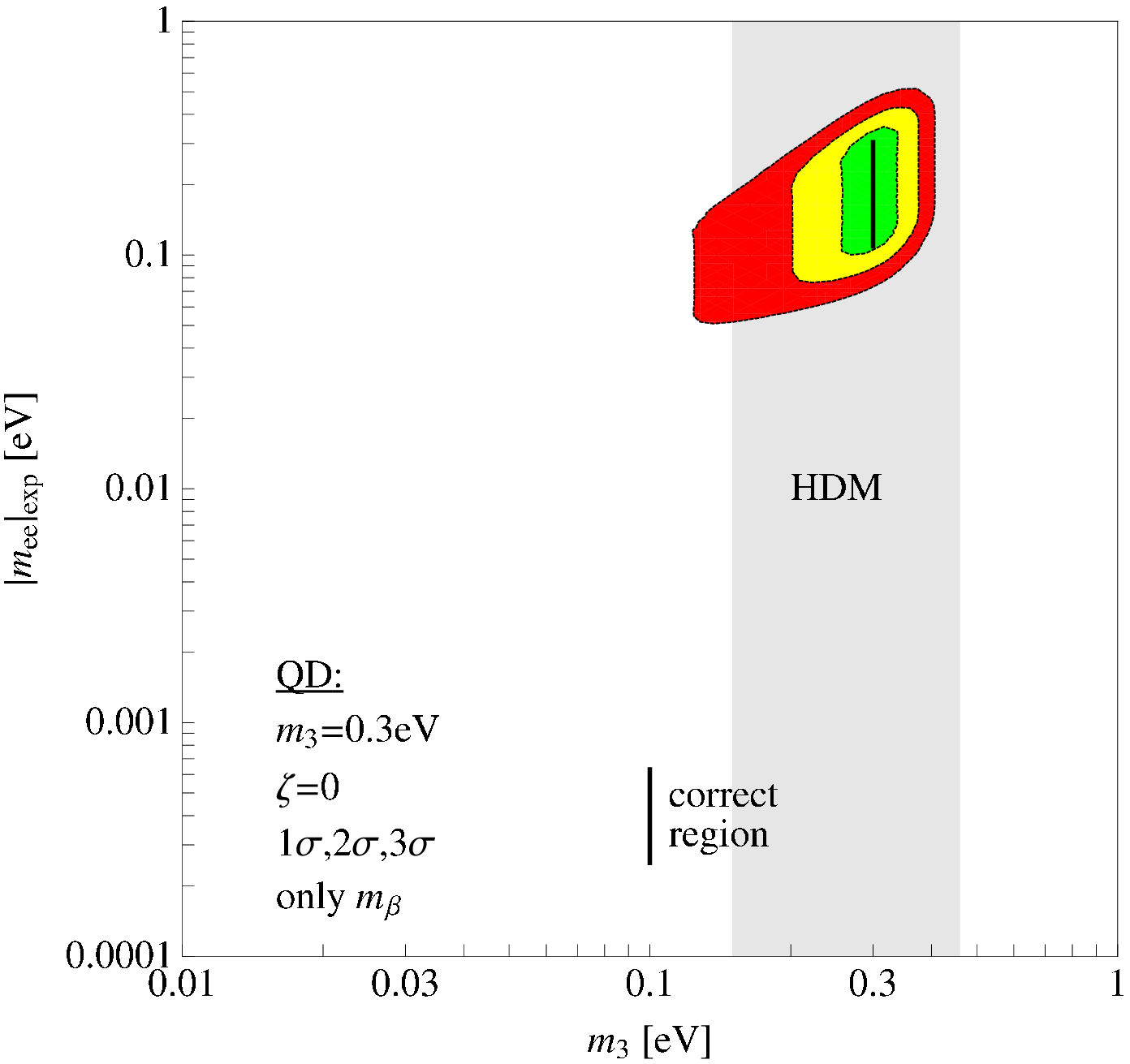,height=5cm,width=6cm} }

\centerline{
\psfig{file=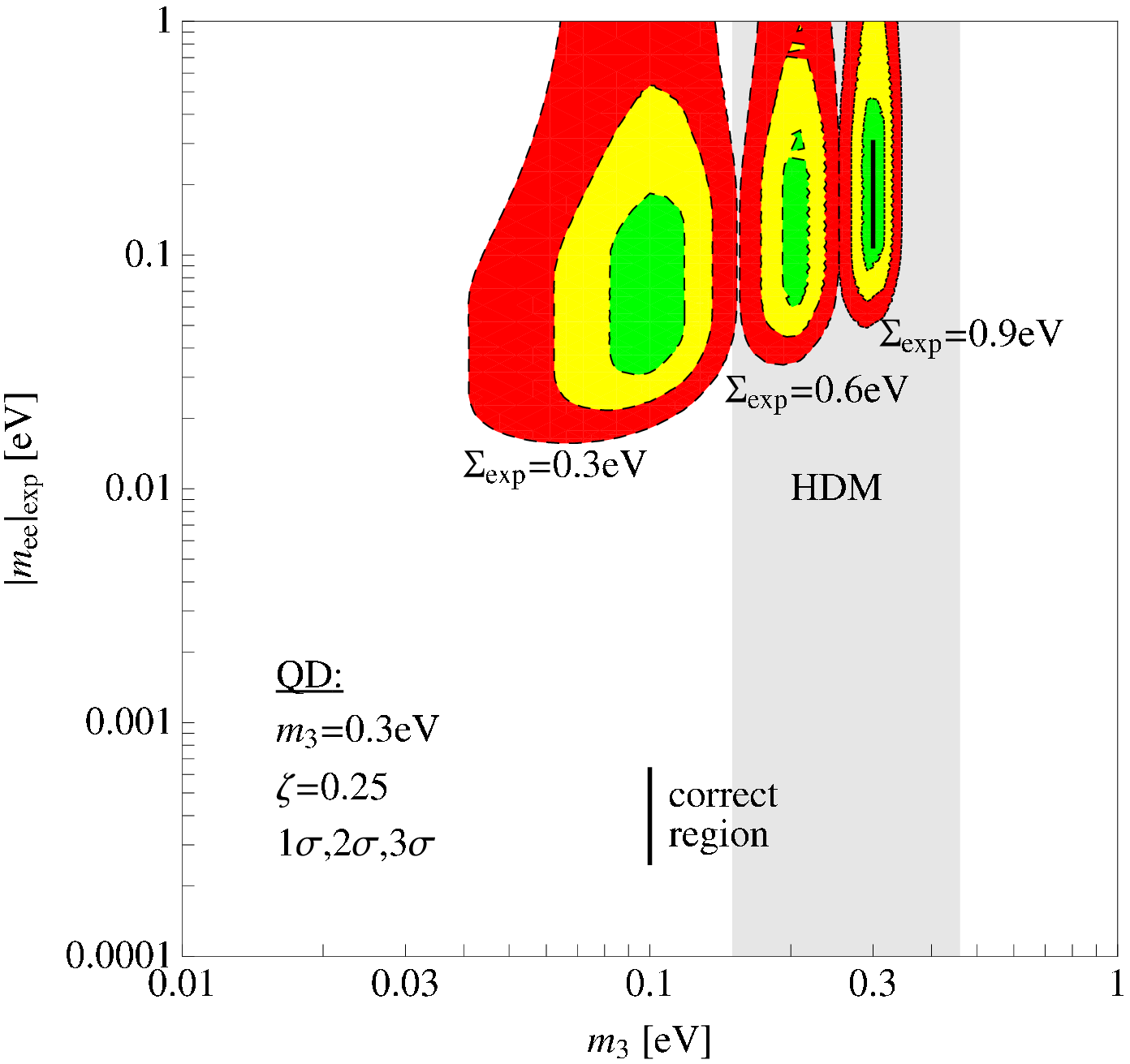,height=5cm,width=6cm}
\psfig{file=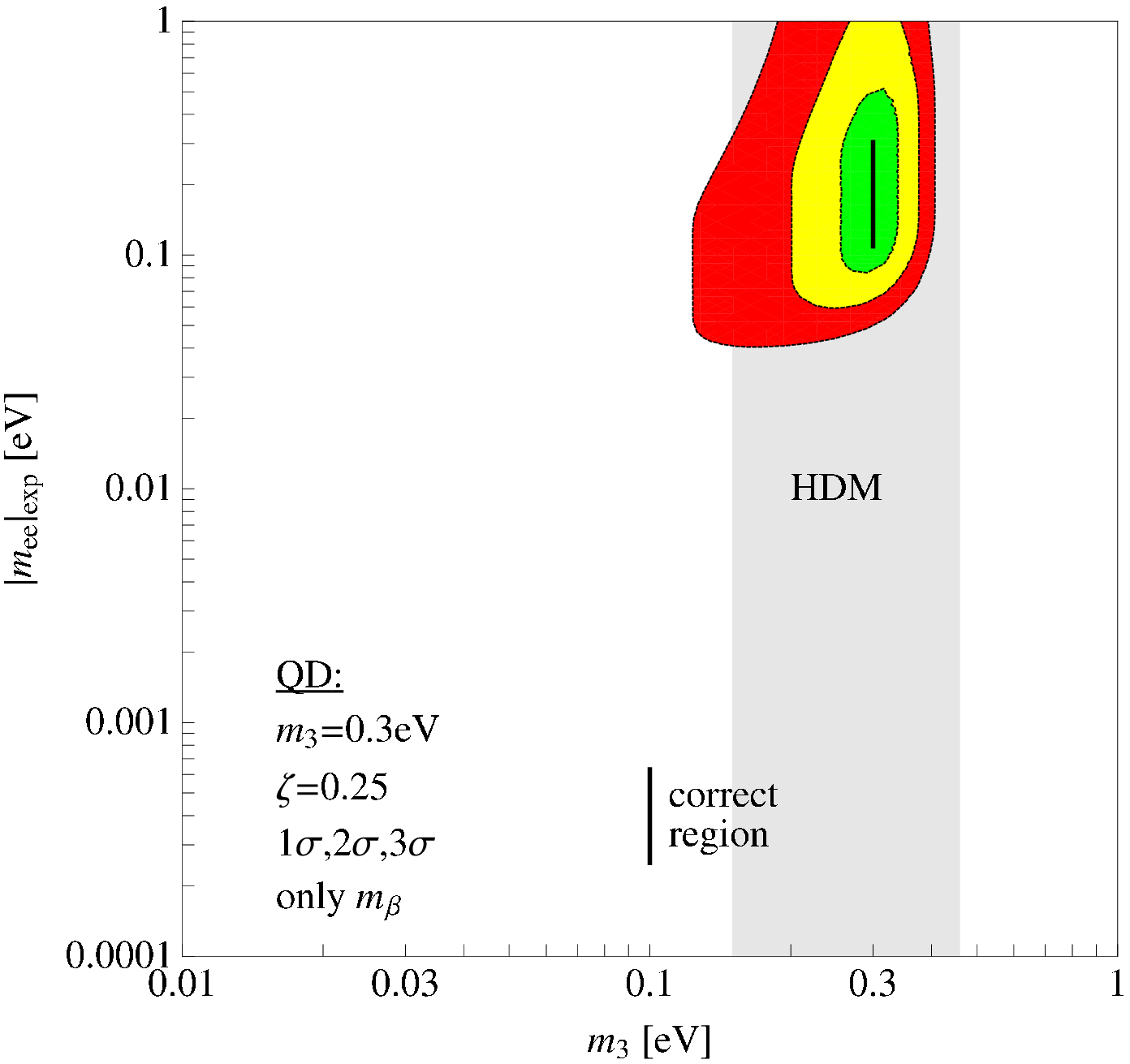,height=5cm,width=6cm} }
\vspace*{8pt}
\caption{\label{fig:MMR}1$\sigma$, 2$\sigma$ and 
3$\sigma$ regions in the $m_3$-$\meff_{\rm exp}$ plane for a
quasi-degenerate neutrino mass scenario. The upper plots are for no
NME uncertainty, the lower plots assume 25\% uncertainty. The left
plots show the correct (solid line) as well as two possible incorrect cosmological 
measurements (dashed lines). The right plots leave $\Sigma$ out of the
fit. The area denoted HDM is the range of \meff~from the claim of 
part of the Heidelberg-Moscow collaboration. Taken from\protect\cite{Maneschg:2008sf}.}
\end{figure}

Perhaps more interesting is the determination of the neutrino mass
scale in future experiments if information from complementary 
neutrino mass observables is combined. For instance\cite{Maneschg:2008sf}, consider the
scenario defined by 

\begin{tabular}{cccc}
$m_3$ [eV] & $\meff$ [eV] & $m_\beta$ [eV] & $\Sigma$
[eV] \\ \hline
0.3    & $0.11-0.30$    &  0.30   &  0.91\\
\end{tabular}

The prospective errors one can use are $\sigma(m_\beta^2) = 0.025$ eV$^2$ 
and $\sigma(\Sigma) = 0.05$ eV, and an ``experimental error'' 
$\sigma(\meff_{\rm exp}) = \frac 12 \, \meff_{\rm exp} \, 
\sigma(\Gamma_{\rm obs}) /\Gamma_{\rm obs}$, where $\sigma(\Gamma_{\rm
obs}) /\Gamma_{\rm obs}$ is
motivated by the GERDA proposal\cite{gerda} to be $\simeq  23
\%$. The ``theoretical error'' from the NME uncertainty was defined as
$ \sigma(\meff) = (1 + \zeta) \, \left(\meff + \sigma(\meff_{\rm exp}) 
\right) - \meff $. 
Depending on the measured effective mass $\meff_{\rm exp}$ one
can now obtain the values of $m_0$ which can be
reconstructed. Fig.~\ref{fig:MMR} shows the results of the analysis.  If
$\zeta=0$ one finds $\sigma(m_3) \simeq 15\%$ at 3$\sigma$, while for
$\zeta=0.25$ it holds that $\sigma(m_3) \simeq 25\%$. If one includes
a wrong cosmological input the reconstruction of $m_3$ can be wrong by up to one order of
magnitude. Leaving $\Sigma$ out of the analysis yields $\sigma(m_3)
\simeq 50\%$, showing that the precision is largely determined by
cosmology.

\begin{figure}[t]
\centerline{
\psfig{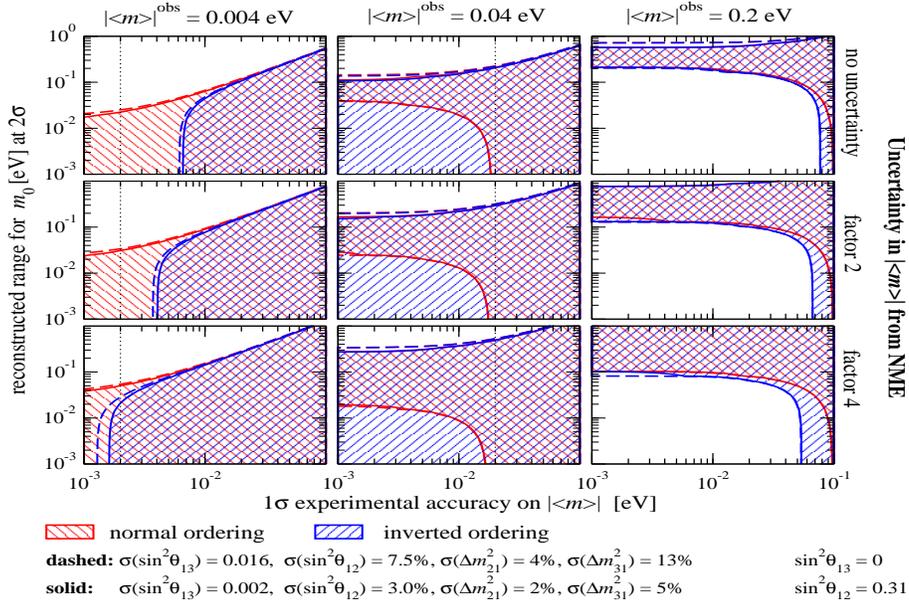}
}
\vspace*{8pt}
\caption{\label{fig:2005zb}
The reconstructed range for the lightest neutrino mass
      at 2$\sigma$~C.L.\ for normal and inverted mass ordering as a function of
     the 1$\sigma$ experimental error on $\meff_{\rm exp}$ (here
called $|\!<\!m\! >\!|^\mathrm{obs}$). The
     results are shown for three representative values
     $\meff^\mathrm{obs} = 0.004, 0.04, 0.2$~eV, 
     and for fixed NME (first row), and an uncertainty of a factor of
     $F=2$ and $F=4$ in the NME (second and third rows).  The dashed (solid) lines correspond to
     the present uncertainties in the oscillation
     parameters.  To the left of the dotted lines, a positive signal
     is obtained at $2\sigma$, whereas to the right only an upper
     bound can be set. Taken from\protect\cite{Pascoli:2005zb}.}
\end{figure}

A detailed analysis was performed in Ref.\cite{Pascoli:2005zb}, from
where we have taken Fig.~\ref{fig:2005zb}. 
As was noted in that paper, the uncertainty of
the oscillation parameters is of little importance in determining
$m_0$. 
To take into account the NME uncertainty the following procedure was proposed: 
${\cal M}$ is the unknown true NME and ${\cal M}_0$ is the NME used to
obtain $\meff_{\rm exp}$, which denotes the effective mass extracted
from an experiment. 
The parameter $F$ is connected to the ratio $\xi = {\cal M}/{\cal
M}_0$ in the sense that $\xi $ ranges from $1/\sqrt{F}$ to $F$. 
If the
experimental error on $\meff_{\rm exp}$ is sufficiently small 
($\ls 0.06$~eV for NME uncertainty $F \ls 3$), 
the neutrino mass spectrum will be shown to be QD, and 
$m_0$ will be constrained to lie in a rather
narrow interval of values limited from below by $m_0 \gs 0.1$~eV. 
The uncertainty in the NME directly translates into an
uncertainty in $m_0$, in analogy to Eq.~(\ref{eq:m0_lim}). 
In the case of an intermediate value of $\meff_{\rm exp} =
0.04$~eV, shown in the middle column of Fig.~\ref{fig:2005zb}, an
allowed  range
of $0.01$~eV~$\ls m_0 \ls 0.1$~eV could be established for 
precise measurements. 
In the case of an inverted ordering only an upper bound $m_0 \ls
0.1$~eV will be obtained. This
result can be easily understood from the usual \meff~vs.~smallest 
mass plots, which are basically flat for the inverted ordering and $m_3
\ls 0.1$ eV. 

Other analyses on neutrino mass extraction from different neutrino
mass experiments including \obb~can be found in Refs.\cite{steen_mass_ana,errors,Bergstrom:2011dt}.

\subsubsection{\label{sec:meff_NHIH}Mass ordering: testing the
inverted hierarchy}

From Fig.~\ref{fig:meff_mass} the interesting possibility of ruling out
the inverted mass ordering becomes obvious. 
The minimal value of the effective mass, repeated here for
convenience, is non-zero and given by  
\be
\meff^{\text{inv}}_{\text{min}} = 
\left(1 - |U_{e3}|^2 \right)  \sqrt{\dma} \left(1 - 2 \, \sin^2
\theta_{12} \right) . 
\ee
If a limit on the effective mass below this value is obtained, the
inverted ordering is ruled out if neutrinos are Majorana particles. In
case the mass ordering is known to be inverted (e.g.~by a
long-baseline experiment or by observation of a galactic supernova)
then the Majorana nature of neutrinos would be ruled out. 
\begin{figure}[t]
\centerline{
\psfig{file=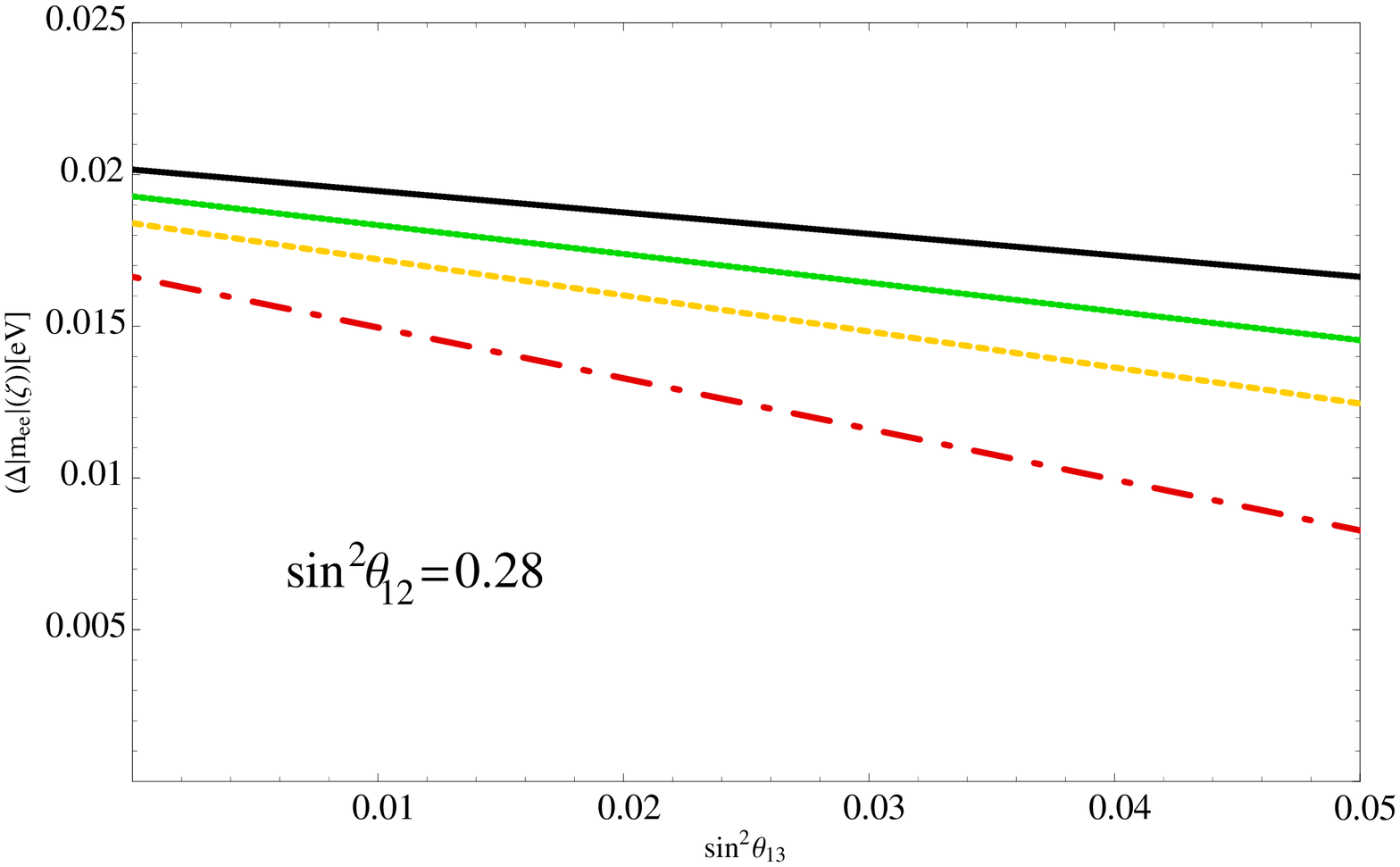,width=5cm,height=4cm} 
\psfig{file=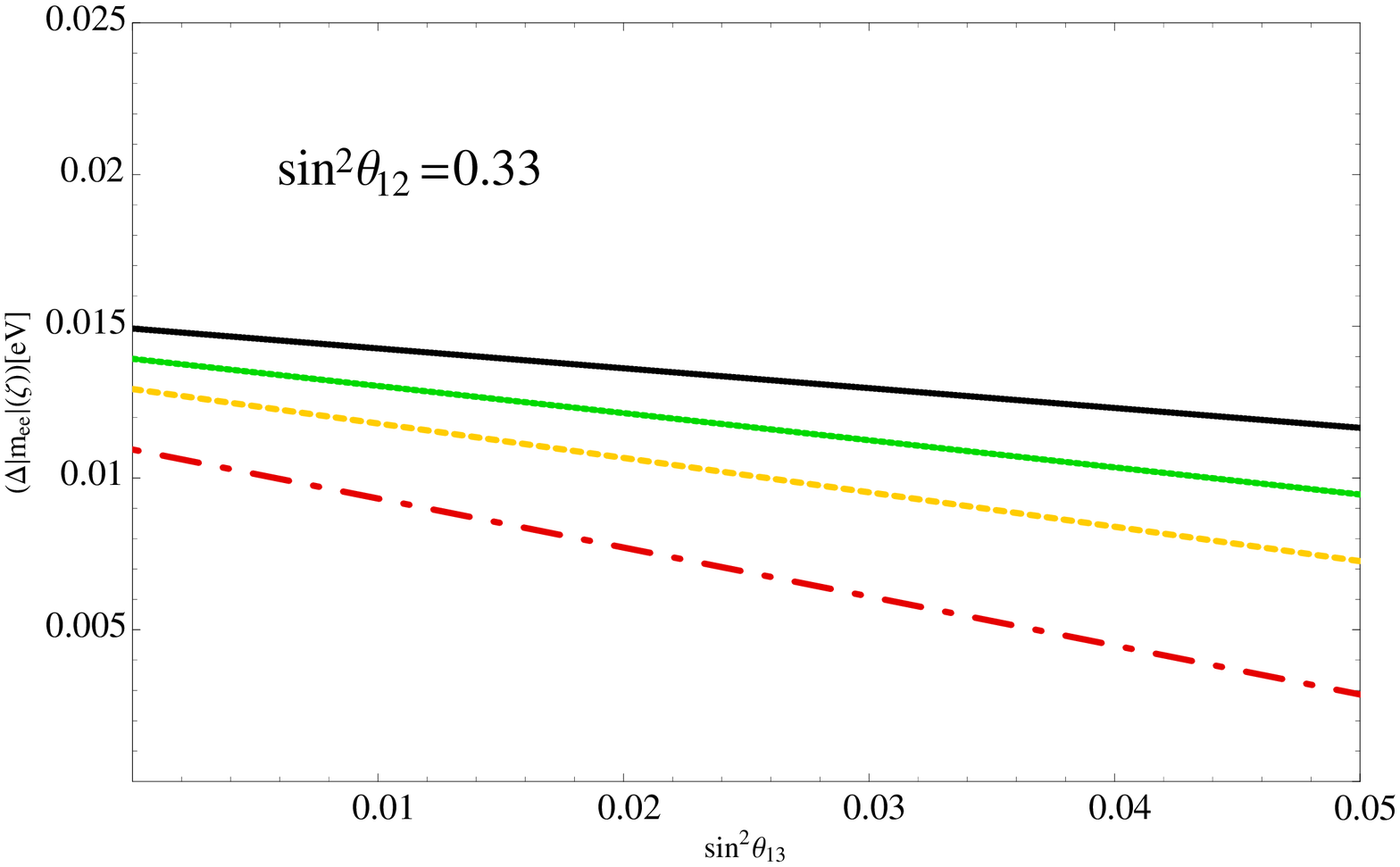,width=5cm,height=4cm}
}
\vspace*{8pt}
\caption{\label{fig:gaps}The difference $\Delta \meff$ of $\meff^{\rm inv}_{\rm min}$ and  
$\zeta \, \meff^{\rm nor}_{\rm max} $ as a function of 
$\sin^2  \theta_{13}$ for 
different nuclear matrix element uncertainty factors $\zeta = 1, 1.5,
2$ and $3$ (from top to bottom). 
We have chosen an illustrative value of the smallest mass
of $0.005$ eV and $\sin^2 \theta_{12} = 0.28$ (left) and $\sin^2
\theta_{12} = 0.33$ (right). }
\end{figure}

\begin{figure}[th]
\centerline{
\psfig{file=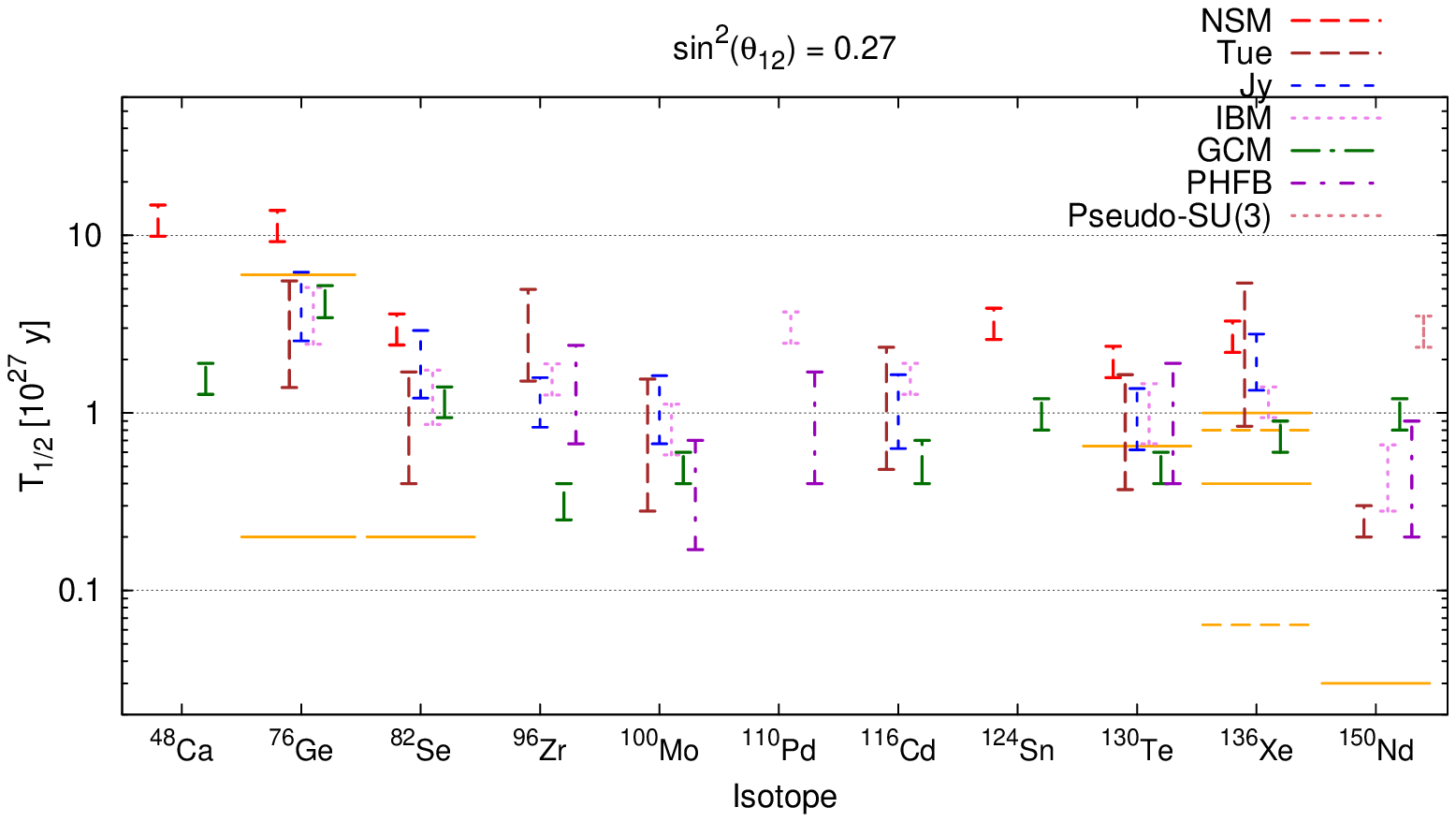,width=7cm,height=5cm} \quad
\psfig{file=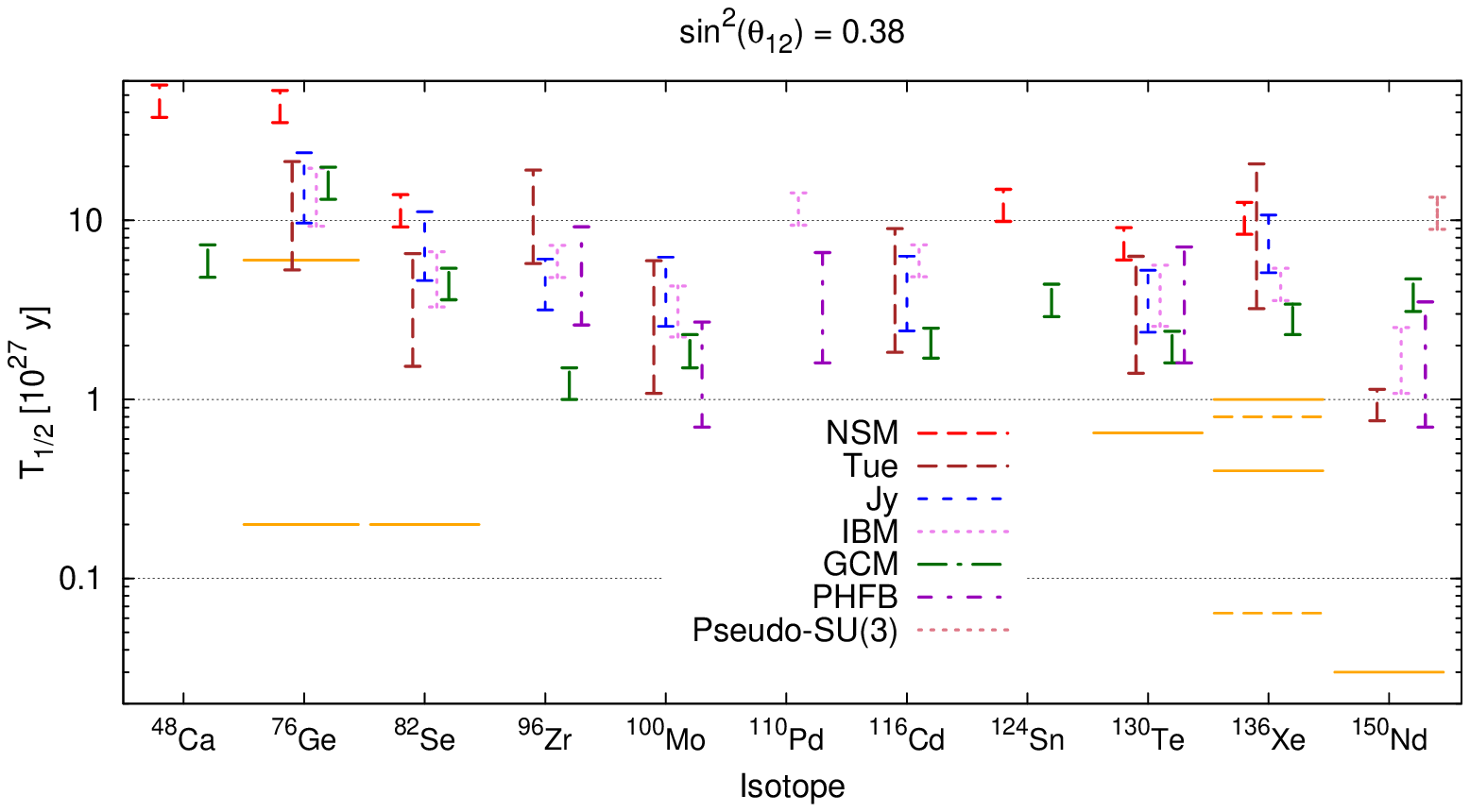,width=7cm,height=5cm} }

\centerline{
\psfig{file=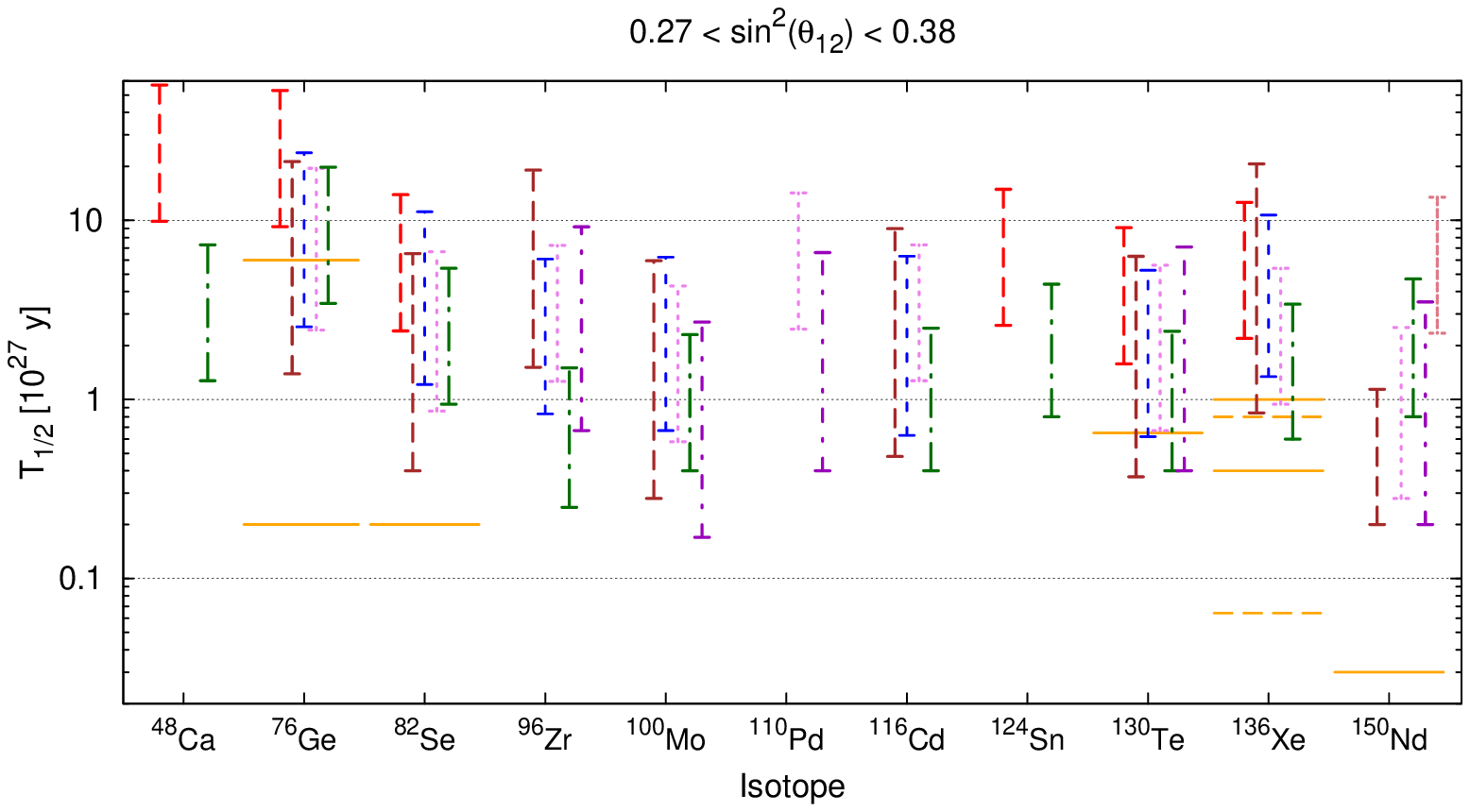,width=7cm,height=5cm} 
\psfig{file=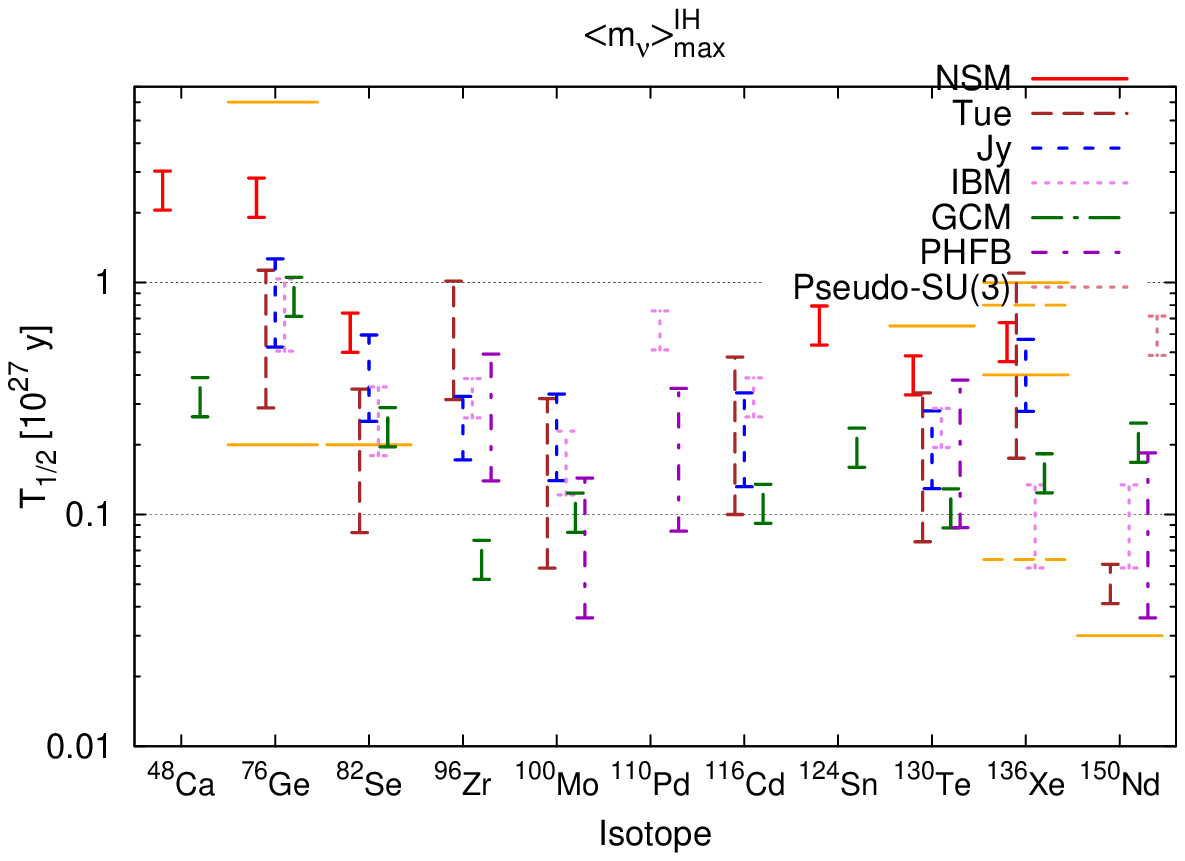,width=7cm,height=5cm} 
}
\vspace*{8pt}
\caption{\label{fig:isohl}Required half-life sensitivities to exclude
and touch the inverted hierarchy for different values of
$\theta_{12}$. The upper plots show the necessary half-lifes for $\sin^2 
\theta_{12} = 0.27$ (upper left) and $\sin^2 \theta_{12} = 0.38$
(upper right). The lower left plot includes the current $3\sigma$
uncertainty for $\theta_{12}$. The lower right plot shows the
necessary half-lifes in order to touch the inverted ordering, which is
independent on $\theta_{12}$.  The small horizontal lines show
expected half-life sensitivities at 90\% C.L.~of running and planned
\obb-experiments. When two sensitivity expectations are
given for one experiment they correspond to near and far time
goals, see Table \ref{tab:fut_exp}. Taken from\protect\cite{DRZ}.}
\end{figure}

As a rough requirement for experiments we 
calculate the difference between the minimal effective mass 
for the inverted ordering and the maximal effective mass for the 
normal ordering multiplied with the nuclear matrix element uncertainty
factor $\zeta$\cite{Pascoli:2002ae,Choubey:2005rq,Lindner:2005kr}
\be \label{eq:Delta}
\Delta \meff \equiv \meff^{\rm inv}_{\rm min} - \zeta \, 
\meff^{\rm nor}_{\rm max} \,.
\ee 
We plot this difference as a function of $\sin^2  \theta_{13}$ in
Fig.~\ref{fig:gaps}. Obviously, the largest dependence stems from
$\theta_{12}$, and the smaller $\theta_{12}$ is, the better. This is clear
from the previous discussion and Fig.~\ref{fig:lovely_isn't_it?}, 
because the smaller $\theta_{12}$ is, the larger is
$\meff^{\text{inv}}_{\text{min}}$. The effect\cite{Lindner:2005kr} of 
non-zero $\theta_{13}$ is to slightly decrease 
$\meff^{\text{inv}}_{\text{min}}$ 
and to slightly increase $\meff^{\text{nor}}_{\text{max}}$. 

One can translate the effective mass necessary to rule out (or touch)
the inverted hierarchy into half-lifes. The very important dependence on
$\theta_{12}$ has recently been discussed in Ref.\cite{DRZ}. 
The plots in Fig.~\ref{fig:isohl} are generated using the compilation 
of NMEs from Table \ref{tab:nme}. The current $3\sigma$ range
corresponds to an uncertainty of a factor 2 in the minimal value of
the effective mass, which is of the same order as the current uncertainty in
the NMEs.  
The factor 2 due to $\theta_{12}$ corresponds to a factor of $2^2 = 4$ in half-life. 
In experiments with background, see Eq.~(\ref{eq:Texp}), this means a
rather non-trivial combined factor of $2^4 = 16$ in the product of 
measuring time, energy resolution, background index and detector
mass. Therefore, a precision determination of the
solar neutrino mixing angle would be very desirable to
evaluate the requirements and physics potential of upcoming
\obb-experiments in order to test the inverted ordering\cite{DRZ}.

\subsubsection{\label{sec:meffCP}Majorana CP phases}
Apart from measuring the effective mass in case of a normal hierarchy,
determination of a Majorana CP phase from \onbb~is probably the most
difficult physics goal related to this process\footnote{Though
$|U_{e3}|$ has some influence on \obb\cite{Lindner:2005kr}, extracting
it from a measurement is also inpractical\cite{ue3_0vbb}.}. One general point to
be made here is that there is only one observable, \meff, and thus
only one of the two Majorana phases (or a combination of the two
phases) can be extracted. In addition, complementary information on the neutrino
mass scale has to be put in for such a measurement. A final remark is
that the process is not CP violating, i.e.~the rate of the $0\nu\beta^+\beta^+$
process depends on the same quantity as the 
\obb~process\footnote{Manifest CP violation from Majorana phases is discussed
e.g.~in\cite{deGouvea:2002gf}.}. 

The prospects of measuring the CP phase in \onbb~have been discussed
in several papers\cite{CP}. A somewhat pessimistic conclusion has been
drawn in Ref.\cite{Barger:2002vy}, whereas the requirements for such a
measurement have been discussed in\cite{nonogo}, and found to be
not too unrealistic.

The requirement for determining the phases is clear from
Figs.~\ref{fig:meff_mass} and \ref{fig:meff_obs}. Experimentally one
should find results lying in the areas indicated with
``CPV'', which are however smeared by experimental and theoretical
uncertainties. This is realistic only for the
inverted ordering or the quasi-degenerate scheme. Neglecting
$\theta_{13}$, the effective mass is in these cases is proportional to 
\be
\meff \propto \left|\cos^2 \theta_{12} + e^{2 i \alpha} \, \sin^2 
\theta_{12}  \right| = \sqrt{1 - \sin^2 2 \theta_{12} \, \sin^2 \alpha} \, . 
\ee
Therefore, the larger $\theta_{12}$ is, the more promising it is to
extract $\alpha$ from measurements. Recall that ruling out the inverted
mass ordering is easier if $\theta_{12}$ is small. 

\begin{figure}[t]
\centerline{
\psfig{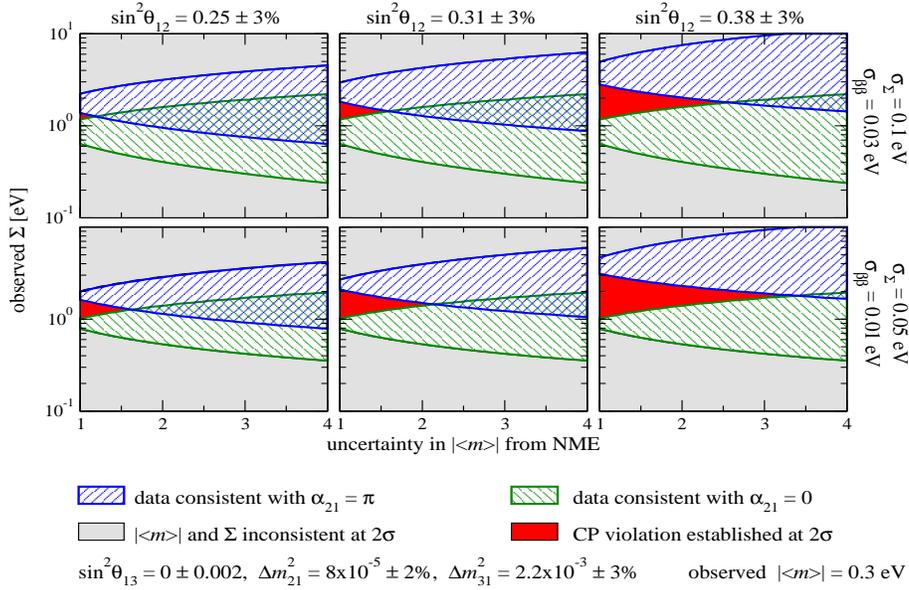}}
\vspace*{8pt}
\caption{\label{fig:2005zbCP}
Constraints on the Majorana phase $2\,\alpha$ (here called $\alpha_{21}$) at 95\%~C.L.\
   from an observed $\meff_\mathrm{exp} = 0.3$~eV and prospective 
   data on $\Sigma$, as a function of the NME uncertainty factor
$F$. Shown are the regions in which the
   data are consistent with one of the CP-conserving values 
    (hatched), observed $\Sigma$ is
   inconsistent with $\meff_\mathrm{exp}$ (light-shaded), and 
   Majorana CP-violation is established (red/dark-shaded).
Taken from\protect\cite{Pascoli:2005zb}.}
\end{figure}

A detailed statistical analysis has been performed
in\cite{Pascoli:2005zb}, from which we present
Fig.~\ref{fig:2005zbCP}. 
One can see that, as expected, for larger values of
$\theta_{12}$ the areas in parameter space become larger. 
For instance, if $\sin^2\theta_{12} \gs 0.3$ 
and $\simeq 10\%$ errors in the measured $\meff_\mathrm{exp}$ and
$\Sigma$ are present, the NME has to been known to better than within a factor of 1.5.
For smaller values of the errors, $\sigma_{\beta\beta} \simeq 0.01$~eV
and $\sigma_\Sigma \simeq 0.05$~eV, Majorana CP-violation could be 
established even for $F \simeq 2$ See Section \ref{sec:meff_mass} for
the definition of $F$). Finally, the
Majorana phase $2\alpha$ has to have a value approximately 
in the interval $\sim (\pi/4 - 3\pi/4)$. In the inverted hierarchy the
required errors have to be smaller, and the determination of the
phase is more challenging.

\subsubsection{\label{sec:meff0}Vanishing effective mass}
Unfortunately, the normal mass hierarchy can allow for 
complete cancellation of the effective mass (see
e.g.\cite{Xing:2003jf}). In terms of 
Fig.~\ref{fig:FD_mass_mech}, this ``cancellation regime'' means that a triangle
can be formed. If $|U_{e3}|=0$ then the requirement is 
\be
\frac{m_1}{m_2} = \tan^2 \theta_{12} \simeq \frac 12 \, , 
\ee
while for $m_1 = 0$ one needs
\be
\frac{m_2}{m_3} = \frac{\tan^2 \theta_{13}}{\sin^2 \theta_{12}} \simeq
3 \, \tan^2 \theta_{13} \, . 
\ee
In both cases the Majorana phases need to be such that the two surviving terms have
opposite sign. For the case of arbitrary $\theta_{13}$ one finds\cite{Dev:2006if}
\bea \D
\cos 2\alpha =\frac{
m_{3}^{2} \,s_{13}^{4}-c_{13}^{4}( m_{1}^{2} \, c_{12}^{4}+m_{2}^{2}
\, s_{12}^{4})}{
2 \,m_{1} \,m_{2} \,s_{12}^{2} \,c_{12}^{2} \,c_{13}^{4}} \, , \\ \D
\cos 2\beta =-\frac{
m_{3}^{2} \,s_{13}^{4}+c_{13}^{4} \,(m_{2}^{2} \,s_{12}^{4}-m_{1}^{2} \,c_{12}^{4})}{
2 \,m_{2} \,m_{3} \,s_{12}^{2} \,s_{13}^{2} \,c_{13}^{2}} \, . 
\eea
It may seem unnatural that the 7 parameters on which \meff~depends
conspire in such a way that the effective mass vanishes. However, 
recall that the effective mass is the $ee$ element of the Majorana
neutrino mass matrix\footnote{Zeros of the remaining elements of $m_\nu$ have been studied
in\cite{Merle:2006du}, the presence of two zeros in \cite{2zero}.}. This matrix is 
generated by the underlying theory of mass generation, and texture
zeros occur frequently in such (flavor) models, see
Ref.\cite{Grimus:2004hf} for a general analysis 
and\cite{Jenkins:2008ms} for symmetries leading to $\meff=0$. 

One may ask whether the effective mass remains zero, or whether
corrections lead to small but non-zero \meff. In fact, there are
several possibilities for non-zero \obb-rates, even if $\meff = 0$: 
\begin{itemize}
\item the first point to make
here is that the dependence of the amplitude on the neutrino parameters
goes as (see Eq.~(\ref{eq:am_SI})) 
\[ 
U_{ei}^2 \frac{m_i}{q^2 - m_i^2 } \simeq 
U_{ei}^2 \, m_i \left(1 + \frac{m_i^2}{q^2} \right) 
= \meff + U_{ei}^2 \, m_i^3 \frac{1}{q^2 } \, . 
\] 
While the second term is very much suppressed by $m_i^2/q^2 \ls 
10^{-12}$ with respect to the usual effective mass term, 
it is in general non-zero, even when \meff~is zero; 

\item another source of correction arises from radiative
corrections. While in the effective theory the renormalization of the
mass matrix is multiplicative (see Section \ref{sec:meffRG}), this 
may no longer be true in the theory beyond the effective one. For instance,
if $m_\nu$ is generated via the type I seesaw mechanism, then
threshold corrections from integrating out the heavy neutrinos one by
one will in general lead to non-zero \meff, even if the $ee$ entry of
$m_D^T M_R^{-1} \, m_D$ is zero;  

\item in the type I seesaw mechanism, $m_D^2/M_R$ is
actually only the leading order term. ``Next-to-leading order'' corrections
$m_D^4/M_R^3$ are present and in general induce non-zero terms in the
$ee$ entry, even if $(m_D^T \, M_R^{-1} \, m_D)_{ee}$ is zero\cite{HLR};

\item if another source of lepton number violation is present, then a
non-zero $ee$ entry of the neutrino mass matrix will be induced via
the Schechter-Valle diagrams from Fig.~\ref{fig:blackbox}, even if
$\meff = 0$; 

\item finally, it is plausible that a flavor-blind Planck scale term
is present, which induces an effective mass of order $v^2/M_{\rm Pl}
\simeq 10^{-5}$ eV. This term arises from the Weinberg operator Eq.~(\ref{eq:Leff}) with
the Planck scale inserted as $\Lambda$.  

\end{itemize}

All these sources give of course very small but in general non-zero  
contributions to the effective mass. One might ask whether one can determine experimentally by
other means if the effective mass vanishes. While this is not
possible, one can show however that the effective mass cannot
vanish: from Fig.~\ref{fig:meff_obs} note that 
$\meff \simeq 0$ corresponds to $m_\beta \ls 0.02$ eV and $\Sigma
\ls 0.1$ eV. Thus, finding these quantities above such values
immediately rules out the possibility of vanishing \meff. Of course,
determining experimentally that the inverted mass ordering is realized
also implies that $\meff \neq 0$.

\subsubsection{\label{sec:meffRG}Renormalization}
\begin{figure}[t]
\centerline{
\psfig{file=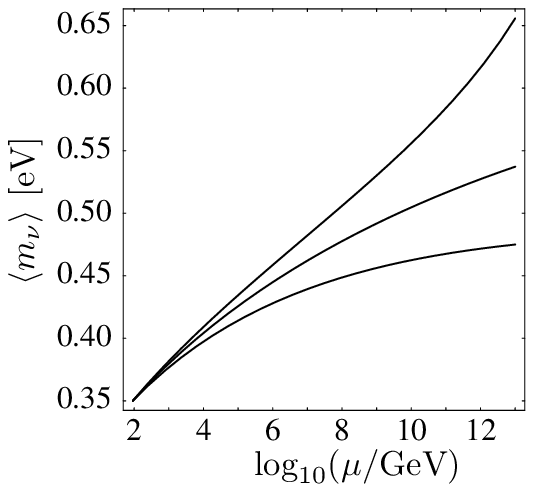,height=5cm,width=6cm}
\psfig{file=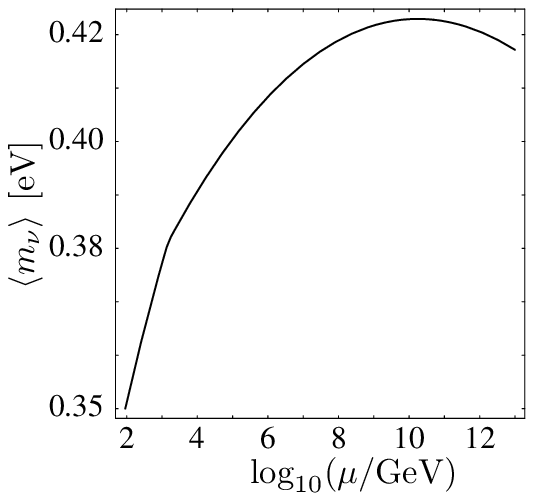,height=5cm,width=6cm}
}
\vspace*{8pt}
\caption{\label{fig:RG}Extrapolation of the effective mass from 0.35
eV at low scale  to higher energies.  
  The SM curves in the left plot correspond to Higgs masses of
  \(114\,\mathrm{GeV}\), \(165\,\mathrm{GeV}\) and 
  \(190\,\mathrm{GeV}\) (from bottom to top).  In the MSSM plot on the
right, the Higgs mass is \(120\,\mathrm{GeV}\), 
$\tan\beta=50$, $M_\mathrm{SUSY}=1.5\,\mathrm{TeV}$. Taken
from\protect\cite{Antusch:2003kp}.} 
\end{figure}

The renormalization group (RG) evolution of neutrino parameters has recently been reviewed
in\cite{Ray:2010rz}. If some unknown high energy theory at a
scale $\Lambda$ leads  to a mass matrix $m^0_\nu$, then in the effective
theory one has the following mass matrix at low scale $\lambda$, where
measurements take place: 
\begin{eqnarray}
m_\nu = I_{\alpha_\nu}\left( 
\begin{array}{ccc}
(m^0_\nu)_{ee} \, I_e^2  & (m^0_\nu)_{e\mu} \, 
I_e \, I_\mu  & (m^0_\nu)_{e\tau} \, I_e \, I_\tau \\ 
\cdot & (m^0_\nu)_{\mu\mu} \, I_\mu^2  & (m^0_\nu)_{\mu\tau} \, I_\mu \, I_\tau \\ 
\cdot & \cdot & (m^0_\nu)_{\tau\tau} \, I_\tau^2 
\end{array} 
\right) , \quad \label{eq:mnu1}
\end{eqnarray}
where 
\be
I_\alpha  \simeq 1 +  \frac{C}{16 \pi^2} \, y_\alpha^2 \,
\ln \frac{\lambda}{\Lambda} \mbox{ and } I_{\alpha_\nu} 
\simeq 1 + \frac{1}{16 \pi^2} \alpha_\nu \, \ln
\frac{\lambda}{\Lambda} \, , 
\ee
with $\alpha, \beta \in \{ e, \mu, \tau \}$, $C = 1$ in the MSSM
and $C = -\frac 32$ in the SM. One can safely drop $y_e$ and $y_\mu$
from the above expression and describe the RG evolution with $I_\tau$
and $I_{\alpha_\nu}$ only. We furthermore have 
\bea  
\alpha^\mathrm{SM}_\nu =
 -3 g_2^2 + 2 (y_\tau^2+y_\mu^2+y_e^2) +
        6 \left( y_t^2 + y_b^2 + y_c^2 + y_s^2 + y_d^2 + y_u^2 \right) 
        + \lambda_{\rm H} \, , 
\\
        \alpha^\mathrm{MSSM}_\nu =
         -\frac{6}{5} g_1^2 - 6 g_2^2 + 6 \left( y_t^2 + y_c^2 + y_u^2 \right).
        \label{eq:alphabarMSSM}
\eea
Here $g_{1,2}$ are the electroweak gauge couplings, $y_x$ the Yukawa coupling of
fermion $x$, and $\lambda_{\rm H}$ the Higgs self-coupling. 
The RG evolution of \meff~is therefore basically a
rescaling of the effective mass with $I_{\alpha_\nu}$. In contrast to the running of
the individual parameters of $\meff$ ($\theta_{12}$, $\theta_{13}$,
$m_1$, $m_2$, $m_3$, $\alpha$ and $\beta$), which can be very
dramatic, the RG evolution of \meff~is modest. Its running does
basically not depend
on the mass ordering or any of the other neutrino mass and mixing
parameters. It is an interesting exercise to consider the $\beta$
functions of the 7 parameters of \meff~and to show that at the end all
dependence on $\theta_{12}$, $\theta_{13}$,
$m_1$, $m_2$, $m_3$, $\alpha$ and $\beta$ drops
out. The effective mass typically increases from 
low to high scale, Fig.~\ref{fig:RG} shows an example for its running\cite{Antusch:2003kp}.

\subsubsection{\label{sec:meff_mod}Distinguishing neutrino models}

\begin{figure}
 \centerline{
 \psfig{file=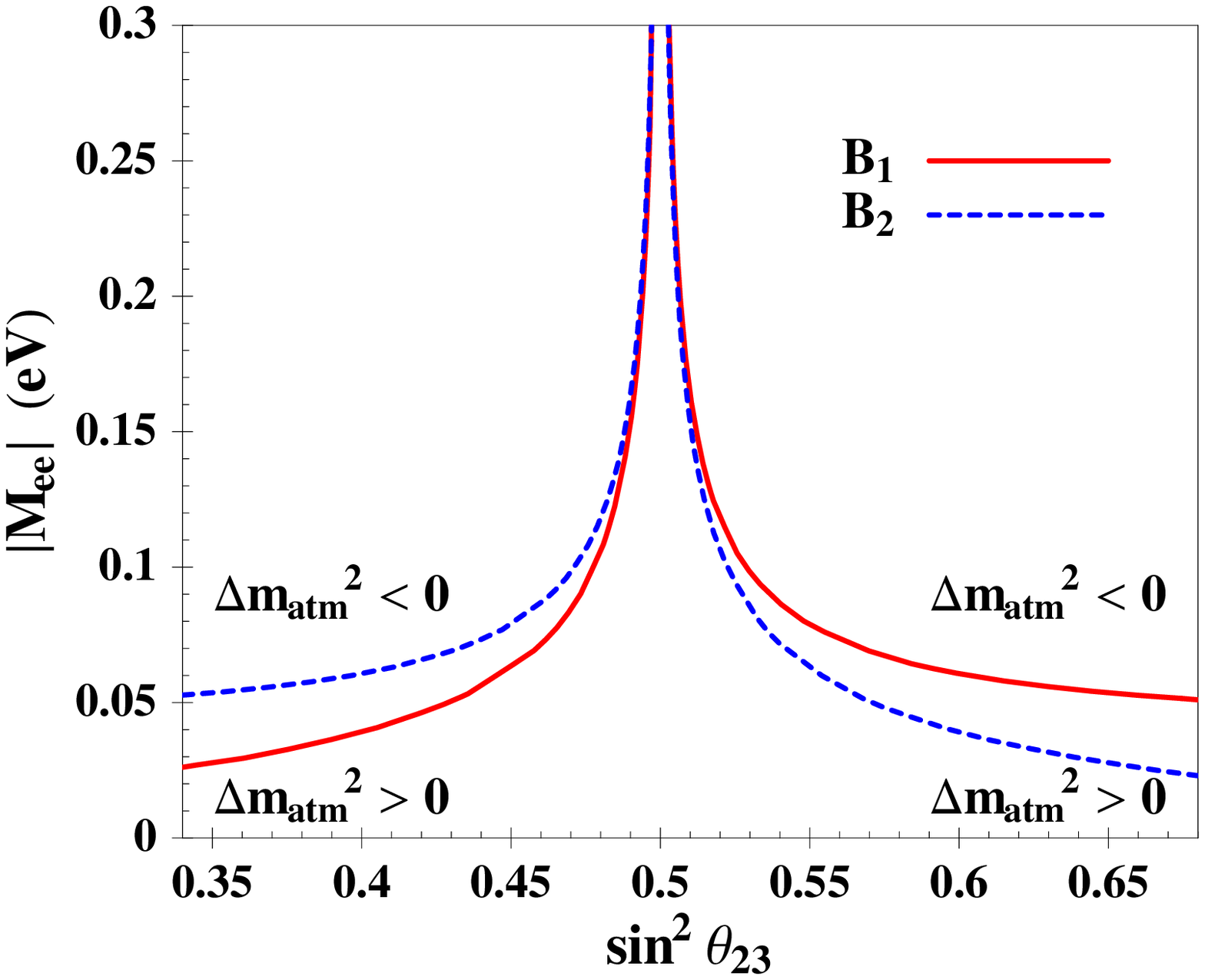,height=7cm,width=6cm}
\psfig{file=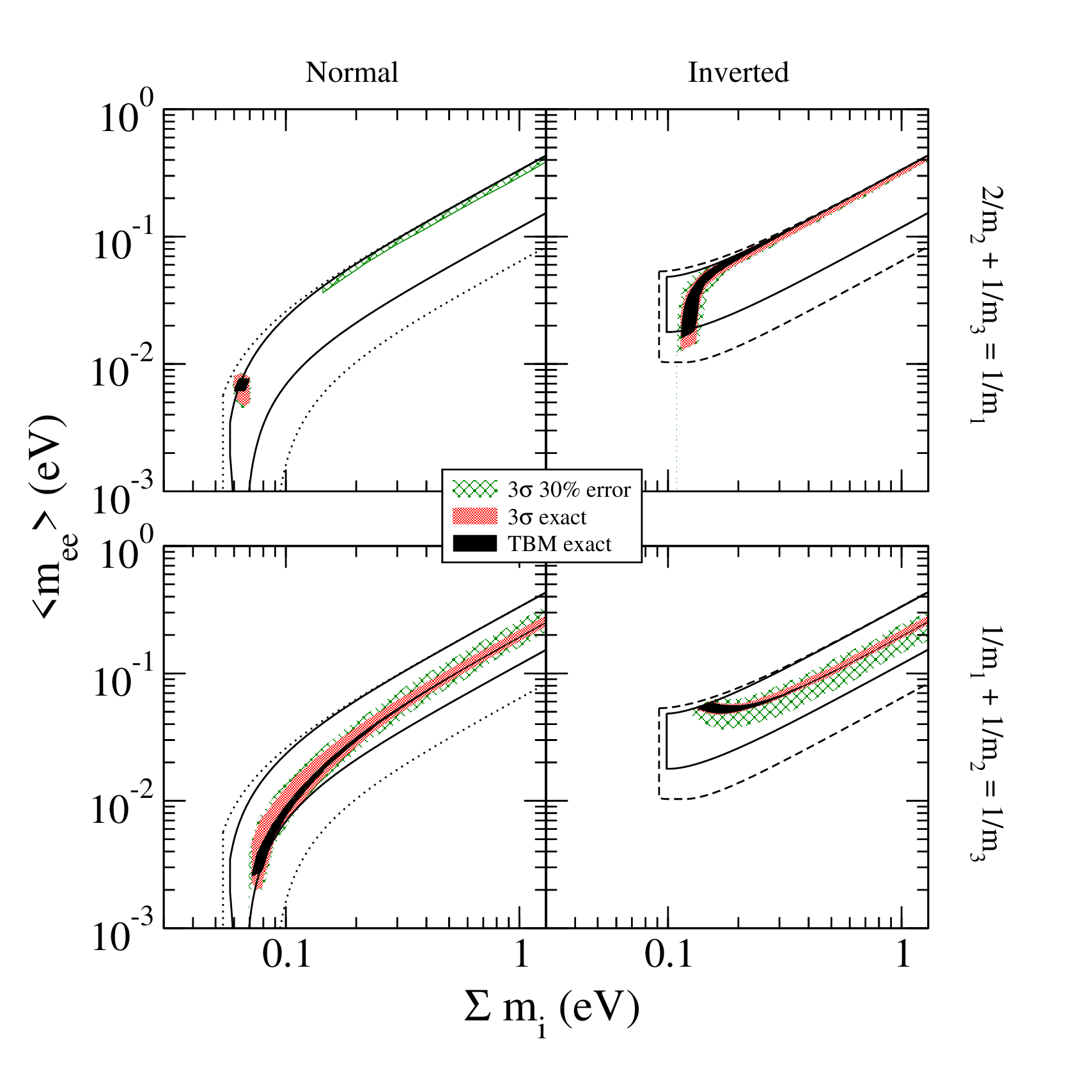,height=7cm,width=9cm}
}
\vspace*{8pt}
 \caption{Left: 
correlation between the effective mass and the atmospheric neutrino
mixing angle in a specific flavor symmetry model.  Taken
from\protect\cite{Hirsch:2007kh}. Right: 
allowed regions in $\meff-\Sigma$ parameter space for the
sum-rules $\frac{2}{m_2}+\frac{1}{m_3}=\frac{1}{m_1}$ (top) and
$\frac{1}{m_1}+\frac{1}{m_2}=\frac{1}{m_3}$ (bottom), for both the
TBM (black)
and $3\sigma$ values (light red) of the oscillation data, as well as
for the sum-rules violated by 30\% (green hatches). Taken
from\protect\cite{Barry:2010yk}.
}
\label{fig:mass_flav}
\end{figure}

We have mentioned in Section \ref{sec:mnu_obs} that the peculiar and
unexpected form of lepton mixing (see Eq.~(\ref{eq:tbm})) is assumed
to have its origin in the presence of flavor
symmetries\cite{flavsym}. 
There is a large abundance of such models, many leading to the same
neutrino mixing scheme, for instance tri-bimaximal mixing (TBM). The question
arises how to distinguish them from one another. It turns out that
neutrino mass observables can help in disentangling the vast amount of
flavor symmetry models. One example is that the flavor symmetry leads
to correlations of the mass matrix elements, which imply correlations
of observables. For instance, the effective mass could be correlated
with the atmospheric neutrino parameter $\sin^2 \theta_{23}$, which
was obtained in a model in Ref.\cite{Hirsch:2007kh}, see
Fig.~\ref{fig:mass_flav}. Recall that in general $\theta_{23}$ has 
no influence on \meff. 

Another point are ``sum-rules'': 
the most general neutrino mass matrix giving rise to TBM is 
\be \label{eq:mnu_tbm}
m_\nu = 
\left(
\bad 
A & B & B \\[0.2cm]
\cdot & \frac{1}{2} (A + B + D) & \frac{1}{2} (A + B - D)\\[0.2cm]
\cdot & \cdot & \frac{1}{2} (A + B + D)
\ea 
\right) .
\ee
As such, the (complex) eigenvalues $ A-B$, $A + 2 \, B$ and
$D$ are independent of the mixing angles: no matter what $A,B,D$
are, the PMNS mixing is given as Eq.~(\ref{eq:tbm}). However, very often the
structure of the mass matrix is simpler than in
Eq.~(\ref{eq:mnu_tbm}), and ``sum-rules'' between the neutrino masses
arise. Examples are\cite{Altarelli:2005yx} $2/m_2 + 1/m_3 = 1/m_1$ or $1/m_2 + 1/m_3 =
1/m_1$, and detailed studies of the predictions can be found
in Ref.\cite{Barry:2010yk}, from which we took the right plot in 
Fig.~\ref{fig:mass_flav}. Other discussions on mass-related
phenomenology of flavor symmetry models can be found in\cite{mass_flav_others}.

\subsubsection{\label{sec:sterile}Light sterile neutrinos}
The easiest way to depart from the standard 3 neutrino picture discussed
so far is to add light sterile neutrinos. In fact,
we have mentioned in Section \ref{sec:neutrinos} several hints which point to the existence of
additional radiation in the Universe, as well as for one or two
additional mass-squared differences in the eV regime. With one or two
light sterile neutrinos the PMNS matrix becomes a unitary $4\times4$
or $5\times5$ matrix. Short-baseline oscillations depend on $U_{e4}$
(and $U_{e5}$) as well as $U_{\mu 4}$ (and $U_{\mu 5}$).  Only 
eV-like mass-squared differences play a role for such experiments, and
only if two sterile states are added does the possibility of CP violation
arise. This can explain the different neutrino and anti-neutrino
results from MiniBooNE and MiniBooNE plus LSND, respectively. 

Table \ref{tab:global-bfp} shows the results\cite{Kopp:2011qd} from a global fit to
the world's short-baseline data, taking into account the recent
re-evaluation of reactor fluxes\cite{thierry}. The data are not
sensitive to whether the
two sterile neutrinos are above or below the three active ones (2+3 or
3+2 scenarios), but are sensitive to whether the active neutrinos are sandwiched
between two sterile ones (1+3+1). Note that the fit to 1+3+1 scenarios
is slightly better than for 3+2/2+3 scenarios. Fitting only the
reactor experiments is possible in a 3+1 or 1+3 scenario, and  
gives\cite{Kopp:2011qd} $|U_{e4}| = 0.151$ and $\Delta m^2_{41} =
1.78$ eV$^2$.

If there are two sterile neutrinos, the nomenclature for the 8 possible mass orderings is as
follows: 
\begin{itemize} 
\item[(i)] SSX, where X = N for a normal and 
X = I for an inverted ordering of the mostly active neutrinos. 
In these schemes the two
 predominantly sterile neutrinos are heavier 
than the three predominantly active neutrinos (2+3 scenarios); 
\item[(ii)] XSS (X = N or I as before), where the two 
 predominantly sterile neutrinos are lighter than the three 
predominantly active neutrinos (3+2 scenarios); 
\item[(iii)] SXS with X = N or I, where the three active neutrinos 
are sandwiched between the sterile ones (1+3+1 scenarios). In 
this class there can be four possible 
scenarios, which we denote as SXSa and SXSb. 
The scheme SXSa corresponds to the state $\nu_5$ higher 
than the three active states 
and SXSb corresponds to the state 
$\nu_5$ lower than the three active states. 
 
\end{itemize}
The individual masses, expressed in terms of smallest mass and the four
mass-squared differences, can be found in Ref.\cite{131}. 
The new sterile neutrinos will contribute to the sum of masses in
cosmology, to the kinematic mass in KATRIN, and, 
if they are Majorana particles, to
\obb. The effects of sterile neutrinos on neutrino-less double beta decay
have been studied by various 
authors\cite{meff_ster,KlapdorKleingrothaus:2000gr,Farzan:2001cj,131,BRZ}.
One simply extends the sums in the definitions of $\Sigma$, $m_\beta$
and $\meff$ from $i=3$ to $i=5$. 
The interpretation of reactor experiments actually observing
oscillations, which is possible after the new reactor
fluxes\cite{thierry} are taken into account, makes the
application of the results to \obb~easier. If only LSND and MiniBooNE
supply oscillation data, the $\stackrel{(-)}{\nu_\mu} \to \stackrel{(-)}{\nu_e}$ 
transitions depend on 
$U_{e i}$ and $U_{\mu i}$ ($i = 4,5$). For $m_\beta$ and $\meff$ only
$U_{e i}$ is required, and one needs to assume something about $U_{\mu
i}$ to extract $U_{e i}$ from the fit results. However, reactor oscillation
survival probabilities depend only on $U_{e i}$. 
There are two more Majorana phases which show up in the 
modified effective mass, which is the sum of the contribution
considered so far ($\meff^{\rm act}$) plus new terms from the sterile
states ($\meff^{\rm st}$):  
\[ 
\meff' = | \underbrace{|U_{e1}|^2 \, m_1 + |U_{e2}|^2 \, m_2 \, e^{2 i
\alpha} + |U_{e3}^2| \, m_3 \, e^{2 i \beta} 
}_{\meff^{\rm act}} + 
\underbrace{|U_{e4}|^2 \, m_4 \, e^{2 i \Phi_1} + |U_{e5}|^2 \, m_5 \, 
e^{2 i \Phi_2} }_{\meff^{\rm st}} | . 
\] 
The usual phase space factors and matrix elements
as in the standard interpretation apply (nuclear physics is 
the same as for the standard case, as long as the
masses do not exceed $q \simeq 100$ MeV). The additional contribution
$\meff^{\rm st}$ from the sterile neutrinos could be leading, 
sub-leading, or of the same order of magnitude as the active neutrino 
part $\meff^{\rm act}$.

\begin{table}[t]
\tbl{\label{tab:global-bfp}Parameter values and $\chi^2$ at the global
    best-fit points for 3+2 and 1+3+1 oscillations. Taken from\protect\cite{Kopp:2011qd}.}
{\begin{tabular}{cccccccc}
  & $\Delta m^2_{41} [\rm eV^2]$ & $|U_{e4}|$ & $|U_{\mu 4}|$ & 
    $\Delta m^2_{51} [\rm eV^2]$ & $|U_{e5}|$ & $|U_{\mu 5}|$ 
& $\chi^2$/dof\\
    \toprule 
    3+2/2+3 &
    0.47 & 0.128 & 0.165 & 
    0.87 & 0.138 & 0.148 
& $110.1/130$\\
    1+3+1 &
    0.47 & 0.129 & 0.154 & 
    0.87 & 0.142 & 0.163 
& $106.1/130$\\ \botrule 
  \end{tabular} }
\end{table}

Neglecting the smallest neutrino mass and using the best-fit values
from Table \ref{tab:global-bfp} gives the following predictions for
the mass observables\cite{BRZ}: 
\begin{itemize}
\item[SSN:]
the active neutrinos give the same contribution as in
the standard case for NH. The contribution to the sum of masses from the
sterile states dominates and is given by $\Sigma = \sqrt{\Delta m^2_{41}} +
\sqrt{\Delta m^2_{51}} \simeq 1.62$ eV. The contribution to $m_\beta$
is $\sqrt{\Delta m^2_{41} |U_{e 4}|^2 + \Delta m^2_{51} |U_{e 5}|^2 }
\simeq 0.16$ eV. The contribution to the effective mass is 
$\left| |U_{e 4}|^2 \, \sqrt{\Delta m^2_{41}} + e^{2i (\Phi_1 - \Phi_2)} \, |U_{e 5}|^2 \,
\sqrt{\Delta m^2_{51}} \right|$, which is between $0.007$ and $0.029$ eV, and
hence larger than the typical value for NH. Thus, the effective mass cannot
vanish (for the best-fit point), in contrast to the standard case.  

\item[SSI:] the active neutrinos give the same contribution as in
the standard case for IH, and the sterile states give the same
predictions for the mass observables as in SSN. The effective mass can
therefore vanish, in contrast to the standard three neutrino case.  

\item[NSS:] the active neutrinos are QD (normal ordering) with a mass scale $\sqrt{\Delta
m^2_{51}} \simeq 0.93$ eV, and this governs the predictions
for $\meff$ and $m_\beta$. For cosmology, $\Sigma = 3 \, \sqrt{\Delta
m^2_{51}} + \sqrt{\Delta m^2_{51} - \Delta m^2_{41}} \simeq 3.4$ eV. 

\item[ISS:] same as NSS, except for inversely ordered active neutrinos. 

\item[SNSa:] the active neutrinos are QD (normal ordering) with a mass scale $\sqrt{\Delta
m^2_{41}} \simeq 0.69$ eV, and this defines the predictions
for $\meff$ and $m_\beta$. The sum of masses is $\Sigma = 3 \, \sqrt{\Delta
m^2_{41}} + \sqrt{\Delta m^2_{41} + \Delta m^2_{51}} \simeq 3.2$ eV.

\item[SNSb:] the active neutrinos are QD (normal ordering) with a mass scale $\sqrt{\Delta
m^2_{51}} \simeq 0.93$ eV, and this defines the predictions
for $\meff$ and $m_\beta$, up to a small correction of order
$|U_{e4}|^2 \, \sqrt{\Delta m^2_{51} + \Delta m^2_{41}} \simeq 0.03$
eV e.g.~for \meff. The sum of masses is $\Sigma = 3 \, \sqrt{\Delta
m^2_{51}} + \sqrt{\Delta m^2_{41} + \Delta m^2_{51}} \simeq 4.0$ eV. 

\item[SISa:] same as SNSa, except for inversely ordered active neutrinos. 

\item[SISb:] same as SNSb, except for inversely ordered active neutrinos. 

\end{itemize} 

\begin{figure}
 \centerline{
 \psfig{file=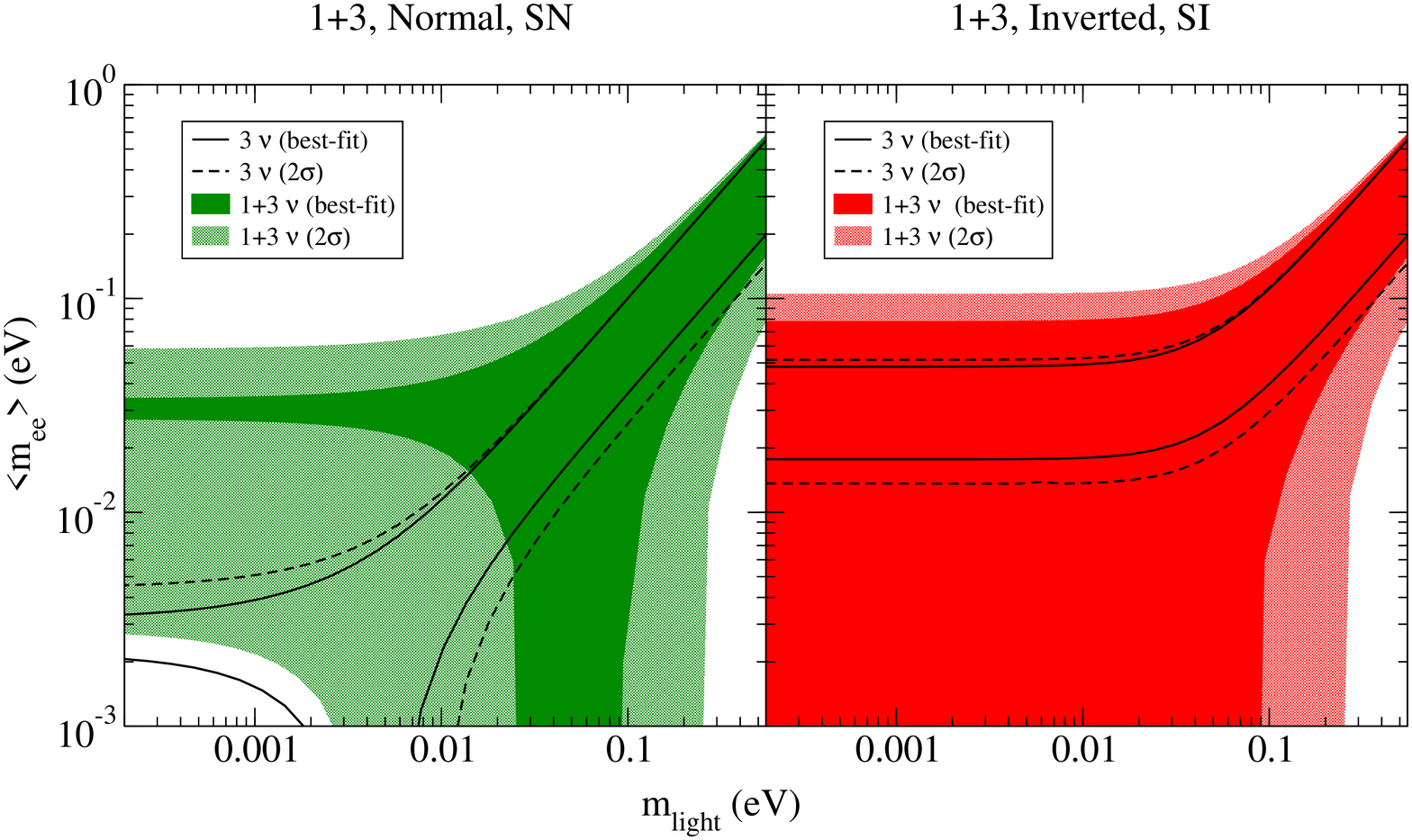,height=6cm,width=10cm} }

\centerline{
\psfig{file=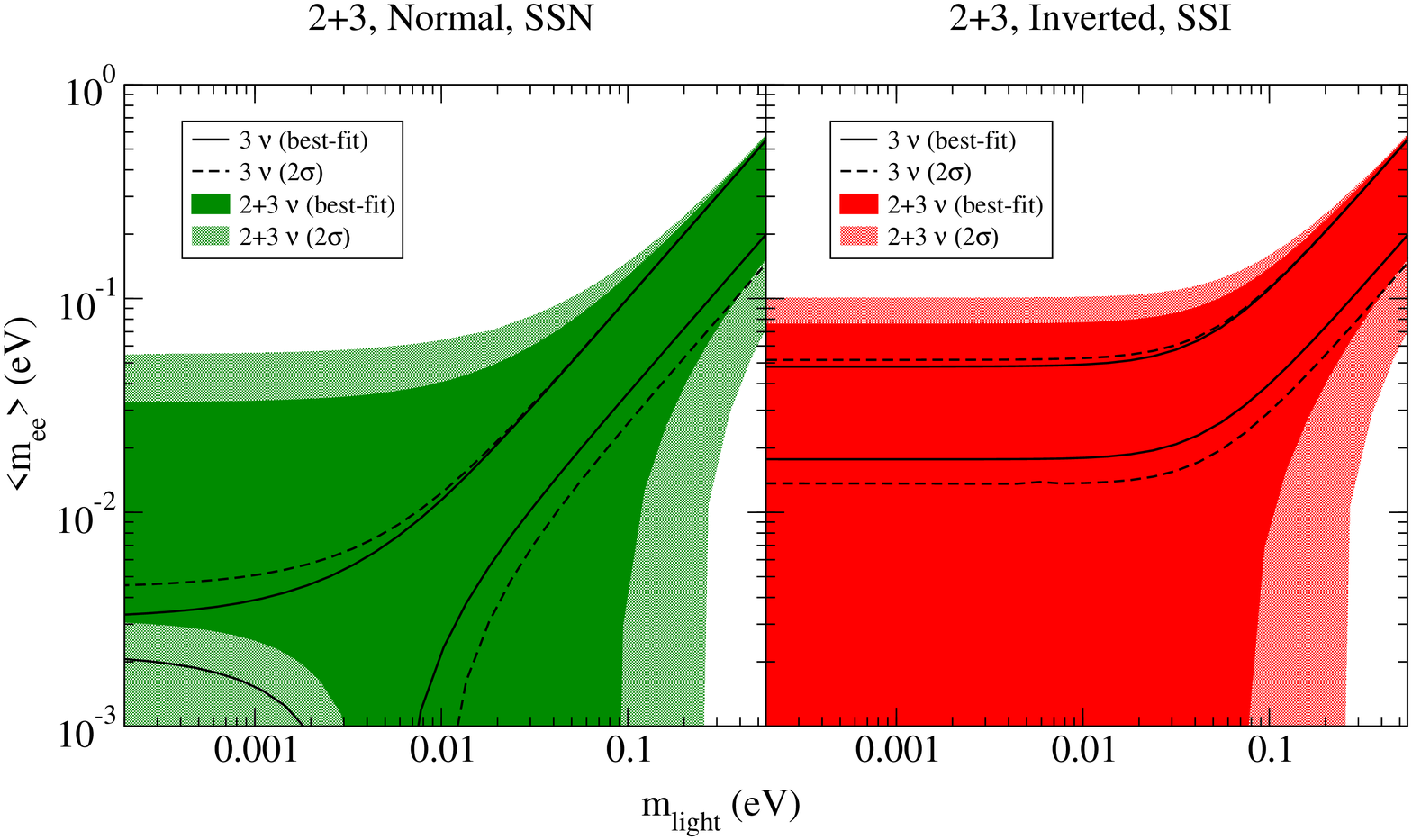,height=6cm,width=10cm}
}
\vspace*{8pt}
 \caption{Top: Effective mass against the smallest mass for the 
1+3 scheme, with $\Delta m^2_{41} = 1.78$ eV$^2$, $|U_{e4}|^2 =
0.023$ (dark) and the $2\sigma$ range $\Delta m^2_{41} = (1.61-2.01)$ eV$^2$, $|U_{e4}|^2
=  0.006 - 0.040$ (light). Bottom: same as above for 2+3 scheme, with 
$\Delta m^2_{41} = 0.47$ eV$^2$, $\Delta m^2_{51} = 0.87$ eV$^2$,
$|U_{e4}|^2 = 0.016$, $|U_{e5}|^2 = 0.019$ (dark) 
and  $\Delta m^2_{41} = (0.42-0.52)$ eV$^2$, $\Delta m^2_{51} = (0.77
- 0.97)$ eV$^2$, $|U_{e4}|^2 = 0.004 - 0.029$, $|U_{e5}|^2 = 0.005 -
0.033$ (light). The black solid and dashed lines correspond to the
standard 3 neutrino best-fit and $2\sigma$ cases. }
\label{fig:meff_st}
\end{figure}

Note that these are all lower limits, because we have put the smallest
neutrino mass to zero. In case of SSI the effective mass can vanish
when the 3 active neutrinos are inversely ordered, in contrast to the
three neutrino case in Section \ref{sec:meff_inverted}. If future
\obb-experiments measure a tiny effective mass, or obtain a limit
below $\meff_{\rm min}^{\rm IH}$ given in Eq.~(\ref{eqn:meffminih}),  and the neutrino mass
ordering is confirmed to be inverted from long-baseline neutrino oscillations, the sterile
neutrino hypothesis would be an attractive explanation for this
inconsistency. This is the first example in which a deviation from the
standard picture of 3 active neutrinos shows up and influences the
interpretation of \obb. We show in Fig.~\ref{fig:meff_st} the 
effective mass against the smallest mass for the 1+3 and 2+3 cases.

Obviously all schemes have difficulties with cosmology, the
contribution to the sum of masses exceeds 1.5 eV in all cases. KATRIN 
and next generation \obb-experiments 
will see a signal in all cases except for SSI and SSN, unless the
masses and mixings take values at the very high end of their currently
allowed ranges. An analysis of
KATRIN's potential to separate one or more sterile 
neutrino component from the active neutrino component has been
performed in\cite{anna}. It was shown that KATRIN will definitely be
able to separate one or more sterile 
neutrino components from the active neutrino ones, if they do in fact have 
mass and mixing in the range considered here. 
With a limit
on the effective mass being around 0.5 eV (see Table
\ref{tab:limits}), the schemes with QD neutrinos with mass scale
$\sqrt{\Delta m^2_{41}}$ or $\sqrt{\Delta m^2_{51}}$ are in fact 
already tested, giving constraints on the Majorana phases
$\alpha$ already at the current stage\cite{131}.

If the ``reactor only'' results of  $|U_{e4}| = 0.151$ and $\Delta
m^2_{41} = 1.78$ eV$^2$ are used, then $\Sigma \gs \sqrt{\Delta
m^2_{41}} \simeq 1.3$ eV or $\Sigma \gs 3 \, \sqrt{\Delta m^2_{41}}
\simeq 4.0$ eV, depending on whether a 1 + 3 or 3 + 1 scheme is
realized. The contribution to KATRIN is either 0.52 or 1.3 eV, and the
effective mass either receives  a contribution of $0.03$ eV, or
corresponds to QD neutrinos with a mass scale $\sqrt{\Delta
m^2_{41}}$. For the 1+3 case, again, the effective mass can vanish if
the active neutrinos are inversely ordered.

The existence of sterile neutrinos can also be tested in upcoming
oscillation experiments and via cosmological observations, as
discussed in Section \ref{sec:mnu_obs}.

\subsubsection{\label{sec:mnuNSI}Exotic modifications of the three neutrino picture}
Exotic modifications of the 3-neutrino framework are possible and may
spoil the discussion presented so far in this Section.

The most obvious modification is that neutrinos are 
{\bf Dirac particles}, in which case there is no \onbb~and writing
this review was all in vain. A useful way to show this in the
effective mass is to note that one Dirac neutrino can be written as
two maximally mixed Majorana neutrinos with common mass $m_i$ and 
opposite CP parity. The effective mass is then 
\be
\sum\limits_i \sqrt{\frac 12} \, 
|U_{ei}|^2 \, \left( m_i + m_i \, e^{i \pi} \right) = 0 \, . 
\ee
A small splitting of the degeneracy can be described with the mass
matrix
\be
m_i 
\left( \baz 
\epsilon & 1 \\
1 & 0 
\ea \right) \rightarrow 
U = \sqrt{\frac12 } \left( 
\baz 
1 + \frac \epsilon 4 & -1 + \frac \epsilon 4 \\
1 - \frac \epsilon 4 & 1 + \frac \epsilon 4
\ea
\right) \mbox{ and } 
m_i^{\pm} = m_i \left(\pm 1 + \frac \epsilon 2 \right) , 
\ee
with the indicated new eigenstates and mixing matrix. 
These {\bf Pseudo-Dirac neutrinos} lead to a contribution to the effective
mass of about $\epsilon \, m_i  = \frac 12 \, \delta m^2 /m_i$, with $\delta m^2 =
(m_i^+)^2 - (m_i^-)^2$. Regarding limits on such splitting, 
roughly speaking, values larger than $\delta m^2 \simeq 10^{-11}$
eV$^2$ for $m_1$ and $m_2$ are forbidden by solar neutrino data, and $\delta m^2 \gs
10^{-3}$ eV$^2$ for $m_3$ by atmospheric data\cite{deGouvea:2009fp}.  If all three states
are Pseudo-Dirac the effective mass is basically zero\cite{Maalampi:2009wm}, 
while if one or two are Pseudo-Dirac interesting predictions for the
effective mass arise. This can happen in ``bimodal'' or {\bf ``schizophrenic''}
scenarios\cite{schizo}, in which at leading order one or two mass
states are Dirac particles while the other one is Majorana. Because lepton number is not
conserved, loop corrections imply small Pseudo-Dirac terms for the
Dirac states. For instance, if $\nu_2$ is a Dirac particle then the
effective mass in the inverted hierarchy is\cite{schizo} $\meff \simeq \sqrt{\dma}
\, c_{12}^2 \, c_{13}^2$, roughly a factor of two larger than the
minimal  value in the standard case, see Eq.~(\ref{eq:meeIHleft}). A
generalization to all possibilities can be found in\cite{schizo1}. 

A possible modification of the three neutrino picture mentioned 
before is the possible {\bf non-unitarity of the PMNS matrix},
which has however negligible effect on the effective mass\cite{Rodejohann:2009ve}. 

Another exotic property is {\bf CPT violation}. Interesting consequences
for \obb~have been considered in\cite{Barenboim:2002hx}, where a simple
one family example is discussed. In the $(\nu, \bar{\nu})$ basis, 
where CPT transforms $\nu$ into $\bar\nu$ up to a phase, the 
mass matrix can be written as 
\be
M = \left( 
\baz 
\mu + \Delta & y^\ast \\
y & \mu - \Delta 
\ea 
\right) . 
\ee
Here $y$ mixes $\nu$ and $\bar\nu$, while $\Delta$ leads to different
masses for  $\nu$ and $\bar\nu$. 
The eigenstates $\nu_\pm$ with masses $m_\pm = \mu \pm \sqrt{|y|^2 +
\Delta^2}$ can be shown to be Majorana neutrinos (i.e.~CPT transforms
$\nu_\pm$ into $\nu_\pm$ up to a phase) only if $\Delta = 0$. This in
turn would imply however that CPT is conserved. CPT is violated for $\Delta \neq
0$, in which case neutrinos cannot be Majorana particles.  The
amplitude for \obb~sums over $m_+$ and $m_-$ and is
non-zero\cite{Barenboim:2002hx}. Therefore, \onbb~takes place even if
neutrinos are strictly speaking not Majorana particles. The neutrino-less double
positron decay proceeds with the same ``effective mass''.

In principle \obb~can also provide limits on parameters associated
with violation of {\bf Lorentz invariance} or the {\bf equivalence
principle}. The constraints\cite{KlapdorKleingrothaus:1998hm} on the
difference of maximal velocities of
mass states or on non-universal couplings of neutrinos to the 
gravitational potential are in general weaker than the ones from neutrino
oscillations\cite{Fogli:1999fs}.

It should be mentioned here that \zbb~constrains violation of  
the {\bf spin-statistics theorem} for neutrinos. With two
identical particles in the final state there are two diagrams with
exchanged  momenta $p_1 \leftrightarrow  p_2$. Their relative sign 
depends on whether Fermi-Dirac or Bose-Einstein statistics applies. 
By writing the amplitude ${\cal A}$ as 
$\cos^2 \phi \, {\cal A}_{\rm fermionic} + \sin^2 \phi \, {\cal
A}_{\rm bosonic}$, conservative limits of  $\sin^2 \phi \ls 0.5$ can be
set\cite{WTF?}.

\section{\label{sec:non-standard}Non-Standard Interpretations}
After discussing is some detail the standard interpretation of \onbb,
we turn to non-standard interpretations, repeated here for
convenience: 

{\it Neutrino-less double beta decay is mediated by some other lepton number
violating physics, and light massive
Majorana neutrinos (the ones which oscillate) potentially leading 
to \obb~give negligible or no contribution.}

It is convenient to express the decay width of \onbb~in the following
form (see Eq.~(\ref{eq:fact}))
\be \label{eq:NS}
\Gamma^{0\nu} = \sum\limits_x G_x (Q,Z) \,  
\left| {\cal M}_x \, \eta_x  \right|^2 .
\ee
Here we sum over all possible mechanisms which are denoted by a
subscript $x$, with matrix element ${\cal M}_x$ and a dimensionless particle
physics parameter $\eta_x$. For
the standard interpretation of light neutrino exchange,  
\be \label{eq:eta_l}
\eta_{\rm l} = \meff/m_e \ls 9.9 \times 10^{-7} \, . 
\ee
Note that different mechanisms can
interfere coherently, a case we will discuss in Section
\ref{sec:distinguish_simul}. 
Most of the times the alternative mechanism is
connected to a high energy scale. The corresponding 
particle physics amplitude, which has to be compared with the standard one $G_F^2 \, \meff/q^2$ from
Eq.~(\ref{eq:am_SI}), could be written as   
\be \label{eq:am_naive_heavy}
{\cal A}_{\rm heavy} \simeq \frac{c}{\Lambda^5} \, . 
\ee
Here $c$ contains new Yukawa and/or gauge couplings and $\Lambda$ is the
new physics scale. This is a helpful but crude approximation, which is in fact not
fulfilled by several mechanisms to be discussed in the following. 
However, the current limit $\meff = 0.5 $ eV corresponds to $\Lambda
\simeq $ TeV, by all means an interesting energy scale. In fact, we
will encounter in what follows some alternative mechanisms with LHC phenomenology.  On
the other hand, it means that if the new physics scale exceeds, say,
10 TeV, then it will not contribute significantly to \obb. In what follows we will
aim at a complete list of non-standard realizations of \onbb, for
earlier reviews see\cite{KlapdorKleingrothaus:1998yy,Mohapatra:1998ye}. 

An ideal experimental signature for drawing the conclusion that a
mechanism different from active neutrino exchange is present would be
that KATRIN and cosmological observations do not see a signal, but
\obb~is observed with a half-life corresponding to, say, $\meff \simeq
0.5$ eV. To put in another way, in plots of neutrino mass observables,
such as in Fig.~\ref{fig:meff_obs}, one ends at points outside the
allowed areas. 
On the other hand, if one ends in those plots in the
allowed areas, then it is not necessary to consider non-standard
interpretations, except for setting limits on the associated
parameters.

\subsection{\label{sec:heavy}Heavy neutrinos}
\begin{figure}[t]
 \centerline{
\psfig{file=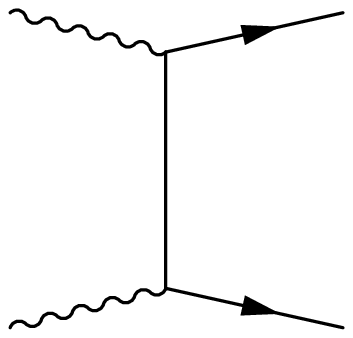,height=4cm,width=4cm}\quad\quad\quad\quad
 \psfig{file=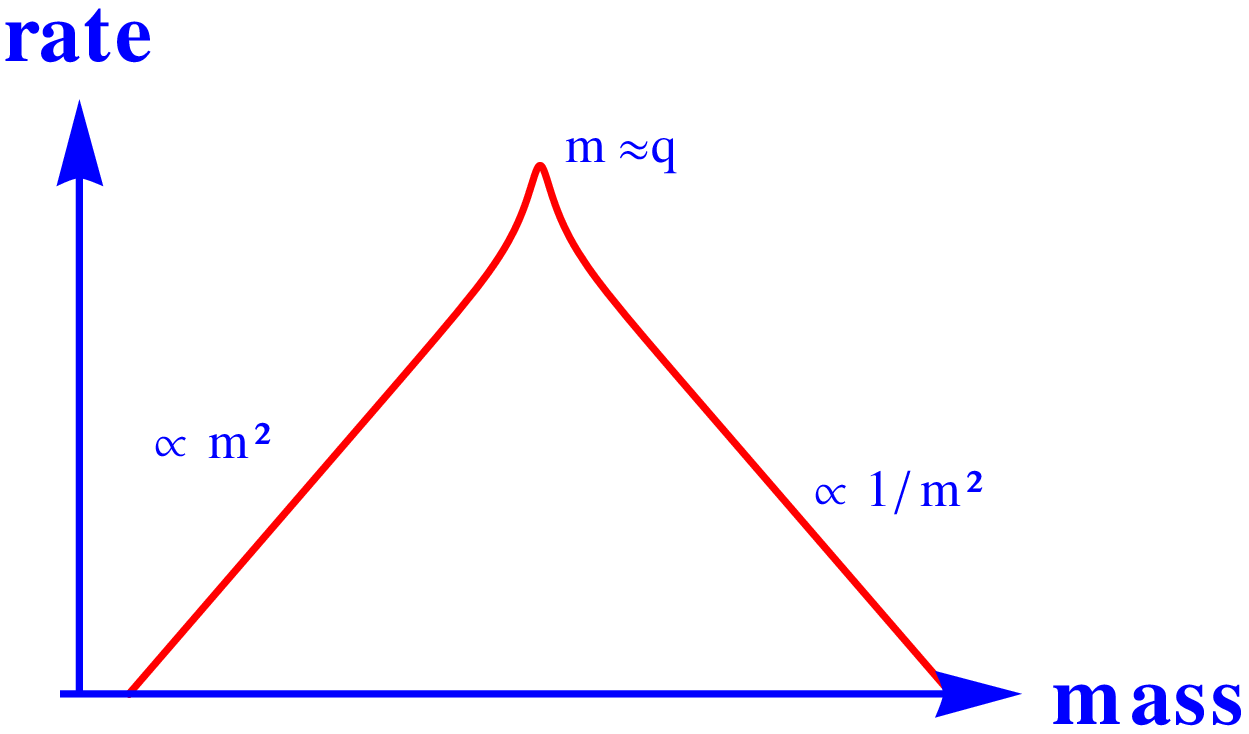,height=5cm,width=6cm}
}
\vspace*{8pt}
 \caption{Left: ``lobster'' diagram for lepton number violating
processes with Majorana neutrino exchange. Right: typical behavior of
\obb-like processes with free Majorana
mass $m$: for small masses $m^2 \ll q^2$ the rate increases with $m^2$, while for
large masses  $m^2 \gg q^2$ it decreases with $m^{-2}$. Here ``small'' and ``large'' are
defined relative to the energy scale $q^2$ of the process, which could be
the mass of a decaying particle/nucleus, or the center of mass energy
of a collider process. The maximum rate can be expected when the mass
corresponds to the available energy. }
\label{fig:heavy_sym}
\end{figure}

An interesting way of realizing \obb~is through the exchange of heavy
Majorana neutrinos\cite{heavy,Halprin:1976mr,Halprin:1983ez}. The Feynman diagram is the same as in
Fig.~\ref{fig:FD_mass_mech}, with the neutrinos not being the ones
whose oscillations are observed, and with the 
PMNS matrix elements $U_{ei}$
replaced by $S_{ei}$, where $S$ is the matrix describing the mixing of
the heavy neutrinos with the SM charged leptons in the charged current
term. Recall the form of
the \obb-amplitude on the particle physics level:  
\be \label{eq:am_gen}
{\cal A} \propto  \frac{m_i}{q^2 - m_i^2} 
\propto \left\{ 
\baz 
m_i & \mbox{ for } q^2 \gg m_i^2 \, , \\ \D
\frac{1}{m_i} &  \mbox{ for } q^2 \ll m_i^2 \, . 
\ea 
\right.  
\ee
As mentioned before, due to the typical structure of the diagram,
symbolically displayed in Fig.~\ref{fig:heavy_sym}, one sometimes
calls it ``lobster diagram''.  
With \obb~being a $t$ (and $u$-) channel process there is no
resonance. In Fig.~\ref{fig:heavy_sym} we show the typical behavior of \obb-like
processes as a function of the Majorana mass. 
Let us stress that the maximum rate can be expected when the mass
corresponds to the available energy, i.e.~about 100 MeV for \obb. In
analogous processes of \onbb~(see Section \ref{sec:alt}) the energy
scale and therefore the range in which the strongest limits on the
mass and mixing arise, may be different. In addition, there could be $s$-channel
processes in which a resonance could be hit, leading to even stronger 
constraints. 
If $m_i \gs 100$ MeV, the rate is proportional to 
\be
\sqrt{\Gamma^{0\nu}} \propto \langle \frac 1 m \rangle \equiv \sum\limits_i
\frac{S_{ei}^2}{m_i}  \, . 
\ee
We will focus on this case of heavy neutrinos. 

Turning to nuclear physics, the neutrino potential in Eq.~(\ref{eq:nupot}) is modified because the
neutrino energy and momentum are dominated by its heavy mass. A
dependence on the axial mass $M_A \simeq 0.9$ GeV, which appears in
the nuclear form factors and which takes into account the finite size of the
nucleons, is introduced because of the short-range nature of the
process. Without the form factor the process would be exponentially
suppressed due to the repulsion of the two decaying nuclei. 
Note that this is the first diagram for
\obb~which is purely short-range or point-like. Details of the nuclear physics 
can be found in Refs.\cite{Vergados:1982wr,haxton,Gnu2,Benes:2005hn,Blennow:2010th}. 
Often one writes the contribution of heavy neutrinos as 
$ \sqrt{\Gamma^{0\nu}} \propto \langle \frac 1 m \rangle \, M_A^2 \,
F(A,m_i)$,   
where $F(A,m_i) = {\cal O}(0.1)$ is a mildly varying function. One can
write the decay width as $\Gamma^{0\nu}_{\rm h} = G_{\rm h}(Q,Z) \, |\eta_{\rm h}
\, {\cal M}_{\rm h}|^2$, where in the context of Eq.~(\ref{eq:NS}) one can write the LNV parameter
for heavy neutrino exchange as $m_p \, \imeff$, with $m_p$ the proton
mass. 
The same phase space factors as in the
standard case apply, and the matrix elements absorb now various factors
such as the dependence on
$M_A$ or $F(A,m_i)$. The particle physics parameter $\langle \frac 1 m
\rangle$ contains all singlet fermions coupling with SM charged
leptons in the charged current terms, including heavy neutrinos from
the type I and III seesaw mechanisms, as well as generalizations
thereof, such as inverse seesaw\cite{Ibarra:2010xw}.

Ref.\cite{Faessler:2011rv} has recently calculated within the QRPA
approach the NMEs 
${\cal M}_{\rm h}$ in the
above convention and found a range of roughly a factor
of two: 172 -- 412
for $^{76}$Ge, 165 -- 408 for $^{82}$Se, 185 -- 404 for $^{100}$Mo and
171 -- 384 for $^{130}$Te. 
The spread originates from variation of
$g_A$, the nucleon-nucleon potential and the model space size. 
The NMEs seem to be much larger than the ones for
the standard case, but as mentioned above they absorb several
parameters. 
With the current limits on the half-lifes from Table \ref{tab:limits}
and the phase space factors from Table \ref{tab:Gnu}, we 
find 
\be \label{eq:heavy}
\langle \frac 1 m \rangle \le \left\{ 
\baz 
(0.75 - 1.8) \times 10^{-8} \, \,{\rm
GeV}^{-1} & \mbox{ for } ^{76}{\rm Ge} \, , \\
(2.8 - 6.9) \times 10^{-8} \, \,{\rm
GeV}^{-1} & \mbox{ for } ^{82}{\rm Se} \, , \\
(1.3 - 2.8) \times 10^{-8} \, \,{\rm
GeV}^{-1} & \mbox{ for } ^{100}{\rm Mo} \, , \\
(0.82 - 1.8) \times 10^{-8} \, \,{\rm
GeV}^{-1} & \mbox{ for } ^{130}{\rm Te} \, , \\
 \ea\right.
\ee
The dimensionless LNV parameter for heavy neutrino exchange is 
\be  \label{eq:eta_h}
\eta_{\rm h} = m_p \, \langle \frac 1 m \rangle  \le 1.7 \times
10^{-8} \, , 
\ee
where $m_p$ is the proton mass.
Interestingly the best limit $\langle \frac 1 m \rangle \le 1.8 \times 10^{-8} \, \,{\rm
GeV}^{-1}$ stems jointly  from $^{76}{\rm Ge}$ and
$^{130}{\rm Te}$. For heavy neutrinos the limit $\sum_i |S_{ei}|^2 \le 0.0052$ from global
fits applies, and this constraint on $|S_{ei}|^2$ is stronger for $m_i
\gs 2.9 \times 10^5$ GeV.  Naively, one can simply compare the particle physics
amplitudes $G_F^2 \, \meff/q^2$ and $G_F^2 \, \langle \frac 1 m \rangle$. With $\meff
\ls 0.5$ eV and $q \simeq 100$ MeV it follows that $\langle \frac 1 m
\rangle \ls 5 \times 10^{-8}$ GeV$^{-1}$, which is basically the same number as 
Eq.~(\ref{eq:heavy}), given the NME uncertainty. As we will see in the
following, the comparison of an alternative mechanism of \obb~to the
standard mechanism {\it on the amplitude level} 
is remarkably successful, and gives constraints which are consistent
with literature values taking the onerous nuclear
physics aspects into account. Formulated provocatively, matrix elements are order one numbers
with a corresponding uncertainty, and comparing the particle physics
amplitudes introduces an order one factor uncertainty, which often
is good enough to understand the particle physics implications of \obb.

 \begin{figure}[b]
 \centerline{
 \psfig{file=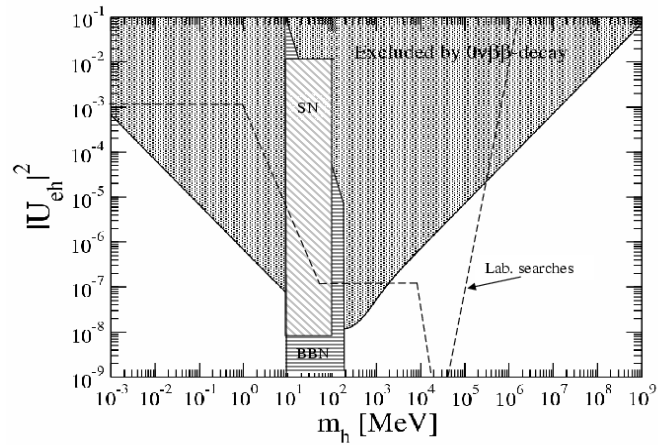,height=6cm,width=9cm}
}
\vspace*{8pt}
 \caption{Exclusion plot in the $|U_{eh}|^2$--$m_h$ plane, where 
$m_h$ and $|U_{eh}|$ are heavy neutrino masses and their mixing with the SM
electron doublet. 
The shaded regions are excluded by $0\nu\beta\beta$-decay, by 
Big Bang nucleosynthesis and by SN1987A neutrino observations. Taken
from\protect\cite{Benes:2005hn}.
}
\label{fig:heavy}
\end{figure}

Fig.~\ref{fig:heavy} shows the exclusion limits on mass and mixing of
heavy sterile neutrinos from Ref.\cite{Benes:2005hn}. The calculation
covers all masses from keV to $10^{15}$ GeV. As expected, the limits
are strongest when the neutrino mass corresponds to the energy scale $q \simeq 100$ 
MeV.\\

There is an obvious source of heavy neutrinos, namely the ones from
the type I seesaw mechanism, $N_{R i}$, with masses $M_i$. Note that these
particles provide two sources of \obb: a {\it direct} one realized by
exchange of $N_{R i}$ and an {\it indirect} one by light neutrino
exchange. However, the $N_{R i}$ are
typically very heavy and have suppressed mixing $S \simeq m_D/M_R
\simeq m_\nu/m_D  \simeq \sqrt{m_\nu/M_R}$, 
therefore they lead to basically vanishing \imeff. Without any strong,
instable and fine-tuned cancellations\cite{Ibarra:2011xn}, the direct contribution from \meff~is larger
in seesaw scenarios\cite{Xing:2009ce,Rodejohann:2009ve,Blennow:2010th}. 
Within type I seesaw there is an exact relation 
\be \label{eq:exactseesaw}
\sum\limits_i N_{ei}^2 \, m_i + S_{ei}^2 \, M_i  = 0 \, , 
\ee
where $|\sum N_{ei}^2 \, m_i |$ is the effective mass \meff~in type I
seesaw scenarios in which the PMNS matrix is strictly speaking 
not unitary and thus denoted here by $N$.  
The zero on the rhs of the above equation is nothing but the upper
left zero in the full seesaw mass matrix in Eq.~(\ref{eq:Lseesaw}). 
Therefore, with Eq.~(\ref{eq:exactseesaw}), 
the limit on $\meff \ls 0.5$ eV directly translates to $|\sum S_{ei}^2 \,
M_i|\ls 0.5$ eV, which in the absence of cancellations 
is much more stringent than $|\sum S_{ei}^2/M_i|
\ls 1.8 \times 10^{-8}$ GeV$^{-1}$. If a low scale seesaw mechanism is
applied and both the $m_i$ and the $M_i$ are below 100 MeV, then
there will be no \onbb~because\cite{deGouvea:2006gz} of the exact seesaw relation 
Eq.~(\ref{eq:exactseesaw}). 

In type III seesaw scenarios the neutral component of the triplet
plays the role of the heavy neutrino in the type I seesaw and the
discussion is analogous.

\subsection{\label{sec:Delta}Higgs triplets}
\begin{figure}[t]
 \centerline{
\psfig{file=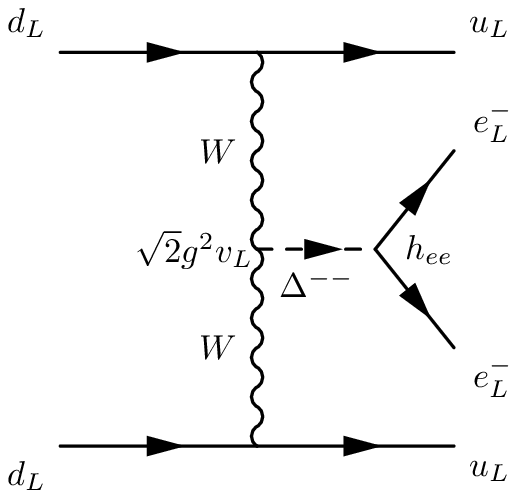,width=6cm,height=4cm}
}
\vspace*{8pt}
\caption{\label{fig:FD_triplet_mech}Quark level Feynman diagram for
the triplet realization of \onbb.}
\end{figure}

Higgs triplets contain a doubly charged scalar and can directly couple
to two electrons and to two $W$-bosons, giving rise to the quark level Feynman diagram shown in
Fig.~\ref{fig:FD_triplet_mech}, first noted
in\cite{Mohapatra:1981pm}. This is the first diagram for \obb~which
does not contain a neutrino line. 
We will show in this Section that in
the simple version based solely on $SU(2)_L \times U(1)_Y$ the triplet
does not play a significant role in \obb. In left-right symmetric
theories this changes, and we will deal with this class of theories in the next
Section. 

The $SU(2)_L$ triplet can be written as 
\begin{eqnarray}
\Delta = \left( \begin{matrix} \Delta^-/\sqrt{2} & \Delta^{--} \cr
\Delta^0 & -\Delta^-/\sqrt{2} \end{matrix} \right)  , 
\end{eqnarray}
and the neutral component receives a vev $\langle \Delta^0 \rangle =
v_L/2$, which induces from the Lagrangian 
${\cal L}_\Delta = h_{\alpha \beta} \overline{L^c}_{\alpha} i\tau_2 
\Delta L_\beta$, where $L_\alpha$ are Lepton doublets of flavor $\alpha$, 
the neutrino mass matrix $m_\nu =  h \, v_L$. The vev $v_L$ is constrained from the electroweak $\rho$
parameter to be less than about 8 GeV, and current limits on the
triplet masses are around 100 GeV\cite{PDG}. The particle physics amplitude for
\obb~can be read off from Fig.~(\ref{fig:FD_triplet_mech}) as 
\be
{\cal A}_\Delta \simeq G_F^2 \, \frac{h_{ee} \, v_L}{m_\Delta^2} 
\ls G_F^2 \, \frac{(m_\nu)_{ee}}{m_\Delta^2} = 
G_F^2 \, \frac{\meff}{m_\Delta^2}
\, . 
\ee
If the triplet was responsible for neutrino mass (type II seesaw) then
$h_{ee} \, v_L = (m_\nu)_{ee}$, which is the largest possible value
of $h_{ee} \, v_L$, unless unnatural cancellations of different seesaw terms
take place.  Comparing with the standard amplitude in
Eq.~(\ref{eq:am_SI}) we see that the rate for triplet exchange is
suppressed with respect to the standard mechanism by at least a factor
$(q/m_\Delta)^4 \ls 10^{-12}$ and hence not of
relevance\cite{Schechter:1981cv,Wolfenstein:1982bf}. Nuclear physics
details add some additional suppression\cite{Haxton:1982ff}.

There are additional diagrams in which one or two of the $W$-bosons 
in Fig.~\ref{fig:FD_triplet_mech} are replaced by
singly charged scalars $\Delta^-$ (see e.g.\cite{Petcov:2009zr}) but
those amplitudes are suppressed by a factor $v_L/v$ for each
$\Delta^-$-quark vertex and by $(m_W/m_{{\Delta^-}})^2$ for each
$\Delta^-$ propagator. 
In principle one can evade these constraints by adding exotic scalars
with specific hypercharge and isospin quantum numbers\cite{Schechter:1981cv,Chen:2006vn}. 

\begin{figure}[t]
 \centerline{
\psfig{file=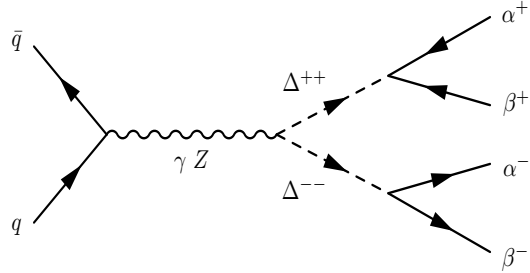,width=7cm,height=4cm}
}
\vspace*{8pt}
\caption{\label{fig:FD_DELTA}Drell-Yan production of Higgs
triplets at the LHC with subsequent decay into like-sign lepton
pairs. In left-right symmetric theories this process is possible as
well.}
\end{figure}

Finally, we should mention the possibility of Higgs triplet production
at the LHC, which is possible up to masses of about 800
GeV\cite{triplet_LHC}. The relevant diagram is shown in
Fig.~\ref{fig:FD_DELTA}. If their branching ratio into leptons is larger
than into a $W$ boson pair, then their decay can give information on the
neutrino mass matrix if in addition the pure type II seesaw is
realized. In fact, BR$(\Delta^{--} \to \alpha^- \, \beta^-) \propto
(m_\nu)_{\alpha \beta}$, and an alternative method to probe Majorana
neutrino properties was possible, as studied 
e.g.~in\cite{triplet_LHC,triplet_LHC1,Petcov:2009zr}. Note in particular that the
branching ratio for decays into two electrons is proportional to the
effective mass. The other entries of the mass matrix could be 
directly studied as well, which is not possible with other processes,
see Section \ref{sec:alt}. In case both the triplet and the type I
seesaw are at work, the exact seesaw relation in
Eq.~(\ref{eq:exactseesaw}) is modified to 
\be \label{eq:exactseesaw1}
\sum\limits_i N_{ei}^2 \, m_i + S_{ei}^2 \, M_i  = h_{ee} \, v_L \, , 
\ee
which links in principle light and heavy neutrino parameters with
triplet parameters. 

\subsection{\label{sec:LR}Left-right symmetric theories}
Left-right (LR) symmetric theories are a popular and appealing extension
of the Standard Model, in which $SU(2)_L \times SU(2)_R \times U(1)_{B
- L}$ is the extended gauge group. Such a gauge symmetry can be arranged in 
breaking patterns of larger groups such as $SO(10)$ or the Pati-Salam
group. It is a natural framework to justify the type I + II seesaw
terms in Eq.~(\ref{eq:mnuseesawI+II}). 
The Higgs sector of the
theory contains\cite{Mohapatra:1980yp} a ``left-handed triplet'' $\Delta_L$ with quantum
numbers $(3,1,2)$, a ``right-handed triplet'' $\Delta_R$ with quantum
numbers $(1,3,2)$, and a bi-doublet $\Phi$ with $(2,2,0)$, which is
not important for \obb. At low energies the potential consequences of LR
symmetry are mainly right-handed currents mediated by a $W_R$ with
coupling strength $g_L = g_R = g$, and the
presence of $\Delta_L$ and $\Delta_R$. Here $\Delta_L$ can be
responsible for a contribution $M_L = h \, v_L$ to neutrino mass. It
can give a direct contribution to \obb, as discussed 
in Section \ref{sec:Delta}, where we have learned that it is
suppressed. The $\Delta_R$ gives mass to the right-handed neutrinos 
$M_R = f \, v_R$, where $v_R$ is the vev of its neutral component and
$f$ a Yukawa coupling matrix. Often is is assumed that a discrete LR
symmetry holds in addition, in which case $h = f^\ast$, or $M_L =
M_R^\ast$. Consequently\footnote{Often one considers $h = f$, or
$M_L = M_R$, which happens when the discrete LR symmetry is 
connected to charge conjugation instead of parity. The limits from LFV
and CP violation in the quark sector case are stronger in this case\cite{Maiezza:2010ic}.}, 
with writing $m_D = y \, v$ it follows 
\be
m_\nu = M_L - m_D^T \, M_R^{-1} \, m_D = v_L \left(h - \frac{v^2}{v_R \,
v_L} \, y^T \, f^{-1} \, y \right) = v_L \left(h - \frac{v^2}{v_R \,
v_L} \, y^T \, {h^\ast}^{-1} \, y \right) . 
\ee 
From the analysis of the scalar potential it follows $v_L \propto
v^2 /v_R$ and therefore neutrino mass is zero in the limit $v_R \to 
\infty$, in which case there are no RH currents, because $M_{W_R}
\simeq g_R \, v_R$.  This connection of small neutrino mass and almost
maximal parity violation makes LR symmetric theories very interesting. 
 
In what regards \obb, there are now several diagrams which allow for
it. In certain variants of LR symmetric models one of the diagrams will dominate over
the other, but we will not enter discussion of the details, and simply
give the limits arising from each diagram individually. \\

First of
all, the $\Delta_R$ can mediate the process in analogy to the diagram
in Fig.~\ref{fig:FD_triplet_mech}. It couples to the $W_R$ instead of
the $W$, and the two emitted electrons (as well as the quarks) 
are right-handed instead of left-handed. The amplitude goes as 
\be
{\cal A}_{\Delta_R} \simeq G_F^2 \left( \frac{m_W}{M_{W_R}}\right)^4 
\frac{f_{ee} \, v_R}{m_{\Delta_R}^2} = G_F^2 \left( \frac{m_W}{M_{W_R}}\right)^4 
\frac{(M_R)_{ee}}{m_{\Delta_R}^2} \, ,
\ee
where $(M_R)_{ee}$ can be written as $\sum V_{ei}^2 \, M_i$, with
$M_i$ the right-handed neutrino masses whose mass matrix $M_R$ is
diagonalized with $V$. Comparing with the naive amplitude in 
Eq.~(\ref{eq:am_naive_heavy}) gives 
$\Lambda^5 \simeq (m_{\Delta_R}^2 \, M_{W_R}^4)/|(M_R)_{ee}|$, and from
the standard amplitude (\ref{eq:am_SI}) it follows 
\be
\frac{|(M_R)_{ee}|}{m_{\Delta_R}^2 \, M_{W_R}^4 } \ls 10^{-15} \, {\rm
GeV}^{-5} \, . 
\ee
Expressing it with a dimensionless quantity is possible by defining 
\be \label{eq:eta_DR}
\eta_{\Delta_R} = \frac{|(M_R)_{ee}|}{m_{\Delta_R}^2 \, M_{W_R}^4 }
\frac{m_p}{G_F^2} \ls 6.9 \times 10^{-6} \, .  
\ee
The limit is compatible with TeV-scale left-right symmetry, one could for
instance rewrite it as 
\be
M_{W_R} \gs 1.9 \, \left( \frac{|(M_R)_{ee}|}{500 \, {\rm GeV}}
\right)^{1/4}  \left( \frac{200 \, {\rm GeV}}{m_{\Delta_R} }
\right)^{1/2}  ~\rm TeV \, . 
\ee
In fact, limits from \obb~are competitive to other means of probing the
parameters associated to LR
symmetry\cite{Guadagnoli:2010sd,Maiezza:2010ic}. Higgs triplets can be
produced at the LHC, in the same way
as shown in Fig.~\ref{fig:FD_DELTA}. Their decay into electrons or
positrons probes $f_{ee} = (M_R)_{ee}/v_R$.  
An interesting possibility in these models is that $M_L$ dominates the
type I + II seesaw formula: $m_\nu = M_L$. In this case $M_R \propto
m_\nu$, i.e.~the heavy neutrinos get diagonalized by the PMNS matrix 
and $(M_R)_{ee}$ becomes less arbitrary. However, in those cases
it turns out that constraints from LFV, in particular $\mu \to 3e$,
which can be mediated by triplets at tree level, force 
$m_\Delta \ll M_i$. This in turn implies that heavy neutrino exchange
in connection with RH currents gives a larger contribution to
\obb\cite{Tello:2010am}.

\begin{figure}[t]
 \centerline{
\psfig{file=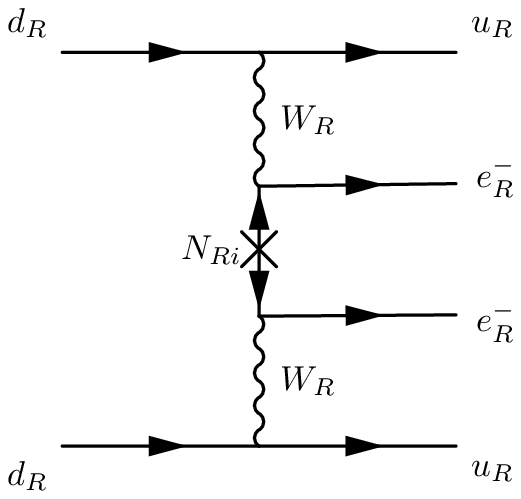,width=5cm,height=4cm} \quad
\psfig{file=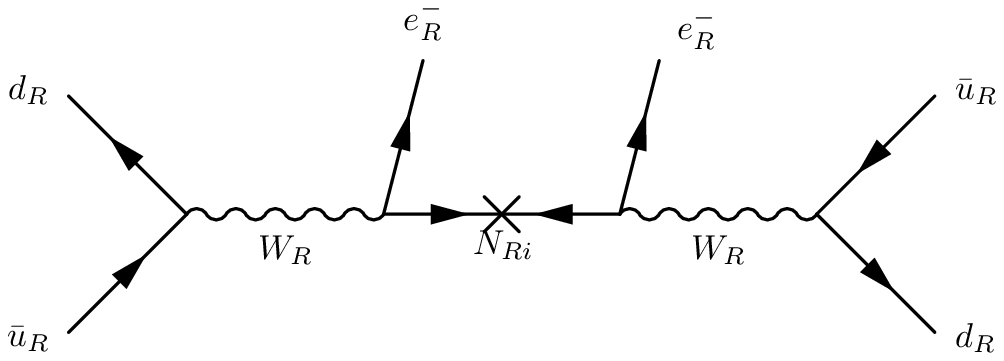,width=7cm,height=4cm}
}
\vspace*{8pt}
\caption{\label{fig:FD_heavy_RH}Left: quark level Feynman diagrams for left-right
symmetric realizations of \onbb~with heavy neutrino exchange. Right:
the corresponding diagram at LHC.}
\end{figure}

Recall that heavy neutrino coupling to the usual LH currents is
suppressed by small mixing $m_D/M_R$. The diagram to study is
therefore the standard one from Fig.~(\ref{fig:FD_mass_mech}) with
$W_R$ exchange\cite{Mohapatra:1986pj}, shown in
Fig.~\ref{fig:FD_heavy_RH}. 
The amplitude goes as
\be \label{eq:A_NR}
{\cal A}_{N_R} \simeq G_F^2 \left( \frac{m_W}{M_{W_R}}\right)^4 
\sum \frac{V_{ei}^2}{M_i}\, . 
\ee
If $f = h$ (or $f = h^\ast$) and type II dominance holds, then $V = U$ ($V = U^\ast$) 
and the PMNS matrix appears in this expression. 
By noting that the NMEs are the same as for the heavy neutrino
exchange discussed in Section \ref{sec:heavy}, we can use the 
limit from Eq.~(\ref{eq:heavy}) to find 
\be \label{eq:NR}
\left|\sum  \frac{V_{ei}^2}{M_{W_R}^4 \, M_i} \right|
 \le \left\{ 
\baz 
(1.8 - 4.3) \times 10^{-16} \, \,{\rm
GeV}^{-5} & \mbox{ for } ^{76}{\rm Ge} \, , \\
(6.7 - 16.6) \times 10^{-16} \, \,{\rm
GeV}^{-5} & \mbox{ for } ^{82}{\rm Se} \, , \\
(3.1 - 16.6) \times 10^{-16} \, \,{\rm
GeV}^{-5} & \mbox{ for } ^{100}{\rm Mo} \, , \\
(2.0 - 4.3) \times 10^{-16} \, \,{\rm
GeV}^{-5} & \mbox{ for } ^{130}{\rm Te} \, , \\
 \ea\right.
\ee
and hence the dimensionless particle physics parameter has the same limit as
$\eta_{\rm h}$ in Eq.~(\ref{eq:eta_h}):  
\be \label{eq:eta_NR}
\eta_{N_R} = m_p \, \left|\sum \frac{V_{ei}^2}{M_i} \right|\left( \frac{m_W}{M_{W_R}}\right)^4
\le 1.7 \times 10^{-8} \, . 
\ee
This limit again corresponds to TeV scale, which can be seen by
rewriting it  as 
\be 
M_{W_R} \gs 1.5   \left( \frac{500 \, {\rm GeV}}{V_{ei}^2 /M_i}
\right)^{1/4} ~\rm  TeV \, . 
\ee
A straightforward phenomenological LHC aspect of heavy neutrino exchange in
left-right symmetric theories is seen in Fig.~\ref{fig:FD_heavy_RH}.  
Like-sign lepton production\cite{Keung:1983uu} is possible
and allows to directly test this mechanism. 
The current limit on $M_{W_R}$ set by LHC data is 1.4 
TeV, both for very light right-handed neutrino
mass\cite{Chatrchyan:2011dx}, as well as for masses
between\cite{Nemevsek:2011hz} 
100 GeV and $M_{W_R}$. 
In the future, LHC can detect masses up
to a few TeV, and right-handed neutrinos up to TeV. This will test
contributions of right-handed neutrino exchange to \obb. 

\begin{figure}[t]
 \centerline{
\psfig{file=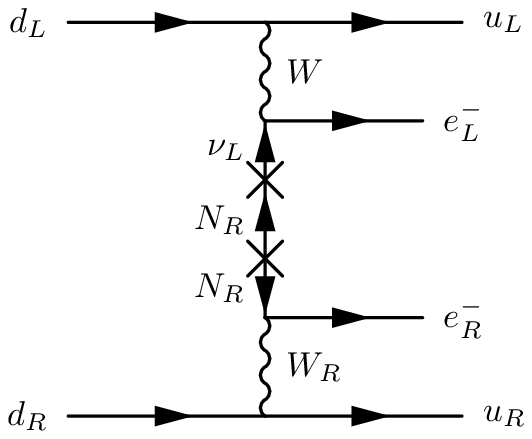,width=6cm,height=4cm} \quad
\psfig{file=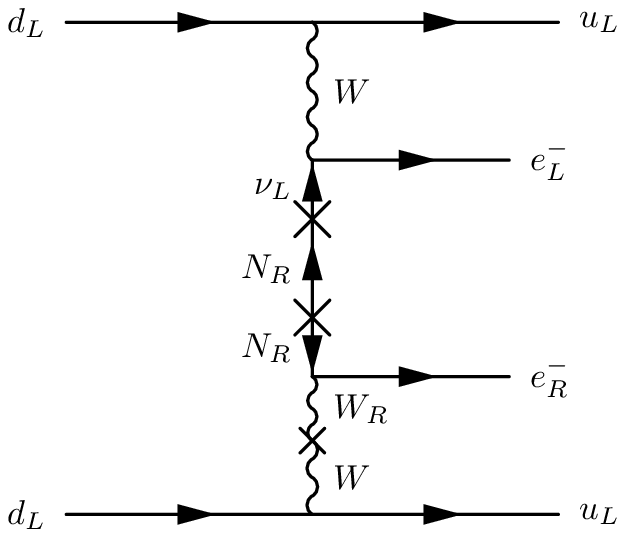,width=6cm,height=4cm}
}
\vspace*{8pt}
\caption{\label{fig:FD_RH}Quark level Feynman diagrams for left-right
symmetric realizations of \onbb. Left is the $\lambda$-mechanism, right the
$\eta$-mechanism.}
\end{figure}

The remaining two diagrams for LR symmetry are stemming from mixing of
the left- and right-handed sectors. First of all, one of the $W$ 
bosons in the standard diagram could be right-handed, leading to the left diagram in
Fig.~\ref{fig:FD_RH}. Its amplitude is  
\be \label{eq:am_LR_lam}
{\cal A}_\lambda \simeq G_F^2 \left(\frac{m_W}{M_{W_R}} \right)^2 \,
\sum U_{ei} \, \tilde{S}_{ei} \, \frac{1}{q} \, , 
\ee
where $\tilde{S}$ is the matrix which quantifies the mixing of the SM
leptons with RH currents. 
Note that one of the hadronic currents is right-handed. 
The dependence on $1/q$ can be understood from the RH
nature of one of the vertices (see the comments after
Eq.~(\ref{eq:SA_calc})). The dimensionless particle physics parameter is
\be
\langle \lambda \rangle \equiv \eta_\lambda = \left(\frac{m_W}{M_{W_R}} \right)^2 \,
\left| \sum U_{ei} \, \tilde{S}_{ei} \right| \, . 
\ee 
The other contribution takes $W$-$W_R$ mixing, quantified by $\tan
\zeta$, into account and has an amplitude given by 
\be\label{eq:am_LR_eta}
{\cal A}_\eta \simeq G_F^2 \tan \zeta  \,
\sum U_{ei} \, \tilde{S}_{ei} \, \frac{1}{q} \, . 
\ee
Here both hadronic currents are left-handed. 
The dimensionless particle physics parameter is
\be
\langle \eta \rangle \equiv \eta_\eta = \tan \zeta \,
\left| \sum U_{ei} \, \tilde{S}_{ei} \right| \, . 
\ee
Note that the usual way in which both diagrams in Fig.~\ref{fig:FD_RH} are drawn may
be confusing. They are actually long-range diagrams with light neutrino
exchange, and the lower vertex receives due to mixing with the RH
current a (small) factor $\tilde{S}_{ei} \simeq m_D/M_R$. 
This term requires non-zero $m_D$ and $M_R$ less than infinitely
heavy. Without $M_R$ and hence without lepton number violation  
it would obviously not be there. The implicit lepton number violation
necessary for the existence of the diagram is illustrated by a Majorana
mass term and a Dirac mass term, which gives a total contribution $m_D/M_R$.

In both the $\lambda$ and the $\eta$ diagrams one of the emitted
electrons is right-handed. The nuclear physics becomes more
complicated now, because the momentum dependence of the amplitudes,
${\cal A}\propto q^\mu = (\omega, \vec q)$, which introduces matrix elements corresponding to
the time and space components of $q^\mu$. In fact, the space components can give
rise to $0^+ \to 2^+$ transitions, whose observation would therefore
be a clear signal\cite{excited2} of the presence of right-handed currents in
\obb. The main point here is that the time component ($\omega$) parts turn out to be
suppressed by order $(E_1 - E_2)/\omega \sim 10^{-2}$ due to
cancellation of the two diagrams with interchanged electron lines,
whose energies are $E_1$ and $E_2$.  This suppression makes the time
component parts of the same order as the space component ($\vec q$)
parts. The latter contain for $\langle \eta \rangle$ (not for $\langle
\lambda \rangle$) two extra matrix elements,
one of which stems from the nuclear recoil $\vec Q \sim \vec q$ 
(with the electrons emitted as $s$-waves). This contribution dominates
and compensates the $(E_1 - E_2)/\omega$ suppression.  These
features are explained in detail e.g.~in Ref.\cite{doi}.

We are not aware of any recent comparative study of the relevant NMEs
for these processes. Ref.\cite{diff_nme2} has recently summarized
the calculation from\cite{Muto:1989cd}. We add to these results
the ones from\cite{Pantis:1996py} (which do not contain $^{150}\rm Nd$) 
 and take the two sets of calculations as a span of NMEs. The result is 
\be
\langle \eta \rangle = \eta_\eta \le \left\{ 
\baz 
(4.0 - 11) \times 10^{-9} & \mbox{ from } ^{76}\rm Ge \, , \\
(1.4 - 4.4)\times 10^{-8}  & \mbox{ from } ^{82}\rm Se \, , \\
(5.4 - 100.6)\times 10^{-9} & \mbox{ from } ^{100}\rm Mo \, , \\
(4.0 - 6.2)\times 10^{-9}  & \mbox{ from } ^{130}\rm Te \, , \\
(1.2 - 1.6)\times 10^{-8} & \mbox{ from } ^{136}\rm Xe \, , \\
1.4 \times 10^{-8}  & \mbox{ from } ^{150}\rm Nd \, .
\ea 
\right. 
\ee
The best limit from $^{130}\rm Te$ of about $6 \times 10^{-9}$ corresponds to the naive 
result obtained by comparing the standard amplitude (\ref{eq:am_SI}) with 
${\cal A}_\eta$, from which $\langle \eta \rangle \ls 5 \times
10^{-9}$ is found. Again, comparing on the particle physics amplitude
works amazingly well. However, this is here somewhat accidental, because 
the usually dominating time component part is suppressed by a factor
$10^{-2}$. This suppression in turn is compensated by terms from the space
components such as the nuclear recoil term. 

However, for $\langle \lambda \rangle $ one
would be two orders of magnitude off. While the naive estimate
would give $\langle \lambda \rangle \ls 5 \times 10^{-9}$, the
more correct procedure described above gives 
\be
\langle \lambda \rangle = \eta_\lambda \le \left\{ 
\baz 
(6.1 - 15)\times 10^{-7} & \mbox{ from } ^{76}\rm Ge \, , \\
(1.8 - 3.8)\times 10^{-6}  & \mbox{ from } ^{82}\rm Se \, , \\
(9.8 - 54.5)\times 10^{-7} & \mbox{ from } ^{100}\rm Mo \, , \\
(5.8 - 8.9)\times 10^{-7}  & \mbox{ from } ^{130}\rm Te \, , \\
(2.1 - 3.2)\times 10^{-6} & \mbox{ from } ^{136}\rm Xe \, , \\
1.4\times 10^{-6}  & \mbox{ from } ^{150}\rm Nd \, .
\ea 
\right. 
\ee
Again $^{130}\rm Te$ dominates the constraints and gives $\langle
\lambda \rangle \ls 9 \times 10^{-7}$. The two orders of magnitude
difference with respect to the naive limit originate from the suppression 
of the dominating time component part of the amplitude 
by a factor of electron energy
$\sim$ MeV divided by neutrino momentum $\sim 100$ MeV.\\

To sum up, the full glory of left-right symmetric theories provides
several possible diagrams for \obb: standard, heavy neutrino exchange,
heavy neutrino exchange with RH currents, left-handed triplet,
right-handed triplet, $\lambda$ and $\eta$. In principle, all should
be considered at the same time, yielding correlated
constraints\cite{Vergados:1982wr,Mohapatra:1986pj,Suhonen:1993rd,Hirsch:1996qw} in a
multi-dimensional parameter space spanned by 
parameters $M_{W_R}$, $\tan \zeta$, $m_{\Delta_R}$,
$(M_R)_{ee}$, $\sum V_{ei}^2/M_i$ and $\sum U_{ei} \tilde{S}_{ei}$.   
One can expect that left-handed 
triplet and heavy neutrino exchange with LH currents can be
neglected, but the remaining diagrams could give observable \obb~if the relevant
masses and scales do not exceed TeV  too much. These scales
correspond to values testable at the LHC, via lepton flavor violation or
rare processes in the quark sector, and interesting works analyzing this  
interplay have recently been published\cite{Maiezza:2010ic,Tello:2010am}.

\subsection{\label{sec:SUSY}Supersymmetric theories}
In the context of supersymmetric theories $R$-parity often plays an
important role. It is defined as $(-1)^{3 \, B + L + 2 \, s}$, where
$B$ ($L$) is baryon (lepton) number and $s$ spin. For particles $R =
1$ while for sparticles $R = -1$. The usual MSSM Lagrangian\cite{MSSM}
conserves $R$. If $R$ is violated, the following renormalizable and gauge invariant
Lagrangian is allowed: 
\be \label{eq:Rparity}
{\cal L}_\slashed{R} = \lambda_{ijk} \, \hat L_i \, \hat L_j \, \hat
e_k^c + \lambda'_{ijk} \, \hat L_i \, \hat Q_j \, \bar d_k^c
+ \lambda_{ijk}'' \hat u_i^c \, \hat d_j^c \, \hat d_k^c + \epsilon_i
\, \hat L_i \hat H_u \, .  
\ee
Here the $\hat L_i$ ($\hat Q_i$)
are superfields which contain the SM lepton (quark) 
doublets as well as the corresponding slepton (squark) doublets,
$u_i^c, d_i^c, e_i^c$ are superfields containing the singlets of
particles and sparticles, while $\hat H_u$ contains the Higgs and
Higgsino doublets; $i = 1,2,3$ is the family index. The terms
proportional to $\lambda''_{ijk}$ violate baryon number, and have to be tiny
to forbid too rapid proton decay. The remaining three terms in
Eq.~(\ref{eq:Rparity}) violate lepton number by one unit. If
$R$-parity is broken\footnote{In principle also the case of $R$-parity conservation can via box
diagrams lead to \obb, if $L$ violating sneutrino mass terms are
present\cite{Faessler:1997db}. 
These are connected to the amplitude of \obb~and $L$ violating Majorana neutrino
mass terms in analogy to the black-box theorem: 
the presence of one of the three implies the presence of the other
two\cite{Hirsch:1997vz,Hirsch:1997is}. However, the constraints from the sneutrino
contribution to neutrino mass are stronger than the ones from
\obb\cite{Hirsch:1997dm}.} 
(see\cite{Barbier:2004ez} for a review), diagrams
with two such vertices can therefore lead to 
\onbb\cite{Mohapatra:1986su,Vergados:1986td,Hirsch:1995ek}. Note that
now there are two $\Delta L = 1$ vertices instead of one explicit $\Delta
L = 2$ mass term.

There is again an interesting possible interplay here, namely that
the $R$-parity violating (RPV) terms are responsible for the neutrino mass
itself. 
It is well-known that loop-induced Majorana neutrino masses can be 
generated by the $\lambda$ and $\lambda'$ terms. 
This would be their indirect contribution to \obb, to be
compared with the direct contributions to be discussed in this
Section.  A systematic analysis of the interplay of direct and
indirect contributions to \obb~is still lacking, but most often the
neutrino mass constraints are weaker than the ones from \obb, or the 
parameter space is chosen such that the RPV contributions to 
\obb~dominate.

\begin{figure}[t]
 \centerline{
\psfig{file=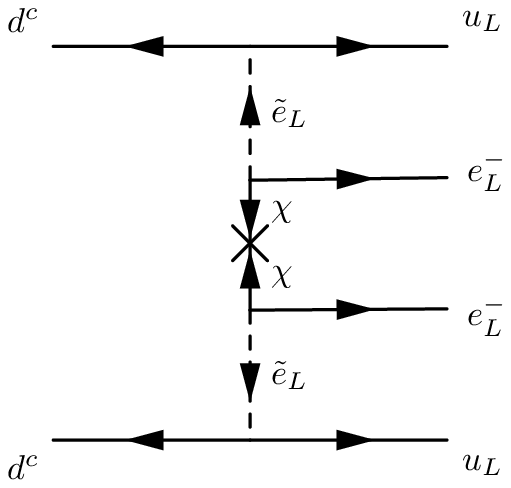,width=4cm,height=4cm} \quad
\psfig{file=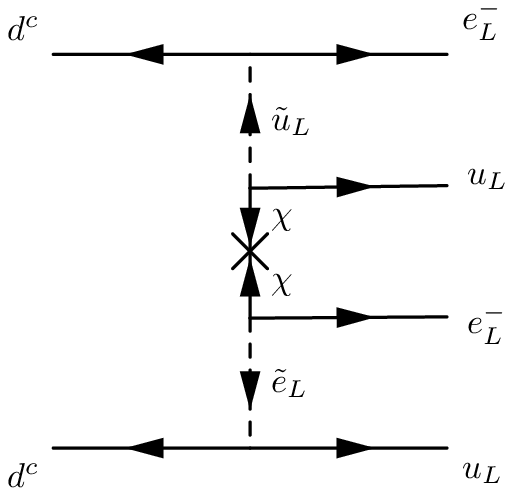,width=4cm,height=4cm} \quad
\psfig{file=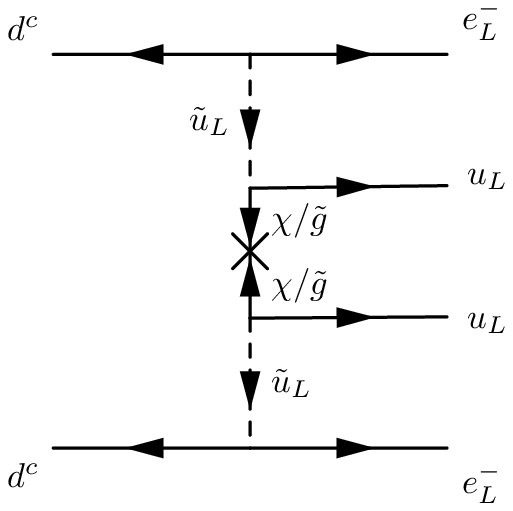,width=4cm,height=4cm} } \vspace{.3cm}

 \centerline{
\psfig{file=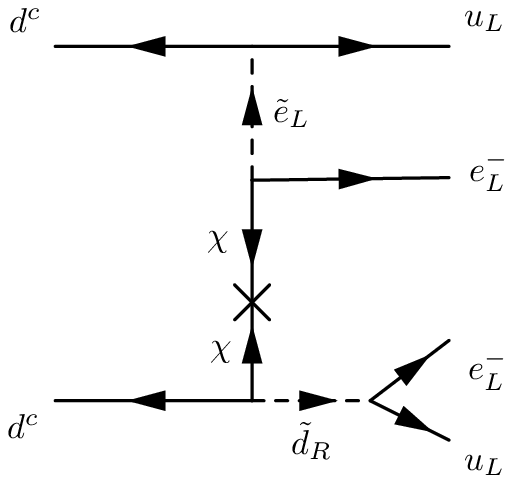,width=4cm,height=4cm} \quad
\psfig{file=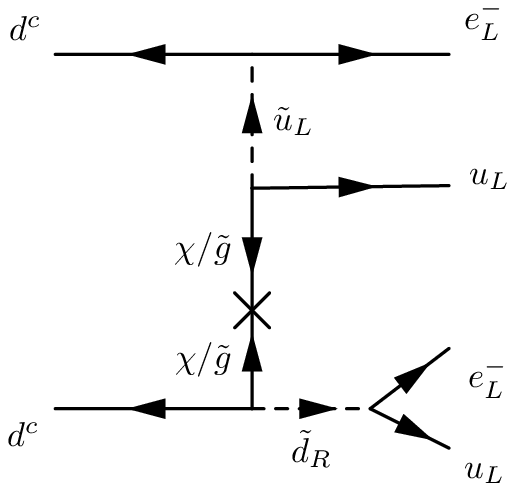,width=4cm,height=4cm} \quad
\psfig{file=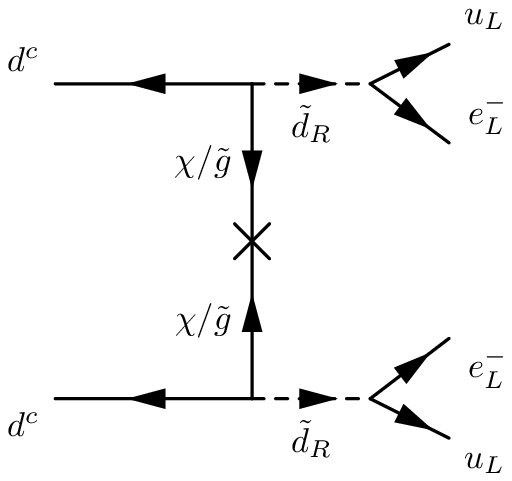,width=4cm,height=4cm} 
}
\vspace*{8pt}
\caption{\label{fig:FD_RPV1}Quark level Feynman diagrams for
short-range $R$-parity violating SUSY contributions to \obb, which are
proportional to $\lambda'^2_{111}$. }
\end{figure}

The most simple possibility here is ``bilinear $R$-parity violation'', in which only
the term $\epsilon_i \, \hat L_i \hat H_u$ is present. This is a
realization of the type I seesaw, with the Higgsino playing the role
of a single (TeV scale) heavy neutrino, which means that only one
light neutrino is massive. Radiative corrections can generate the
necessary other neutrino masses\cite{Hirsch:2000ef}. 
Bilinear $R$-parity violation leads to mixing of neutrinos with
neutralinos and of charged leptons with charginos. 
Effective $\tilde d_L$-$u$-$e^c$, $\tilde e_L$-$e^c$-$\nu$, 
$\tilde d_R$-$d$-$\nu$ and $\tilde u_L$-$u^c$-$\nu$ vertices arise,
and can lead to \obb~in diagrams\cite{Faessler:1997db} which are
similar to the ones in Fig.~\ref{fig:FD_RPV1}. For instance, one could
have the upper
left diagram with $W$ instead of $\tilde e_L$ exchange, or the lower
left one with neutrino and $W$ exchange, instead of the $\tilde e_L$
and $\chi$, respectively. 
Those diagrams have been found to be suppressed with respect to the
standard mass mechanism\cite{Hirsch:1998kc}. We will concentrate on
the trilinear terms from now on.

\begin{figure}[t]
 \centerline{
\psfig{file=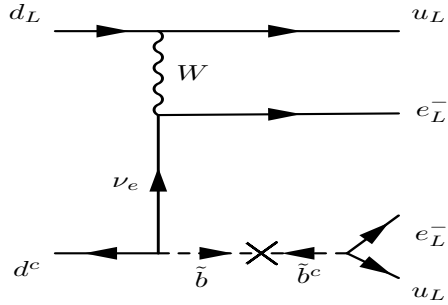,width=6cm,height=4cm} 
}
\vspace*{8pt}
\caption{\label{fig:FD_RPV2}Quark level Feynman diagram for
long-range $R$-parity violating SUSY contribution to \obb, which is 
proportional to $\lambda'_{131} \, \lambda'_{113}$. }
\end{figure}

The two RPV contributions to \obb~are shown in Figs.~\ref{fig:FD_RPV1}
(neutralino/gluino exchange, or $\lambda'_{111}$ mechanism) 
and \ref{fig:FD_RPV2} (squark exchange, or $\lambda'_{131} \,
\lambda'_{113}$ mechanism). Here the pion exchange 
dominance\cite{Vergados:1981bm,Faessler:1996ph} mentioned in Section
\ref{sec:NME_NSI} is realized. Limits on RPV SUSY parameters from
\obb~have been derived in\cite{RPV_limits,Faessler:2007nz}.

The short distance diagrams are shown in Fig.~\ref{fig:FD_RPV2}. The
naive estimate for the amplitude is 
\be
{\cal A}_{\slashed{R}_1} \simeq \frac{\lambda'^2_{111}}{\Lambda_{\rm SUSY}^5} \, , 
\ee
where we set all sparticle masses to the same SUSY scale $\Lambda_{\rm
SUSY}$ and the only relevant coupling is $\lambda'_{111}$, because
the other vertices are order one gauge couplings. Comparing with the
standard amplitude (\ref{eq:am_SI}) 
gives $ \lambda'^2_{111}/\Lambda_{\rm SUSY}^5 \ls 
7 \times 10^{-18}$ GeV$^{-5}$. The scale $\Lambda_{\rm SUSY}^5 $
differs for the six diagrams in Fig.~\ref{fig:FD_RPV1}, for instance
it is related to $m_\chi \, m_{\tilde e_L}^4$ in the upper left and 
$m_{\tilde g} \, m_{\tilde u_L}^4$ in the upper right, etc.

Note that in the diagrams $\chi$ denotes all four neutralinos, which are linear
combinations of neutral gauginos and Higgsinos. In case that gluinos
and/or squarks are exchanged, Fierz transformations have to be performed to obtain
colorless operators. As a result\cite{vergados}, scalar, pseudoscalar
and tensor matrix elements arise. At the end of the day, one can
express the matrix element as\cite{Hirsch:1995ek,Faessler:1996ph,Allanach:2009xx}
\bea \label{eq:am_RPV1}
{\cal M}_{{\slashed R}_1} = \eta_{\tilde{g} }
\left( 
{\cal M}_{\tilde g}^{2N} +  {\cal M}^{\pi} 
\right) 
+ \eta_\chi  
\left( 
{\cal M}_{\tilde g}^{2N} +  {\cal M}^{\pi} 
\right) 
+ \eta_{\tilde{g}}' 
\left( 
{\cal M}_{\tilde f}^{2N} +  \frac 58 \, {\cal M}^{\pi} 
\right) \\
+ \eta_{\chi \tilde{e}}  
\left( 
{\cal M}_{\tilde f}^{2N} +  \frac 58 \, {\cal M}^{\pi} 
\right) 
+ \eta_{\chi \tilde{f}}  
\left( 
{\cal M}_{\tilde f}^{2N} +  \frac 58 \, {\cal M}^{\pi} 
\right) , 
\eea
where ${\cal M}_{\tilde g}^{2N} = {\cal O}(100)$, 
${\cal M}_{\tilde f}^{2N} = {\cal O}(10)$ are 2 nucleon NMEs and
${\cal M}^{\pi} = {\cal O}(100)$ 
pion exchange NMEs (their absolute magnitude exceeds the one of ${\cal
M}_{\tilde g}^{2N}$). Whether neutralino or gluino exchange dominates
depends on the SUSY parameters. 

The particle physics parameters in Eq.~(\ref{eq:am_RPV1})
are\cite{Hirsch:1995ek,Faessler:1996ph,Allanach:2009xx}  
\begin{eqnarray} \label{eq:pp_RPV1}
\eta_{\tilde{g}} &=& \frac{\pi\alpha_3}{6}
\frac{\lambda'^2_{111}}{G^2_F} \frac{m_p}{m_{\tilde{g}}}
\left(\frac{1}{m^4_{\tilde{u}_L}} + \frac{1}{m^4_{\tilde{d}_R}} -
\frac{1}{2m^2_{\tilde{u}_L}m^2_{\tilde{d}_R}} \right) , \nonumber \\
\eta_{\chi} &=& \frac{\pi\alpha_2}{2} \frac{\lambda'^2_{111}}{G^2_F}
\sum_{i=1}^4 \frac{m_p}{m_{\chi_i}}
\left(\frac{V^2_{L_i}(u)}{m^4_{\tilde{u}_L}} +
\frac{V^2_{R_i}(d)}{m^4_{\tilde{d}_R}} -
\frac{V_{L_i}(u)V_{R_i}(d)}{m^2_{\tilde{u}_L}m^2_{\tilde{d}_R}}\right),
\nonumber \\
\eta'_{\tilde{g}} &=& \frac{2\pi\alpha_3}{3}
\frac{\lambda'^2_{111}}{G^2_F} \frac{m_p}{m_{\tilde{g}}}
\frac{1}{m^2_{\tilde{u}_L}m^2_{\tilde{d}_R}} \, ,   \\
\eta_{\chi\tilde{e}} &=& 2\pi\alpha_2 \frac{\lambda'^2_{111}}{G^2_F}
\sum_{i=1}^4 \frac{m_p}{m_{\chi_i}}
\frac{V^2_{L_i}(e)}{m^4_{\tilde{e}_L}} \, , \nonumber  \\
\eta_{\chi\tilde{f}} &=& \pi\alpha_2 \frac{\lambda'^2_{111}}{G^2_F}
\sum_{i=1}^4 \frac{m_p}{m_{\chi_i}}
\left(\frac{V_{L_i}(u)V_{R_i}(d)}{m^2_{\tilde{u}_L}m^2_{\tilde{d}_R}}
-
\frac{V_{L_i}(u)V_{L_i}(e)}{m^2_{\tilde{u}_L}m^2_{\tilde{e}_L}}
-
\frac{V_{L_i}(e)V_{R_i}(d)}{m^2_{\tilde{e}_L}m^2_{\tilde{d}_R}}
\right),\nonumber
\end{eqnarray}
where $\alpha_3$, $\alpha_2$ are the $SU(3)_C$ and $SU(2)_L$ fine
structure constants, respectively, and $V$ are rotation matrices to go
from the gaugino/Higgsino basis to the neutralino basis. As an
example, consider gluino and pion exchange dominance, in which case
the product of matrix elements and particle physics parameters in 
Eqs.~(\ref{eq:am_RPV1},\ref{eq:pp_RPV1}) simplifies to 
\be
\eta_{\rm \slashed{R}_1}^{\tilde g} \, {\cal M}_{{\slashed R}_1}^{\tilde g} \simeq 
\frac{\pi\alpha_3}{6}
\frac{\lambda'^2_{111}}{G^2_F} \frac{m_p}{m_{\tilde{g}} \,
m^4_{\tilde{d}_R}} \left(1 +
\left(\frac{m_{\tilde{d}_R}}{m_{\tilde{u}_L}}\right)^2 
\right)^2  {\cal M}^{\pi} \, . 
\ee
The relevant NMEs are\cite{Faessler:2011rv} between 387 -- 569 
for $^{76}$Ge,  375 -- 594 for $^{82}$Se, 412 -- 589 for $^{100}$Mo and
 385 -- 540 for $^{130}$Te. Hence, the current limits on $\eta_{\rm
\slashed{R}_1}^{\tilde g}$ are 
\be \label{eq:eta_RPV1}
\eta_{\rm \slashed{R}_1}^{\tilde g} \le \left\{ 
\baz 
(4.9 - 7.5) \times 10^{-9} & \mbox{ for } ^{76}{\rm Ge} \, , \\
(0.9 - 1.8) \times 10^{-8} & \mbox{ for } ^{82}{\rm Se} \, , \\
(0.8 - 1.1) \times 10^{-8} & \mbox{ for } ^{100}{\rm Mo} \, , \\
(5.5 - 7.7) \times 10^{-9} & \mbox{ for } ^{130}{\rm Te} \, . \\
 \ea\right.
\ee
The value of $\eta_{\rm \slashed{R}_1}^{\tilde g} \ls 7.5 \times
10^{-9}$ translates into $\lambda'^2_{111}/(m_{\tilde{g}} \,
m^4_{\tilde{d}_R}  (1 + m_{\tilde{d}_R}^2 /m_{\tilde{u}_L}^2)^2 ) \ls
1.8 \times 10^{-17}$ GeV$^{-5}$, in very good agreement with the naive limit 
$ \lambda'^2_{111}/\Lambda_{\rm SUSY}^5 \ls 
7 \times 10^{-18}$ GeV$^{-5}$. 

\begin{figure}[t]
 \centerline{
\psfig{file=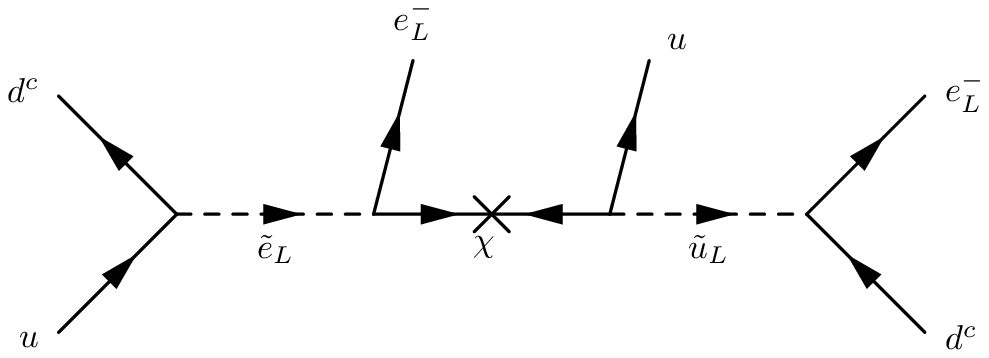,width=7cm,height=4cm} \quad
\psfig{file=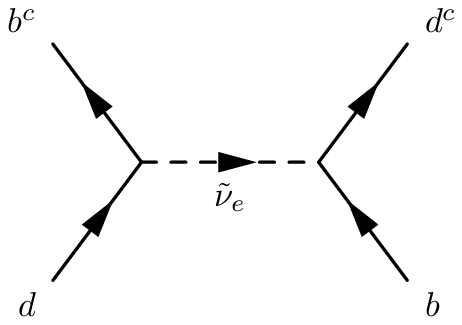,width=4cm,height=4cm}
}
\vspace*{8pt}
\caption{\label{fig:FD_SUSY_test}Left: resonant selectron production
at the LHC as test of the short-range $\lambda'_{111}$ RPV diagrams in
Fig.~\ref{fig:FD_RPV1}. Right:
$B^0$-$\bar{B}^0$ mixing as test of the long-range $\lambda'_{131} \,
\lambda'_{113}$ RPV diagram in Fig.~\ref{fig:FD_RPV2}.  }
\end{figure}

\begin{figure}[t]
 \centerline{
\psfig{file=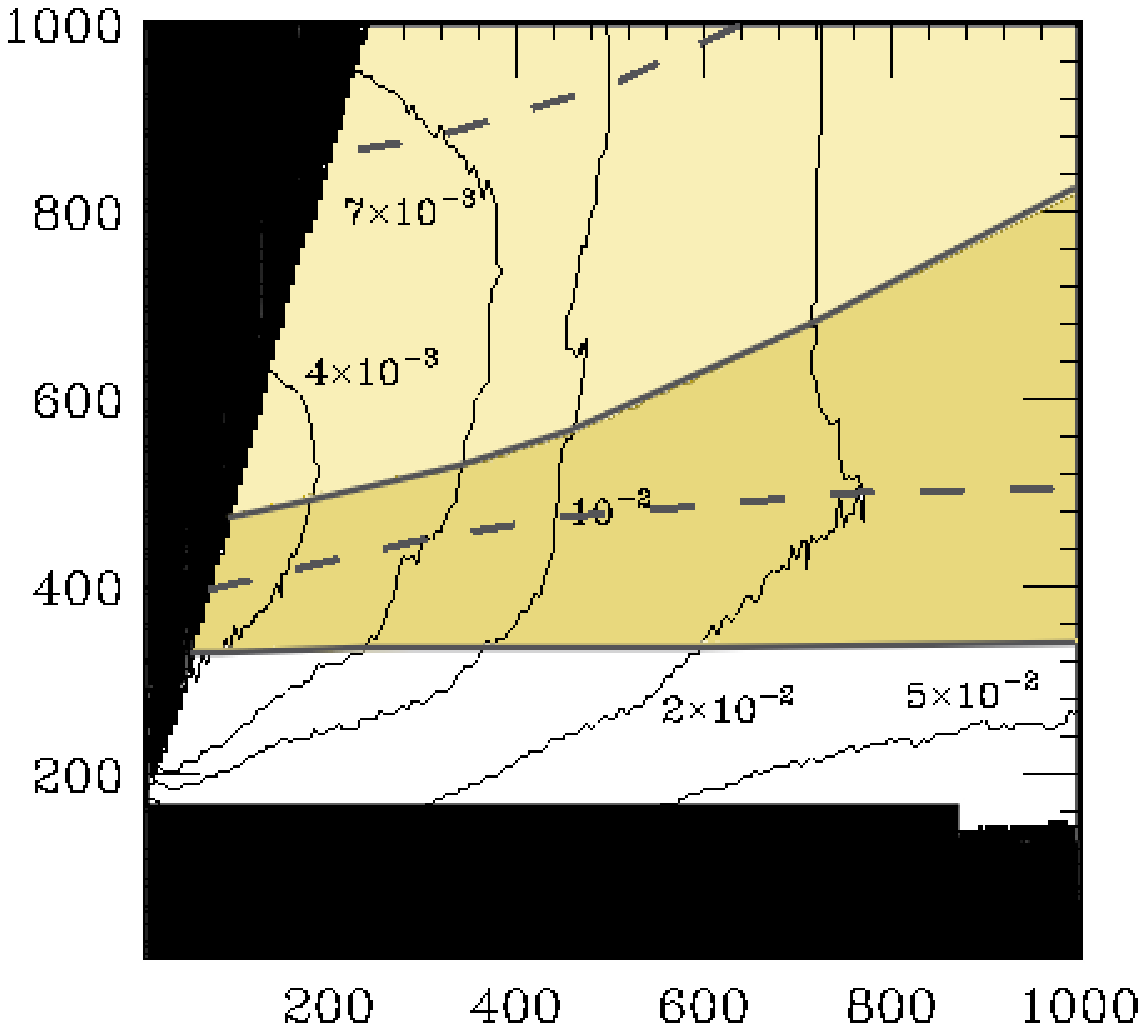,width=7cm,height=6cm}
\psfig{file=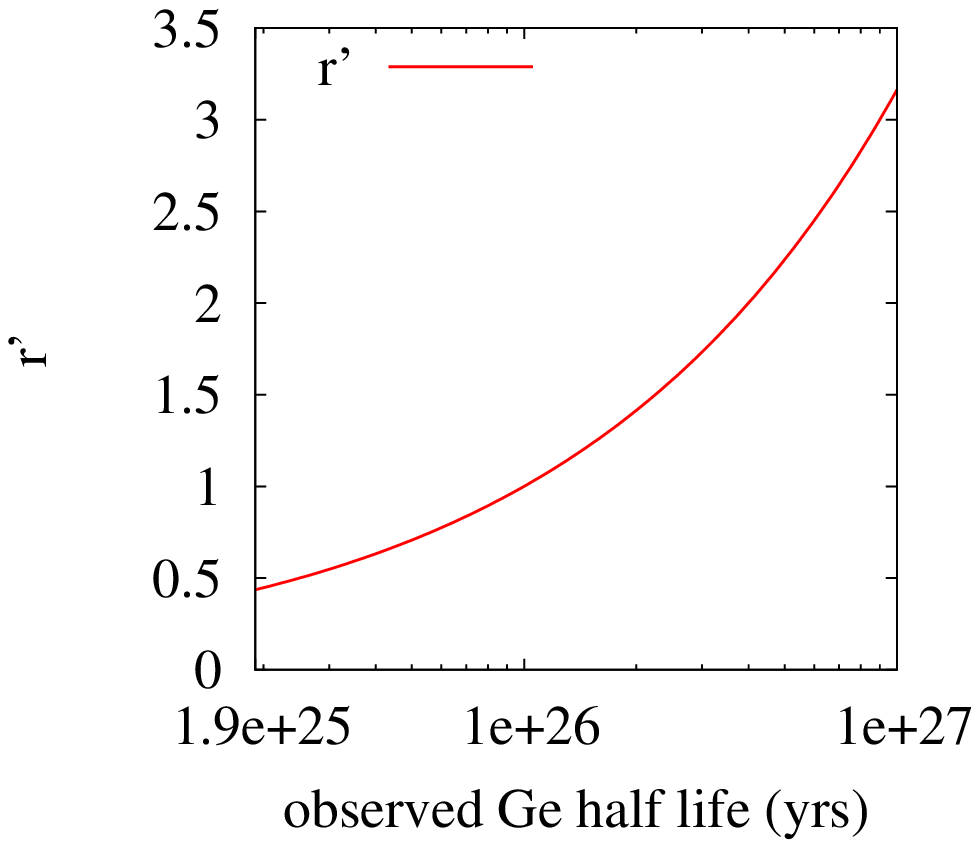,width=6.6cm,height=5.6cm}
}
  \caption{Left: mSUGRA parameter space ($m_0$ vs.~$m_{1/2}$) in which
  single slepton production may be observed at the LHC with $\sqrt{s}
= 14$ TeV and 10 fb$^{-1}$ of integrated luminosity. 
  The labelled contours show the search reach given by the labelled
  value of $\lambda'_{111}$. The white, dark-shaded and light-shaded regions
  show for $^{76}$Ge that observation of single slepton production at the $5\sigma$ level
  would imply $\bhalflife < 1.9 \times 10^{25} ~\textrm{yrs}$,  
$10^{27}~{\rm yrs} > \bhalflife > 1.9 \times 10^{25}$ yrs and
  $\bhalflife > 10^{27} ~\textrm{yrs}$, respectively. 
The upper and lower dashed curves
show where the contour between the dark-shaded and light-shaded regions would
move to if $\meff=0.05$ eV were included 
with constructive or destructive interference, respectively. 
Right: ratio of the RPV amplitude (\ref{eq:am_RPV1}) and the total
amplitude of \obb~vs. the half-life if $m_0 = 680$ GeV and $m_{1/2} =
440$ GeV. Taken from\protect\cite{Allanach:2009xx}.
  \label{fig:lamp111_Disc} }
\end{figure}

Of course, supersymmetric particles are expected to be produced at the
LHC, and Refs.\cite{Allanach:2009iv,Allanach:2009xx} have recently
analyzed the interplay of RPV contributions to \obb~and collider
physics. In particular, resonant selectron production\cite{dreiner}, $u \, d^c \to
\tilde{e}_L \to e \, \tilde{\chi} \to e \, u \tilde{u}_L \to e \, u  \, e
\, d^c$, was studied. The Feynman diagram is sketched in
Fig.~\ref{fig:FD_SUSY_test}; note the typical like-sign dilepton structure. 
The first and last reactions in the chain involve
$\lambda'_{111}$ and the parton level cross section is proportional to
$\lambda'^2_{111}/\hat s$. A numerical scan of a mSUGRA-like breaking
scenario with $m_0$ and $m_{1/2}$ between $40$ and $10^3$ GeV, 
vanishing trilinear coupling $A_0$ and $\tan \beta = 10$ was
performed, the result of which is given in
Fig.~\ref{fig:lamp111_Disc}. 
In the white region, resonant selectron production is forbidden by the
current \obb-limits on $^{76}$Ge. 
Observation in the darker shaded region implies that \obb~should be
observed in GERDA. Hence, if \obb~is discovered, 
searching for resonant selectron production at LHC is a direct test of
the $\lambda'_{111}$ hypothesis. In light-shaded regions one does not
expect observation of \obb, and would hence rule out a possible
contribution to the process. 

There is the possibility that the $R$-parity violating
diagram and the standard one contribute simultaneously 
(see Section \ref{sec:distinguish}). A possible effect of this is  shown
in Fig.~\ref{fig:lamp111_Disc}: constructive interference of $\meff =
0.05$ eV would move the interesting dark-shaded region up, and render 
observation of resonant selectron production very
difficult. Destructive interference would move it down and make it
easier. The right plot in Fig.~\ref{fig:lamp111_Disc} shows the ratio
of the $R$-parity violating amplitude Eq.~(\ref{eq:am_RPV1}) and the total amplitude of
\obb, for a particular point in parameter space. Extracting the value
of $\lambda'_{111}$ from LHC and measuring the half-life of \obb~fixes
this value.\\

The long-range diagram from Fig.~\ref{fig:FD_RPV2}, given first
in\cite{Babu:1995vh}, involves no suppression by neutrino mass, and
the amplitude can be estimated as 
\be
{\cal A}_{ \slashed{R}_2}^b \simeq G_F \, \frac 1q \, U_{ei} \, 
\frac{m_b}{\Lambda_{\rm SUSY}^3} \, \lambda'_{131} \, \lambda'_{113}
\, . 
\ee
Here we have set all SUSY masses to a common scale $\Lambda_{\rm
SUSY}$, and took into account that the $\tilde b$-$\tilde{b}^c$ mixing
is proportional to $m_b/\Lambda_{\rm SUSY}$ (see below). Comparing
with the standard amplitude Eq.~(\ref{eq:am_SI}) gives the constraint 
$ \lambda'_{131} \, \lambda'_{113}/\Lambda_{\rm SUSY}^3 \ls 10^{-14}$
GeV$^{-3}$. A more precise calculation gives constraints on the 
following  quantity 
\be \label{eq:etaRPV2}
\eta_{ \slashed{R}_2}^b = \frac{\lambda'_{131} \, \lambda'_{113}}{2\sqrt{2} \,
G_F} \, \sin 2 \theta^b \left(\frac{1}{m_{\tilde{b}_1}^2} 
- \frac{1}{m_{\tilde{b}_2}^2} \right) . 
\ee
The angle $\theta^b$ and the masses $m_{\tilde{b}_{1,2}}^2$ 
in this expression arise from diagonalization
of the symmetric matrix 
\be
{\cal M}_b^2 = 
\left( \baz 
m_{\tilde{b}_L}^2 + m_b^2 - 0.42 \, M_Z^2 \, \cos 2 \beta & 
- m_b \, (A_b + \mu \, \tan \beta ) \\ 
\cdot & m_{\tilde{b}_R}^2 + m_b^2 - 0.08 \, M_Z^2 \, \cos 2 \beta
\ea \right) , 
\ee
where $\tan \beta$ is the ratio of up- and down-type Higgs vevs, $\mu$ is
the $\mu$-parameter, $A_b$ the trilinear coupling of Higgs scalars and
fermions, and $m_{\tilde{b}_L}^2$
$(m_{\tilde{b}_R}^2)$ the soft masses of the SUSY partners of the
left-handed (right-handed) $b$ quark. Nuclear physics is again dominated
by pion exchange\cite{Faessler:2007nz}, with the relevant NMEs 2 to 3
orders of magnitude larger than the 2 nucleon NMEs. The spread of NMEs
in\cite{Faessler:2011rv} is 396 -- 728
for $^{76}$Ge, 379 -- 720 for $^{82}$Se, 405 -- 691 for $^{100}$Mo and
382 -- 641 for $^{130}$Te. One finds 
\be \label{eq:RPV2}
\eta_{\rm \slashed{R}_2}^b \le \left\{ 
\baz 
(4.0 - 7.3) \times 10^{-9} & \mbox{ for } ^{76}{\rm Ge} \, , \\
(1.5 - 2.8) \times 10^{-8} & \mbox{ for } ^{82}{\rm Se} \, , \\
(0.7 - 1.2) \times 10^{-8} & \mbox{ for } ^{100}{\rm Mo} \, , \\
(4.6 - 7.7) \times 10^{-9} & \mbox{ for } ^{130}{\rm Te} \, . \\
 \ea\right.
\ee
The agreement with the naive limit is very good. We have only
considered here the $b$ squark diagram. There are identical diagrams
with $d$ and $s$ squark mixing, proportional to $m_d \, \lambda'_{111} \,
\lambda'_{111}$ and $m_s \, \lambda'_{121} \, \lambda'_{112}$,
respectively. The first case depends therefore on the same parameters
as the neutralino/gluino diagrams discussed above, but due to its
dependence on $m_d$ it is suppressed. 
The diagram with $s$ squark
mixing can be shown to be sub-leading due to strong limits from
$K^0$-$\bar{K}^0$ mixing\cite{Choudhury:1996ia}, in which at tree
level sneutrino exchange takes place. Those limits are of order 
$\lambda'_{121} \, \lambda'_{112} \ls 
10^{-9} \, (\Lambda_{\rm SUSY}/100 \, \rm GeV)^2$, whose dependence on
the parameters is easy to understand. About the same
order are the limits on $\lambda'_{131} \, \lambda'_{113}$ from
$B^0$-$\bar{B}^0$ mixing (Fig.~\ref{fig:FD_SUSY_test} sketches the
relevant Feynman diagram), which have to be compared with 
$ \lambda'_{131} \, \lambda'_{113} \ls 10^{-8} \, (\Lambda_{\rm SUSY}/100
\, \rm GeV)^3$ from \onbb.  
This implies an interesting interplay of 
$B$ physics and \obb: as long as the SUSY breaking scale does not
exceed TeV, the limits are similar. However, as the $B^0$-$\bar{B}^0$
mixing diagram proceeds with sneutrino exchange and the \obb-diagram
with $b$ squarks, a more detailed analysis is in order, which has been
performed in Ref.\cite{Allanach:2009xx}. As a result, the
$B^0$-$\bar{B}^0$ constraint is currently stronger than the one 
from \obb, but can be responsible for observable half-lifes of $10^{26}
$ -- $10^{27}$ yrs for $^{76}$Ge, which was the isotope 
studied in\cite{Allanach:2009xx}. In analogy to the right plot of
Fig.~\ref{fig:lamp111_Disc} one could again define a ratio of matrix
elements and study its range as a function of the half-life\cite{Allanach:2009xx}.

\subsection{\label{sec:Majorons}Majorons}
\begin{figure}[t]
 \centerline{
\psfig{file=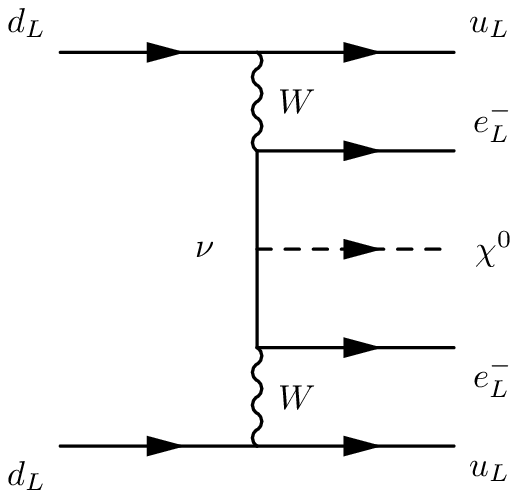,width=6cm,height=4cm}
\psfig{file=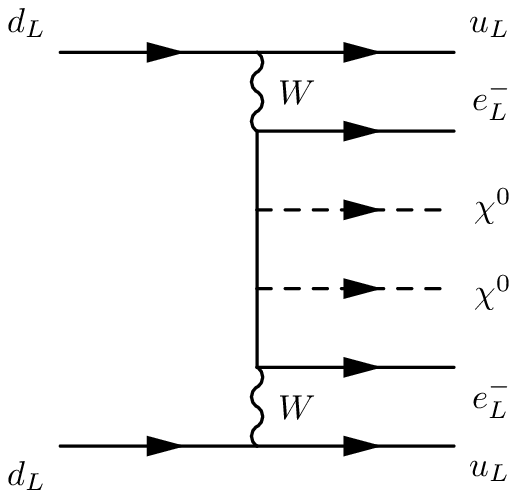,width=6cm,height=4cm}
}
\vspace*{8pt}
\caption{\label{fig:FD_Majoron}Quark level Feynman diagram for a one
and two Majoron realizations of \onbb.}
\end{figure}

The term Majoron denotes very light or massless particles $\chi^0$ which can
couple to neutrinos. Originally Majorons were Goldstone bosons of
spontaneously broken global lepton number. This Majoron could be 
part of a weak singlet\cite{Chikashige:1980ui}, doublet or 
triplet\cite{Gelmini:1980re}, the latter two cases 
being ruled out by their unacceptable contribution to the 
$Z$ width. Another set of important constraints stems from
astrophysics\cite{Raffelt:1990yz,Tomas:2001dh}. 
In the context of triplet Majorons it has been noted 
that the decay mode\cite{Georgi:1981pg} 
\be
(A,Z) \ra (A,Z+2) + 2 \, e^- + \chi^0 \, 
\ee
is induced, see Fig.~\ref{fig:FD_Majoron}. Several different approaches of (almost) massless 
scalar particles coupling to neutrinos and their impact on \obb~have
been made in the 
past\cite{Aulakh:1982yn,Mohapatra:1988fk,Masiero:1990uj,Berezhiani:1992cd,Burgess:1993xh,Carone:1993jv,Bamert:1994hb,Montero:2000ar,Mohapatra:2000px}.
Those include scenarios in which Majorons are not Goldstone bosons,
or carry lepton number, such that lepton number is actually conserved 
and \obb~is forbidden. Other examples are when Majorons are vector 
particles\cite{Carone:1993jv}, or doublet
Majorons\cite{Aulakh:1982yn} in which the Majoron is the SUSY partner
of the neutrino. Extra-dimensional Majorons with a set of Kaluza-Klein
modes was also proposed\cite{Mohapatra:2000px}. 
Simple singlet Majoron models allow coupling of
$\chi^0$ to right-handed neutrinos   
with strength $m_\nu/M$, where $M$ is the scale of spontaneous lepton
number breaking, hence $M \simeq M_R$.  Thus one does not expect sizable
coupling and \obb-rates. This is different in more complicated models. 
It was also realized that decays with two Majorons in
the final state are possible\cite{Mohapatra:1988fk}: 
\be
(A,Z) \ra (A,Z+2) + 2 \, e^- + 2 \, \chi^0 \, , 
\ee
see Fig.~\ref{fig:FD_Majoron}. At the end of the day, one writes the decay rate as 
\be
\Gamma^{0\nu} = \left|\langle g_\chi \rangle\right|^{(2~\rm or ~4)}
\left|{\cal M}_\chi \right|^2 \, G_\chi(Q,Z) \, , 
\ee
where $\left|\langle g_\chi \rangle\right|$ is an averaged and
model-dependent coupling constant, its power obviously depending on 
single or double emission. The phase space factor $G_\chi(Q,Z)$ depends also on
 the number of final state particles, but also on the model, in
particular on the nature of the Majoron. The experimental quantity to
distinguish Majoron modes from \obb~is of course the energy spectrum
of the two emitted electrons. In the original triplet model, with
$g_\chi \, \bar \nu \, \chi \, \nu$
the coupling of the Majoron with two neutrinos, the amplitude can be
written as ${\cal A} \simeq G_F^2 \, \langle g_\chi \rangle/q^2 $, which has
one dimension of energy less than the amplitudes considered before,
because the phase space integration for one additional final state
particle implies two powers of energy. Hence, the decay width goes as
$Q^7$ ($n=1$) instead of $Q^5$ for \obb. Similar models with double Majoron
emission have consequently a decay width proportional to $Q^9$
($n=3$). 
\begin{table}[t]
\tbl{\label{tab:Majorons}Categories of Majoron models as first proposed
in\protect\cite{Bamert:1994hb}. Given are whether \obb~can take place,
if single or double Majoron emission is predicted, the spectral index
and the lepton number of the Majoron, whether it is a Goldstone boson,
and the limit on its coupling, taken
from\protect\cite{Arnold:2006sd}. The limit for
IF is estimated here.}
{\begin{tabular}{ccccccc}
category & \obb & mode & $n$ & $L_\chi$ & GB? & $\langle g_\chi
\rangle $ \\     \toprule 
IB & \checkmark & $\chi^0$ & 1 & 0 & -- & $1.7 \times 10^{-4} $\\  
IC & \checkmark & $\chi^0$ & 1 & 0 & \checkmark & $1.7 \times 10^{-4} $\\
ID & \checkmark & $\chi^0\chi^0$ & 3 & 0 & -- & 1.5 \\
IE & \checkmark &  $\chi^0\chi^0$ & 3 & 0 &  \checkmark & 1.5 \\ 
IF (bulk) & \checkmark & $\chi^0$ & 2 & 0 & \checkmark & $\sim
10^{-4}$ $^\ast$ \\ 
\hline 
IIB & -- & $\chi^0$ & 1 & -2 & -- & $1.7 \times 10^{-4} $\\ 
IIC & -- & $\chi^0$ & 3 & -2 & \checkmark & 0.024 \\ 
IID & -- & $\chi^0\chi^0$ & 3 & -1 & -- & 1.5 \\ 
IIE & -- & $\chi^0\chi^0$ & 7 & -1 & \checkmark & 1.3 \\ 
IIF (vector) & -- & $\chi^0$ & 3 & -2 & -- & 0.024 \\  \botrule 
  \end{tabular} }
\begin{tabnote}
$^\ast$this is a limit on $g^{2/5}/M$ in units of 
GeV$^{-1}$, where $g$ is the
$\chi^0 \nu \nu$ coupling and $M$ the low energy string scale in the extra-dimensional
framework studied in\cite{Mohapatra:2000px}. 
\end{tabnote}
\end{table}
With the same logic it
follows that single Majoron decays where
the coupling goes with $\partial^\mu \chi$ have a width proportional
to $Q^9$ ($n=3$), while double Majoron decays go with $Q^{13}$  
($n=7$). The integer number $n$ in the above considerations indicates
the ``spectral index'' of the two electron spectrum\cite{Bamert:1994hb}
\be
\frac{d\Gamma^{0\nu}}{dE_1 \, dE_2} \propto 
(Q - E_1 - E_2)^n \, \sqrt{E_1^2 - m_e^2} \, \sqrt{E_2^2 - m_e^2} \,
E_1 \, E_2  \, , 
\ee
neglecting Fermi functions and prefactors. Reasonable estimates could
now be made, again by comparing the amplitudes, and taking into
account the different phase space dependence and a factor $2 (2 \pi)^3$ for each additional phase space
integration. In this way one finds for instance that for single Majoron
modes with $n = 1$ the standard contribution $(G_F^2 \, \meff/q^2)^2
\, Q^5$ has to be compared with $(G_F^2 \, \langle g_\chi
\rangle/q^2)^2 \, Q^7 /(2(2\pi)^3)$, from which it follows 
$\langle g_\chi \rangle \ls 10^{-5}$, and for $n = 3$ that
$\langle g_\chi \rangle \ls 1$.  
Nuclear physics aspects are dealt with
in\cite{Doi:1987rx,Hirsch:1995in}, 
and we will not go into detail
here. For single Majoron and $n = 1$ cases the NMEs from the standard
interpretation can be used, while for the other cases different NMEs
need to be calculated, similar to the situation for the $\langle
\lambda \rangle$ and $\langle \eta \rangle $ terms in the presence of
right-handed currents, discussed in Section \ref{sec:LR}. 
We rather summarize the limits on the various model
categories, which first have been described
in\cite{Bamert:1994hb}. This is shown in Table \ref{tab:Majorons}.

\subsection{\label{sec:NS_other}Other mechanisms}

We will discuss other proposed realizations of \obb~in this Section.

{\bf Non-renormalizable effective operators} ${\cal O}^{4 + d}$ in the
Lagrangian ${\cal L}_{\rm eff} = {\cal O}^{4 + d} /\Lambda^d$ 
can generate neutrino Majorana masses and/or
lepton number violation, the most simple example being the Weinberg 
operator of dimension $4 + d = 4 + 1 = 5$ in Eq.~(\ref{eq:Leff}). 
Operators with $\Delta L = 2$ have been classified up to dimension 11 
in\cite{Babu:2001ex,deGouvea:2007xp}. They can generate neutrino mass directly
(the Weinberg operator) or via loops, by closing some of the external
legs. It is also possible that those lepton number violating 
operators generate a direct contribution to \obb ~(note that
\obb~is effectively a $\bar u  d \, \bar u d \, e e $ operator, which
has dimension 9). There are five 
dimension 9 and fifteen dimension 11 operators which have this
property\cite{Babu:2001ex}. One example is ${\cal O}^9 =
LLQQd^cd^c$. Closing the external $Q$ and $d^c$ lines with Higgs 
loops gives a neutrino mass term of order $m_\nu \sim m_d^2/\Lambda /(16 \pi^2)^2$. A
limit on $\Lambda$ is estimated from the direct contribution of ${\cal
O}^9$ to \obb, which has an amplitude of order $\Lambda^{-5}$. Hence
one limits $\Lambda \gs 3 $ TeV, and therefore a tiny mass of $m_\nu
\sim 10^{-4}$ eV is generated. All dimension 9 operators generate a
limit of order 3 TeV on their associated suppression scale. Regarding  
dimension 11 operators, their \obb-amplitude can be estimated as 
$v^2/\Lambda^7$, hence $\Lambda \gs$ TeV. In\cite{deGouvea:2007xp} 
all 129 operators up to dimension 11 have been studied and the scale
$\Lambda$ has been fixed by requiring the operator to generate $m_\nu
\simeq 0.05$ eV. This fixes their contribution to \obb. 
Some of the operators can now be disfavored, because their 
direct  contribution to \obb~can be too large\cite{deGouvea:2007xp}.

{\bf Leptoquarks} can couple to quarks and leptons and the SM Higgs 
doublet, and have the potential
to lead to lepton number violation and \obb. Their properties are similar to
$R$-parity violating mechanisms of \obb. In
Ref.\cite{Hirsch:1996ye} the vertices for $(S,V^\mu)$-$d$-$\nu$, 
$(S,V^\mu)$-$d$-$e$,  $(S,V^\mu)$-$u$-$\nu$ and  $(S,V^\mu)$-$u$-$e$ 
interactions have been worked out, where $S$ ($V^\mu$) are scalar 
(vector) leptoquarks with electric charge $-\frac 13$ or $\frac 23$. Effective
$u$-$e$-$\nu$-$d$ vertices arise and the coefficients depend 
on the original leptoquark couplings and masses, the latter obviously
as $M_{S,V}^{-2}$. Writing the amplitude naively as 
${\cal A}_{\rm LQ} \sim G_F \, a/M_{S,V}^2 /q$, where $a$ the typical
coefficient for the effective vertex, one finds limits of $a \ls 10^{-9}$ for
100 GeV leptoquarks, which is within one order of magnitude 
to the actual limits derived in\cite{Hirsch:1996ye}. In that paper the
definition of the coefficients in terms of original parameters can be
found. 
 
In\cite{KlapdorKleingrothaus:2002mk} {\bf scalar bilinears} (coupling to two
fermions) have been considered, and typically one dimensionful
coupling $\mu$ and 3 propagators are present in the \obb-diagrams, leading to
amplitudes of the form $\mu/M^6$, where $M$ is the common mass of the
bilinears (see also\cite{Brahmachari:2002xc}).

Rather surprisingly, given the popularity of scenarios with {\bf extra
spatial dimensions}, there are only few papers discussing its
consequences on \obb. Ref.\cite{ArkaniHamed:1998vp} showed that within ADD
scenarios small Majorana neutrino masses can result if lepton number
is broken on distant branes, with the breaking being communicated 
to our brane by messenger particles. Ref.\cite{extraDim} used this
finding and translated limits on the Majorana mass 
$\meff$ in limits on the number of extra dimensions, compactification 
radius of the extra dimension and messenger mass. 
A generic feature of extra dimensional theories is the
presence of Kaluza-Klein (KK) excitations of particles which feel the
extra dimensions. If Majorana neutrinos do so, the associated tower
contributes in principle to \obb. The case of all excitations being Majorana
neutrinos was discussed within a particular model
in\cite{Bhattacharyya:2002vf}.  
Excitations heavier than 100 MeV will have NMEs with the
characteristic features of heavy neutrino exchange discussed in
Section \ref{sec:heavy}. 
In the model considered in\cite{Bhattacharyya:2002vf} two parameters
had to be chosen, the radius $R$ of the extra dimension and the brain
shift parameter $a$, introduced to make the neutrinos with opposite
CP parity couple to the $W$ bosons with unequal strength. Constraints
on those parameters are possible. Ref.\cite{michael} studied an
extra-dimensional scenario based on a warped Randall-Sundrum model
leading to low scale seesaw, in which also a
tower of order GeV sterile neutrinos is present, which can be
constrained by \obb.  
Models in
which only gauge bosons or Higgs scalars possess KK excitations, such
as in\cite{Chang:2003sx}, are dominated by the usual light neutrino
mass mechanism. 

{\bf Scalar octet seesaw} has been proposed in
Ref.\cite{FileviezPerez:2009ud} to have TeV scale neutrino mass
generation with sizable LHC cross sections. In this mechanism a 
one-loop diagram including a weak scalar triplet $S$ and weak fermion
singlets or triplets $\rho_i$, all octets under $SU(3)_C$, produces a
Majorana neutrino mass. The $\rho$ and $S$ particles could mediate double beta decay via the usual
standard diagram with the $W$ replaced by $S$ and the neutrinos
replaced by $\rho_i$. The amplitude is proportional to $c_{ud}^2 /(m_S^4 \,
m_{\rho_i})$, where $c_{ud}$ is the coupling of $S$ to $u$ and $d$
quarks, which is constrained from flavor violating transitions. 

A {\bf fourth generation} Majorana neutrino with mass $M_4$ behaves exactly as a heavy
neutrino discussed in Section \ref{sec:heavy}. Therefore\cite{Lenz:2010ha}, 
it receives the constraint $|S_{e4}|^2/M_4 \le 1.8 \times 10^{-8}$
GeV$^{-1}$, with $S_{e4}$ being its mixing with the electron.  Pushing
its mass down to collider level would require cancellation with other
contributions to \obb. 

{\bf Composite neutrinos}\cite{comp} lead to heavy neutrinos $N^\ast$ which are excited states
corresponding to a scale $\Lambda_c$ of the SM neutrinos. Their coupling with gauge bosons goes with 
$f /\Lambda_c$, $f$ being a coupling constant, 
and the amplitude for \obb~goes with 
$f^2/\Lambda_c /M_{N^\ast}$ and is sensitive to TeV scale
exciteness\cite{Panella:1997wa}.

So-called {\bf 3-3-1 models} with an initial $SU(3)_L$ gauge symmetry
contain new gauge bosons and scalars, which can contribute to \obb. 
Those cases have been studied in
Refs.\cite{Pleitez:1993gc,Montero:2000ar,Montero:2000ar1}, 
and constraints on the masses and mixings with the SM fermions
have been obtained. 
Majoron emission is also possible in those models, because typically
neutral scalars with lepton number exists, whose vevs induce
spontaneous violation of lepton number, see Section
\ref{sec:Majorons}.


We conclude this section with more exotic proposals.  
Effects of scalar unparticles in \obb~have been discussed
in\cite{Zhang:2008zzy}, and an unusual model with colored scalars
coupling to leptons and quarks, which can mediate \obb, in\cite{Gu:2011ak}. 
Recently it was proposed that a huge number of copies of SM particles 
exists\cite{Dvali:2007hz}, which could solve the hierarchy problem
and, if a permutation symmetry is added, explain also small neutrino
masses. It was shown\cite{Kovalenko:2010qv} that this leads to
basically vanishing amplitudes for \obb.

\section{\label{sec:distinguish}Distinguishing mechanisms for neutrino-less double beta decay}
We have seen in the last two Sections that there are several well
motivated frameworks in which observable \onbb~can be
expected. Obviously, means to distinguish the various possibilities
are necessary. This is a common problem for all experiments looking
for new physics, for instance lepton flavor violation, where
observation of, say, $\mu \to e \gamma$ alone does not prove the
presence of supersymmetry, but could mean a lot of different things
(Higgs triplets, extra dimensions, non-unitary PMNS matrix, etc.). In
this Section we will classify three possible tests of the
underlying mechanism of neutrino-less double beta decay. Mostly 
the dominance of one mechanism is assumed, but we will 
discuss the simultaneous presence of more than one mechanism as well.

\subsection{\label{sec:distinguish_other}Distinguishing via effects in
other observables}

This obvious possibility has been discussed at several occasions in
the last Section. In particular within $R$-parity violating SUSY and left-right symmetric
theories TeV scale particles can lead to observable \obb. Production
of such particles at the LHC is then a check of these mechanisms, in
particular if the like-sign dilepton signature can be used, such as
for heavy right-handed neutrino production\cite{Maiezza:2010ic}, Higgs
triplet decays\cite{triplet_LHC}, 
or resonant selectron production\cite{Allanach:2009iv}. However, checks are also 
possible in processes in which lepton number is not violated, but
instead quark or lepton flavor is not
conserved\cite{Cirigliano:2004tc}. To perform such studies, one can
express the relevant processes in terms of effective operators
suppressed by some high energy scale. The scales of flavor violation and lepton number violation
could be different, but are related or even identical in some cases. The flavor parameters on
which \obb~and flavor violating processes depend can also be
different. 

Examples mentioned above are $B^0$-$\bar{B}^0$ mixing induced by 
$\lambda'_{131} \, \lambda'_{113}$ couplings\cite{Allanach:2009xx}. 
Note that here the parameters
corresponding to flavor $(\lambda'_{131} \, \lambda'_{113})$ are the
same for \obb~and $B^0$-$\bar{B}^0$ mixing, but different particles
are involved: squarks in \obb~and sneutrinos in $B^0$-$\bar{B}^0$
mixing. In left-right symmetric theories an important contribution to lepton
flavor violation stems from Higgs triplet exchange, which can mediate
$\mu \to 3e$ at tree level. Here the flavor physics parameters (also
the ones for $\mu \to e \gamma$) are not
directly related to the ones which govern \obb. If TeV scale physics
generates \obb, and if no special flavor
structures and not too different flavor and lepton number violating scales
are present, one expects\cite{Cirigliano:2004tc} a ratio $R \gg 10^{-2}$ of the rates for
$\mu$-$e$ conversion in nuclei and $\mu \to e \gamma$. Therefore, if
$R \simeq 10^{-2}$ is observed, the standard interpretation of light
neutrino exchange in \obb~is favored\footnote{Massive neutrinos imply
lepton flavor violation in decays 
like $\mu \to e \gamma$ at an unobservably small level.}.

\subsection{\label{sec:distinguish_exp}Distinguishing via decay products} 
\begin{figure}[t]
 \centerline{
\psfig{file=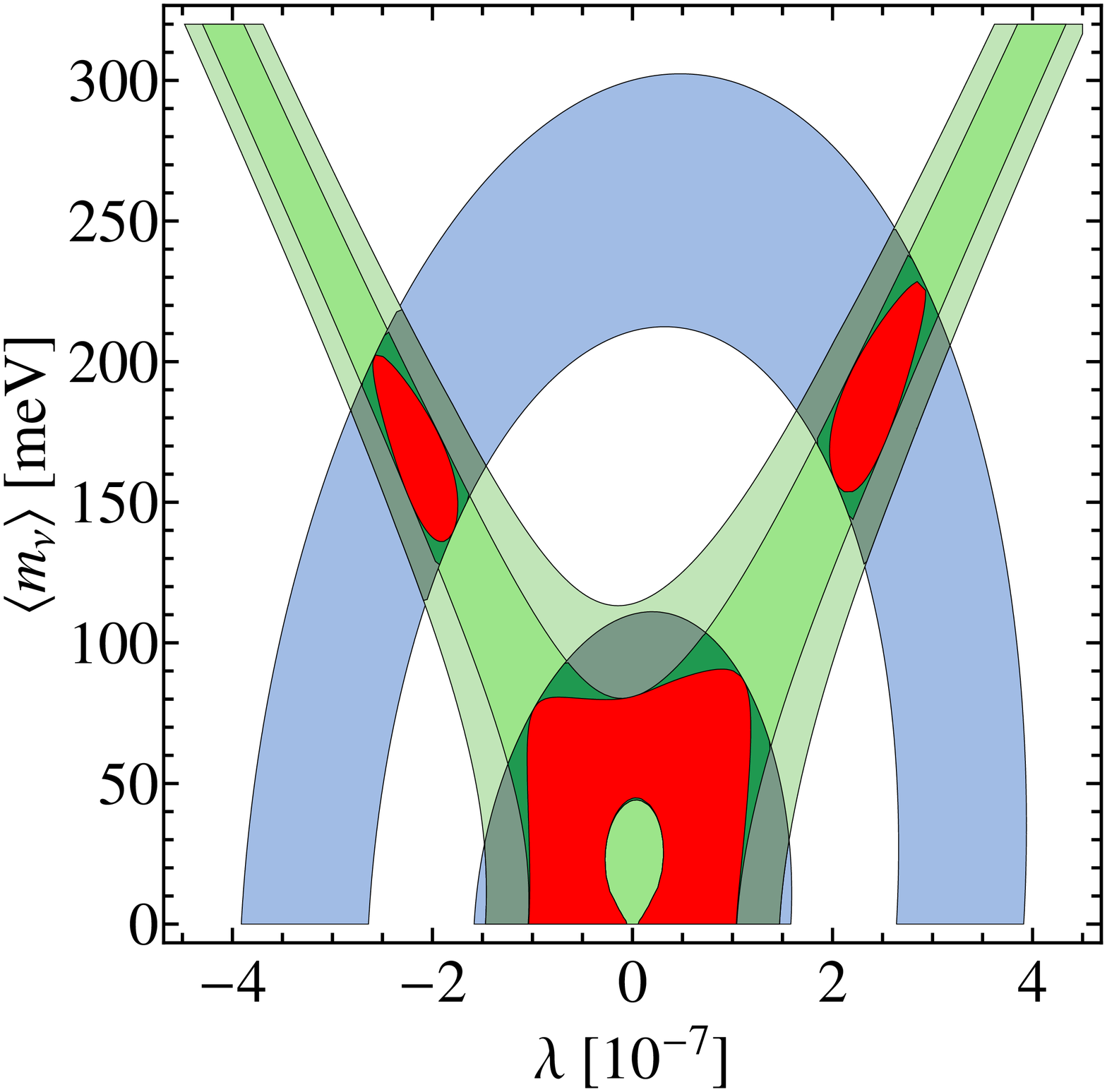,width=7cm,height=5.5cm}
\psfig{file=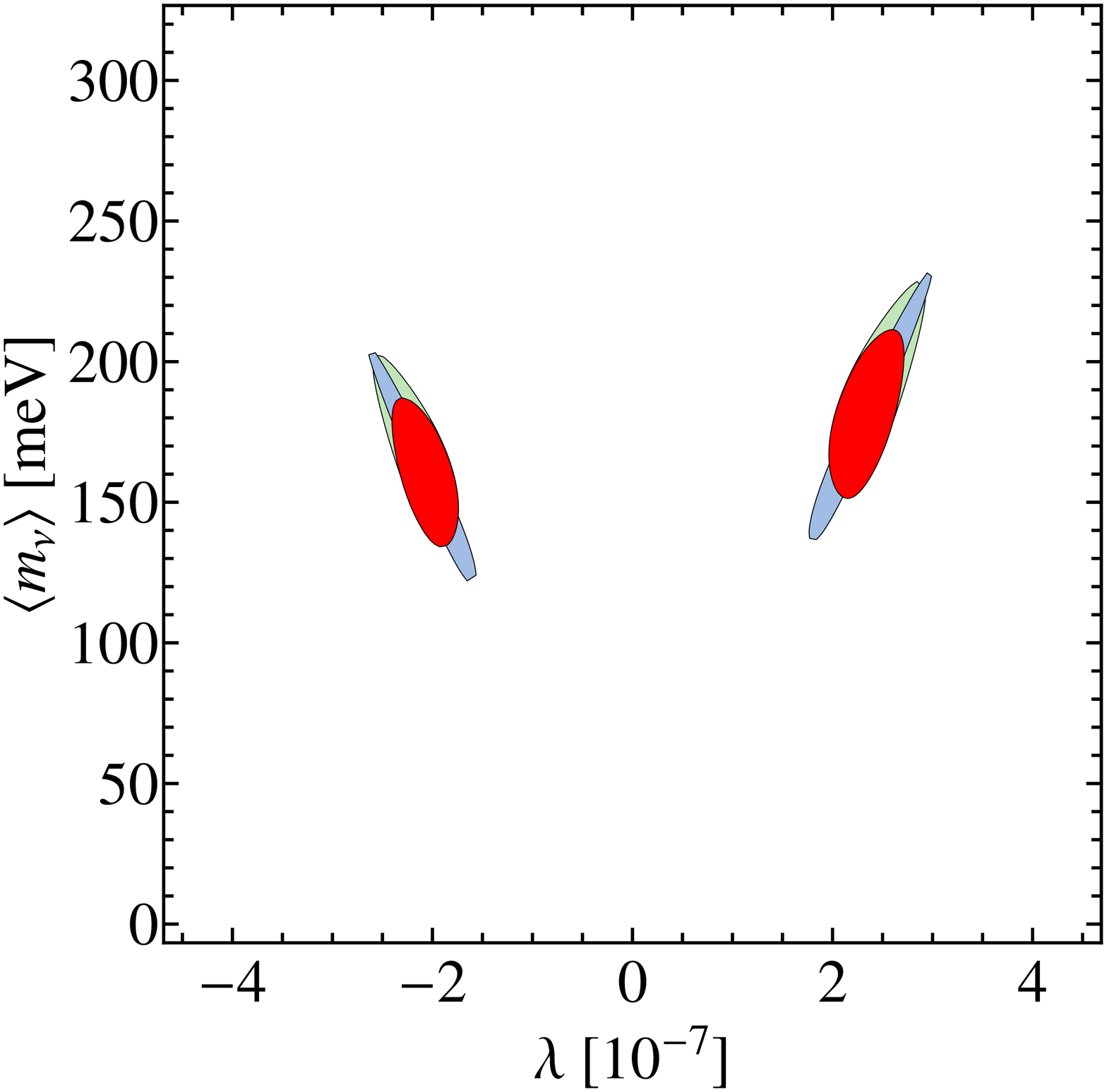,width=7cm,height=5.5cm}
}
\vspace*{8pt}
\caption{\label{fig:nemo}Left: constraints at $1\sigma$ on the model parameters from 
an observation of \obb~of $^{82}$Se at half-life $10^{25}$ yrs 
(outer blue elliptical area) and $10^{26}$ yrs (inner blue elliptical
area). Adding the reconstruction of the angular (outer, lighter green)
and energy difference (inner, darker green) distribution drastically
shrinks the allowed parameter space. Right: adding information from
the decay of $^{150}$Nd. In this example, 30\% admixture of 
the $\lambda$-mechanism is assumed. Taken from\protect\cite{Arnold:2010tu}.}
\end{figure}

We have seen that the Lorentz structure of the different mechanisms of
\obb~can be different. This implies that energy and angular
correlations of the two emitted electrons may be 
different\cite{doi,angular1,angular2}. In particular the SuperNEMO experiment
will be able to perform such measurements, because the set up of foils
with \obb-isotopes in a magnetic field allows tracking of the
individual electrons, instead of ``only'' measuring their total
energy. The design of the detector allows 
direct detection of two electrons from double beta decay by a 
tracking chamber and a calorimeter measuring individual energies and 
times-of-flight. In Ref.\cite{Arnold:2010tu} the collaboration has simulated
the potential discrimination power between the standard mechanism and
the $\lambda$-mechanism within left-right symmetry. The differential
decay width can be written as\cite{doi,angular1,angular2} 
\be
\frac{d\Gamma}{dE_1 \, dE_2 \, d \cos \theta } \propto 
\left\{ \baz 
(1 - \beta_1 \, \beta_2 \, \cos \theta) & \mbox{standard mechanism} \\
(E_1 - E_2)^2 \, (1 + \beta_1 \, \beta_2 \, \cos \theta) & \lambda \mbox{ mechanism} 
\ea \right. \, , 
\ee
where $E_{1,2}$ are the kinetic energies of the electrons, 
$\beta_{1,2}$ their velocities and $\theta$ the angle between them. One
can define an asymmetry $A_\theta = (N_+ - N_-)/(N_+ + N_-)$, where
$N_+$ ($N_-$) is the number of events with $\theta > \pi/2$  ($\theta
< \pi/2$). Another asymmetry is $A_E = (N_> - N_<)/(N_> + N_<)$, where
$N_>$ ($N_<$) is the number of events with $E_1 - E_2 < Q/2$ ($E_1 -
E_2 > Q/2$), where $Q$ is the energy release. Fig.~\ref{fig:nemo}
shows a result from\cite{Arnold:2010tu}, where a 30\% error on the
NMEs and the simultaneous presence of the standard
term and 30\% admixture of the $\lambda$-mechanism has been
assumed. The energy difference distribution turns out to have a
stronger discrimination power.

Another aspect of identifying the \obb-mechanism with the decay
product is when the Majorons as additional particles are emitted, in
which case the energy spectrum of the electrons is different from the
\obb- or the \zbb-spectrum, as discussed in Section
\ref{sec:Majorons}.

\subsection{\label{sec:distinguish_nucl}Distinguishing via nuclear
physics}

\begin{figure}[t]
 \centerline{
\psfig{file=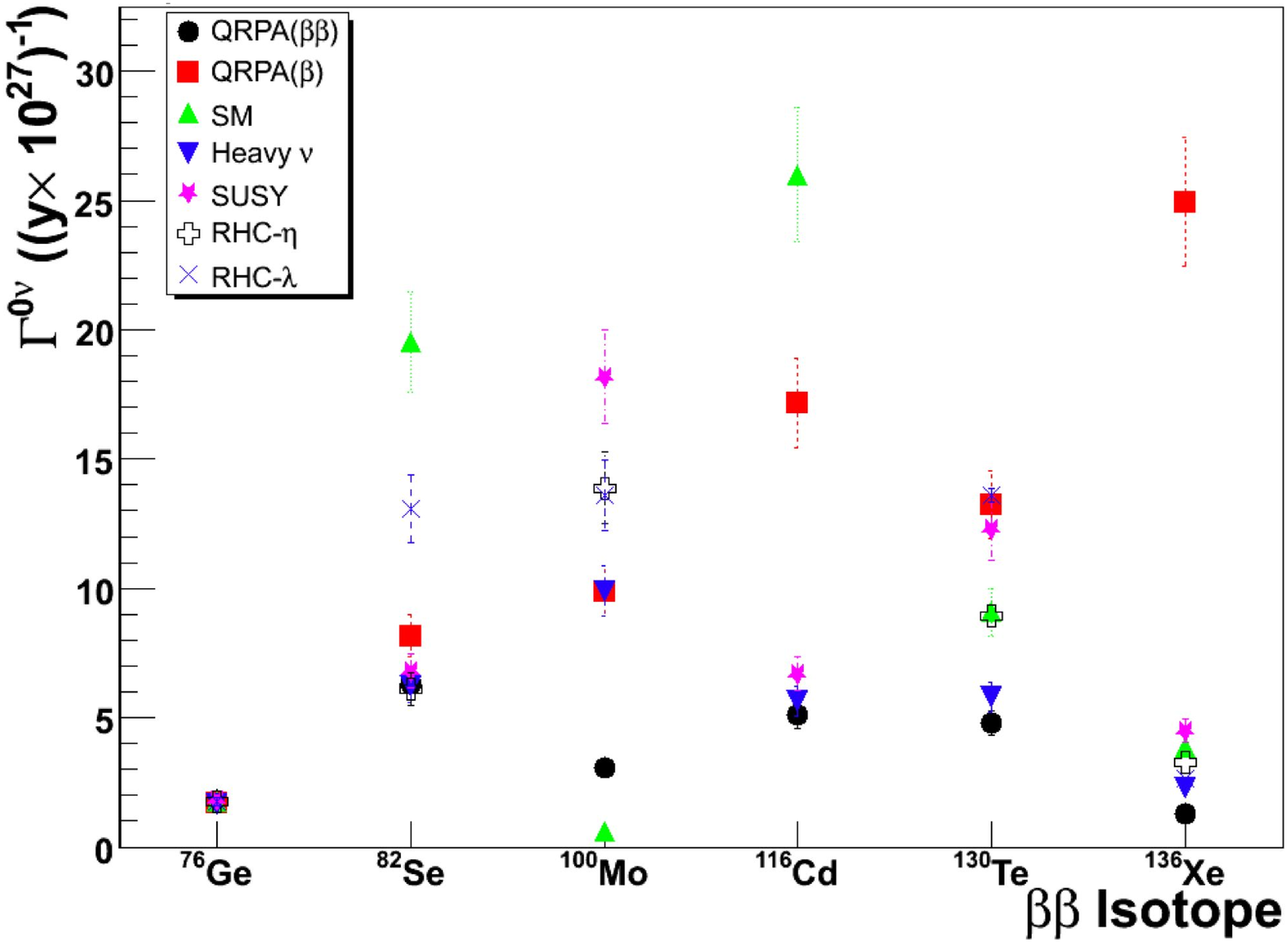,width=9cm,height=5.5cm}
}
\vspace*{8pt}
\caption{\label{fig:diff_NME}Predictions for the inverse half-life for
different NMEs, \obb-mechanisms and isotopes.
Taken from\protect\cite{diff_nme2}.}
\end{figure}

We have not spent much attention on the nuclear physics details of
the \obb-mechanisms, and argued mainly on the particle physics
amplitude level. However, there is nuclear physics involved, and if
its uncertainties can be kept under control, it could in fact be 
helpful\cite{diff_nme1,diff_nme2,diff_nme3} to 
disentangle the various mechanisms of \obb. The use of multi-isotope
determination of \obb~to test NME models for the standard mechanism was discussed
in\cite{BP,errors}. However, it should be
clear that this is a quite challenging task, and in particular
requires that the  spread of the NME calculations is not much larger than 
the experimental error on the half-life. Nevertheless, the 
same strategy can be applied to disentangle different mechanisms of \obb. 
Fig.~\ref{fig:diff_NME} shows the result of Ref.\cite{diff_nme2}, where different isotopes,
NMEs and mechanisms of \obb~were compared. Those were the $\lambda$
and $\eta$ diagrams within left-right symmetry, heavy neutrino
exchange and $R$-parity violating SUSY. The NMEs were two sets of QRPA
calculations (with their parameters fitted to reproduce single beta
decay and \zbb, respectively) and a shell model evaluation. A 10\%
error on the calculations was assumed. The
individual parameters of lepton number violation were chosen such that
for $^{76}$Ge the half-life is the same for all mechanisms.  As can be
seen from Fig.~\ref{fig:diff_NME}, for different isotopes there can be
a significant spread of the half-lifes. By simulating sets of
\obb-rates it was estimated that 3 positive experimental results are required
to pin down the mechanism of \obb, if a total (theoretical,
systematical and statistical) uncertainty of 20\% or
less can be achieved. For 40\% uncertainty four results would be
necessary. Analyses in similar spirit can be found in
Refs.\cite{diff_nme1,diff_nme3}. Obviously, multi-isotope
determination is here crucial. \\

Another possibility to distinguish the mechanisms is the rate 
of the ground-state-to-ground-state transitions
to the rates of decays into excited
states\cite{excited0,excited0a,excited2,Simkovic:2001ft}. The latter
could be $0_1^+$ or $2_1^+$. 
The decay into $0_1^+$ is experimentally easier to identify
because two photons associated with the transition first to $2_1^+$
and then to the ground state are emitted. Transitions to $2_1^+$
states have higher sensitivity to right-handed
currents\cite{excited2}, and observation with large rates would signal
the dominance of these mechanisms, in particular the $\lambda$-contribution. 
The experimental situation is summarized in\cite{Barabash:2007ig}, it
is worth noting that \zbb~to $0_1^+$ excited states has been observed. 
Observation of \obb~into ground states and excited states in
the same experiment could be used to rule out the possibility that an
unidentified background peak mimics the \obb-signal\cite{DLZ}. The development 
of high granularity detectors, large enough \obb-rates and precise
nuclear physics is necessary to realize this consistency test. 
In the standard interpretation the 
decay to excited states occurs with a rate suppressed by a factor 
$R_{\rm ex} \simeq 10^{2 \ldots 3}$ with respect to the $0^+_{\rm g.s.} \to 0^+_{\rm g.s.}$
transition. This factor is a combination of kinematics $((Q - E_{\rm
ex})/Q)^5$, $Q - E_{\rm ex}$ being the energy release to the excited
state, and nuclear physics $|{\cal M}_{\rm g.s.}/{\cal M}_{\rm
ex}|^2$, and is sensitive to the mechanism of \obb.  For instance,
Ref.\cite{Simkovic:2001ft} has found for $^{76}$Ge suppression factors of $R_{\rm ex}
= 96, 48$ and $120$ for the standard mechanism, heavy neutrino exchange
and gluino exchange in $R$-parity violating SUSY, 
respectively. For $^{136}$Xe the factors are 17, 38 and 153, while for
$^{100}$Mo the result was 17, 17 and 59. These differences may be used to distinguish
the mechanism, if one assumes the nuclear physics uncertainties to be under
control. In this respect we compare the NMEs for $^{76}$Ge from
Ref.\cite{Simkovic:2001ft} (QRPA) with the ones from\cite{IBM}
(IBM)\footnote{Given the progress made in recent years 
it may not be fair to compare a 10 year old calculation
with a very recent one. However, we are not aware of any recent QRPA
re-evaluation and expect the ratios to be more stable than the NMEs
themselves.}. 
For $^{76}$Ge the QRPA NMEs for ground state and $0_1^+$ transitions
are 2.80 and 0.994, leading to a factor 7.93 in the relative
half-lifes. The IBM NMEs are 5.465 and 2.479, hence a factor 4.84. For
$^{100}$Mo the QRPA NMEs are 3.21 and 1.76, thus a ratio 3.33. IBM
gives NME values of 3.732 and 0.419, thus a ratio 21.26. Therefore, the notorious
NME uncertainty will again be a problem of the procedure described
here. Nevertheless, important information to the field would
be added by observation of \obb~into excited states.

Another possibility\cite{bbvsEC,Vergados:1982wr} to disentangle the mechanisms is the ratio of
\obb~to $0\nu\beta^+\beta^+$, $0\nu\beta^+\rm EC$ or $0\nu\rm ECEC$,
see Eqs.~(\ref{eq:b+b+},\ref{eq:b+EC},\ref{eq:ECEC}). For instance\cite{bbvsEC},
the ratio of the rates of \obb~of $^{76}$Ge and $0\nu\beta^+\beta^+$ of $^{106}$Cd
are about 2087, 30435 and 1826 for the standard, the $\lambda$- and the
$\eta$-mechanism, respectively. For $0\nu\beta^+\rm EC$ of $^{106}$Cd
the ratios are 148, 12 and 217. The same comments on nuclear physics uncertainties as for excited
states apply here, in addition to the problem of 
even lower rates. Recall however the possibility of resonant
enhancement\cite{ECECenh} of $0\nu\rm ECEC$. Double electron capture
to excited states has recently been discussed as another way to
distinguish mechanisms\cite{Vergados:2011gi}.

\subsection{\label{sec:distinguish_simul}Simultaneous presence of
several mechanisms}
We will now discuss aspects of simultaneous presence of more than one
\obb-mechanism. The different mechanisms can add coherently in the
amplitude, see Eq.~(\ref{eq:NS}), and interference effects are possible. However, at leading
order only terms in which the helicities of the emitted electrons are
identical can interfere. The fact that helicity is not exactly equal to
chirality for the emitted electrons with energy $E_e$ 
leads at the end to a phase space factor of the
interference term suppressed\cite{Faessler:2011rv,Faessler:2011qw} by one order of
magnitude (corresponding very roughly to $(E_e/Q)^2$), and
interference is almost negligible. 
For instance, in the standard mechanism both
electrons are left-handed, while in the $\lambda$-mechanism one is
right-handed, thus these two diagrams do not interfere. 
The chirality of the emitted electrons is indicated in 
the respective quark level Feynman diagrams which are shown in this
review. It is conceivable that destructive interference of several mechanisms
leads to a vanishing rate of \obb~in one or more isotopes. 
Nuclear physics differences of the relevant NMEs could, but do not
have to, lead to a non-vanishing rate in other
isotopes\cite{Faessler:2011rv}.

\begin{figure}[t]
 \centerline{
\psfig{file=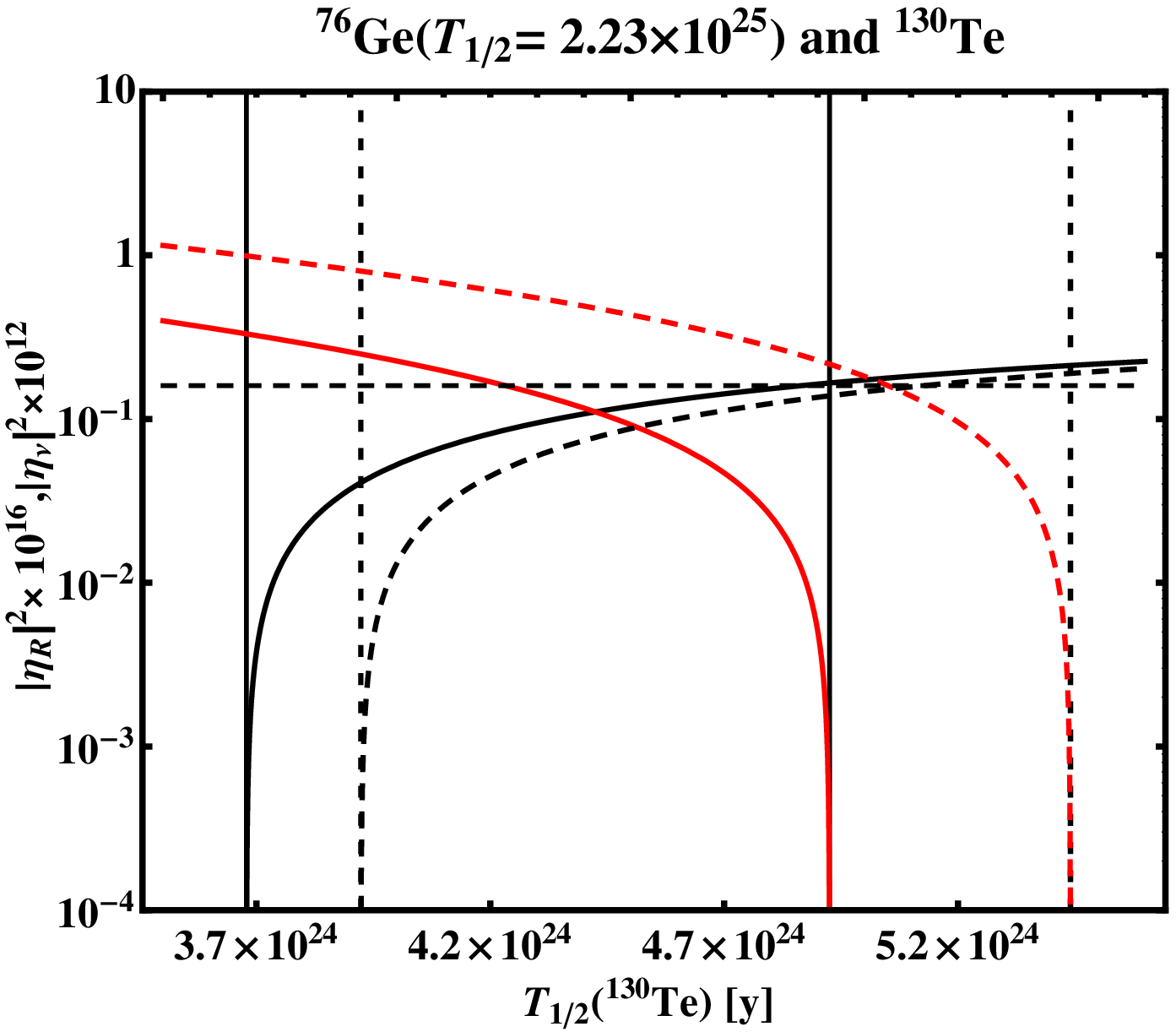,width=7cm,height=5.5cm}
\psfig{file=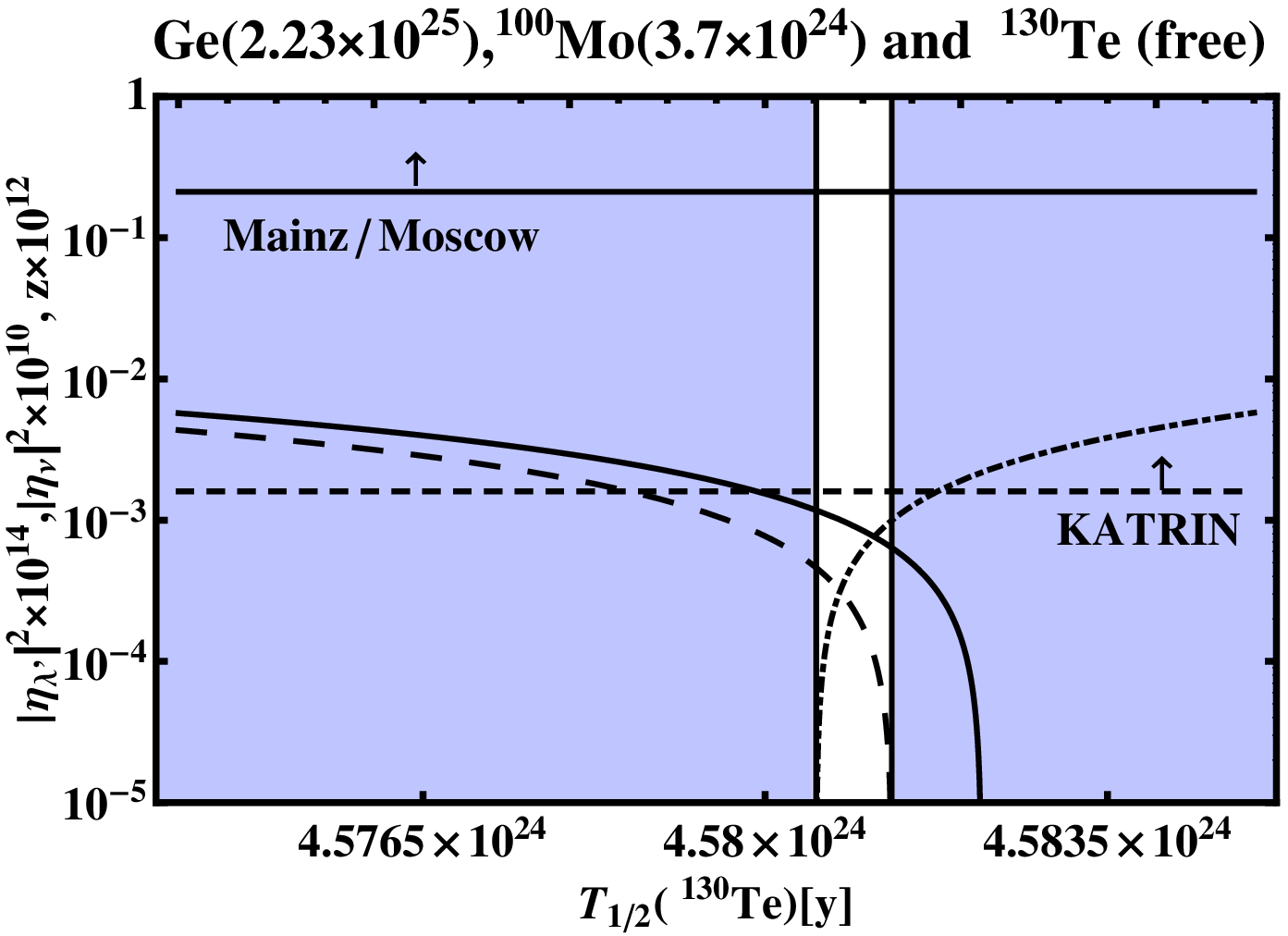,width=7cm,height=5.5cm} 
}
\vspace*{8pt}
\caption{\label{fig:simul_no_in}Left: solving the case of
non-interfering processes in Eq.~(\ref{eq:2nonint})
for the indicated hypothetical half-lifes of $^{76}$Ge and $^{130}$Te. 
The black (red) lines are $|\eta_{\rm l}|^2$ ($|\eta_{N_R}|^2$), the spread
between the solid and dashed lines arises from varying the NMEs
between $5.44 - 5.82$ ($4.18 - 4.70$) and  $411.5 - 264.9$ ($384.5 -
239.7$) for $^{76}$Ge ($^{130}$Te). The horizontal dashed line
corresponds to $\meff \le 0.2$ eV. The physical (positive) solutions for
$|\eta_{\rm l}|^2$ and $|\eta_{N_R}|^2$ are constrained within the solid
(dashed) lines. Right: two interfering processes with constructive
interference $\cos \phi = 1$ and fixed NMEs. The solid line is 
$|\eta_{\rm l}|^2$, the dashed line is $|\eta_{\rm \slashed{R}_1}^{\tilde g}|$, the
dash-dotted line is $2 \,|\eta_{\rm l}| \, |\eta_{\rm \slashed{R}_1}^{\tilde g}| \cos \phi $. 
The blue areas are forbidden. Taken from\protect\cite{Faessler:2011qw}.}
\end{figure}

The procedure to deal with the presence of several mechanisms has been
outlined in\cite{Simkovic:2010ka,Faessler:2011qw}. Consider first
the presence of two essentially non-interfering mechanisms, 
e.g.~light and heavy right-handed neutrino exchange. If two experiments
using different isotopes have found
evidence for \obb, one has (see Eq.~(\ref{eq:fact}))
\bea \label{eq:2nonint}
(T_{1/2}^a)^{-1} = G^a \left(|{\cal M}_l^a|^2 \, |\eta_{\rm l}|^2 
+ |{\cal M}_{N_R}^a|^2 \, |\eta_{N_R}|^2 \right) , \\ 
(T_{1/2}^b)^{-1} = G^b \left(|{\cal M}_l^b|^2 \, |\eta_{\rm l}|^2 
+ |{\cal M}_{N_R}^b|^2 \, |\eta_{N_R}|^2  
\right) , 
\eea
where the superscript $a,b$ denotes the two isotopes and the
subscripts $l$ and $N_R$ denote standard light neutrino exchange and
heavy right-handed neutrino exchange with $W_R$ instead of $W$ 
in left-right symmetric theories, see Eq.~(\ref{eq:A_NR}). Solving for
the particle physics parameters gives 
\bea \D 
|\eta_{\rm l}|^2 = \frac{|{\cal M}_{N_R}^b|^2/(T_{1/2}^a \, G^a) - 
|{\cal M}_{N_R}^a|^2/(T_{1/2}^b \, G^b)}{|{\cal M}_l^a|^2 \,|{\cal
M}_{N_R}^b|^2 - |{\cal M}_l^b|^2 \,|{\cal M}_{N_R}^a|^2 } \, , \\ \D 
|\eta_{N_R}|^2 = \frac{|{\cal M}_l^a|^2/(T_{1/2}^b \, G^b) - 
|{\cal M}_l^b|^2/(T_{1/2}^a \, G^a)}{|{\cal M}_l^a|^2 \,|{\cal
M}_{N_R}^b|^2 - |{\cal M}_l^b|^2 \,|{\cal M}_{N_R}^a|^2 } \, . 
\eea
Recall the present limits of $|\eta_{\rm l}| \ls 9.8 \times 10^{-7}$
and $|\eta_{N_R}| \ls 1.7 \times 10^{-8}$ from
Eqs.~(\ref{eq:eta_l},\ref{eq:eta_NR}).  
Fig.~\ref{fig:simul_no_in}, taken from\cite{Faessler:2011qw} shows an
example solution of Eq.~(\ref{eq:2nonint}). Knowing the half-life of
one isotope constrains the half-lifes of the other ones.

Consider now two interfering diagrams, for instance the standard mechanism and
gluino exchange within $R$-parity violating SUSY, where the emitted electrons are both
left-handed. There is an
unknown phase between the two contributions, and the total half-life 
can be written as 
\be\label{eq:3nonint}
(T_{1/2})^{-1} = G \left(|{\cal M}_l|^2 \, |\eta_{\rm l}|^2 
+ |{\cal M}_{\slashed{R}_1}^{\tilde g}|^2 \, |\eta_{\rm \slashed{R}_1}^{\tilde g}|^2 
+ 2 \, |{\cal M}_l| \, |{\cal M}_{\slashed{R}_1}^{\tilde g}| \, |\eta_{\rm l}| \,
|\eta_{\rm \slashed{R}_1}^{\tilde g}| \cos \phi \right) . 
\ee
Obviously, three positive observations of \obb~in three different
isotopes are required in order to extract the three independent
parameters $|\eta_{\rm l}|$, $|\eta_{\rm \slashed{R}_1}^{\tilde g}|$ 
and $\cos \phi$. An example from\cite{Faessler:2011qw} is presented in
Fig.~\ref{fig:simul_no_in}. The current limit is $|\eta_{\rm
\slashed{R}_1}^{\tilde g}| \le 7.5 \times 10^{-9}$, see
Eq.~(\ref{eq:eta_RPV1}).

\section{\label{sec:alt}Alternative Processes to Neutrino-less Double Beta Decay}
The last section of this review deals shortly with alternative
processes to \obb, i.e.~alternative probes of lepton number
violation. The presence of $\Delta L = 2$ can manifest itself in going from
lepton number $L = 0$ to $L = \pm 2$, typical for a decay process, or from
$L = \pm 1$ to $L = \mp 1$, typical for conversion processes. It could
also be that $|L| = 1$ goes to $|L| = 3$, e.g~in lepton decays or
collisions with initial leptons. Finally, in like-sign lepton
collisions or $0\nu$ECEC one could go from $L = -2$ to $L = 0$.\\

Neutrino oscillation probabilities are not sensitive to the Majorana
nature of neutrinos. However, in principle $\nu_\alpha \to
\bar{\nu}_\beta$ transitions are possible, whose probabilities are
unfortunately suppressed by the factor $(m_i/E)^2$, in analogy to the standard
mechanism of \obb. There are in principle differences between Dirac
and Majorana neutrinos, for instance it is easy to show that Majorana
neutrinos do not have a vector current. Again, in amplitudes the
difference of Dirac and Majorana neutrinos due to the absence of
vector currents for the latter goes with $m_i/E$. This annoying
property is known as the\cite{Kayser:1982br} ``practical Dirac-Majorana confusion
theorem''. 

A recent review on the electromagnetic properties of neutrinos can be
found in\cite{Giunti:2008ve}. In short, Majorana neutrinos cannot
possess diagonal magnetic moments, i.e.~$(\nu_e) \to \bar{\nu}_e$ transitions
would only be possible for Dirac neutrinos. This can be seen by
looking at the magnetic moment operators\cite{mm} $\mu_{\alpha \beta} \, 
\overline{\nu_L}_\alpha \, \sigma_{\mu\nu} \, (\nu_R)_\beta \, F^{\mu \nu}$
for Dirac and $\mu_{\alpha \beta} \, 
\overline{\nu_L}_\alpha \, \sigma_{\mu\nu} \, (\nu_L^c)_\beta\, F^{\mu \nu}$ for Majorana
neutrinos. In $\nu_e \, e$ scattering experiments the helicity and
flavor of the final state neutrino cannot be measured and there is no way to
distinguish Dirac from Majorana in this way. One possibility would be
via spin flavor transitions in supernovae, in which a magnetic field
triggers $(\nu_e)_L \to (\bar{\nu}_\mu)_R$, with subsequent
oscillation of the active $(\bar{\nu}_\mu)_R$ into
$(\bar{\nu}_e)_R$. The usual $\nu_e$ neutronization burst can be heavily
affected by this effect\cite{Akhmedov:1992ea}. While the SM extended
with massive neutrinos does generate too small magnetic moments, $\mu
\propto m_\nu$, in some extensions of the SM it would be possible
to generate the required large magnetic moments\cite{mm_new}. \\

We have mentioned already high energy tests of lepton number violation
such as the like-sign dilepton signature
of heavy Majorana neutrino production\cite{seesaw_LHC,Maiezza:2010ic}, Higgs triplet 
decays\cite{triplet_LHC}, or resonant selectron
production\cite{Allanach:2009iv}. One may wonder whether there are
{\it low energy processes}, in analogy to \obb, which can probe the effective Majorana
mass, maybe even without any nuclear physics complications. 
However, the $(m_i/E)^2$ suppression of the rate together with
Avogadro's number $N_A$
render \obb~the only realistic probe. With order kg of a \obb-isotope one
has order $N_A$ atoms, which compensates the Dirac/Majorana factor
$(m_i/q)^2$. In principle, there are decays like $K^+ \to \pi^- \, e^+
e^+$, which depend on the effective mass in the same way as
\obb~does, and do not suffer from NME uncertainties. However,
calculating the branching ratio yields\cite{Ali:2001gsa}
\be
{\rm BR}(K^+ \to \pi^- \, e^+ e^+) \sim 10^{-33} \,
\left(\frac{\meff}{\rm eV} \right)^2 \, , 
\ee
to be compared with the experimental upper limit\cite{LS_mes,PDG} of ${\rm BR}(K^+ \to
\pi^- \, e^+ e^+) \le 6.4 \times 10^{-10}$. If it was possible to
increase the number of charged kaons by 20 orders of magnitude, one
could go for decays like\cite{Abad:1984gh} $K^+ \to \pi^- \, \mu^+ \mu^+$
(``neutrino-less double muon decay'') and test $\langle m_{\mu\mu}
\rangle$, i.e.~the other entries of the mass
matrix\cite{JapDL2,ichDL2,KaiDL2,Dib:2000wm,Ali:2001gsa,Atre:2005eb}. 
Other decays which have been studied in the past include lepton number
violating decays of $\tau$ leptons\cite{Atre:2005eb,Gribanov:2001vv}, 
top quarks and $W$ bosons\cite{BarShalom:2006bv}, $D$ and $B$
meson decays\cite{ng,Ali:2001gsa,Atre:2009rg,Cvetic:2010rw}, or
hyperons\cite{Barbero:2007zm}. Collider processes such as\cite{Flanz:1999ku} $\nu_\mu \,
N \to X \mu^- \, \alpha^+ \beta^+ $ or\cite{Flanz:1999ah} $e^- p \to X \nu_e \, \alpha^+
\beta^+$ have also been discussed. In addition, searches
for conversion processes such as\cite{doi,Kamal:1979vw,Domin:2004tk} 
$\mu^- \, (A,Z) \to e^+ (A,Z-2)$ or\cite{Missimer:1994xd} $\mu^- \, (A,Z) \to \mu^+ (A,Z-2)$
have been proposed. 

For very light and very heavy Majorana neutrinos the
above processes are not very helpful. Recall however the general property of
Majorana neutrino exchange as displayed in Fig.~\ref{fig:heavy_sym}:
for neutrinos whose masses correspond to the typical energy scale of
the process the sensitivity is largest. Taking $K^+ \to \pi^- \, \mu^+
\mu^+$ as an example, Ref.\cite{Dib:2000wm} has obtained very strong 
limits on masses between 245 MeV and 389 MeV, with $|U_{\mu i}|^2$
down to the $10^{-9}$ regime. The constraints are strong because there
can be ``$s$-channel'' diagrams. Other decays have been analyzed
in\cite{Atre:2009rg,Helo:2011yg}.

Not many works exist which study the above processes in non-standard
mechanisms. Examples include Ref.\cite{Littenberg:2000fg}, where meson
decays such as $K^+ \to \pi^- \, \mu^+
\mu^+$ mediated by $R$-parity violating SUSY were found to provide no
significant limits. The same was shown in\cite{Simkovic:2001fs} for
$(\mu^-, \mu^+)$ conversion, or for $(\mu^-, e^+)$ conversion in various
mechanisms\cite{Leontaris:1982xz}. Doubly charged Higgs
exchange in $K^+ \to \pi^- \, \mu^+ \mu^+$ was also found to generate
negligible rates\cite{Picciotto:1997tk}.\\

A particularly clean probe of lepton number violation is ``inverse
neutrino-less double beta decay''. This is not $(A,Z+2)^{++} + 2 \, e^- \to
(A,Z)$, but 
\be
e^- \, e^- \to W^- \, W^- \, . 
\ee
This reaction can be tested if a future linear collider is run in a 
basically background-free like-sign mode, and has frequently been
proposed as a probe of LNV and new physics in
general\cite{other}. 
The process does not involve any nuclear, hadronic or atomic
uncertainties or difficulties and is presumably the cleanest probe of
lepton number violation. 
\begin{figure}[t]
 \centerline{
\psfig{file=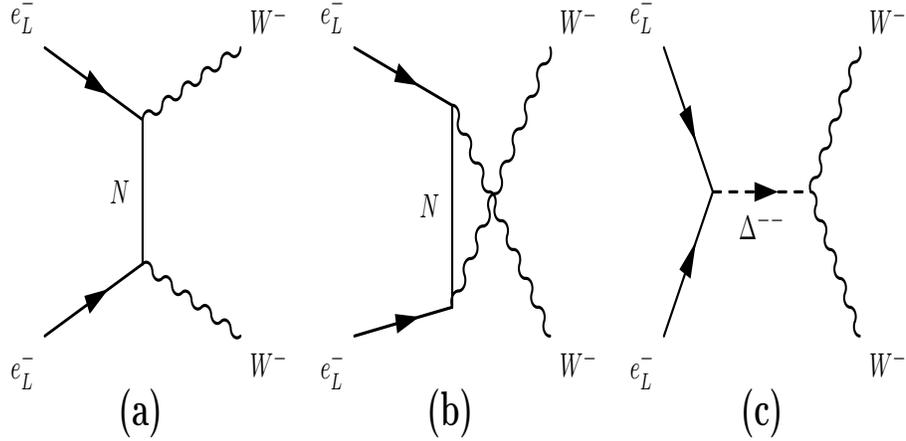,width=12cm,height=6cm}
}
\vspace*{8pt}
\caption{\label{fig:inv}Feynman diagram for inverse \onbb. Diagrams (a)
and (b) are Majorana neutrino $N$ exchange, diagram (c) triplet exchange.}
\end{figure}
If Majorana neutrinos are exchanged, see Fig.~\ref{fig:inv}, and with 
neglecting the mass of the $W$, the cross section reads 
\be \label{eq:sigee}
\frac{ d \sigma}{d \cos \theta} = 
\frac{G_F^2}{32 \, \pi} \left\{ \sum (M_\nu)_i \, V_{ei}^2 \left( 
\frac{t}{t - (M_\nu)_i} + \frac{u}{u - (M_\nu)_i} 
\right) \right\}^2  , 
\ee
where $t$ and $u$ are the usual Mandelstam variables, $(M_\nu)_i$ is the
mass of the neutrinos (including light $m_i$ and heavy $M_i$) 
and $V_{ei}$ their mixing with electrons ($N_{ei}$ and $S_{ei}$). 
There are interesting special cases for the cross 
section: 
\begin{itemize}
\item if only light active Majorana neutrinos contribute to 
the process, then the cross section is  
\bea \D 
\sigma(e^- e^- \ra W^- W^-) = \frac{G_F^2}{4 \, \pi} \, \meff^2 \\ \D 
\le 4.2 \times 10^{-18} \left(\frac{\meff}{1 \, {\rm eV} }
\right)^2 \, {\rm fb} \, ,
\eea  
hence far too small to be observable; 
\item if only heavy Majorana neutrinos contribute to 
the process, then we can bound the cross section 
using the \obb-limit from Eq.~(\ref{eq:heavy}) as 
\be 
\sigma(e^- e^- \ra W^- W^-) = \frac{G_F^2}{16 \, \pi} \, s^2 
\, \imeff^2 \le 2.6 \times 10^{-3} \, \left(\frac{\sqrt{s}}{\rm TeV}
\right)^4 \left(\frac{\imeff}{5 \times 10^{-8} 
\, \rm GeV^{-1}} \right)^2 \, {\rm fb} 
\ee 
again far too small to be observable; 
\item the high energy limit of $\sqrt{s} \to \infty$ is 
\be
\sigma(e^- e^- \ra W^- W^-) = \frac{G_F^2}{4 \, \pi} \, \left(V_{ei}^2
\, (M_\nu)_i \right)^2 \, .
\ee
This seems to violate unitarity, because the cross section for an $s$-wave process should
vanish in the high energy limit. However, recall the exact seesaw
relation $\sum N_{ei}^2 \, m_i + S_{ei}^2 \, M_i  = 0$, as 
discussed in in Eq.~(\ref{eq:exactseesaw}). This relation guarantees
that the cross section vanishes in the high energy limit. 
\end{itemize}
\begin{figure}[t]
 \centerline{
\psfig{file=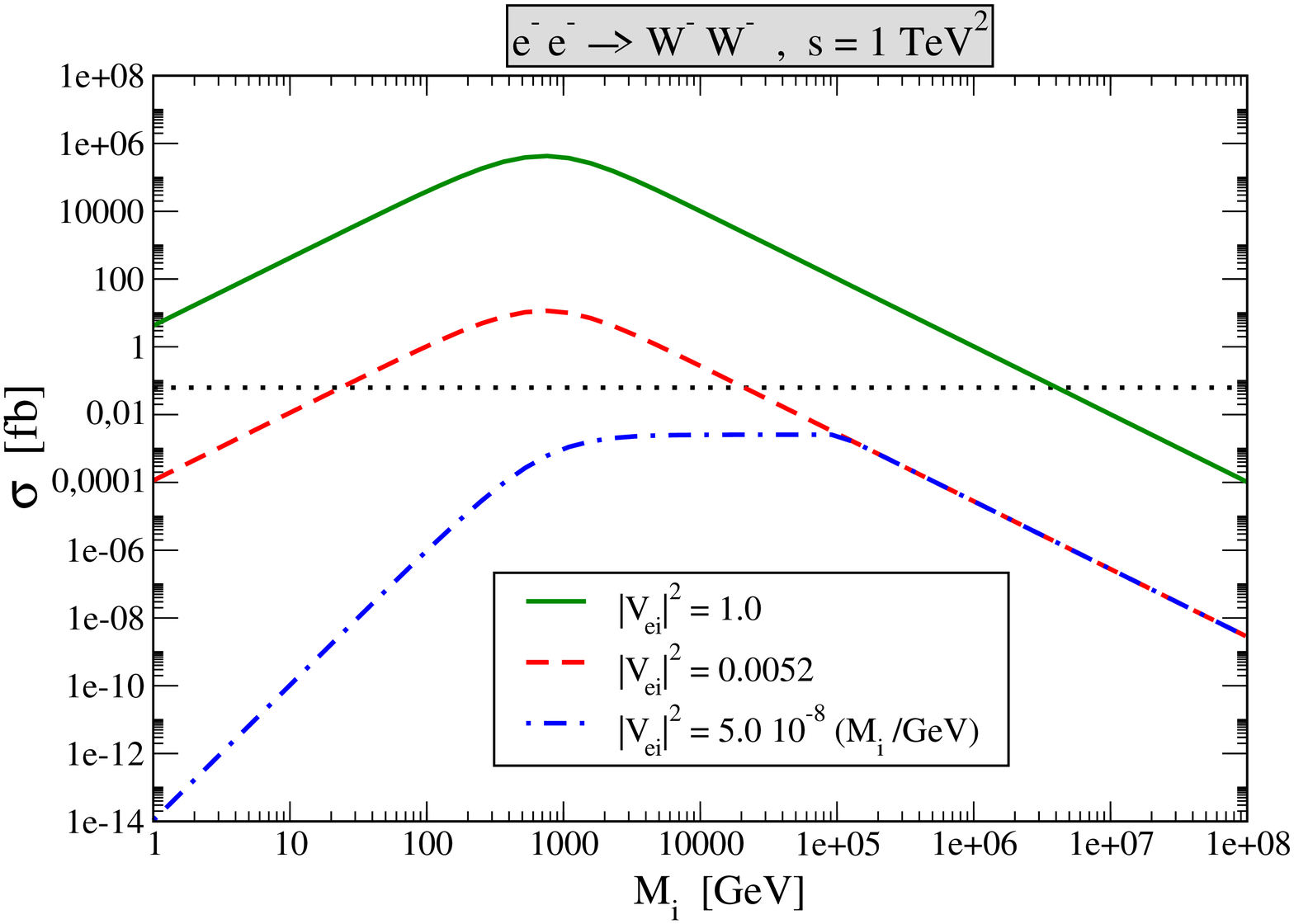,width=7cm,height=5.5cm}
\psfig{file=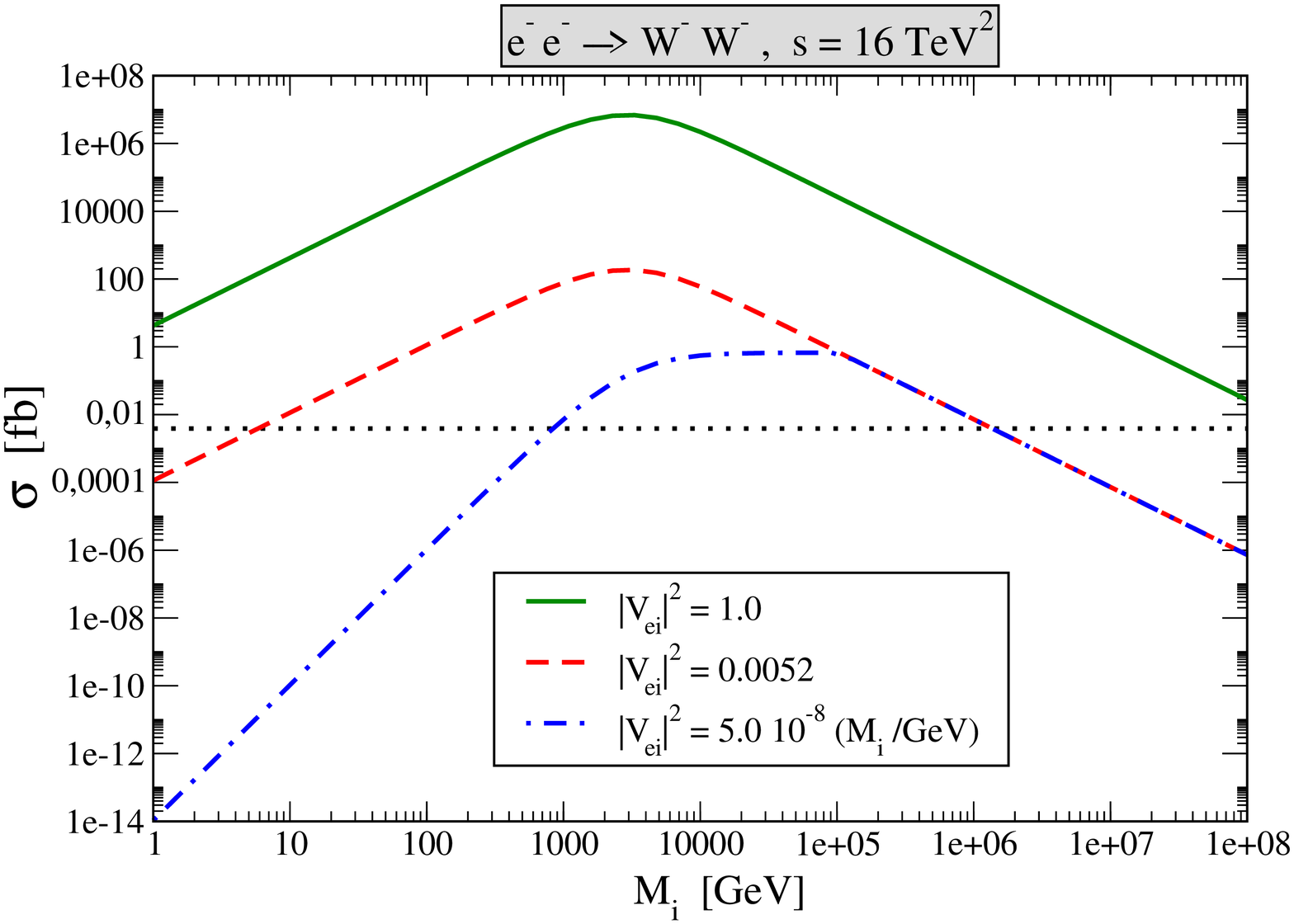,width=7cm,height=5.5cm}
}
\vspace*{8pt}
\caption{\label{fig:inv_xs}Cross section for $e^- e^- \ra W^- W^-$ with 
$\sqrt{s} = 1$ TeV (left) and $\sqrt{s} = 4$ TeV (right) and three  
limits for the mixing parameter $|V_{ei}|^2$. The dotted line 
corresponds to five events for an 
assumed luminosity of 80 $(s/{\rm TeV^2})$ fb$^{-1}$.}
\end{figure}
While small and large masses cannot give sizable cross sections,
intermediate scale neutrino masses $(M_\nu)_i \sim \sqrt{s}$ can give
appreciable event numbers, as expected from the general 
behavior of LNV processes with Majorana neutrinos involved. 
In Fig.~\ref{fig:inv_xs} we show an
example of the cross section as a function of neutrino mass. Different
limits on the mixing $V_{ei}$ are inserted: no limit, the global limit
and the limit as implied from the limit on $\imeff = |V_{ei}|^2/M_i$.
Note how the case of  $|V_{ei}|^2 = 1$ follows the general trend of
Fig.~\ref{fig:heavy_sym}. The above processes can also be searched for
at $e^- \mu^-$ or $\mu^- \mu^- $ machines.

The process $e^- e^- \ra W^- W^-$ can also be mediated by a Higgs
triplet, but due to small neutrino masses has a tiny cross sections
unless a very narrow resonance is met. One can show that if neutrino and
triplet exchange occur simultaneously, the unitarity of the cross
section is saved by the relation in
Eq.~(\ref{eq:exactseesaw1}). Right-handed Higgs triplets or $W_R$ could be
observed, however, if the electron beams
are properly polarized. 

Double chargino production $e^- e^- \ra \chi^- \chi^-$ in 
supersymmetry has been studied in\cite{Hirsch:1998gr}. The diagram is
basically the same as (a) and (b) in Fig.~\ref{fig:inv}, with the $W$
replaced by $\chi$ and the neutrino by a sneutrino. Recall that 
lepton number violation in the sneutrino sector implies Majorana
neutrinos\cite{Hirsch:1997vz}. It turns out that the limits from
\obb~render double chargino production cross sections too small. Other
works on lepton number violating $e^- e^-$ collisions within
supersymmetry can be found in\cite{eeSUSY}. 

As mentioned before, lepton number violating sneutrino mass terms, 
the amplitude of \obb~and Majorana neutrino mass terms imply each
other: if one of the three is present, the other two are there as
well\cite{Hirsch:1997vz,Hirsch:1997is}. This leads to a splitting in 
the $\tilde \nu$-$\bar {\tilde \nu}$ system and therefore to 
lepton number violating 
sneutrino--anti-sneutrino mixing\cite{Grossman:1997is}, whose 
parameters depend heavily on SUSY parameters, and whose 
observation is usually a very challenging task\cite{snusnuobs?}.

\section{\label{sec:concl}Summary}
Neutrino-less double beta decay experiments are much more than
neutrino mass experiments, their importance is much broader and
deeper. The violation of lepton and baryon number is a rather generic feature of
theories beyond the Standard Model,  and searches for \obb~or proton
decay are probes of fundamental physics related to high energies, with
a variety of important consequences in particle physics and cosmology.   
The significance of the decay is underlined by the 
excessive list of references provided in this review. 

In the next 20-30 years \obb~will be the only realistic
probe to test the conservation of lepton number. The existing and
upcoming results, when interpreted in terms of a specific particle
physics scenario, allow to constrain a variety of important parameters, some of which
can only be probed by \obb, others can also be tested in different
and complementary experiments. Best motivated is presumably the 
standard interpretation of light neutrino exchange, where the inverted
mass ordering will begin to be tested within this
decade. Quasi-degenerate neutrinos will generate a signal, and should
in this case be detectable also in direct searches and cosmological
observations. This would be the ideal case to identify the mechanism. 
There are however many non-standard interpretations of \obb, 
the most frequently discussed mechanisms for the decay are
summarized in Table \ref{tab:sum}. The unambiguous determination of
the underlying mechanism is in general less straightforward than for
quasi-degenerate neutrinos. However, naive estimates show that an effective mass of order 
0.1 eV is associated with an amplitude that corresponds to the amplitude
for exchange of TeV scale heavy particles. This energy scale has a
variety of potential effects in currently running particle physics
experiments, such as LHC, lepton flavor violation, FCNC, etc. 
It should be noted that, though some progress was made in recent years, 
high precision physics with \obb~will presumably not be possible: 
nuclear matrix elements are unlikely to be known with more than 20\%
precision. Currently, one has to accept the (shrinking) ${\cal O}(1)$ ranges of NME calculations
and perform analyses of \obb-results keeping them in mind.

An impressive number of upcoming experiments promises an 
exciting future for the field.  
Multi-isotope determination of \obb~with different experimental approaches will be
possible and is crucial in order to make an unambiguous claim of
observation, help clarifying the nuclear matrix element calculations, 
and distinguish the different mechanisms. In order to identify the
origin of \onbb~(the ``inverse problem'' of \obb) three different
possibilities exist: via effects in other observables, via exploring
the decay products, and via nuclear physics effects. 
After the violation of lepton number is established, a highly
interesting physics program of identifying the underlying mechanism 
and its origin will be possible.

\begin{table}[t]
\tbl{\label{tab:sum}Most important mechanisms for \onbb. Given are the
absolute value of the amplitude ($q \simeq 0.1$ GeV is the momentum
exchange for long-range processes) with the particle physics parameter 
written in bold face. The current limits on these quantities are provided,
and tests to identify the mechanism by other means are indicated. RHC
denotes right-handed currents, $\slashed{R}$ stands for $R$-parity
violation and there are several Majoron variants.}
{\begin{tabular}{cccc}
& amplitude and & &  \\
mechanism &  particle physics parameter & current limit & 
test \\     \toprule 
light neutrino exchange & $\frac{G_F^2}{q^2} \boldsymbol{\left|
U_{ei}^2 \, m_i\right|}$ 
& 0.5 eV & $\ba \mbox{oscillations,} \\ \mbox{cosmology,} \\ \mbox{neutrino mass}
\ea $\\ \hline 
heavy neutrino exchange & $ G_F^2 
\boldsymbol{\left| \frac{S_{ei}^2}{M_i}\right|}$ & $2 \times 10^{-8} $ 
GeV$^{-1}$ & $\ba \mbox{LFV,} \\ \mbox{collider} \ea $ \\ \hline 
heavy neutrino and RHC & $G_F^2 \, m_W^4   
\boldsymbol{\left| \frac{V_{ei}^2}{M_i \, M_{W_R}^4}\right|}$ & $4 \times
10^{-16}$ GeV$^{-5}$ & $\ba \mbox{flavor,} \\ \mbox{collider} \ea $ \\ \hline 
Higgs triplet and RHC & $G_F^2 \, m_W^4   
\boldsymbol{\left|\frac{(M_R)_{ee}}{m_{\Delta_R}^2\, M_{W_R}^4}\right|} $ & $10^{-15}$
GeV$^{-1}$ & $\ba \mbox{flavor,} \\ \mbox{collider}  \\ \mbox{$e^-$
distribution} \ea $ \\  \hline 
$\lambda$-mechanism with RHC &  
$G_F^2 \frac{m_W^2}{q} 
\boldsymbol{\left| \frac{U_{ei} \, \tilde{S}_{ei}}{M_{W_R}^2}\right|} $ &
$1.4 \times 10^{-10}$ GeV$^{-2} $ &  $\ba \mbox{flavor,} \\
\mbox{collider,} \\ \mbox{$e^-$ distribution} \ea $ \\ \hline 
$\eta$-mechanism with RHC &  
$G_F^2 \frac{1}{q} \,  
\boldsymbol{\tan \zeta \,\left| U_{ei} \, \tilde{S}_{ei} \right|} $ &
$6 \times 10^{-9}$ &  $\ba \mbox{flavor,} \\
\mbox{collider,} \\ \mbox{$e^-$ distribution} \ea $ \\ \hline 
short-range $\slashed{R}$ & 
$ \ba \boldsymbol{\frac{\left|\lambda'^2_{111}\right|}{\Lambda_{\rm SUSY}^5}} \\ 
\Lambda_{\rm SUSY} =
f(m_{\tilde{g}},m_{\tilde{u}_L},m_{\tilde{d}_R},m_{\chi_i}) \ea $ 
& 
$ \ba 7 \times 10^{-18} ~{\rm GeV}^{-5} \\ \ea $ & $\ba \mbox{collider,} \\
\mbox{flavor}  \ea $\\ 
\hline 
long-range $\slashed{R}$ & $ \ba \frac{G_F}{q} \, 
\boldsymbol{\left|\sin 2 \theta^b \, \lambda'_{131} \, \lambda'_{113} 
\left(\frac{1}{m_{\tilde{b}_1}^2} - \frac{1}{m_{\tilde{b}_2}^2 }\right) \right|} \\ 
\sim  \frac{G_F}{q} m_b \boldsymbol{\frac{\left|\lambda'_{131} \, \lambda'_{113}\right|}
{\Lambda_{\rm SUSY}^3}} \ea $ 
& $\ba  2 \times 10^{-13} ~{\rm GeV}^{-2} \\ \mbox{ } \\
 1 \times 10^{-14} ~{\rm GeV}^{-3} \ea $
& $\ba \mbox{flavor,} \\ \mbox{collider}\ea $\\ 
\hline
Majorons & $\propto \boldsymbol{|\langle g_\chi \rangle|}$ or $\boldsymbol{|\langle g_\chi
\rangle|^2}$ & $10^{-4} \ldots 1 $ &
$\ba \mbox{spectrum,} \\ \mbox{cosmology} \ea $ \\  \botrule
  \end{tabular} }
\end{table}

\section*{Acknowledgements}
I am very grateful for discussions with E.~Akhmedov, M.~Duerr,
M.~Lindner and K.~Zuber, and thank J.~Barry, A.~Dueck and M.~Duerr for
help in producing figures and tables. 
This work was supported by the ERC under the Starting Grant 
MANITOP and by the DFG in the project RO 2516/4-1 as well as in the 
Transregio 27 ``Neutrinos and Beyond''.

\end{document}